\documentclass[12pt,oneside,a4paper]{report}

\usepackage{amssymb}
\usepackage{graphics}
\usepackage{epsfig}
\usepackage{a4wide}
\usepackage{latexsym}
\usepackage[Sonny]{fncychap} 
\usepackage{graphicx}
\usepackage{amsmath}
\usepackage{bm}
\usepackage{xspace}
\usepackage{bbold}
\numberwithin{equation}{section}
\usepackage{enumerate}
\usepackage{doublespace}
\usepackage{multirow}
\usepackage{subfigure}

\def\baselinestretch{1.2}

\newcommand{\clearemptydoublepage}
             {\newpage{\pagestyle{empty}\cleardoublepage}}
\renewcommand{\baselinestretch}{1.2}

\begin{document}

% -- FRONTPAGE --

\pagestyle{empty}
\renewcommand{\baselinestretch}{1}

\begin{titlepage}
\topskip1.3in
\begin{center}
\vspace{2cm}

\begin{center}
\begin{tabular}{l}
\hline
  \\ \\

\huge{{\bfseries\textsc{Alternative Theories}}}\\ \\
\huge{{\bfseries\textsc{$\; \; \; \; \; \;$ of Gravity}}}\\

\\\\  
\hline
\end{tabular}
\end{center}

\vspace{2.5cm}
{\Large
Timothy Clifton\\
\vspace{0.5cm}
King's College\\}
\vspace{2.3cm}
\parbox{20mm}{
    \epsfxsize=20mm
\epsfbox{CU.arms.colour}
}\\
\vspace{2.3cm}
{\large
Dissertation submitted for the degree of Doctor of Philosophy\\
\vskip5pt
Department of Applied Mathematics and Theoretical Physics\\
University of Cambridge\\ 
August 2006}
\end{center}

\clearemptydoublepage

\end{titlepage}

\clearemptydoublepage

% -- PREAMBLE --

\pagestyle{plain}
\pagenumbering{roman}
\begin{center}
\bfseries{\textsc{ACKNOWLEDGEMENTS}}
\end{center}

This work would not have been possible without the help of a lot of
people.  I would like to thank my family and friends for
their encouragement, my collaborators David Mota and Robert Scherrer
for their hard work and my office mate Simon West for innumerable delightful
lunches.  Most of all I would like to thank my supervisor, John Barrow, for his
constant support, his guidance and for his untiring correction of my speling and, grammar.

\clearemptydoublepage

\vspace{5cm}
\begin{center}
\bfseries{\textsc{DECLARATION}}
\end{center}

The research presented in this thesis was performed in the Department
of Applied Mathematics and Theoretical Physics at the University of
Cambridge between October 2003 and July 2006. 
This dissertation is the result of my own work, except as stated below or where
explicit reference is made to the results of others.

The content of this thesis is based on the research papers
\begin{itemize}
\item
\textit{`Spherically Symmetric Solutions to Fourth-Order Theories of
  Gravity'}

Timothy Clifton

gr-qc/0607096
\item
\textit{`Decaying Gravity'}

Timothy Clifton and John Barrow

Physical Review D \textbf{73}, 104022 (2006)
\item
\textit{`Further Exact Cosmological Solutions to Higher-Order Gravity
Theories'}

Timothy Clifton and John Barrow

Classical and Quantum Gravity \textbf{23}, 2951 (2006)
\item
\textit{`The Existence of G\"{o}del, Einstein and de Sitter
  Universes'}

Timothy Clifton and John Barrow

Physical Review D \textbf{72}, 123003 (2005)
\item
\textit{`Exact Cosmological Solutions of Scale-Invariant Gravity
  Theories'}

John Barrow and Timothy Clifton

Classical and Quantum Gravity \textbf{23}, L1-L6 (2006)
\item
\textit{`The Power of General Relativity'}

Timothy Clifton and John Barrow

Physical Review D \textbf{72}, 103005 (2005)
\item
\textit{`Constraints on the Variation of G from Primordial
  Nucleosynthesis'}

Timothy Clifton, John Barrow and Robert Scherrer

Physical Review D \textbf{71}, 123526 (2005)
\item
\textit{`Inhomogeneous Gravity'}

Timothy Clifton, David Mota and John Barrow

Monthly Notices of the Royal Astronomical Society \textbf{358}, 601
(2005)
\end{itemize}

A brief discussion of the number of arbitrary
functions in section 5.2.1 is due to John Barrow.  Section 5.2.3 was written
collaboratively with John Barrow where the analysis of $R^n$ theories
in this section is due primarily to him.  Discussion in the
introduction to section 7.1 is also due to John Barrow.  Section 7.1.3
was written collaboratively with David Mota where his main contribution
was the numerical calculations involved and the production and
discussion of figures 7.6 to 7.15.  In section 9.1.1 the numerical
work and the production of figure 9.1 was performed by Robert
Scherrer.  John Barrow also provided ideas and guidance which were of
use throughout this work.  Use was made of the computer programs Mathematica and Maple.

This dissertation is not substantially the same as any that I
have submitted, or am submitting, for a degree, diploma or other 
qualification at any other university.

\vspace{4cm}
Signed: .......................................
\hspace{1.2cm}  
Dated: ........................................

\clearemptydoublepage

\vspace{5cm}
\begin{center}
\Large{\bfseries{\textsc{Alternative Theories of Gravity}}}
\end{center}

\begin{center}
\textsc{Timothy Clifton}
\end{center}

\bigskip

\begin{center}
\bfseries{\textsc{SUMMARY}}
\end{center}

This work investigates alternative theories of gravity, the solutions
to their field equations and the constraints that can be imposed upon them
from observation and experiment.  Specifically, we consider the
cosmologies and spherically symmetric solutions that can be expected to result from scalar-tensor and
fourth-order theories of gravity.  We find exact cosmological
solutions of various different kinds; isotropic and anisotropic,
homogeneous and inhomogeneous.  These solutions are used to
investigate the behaviour of the Universes at both late
and early times, to investigate the effects of corrections to general
relativity on approach to an initial singularity and to look for
effects which may be observable in the present day Universe.  We use
physical processes, such as the primordial nucleosynthesis of the light
elements, to impose constraints upon any deviations from the standard
model.  Furthermore, we investigate the vacuum spherically symmetric
solutions of these theories.  This environment is of particular interest
for considerations of the local effects of gravity, where the most
accurate experiments and observations of gravitational phenomena can
be performed.  Exact solutions are obtained for this situation and
their stability analysed.  It is found that a variety of new behaviour
is obtainable in these theories that was not previously possible in
the standard model.  This new behaviour allows us an extended
framework in which to consider gravitational physics, and its
cosmological consequences.

\cleardoublepage
\clearemptydoublepage

% -- PAGESTYLE --

\addtolength{\headheight}{\baselineskip}

% -- CONTENTS --

\tableofcontents
\clearemptydoublepage
%\listoffigures
%\clearemptydoublepage
%\listoftables
%\clearemptydoublepage

% -- CHAPTERS --

\pagenumbering{arabic}
\setcounter{page}{1}

\chapter{Introduction}
\label{Introduction}

\bigskip

The standard model of gravitation and cosmology is based on the theory
of general relativity (GR).  GR is one of the cornerstones of modern
theoretical physics and has been shown to be enormously successful not
just in its conceptual ingenuity and mathematical elegance, but also
in its ability to explain real physical phenomena.  This is
particularly astonishing considering that the vast majority of
relativistic gravitational effects were completely unknown at the time
of its conception.  

Whilst in most branches of physics it has been the case that
experiments have been performed and theory developed to fit these
observations, this has not been the case for the development of
relativity theory.  Instead, due to the extreme weakness of gravity,
it has been the situation that the theory of relativity was developed
with only a minimal input from experiment.  The theory then had only
to satisfy the few experiments that were available at the time.  Today
this is no longer the case; an ever increasing body of experimental
and observational evidence is being accumulated.  This wealth of data
has brought gravitational physics up to the status of a real
experimental science, and so it must be confirmed that the theory
which has been developed does in fact adequately model the observations.

In making any comparison between theory and observation it is useful
to have a framework in which to consider the theory.  Such frameworks
have been constructed for gravitational physics by a number of authors
for a variety of different circumstances.  For example, the
parameterised post-Newtonian approach has been developed to provide a
framework in which weak field tests of gravity can be interpreted.
Once the parameters in this very general approach have been
constrained by the appropriate observations, these constraints can
then be compared to the predictions of the relativistic theory.
However, if we have only a single theory it is difficult to interpret
exactly what these observations mean; we can only say if they are
consistent with that one theory or not.  In order to make a proper
evaluation of just how influential these constraints are it is
necessary to have a variety of theories, each differing from the
others in their own particular modifications.  It is the development
of just such alternative theories that is the subject of this work.

As well as providing a foil against which to test GR, alternative
theories of gravity are also interesting to study in their own right.
The very existence of logically consistent alternatives to GR demands
investigation to establish their potential compatibility with
experiment, how they manifest the usual relativistic phenomena and if
any new phenomena can be identified that are not present in the
standard model.  This avenue of research becomes particularly pressing
if one wishes to attempt to explain aspects of gravitational physics
which appear unsatisfactorily accounted for in the standard model.  A
number of these apparent shortcomings present themselves both from
observations of physical phenomena as well from more philosophical
concerns regarding the theory itself.  Examples which are often cited
in the literature are the apparent need for more than 95$\%$ of our
Universe to be made from dark matter and dark energy (neither of which
is satisfactorily understood), the anomalous acceleration of Pioneer,
the incompatibility of GR with quantum field theory and the lack of
application of Mach's principle.  Any one of these provides sufficient
motivation to investigate potential alternatives to GR.

In considering alternative theories of gravity it is necessary to
impose some \textit{a priori} conditions, to limit the number of
candidate theories.  We proceed as Dicke \cite{Dick}, and choose that

\begin{enumerate}
\item{Space-time is a differentiable four-manifold with a metric and a connection.}
\item{The theory must be relativistic.}
\item{General coordinate covariance should be manifest.}
\item{There should exist no a priori geometric structure.}
\item{The field equations should be derivable from an action principle.}
\item{The theory should be simple.}
\end{enumerate}

The first condition is the statement that we are restricting ourselves
to what are often called `metric theories of gravity'.  Associating a
metric with the curvature of space-time is a powerful idea, leading
directly to the equivalence of inertial and gravitational mass as
neutral test-particles (those with negligible self-gravity) follow
geodesics of the space-time.  This equivalence of gravitational and
inertial masses has been shown to very high accuracy by E\"{o}tv\"{o}s
type experiments.  Furthermore, this condition limits our
considerations to four dimensional theories with Riemannian geometry.

The second condition can be well justified from both experimental and
philosophical considerations and is closely related to the first.  It
is known that at any point on a smooth manifold we can make a local
coordinate transformation to normal coordinates.  In such a coordinate
system the effects of gravity should be transformed away (up to tidal
forces) and we should recover special relativity.  This is a
manifestation of the weak equivalence principle which states:
`\textit{a freely falling frame in a gravitational field is equivalent
  to an inertial frame in the absence of gravity}'.

The condition of general coordinate covariance implies that we should
be free to choose any coordinates to mark the positions of physical
events on our space-time manifold.  It is the position of the events
themselves and the geometry of space-time that is important, not the
arbitrary set of coordinates we put on the manifold.  This condition
requires that the theory should be formulated in terms of tensors.

Condition four is the idea that there should be no absolute space.  It
is the requirement that the geometry of space-time should be entirely
determined by its matter content.  This condition is in keeping with
Mach's principle which states that the inertial forces acting on a
body can only be defined with respect to other bodies, and not with
respect to abstract concepts such as absolute space-time.

The condition that the field equations of the theory should be
derivable from an action principle gives us covariantly conserved
quantities, such as energy-momentum, via N\"{o}ther's theorem and
allows us to use Lagrangian mechanics.  Further justification for this
condition may be found in quantum mechanics; if a satisfactory quantum
theory of gravity is to be found this strongly suggests that the
theory should be derivable from an action principle.

The last of the above conditions is perhaps the most tenuous and is
the one which could be most readily relaxed.  We choose to interpret
this condition as meaning that the gravitational Lagrangian should be
a function of simple contractions of the Riemann tensor and one extra
scalar field only.  This still leaves us with a great number of
alternative theories, including the most general formulation of the
scalar-tensor theories as well as the fourth-order theories of
gravity.  Both of these classes of theory contain GR as a special (or
limiting) case and it is these theories which this work will be
concerned with.

It is clear that GR satisfies all of the above conditions.  However,
there are a number of additional conditions that GR also obeys which
will not necessarily be obeyed by the theories we will be considering.
These conditions are:

\begin{enumerate}
    \item{The strong equivalence principle.}
\item{The space and time independence of Newton's constant.}
\item{The linearity of the action in second derivatives of the metric.}
\end{enumerate}

The first of these conditions, the strong equivalence principle, is
the statement that: `\textit{massive self-gravitating objects should
  follow geodesics of the space-time}'.  Whilst the metric formalism
above assures us of this property for neutral test-particles, we are
no longer assured that it should be true in general for extended
massive bodies.  These objects distort the background space-time in
which they exist and whilst it is true that in GR they still follow
geodesics, this is no longer true in general for the alternative
theories we will be considering.  This type of violation of the strong
equivalence principle is often referred to as the Nordvedt effect.

The second condition that is met by GR is that at any point in
space-time one can take the weak-field and slow motion limit to
recover Newtonian gravity, with the same value of Newton's constant as
is measured on Earth.  This condition is explicitly violated by some
of the theories we will consider.  In fact, the Brans-Dicke theory was
devised specifically in order to create a theory of gravitation in
which space-time variations of Newton's constant could occur.
Constraining the parameters of these theories is then equivalent to
constraining the potential variations of Newton's constant which they
model.

The violation of the last of these three conditions is perhaps the
most worrying.  It is the linearity in second derivatives of the
Einstein-Hilbert action that ensures that the field equations of GR
are of no higher than second order in derivatives of the metric.
Whilst this is also true of scalar-tensor theories, it is no longer
true of theories which are constructed from contractions of the
Riemann tensor only (except for the Einstein-Hilbert action with a
cosmological constant).  Generating Lagrangians which are non-linear
in second derivatives yield field equations which are fourth-order in
derivatives of the metric (with the exception of additional terms in
the Lagrangian which are total diverges, such as the Gauss-Bonnet
term).  Also, variation of the action with respect to the metric is
then no longer equivalent to Palatini approach of varying with respect
to the metric and an independent connection.  In fourth-order theories
these two approaches yield two different sets of field equations.  In
this work we will only be concerned with the metric variation
approach.

The investigation of alternative theories of gravity is hampered by
the additional complexities that modifications to GR introduce.  It
seems that GR is unique not only in satisfying all of the conditions
listed above, but also in being the simplest relativistic metric
theory of gravitation that can be conceived of.  Any modification to
GR invariably results in a set of field equations that is considerably
more complicated than Einstein's equations.  In dealing with this
complexity there appear in the literature two different ways of
proceeding.  The first is to look for approximate solutions to a
specific theory of interest.  This approach is particularly useful in
the weak-field and low-velocity regime where frameworks such as the
parameterised post-Newtonian approach can often be employed.  It is
the approach that is most frequently taken up when investigating
theories that are motivated by a desire to overcome the perceived
shortcomings of the standard theory (examples of which were given
above).  The second approach is to look for exact solutions of the
simplest possible modifications to GR.  The idea behind this approach
is to understand as well as possible the effect of modifying the
standard theory.  Once this behaviour is well understood it can then
be used as an approximation to more complicated theories of specific
interest, as well for considerations of that particular modification.
It is the second approach that will be most closely adhered to in this
work.

The program for this study will be to investigate the form of the
solutions to the simplest alternative theories that meet the demands
listed above; these will be the scalar-tensor theories and the
fourth-order theories, which are often collectively referred to as the
extended theories of gravity.  Particular attention will be focussed
on highly symmetric situations.  High symmetry space-times are the
most readily solved for and are often the ones of most physical
interest.  These solutions will then be used to model physical
processes that occur in the Universe.  Comparing these models with
observation will allow constraints to be imposed upon the theory,
which will limit the allowed deviations of this kind from GR.  In
performing this analysis we will be able to note new behaviour that
was not previously obtainable in the general relativistic case.

We begin in chapter 2 by formally introducing the theories that will
be considered.  The generating Lagrangians will be explicitly given
and the field equations derived from them.  The form of these field
equations will immediately be seen to be more complicated than the
general relativistic case.  We will then give a brief discussion of
how these theories can be conformally transformed to look like general
relativity minimally coupled to a scalar field.  This conformal
equivalence will be useful in finding solutions to the field
equations.

In chapter 3 the homogeneous and isotropic solutions to the
scalar-tensor field equations will be given.  For the simplest case of
a spatially flat Brans-Dicke universe containing a single perfect
fluid the general solutions to the problem can be found.  These
general solutions have a late time power law attractor for the
evolution of the scale factor and at early times are generally
dominated for a period by the free component of the gravitational
scalar itself.  For this reason the general solution for the radiation
dominated universe is presented explicitly, as it is this component of
a realistic universe which is expected to dominate at early times, and
the late time attractors are given for dust and vacuum energy
densities.  Vacuum solutions with non-zero spatial curvature are also
given, and will be used in chapter 7 for a discussion of inhomogeneous
cosmologies.  The spatially-flat perfect fluid solutions will be used
in chapter 9 to model the expansion of the Universe during primordial
nucleosynthesis.  The homogeneous and isotropic cosmological solutions
for more general scalar-tensor theories are also given in this
chapter.

In chapter 4 the homogeneous and isotropic spatially flat cosmologies
of a simple class of fourth-order gravity theories are investigated.
A phase plane analysis is performed of the solutions and it is found
that the late time attractor solutions for both vacuum cosmologies and
universes filled with perfect fluids are simple power-law functions of
time.  These solutions are used in chapter 9 to model the expansion of
the Universe during both primordial nucleosynthesis and microwave
background formation.

In chapter 5 a variety of other homogeneous fourth-order cosmologies
are presented.  These include solutions of the G\"{o}del type, as well
as Einstein static, de-Sitter and Bianchi type I Kasner-like
solutions.  For the case of the G\"{o}del solutions the existence of
closed time-like curves, and hence the theoretical possibility of time
travel, is investigated.  The Kasner-like solutions are used to
perform an analysis of the approach towards an anisotropic
cosmological singularity in these theories.  It is found that, unlike
in the general relativistic case, these cosmologies do not undergo an
infinite sequence of chaotic oscillations as the initial singularity is
approached.  It is also confirmed that matter sources do not generally
dominate the evolution of these universes on approach to the singularity.

In chapter 6 static and spherically symmetric vacuum solutions are
presented for both the scalar-tensor and fourth-order cases.  For the
case of the simple class of fourth-order theories that is studied it
is found, using a dynamical systems approach, that the general
solution to the problem is not asymptotically flat.  An explicit $r$
dependent expression is found for the asymptotic attractor and an
exact solution is presented which has the same asymptotic behaviour.
A perturbative analysis is performed which confirms that the exact
solution corresponds to the mode with the appropriate Newtonian limit
and also indicates the existence of extra modes which are damped
oscillatory waves.  This solution is used in chapter 10 to model solar
system gravitational experiments in such a space-time.

Chapter 7 investigates the inhomogeneous cosmologies of these
alternative theories.  In the scalar-tensor case exact solutions are
presented for inhomogeneous universes which asymptotically approach
the homogeneous and isotropic solutions of chapter 3.  A detailed
investigation of the spherical collapse model in these theories is
also performed: firstly for exact flat and closed vacuum solutions
which are matched at a boundary and secondly for flat and closed
universes containing a perfect fluid.  For the case of spherical
collapse with a perfect fluid numerical simulations are performed, as
no exact solutions are know for closed perfect fluid universes.  The
evolution of the scalar field, and hence Newton's constant, are
followed both inside and outside the collapsing region and variations
are tracked.  For the case of fourth-order theories an exact
spherically symmetric vacuum solution is presented which
asymptotically approaches the homogeneous and isotropic vacuum
solution of chapter 4.  This solution is used in chapter 10, as an
alternative to the static solution found in chapter 6, to model
gravitational experiments that are carried out in the solar system.  A
linear perturbative analysis is carried out on the background of this
solution and it is found that the exact solution does indeed reduce to
the appropriate Newtonian limit and that the extra modes present in
the general solution are manifest as modes growing as $r^2$ in the
linearisation.

In chapter 8 an analysis is made of scalar-tensor cosmologies where
energy is allowed to be exchanged between the gravitational scalar
field and a perfect fluid.  In such a theory all of the conditions
required of the theory above are still met.  However, some of the
meaning of condition 1 is lost as test-particles no longer follow
exact geodesics of the space-time geometry.  This is due to the
interaction with the gravitational scalar, which acts as an external
force on the test-particle.  The exchange of energy and entropy
between the scalar field and the perfect fluid is discussed and the
evolution of homogeneous and isotropic universes is then considered.
For a general form of the parameter controlling the energy exchange
the field equations for the scale factor and scalar field can be
decoupled and integrated to two first order ordinary differential
equations.  From this point some simple specific functions of the
energy exchange parameter are proposed and these equations are then
integrated again to give explicit expressions for the evolutions of
the scale factor and scalar field.

In chapter 9 cosmological processes are modelled in the solutions that
were presented in previous chapters.  Specifically, primordial
nucleosynthesis is modelled in the general solutions for radiation
dominated spatially flat scalar-tensor universes.  Previous analyses
have been performed on primordial nucleosynthesis in these theories,
but have always taken the late time power law solutions to describe
the evolution of the Universe.  In this study we consider the general
form of the solution and hence allow for Newton's constant to be
varying whilst nucleosynthesis occurs.  This allows, for the first
time, observational constraints to be imposed upon the free
scalar-dominated period of such a universe's expansion.  In addition, we
study primordial nucleosynthesis and microwave background formation in
the power law solutions of the fourth-order theories found in chapter
4.  These observable processes allow the theory to be constrained from
observation.

Chapter 10 provides an analysis of the weak field constraints
available on the theory.  A brief explanation of the parameterised
post-Newtonian approach, and how it can be used to place constraints
on the scalar-tensor theories of gravity, is given.  The spherically
symmetric vacuum solutions to the fourth order theories found in
chapters 6 and 7 are then analysed.  It is shown that these theories
do not fit so readily into the usual formalism, and that in these
theories one should expect extra gravitational forces in the weak
field limit which are not present in GR.  For the static solution this
extra force takes the form of a force which drops off as $r^{-1}$ and
for the inhomogeneous cosmological solution this force looks like a
friction force, for a slowly moving test-particle.  The $r^{-1}$ force
in the static solution has a dramatic impact on the perihelion
precession of closed orbits in such a space-time and, applying
observational constraints from the orbit of Mercury, it is possible to
constrain the theory dramatically if this solution is used to model
the solar system.  The inhomogeneous cosmological solution fits better
into the parameterised post-Newtonian formalism and, if it is used to
model the solar system, can be used to place tight constraints on this
theory from the observations of the Shapiro time delay of radio
signals from the Cassini space craft as it passes behind the Sun.  The
extra gravitational forces present in these solutions, and their lack
of asymptotic flatness, show explicitly new behaviour in these
theories which is not present in GR.

In chapter 11 we summarise our results and provide concluding
remarks.

\newpage
\textbf{Conventions}
\newline

The Riemann curvature tensor is
\begin{equation*}
{R^a}_{b c d} = {\Gamma^a}_{b d , c}- {\Gamma^a}_{b c , d} +
{\Gamma^a}_{c e} {\Gamma^e}_{b d} - {\Gamma^a}_{d e} {\Gamma^e}_{c b}
\end {equation*}
where Latin indices refer to space-time components and we use Greek
indices to refer to space components.  The signature of the metric is
taken to be $(-,+,+,+)$ and we choose units in which $c=1$.

\clearemptydoublepage
\chapter{Extended Theories of Gravity}
\label{Theories}

\bigskip

`Extended theories of gravity' is the collective term usually applied
to the scalar-tensor and fourth-order theories of gravity.  In this
chapter we will introduce these theories, deriving their field
equations and showing their conformal equivalence to GR.

\section{Actions and Field Equations}

We present in this section the generating Lagrangians for the
theories that will be considered in the remainder of this work.  The
actions obtained from integrating these Lagrangians over all space
will be extremised to obtain the relevant field equations.

Considering gravitational theories derived from Lagrangians has the
advantage of ensuring coordinate independence of the derived field equations,
as well as the covariant conservation of quantities via
N\"{o}ther's constraint.

\subsection{Scalar-Tensor Theories}

A general form of the scalar-tensor theory \cite{Ber68, Nor70, Wag70} can be derived from the
Langrangian density
\begin{equation}
\label{ST}
\mathcal{L} = \frac{1}{16 \pi} \sqrt{-g} \left( f(\phi) R-g(\phi)
\partial_a \phi \partial^a \phi -2 \Lambda(\phi) \right) + \mathcal{L}_m(\Psi, h(\phi) g_{ab})
\end{equation}
where $f, g, h$ and $\Lambda$ are arbitrary functions of the scalar
field $\phi$ and $\mathcal{L}_m$ is the Lagrangian density of the
matter fields $\Psi$.  The function $h(\phi)$ can be absorbed into the
metric by a conformal transformation of the form \cite{dicke62}
\begin{equation*}
h(\phi) g_{ab} \rightarrow g_{ab}.
\end{equation*}
The conformal frame picked out by this choice is the one in
which there is no direct interaction between the scalar field and
matter fields and is usually referred to as the Jordan frame  (see
section \ref{confsec} for a discussion of conformal transformations in
these theories).  Test-particles in this conformal frame follow
geodesics of the metric and the weak equivalence principle is obeyed.
The effect of this transformation on the rest of the Lagrangian can
then be absorbed into redefinitions of $f$, $g$ and $\Lambda$.

By a redefinition of the scalar field $\phi$ we can now set $f(\phi)
\rightarrow \phi$, without loss of generality.  The Lagrangian density (\ref{ST})
can then be written as
\begin{equation}
\label{ST2}
\mathcal{L} = \frac{1}{16 \pi} \sqrt{-g} \left(\phi R-\frac{\omega(\phi)}{\phi}
\partial_a \phi \partial^a \phi -2 \Lambda(\phi)\right) +
\mathcal{L}_m(\Psi, g_{ab})
\end{equation}
where $\omega(\phi)$ is an arbitrary function which has been chosen in analogy with the
Brans-Dicke theory and $\Lambda$ is a $\phi$-dependent generalisation
of the cosmological constant.  This theory reduces to the Brans-Dicke
theory in the limit $\omega \rightarrow$constant and $\Lambda
\rightarrow 0$, and approaches general relativity in the limit $\omega
\rightarrow \infty$, ${\omega}'/\omega^2 \rightarrow 0$ and $\Lambda \rightarrow 0$.

The variation of the action derived from integrating (\ref{ST2}) over
all space is
\begin{equation}
\delta I = \frac{1}{16 \pi} \int d\Omega \sqrt{-g} \left[ \frac{1}{2}
  g^{a b} \delta g_{a b} \mathcal{L} +\phi \delta R
  +\frac{\omega (\phi)}{\phi} \partial^a \phi \partial^b \phi
  \delta g_{a b} +8 \pi T^{a b} \delta g_{a b} \right]
\end{equation}
where use has been made of
\begin{equation}
\label{vars}
\delta g^{a b} = -g^{a c} g^{b d} \delta g_{c d}, 
\qquad \qquad
\delta \sqrt{-g} = \frac{1}{2} g^{a b} \delta g_{a b}
\end{equation}
and
\begin{equation*}
T^{a b} = \frac{2}{\sqrt{-g}} \frac{\delta \mathcal{L}_m}{\delta g_{a
    b}}.
\end{equation*}
Now using the relation \cite{DeW65}
\begin{equation}
\label{Rab}
\delta R_{a b} = \frac{1}{2} g^{c d} (\delta g_{a c ; b d} + \delta g_{b d ; a c} -  \delta
g_{a b ; c d} - \delta g_{c d ; a b})
\end{equation}
we can write
\begin{equation}
\label{field1}
\delta I \simeq \frac{1}{16 \pi} \int d \Omega \sqrt{-g} (-S^{a b}+8 \pi T^{a b}) \delta g_{a b}
\end{equation}
where
\begin{equation*}
S^{a b} = \phi (R^{a b} - \frac{1}{2} g^{a b} R) - (g^{a c} g^{b d}-
g^{a b} g^{c d}) \phi_{;c d}- \frac{\omega}{\phi} (g^{a c} g^{b d}- \frac{1}{2}
g^{a b} g^{c d}) \phi_{,c} \phi_{,d}+ g^{a b} \Lambda
\end{equation*}
and $\simeq$  means equal up to total divergences.  Such terms are
irrelevant here as they can be transformed via Gauss's theorem to
terms on the boundary which are assumed to vanish.

The first variation of the action corresponding to (\ref{ST2}), with respect to
the scalar field $\phi$, is now
\begin{equation}
\label{field2}
\delta I \simeq \frac{1}{16 \pi} \int d\Omega \sqrt{-g} \left(R +2 \omega
\frac{\square \phi}{\phi}- \frac{\omega}{\phi^2} \phi_{,c} \phi_,^c
+\frac{\omega'}{\phi} \phi_{,c} \phi_,^c -2 \Lambda' \right) \delta \phi
\end{equation}
where primes denote differentiation with
respect to $\phi$.  Eliminating $R$ with the trace of
(\ref{field1}) and looking for stationary points of the action, by
setting the first variation to zero, then gives the field equations
\begin{equation}
\label{STfields}
 \phi (R_{a b} - \frac{1}{2} g_{a b} R) = 8 \pi T_{a b} +(\delta^c_a
 \delta^d_b - g_{a b} g^{c d}) \phi_{;c d}+ \frac{\omega}{\phi} (\delta^c_a \delta^d_b- \frac{1}{2}
g_{a b} g^{c d}) \phi_{,c} \phi_{,d}- g_{a b} \Lambda
\end{equation}
and
\begin{equation}
\label{STfields2}
(2 \omega +3) \square \phi = 8 \pi T- \omega' \phi_{,a} \phi_,^a -4
  \Lambda+2 \phi \Lambda'.
\end{equation}

\subsection{Fourth-Order Theories}

The fourth-order theories we shall consider are derived from functions of
the three possible linear and quadratic contractions of the Riemann
curvature tensor: $R$, $R_{ab}R^{ab}$ and $R_{abcd}R^{abcd}$ \footnote{As pointed out by De Witt \cite{DeW65} there is a fourth
possibility, namely $\epsilon^{a b c d} R_{e f a b} {R^{e f}}_{c d}$. However,
this contraction is of no physical interest due to parity considerations.}
. The relevant weight-zero scalar density for this general class of theories
is then given by 
\begin{equation}
\mathcal{L}=\chi ^{-1}\sqrt{-g}f(X,Y,Z)  \label{density}
\end{equation}%
where $f(X,Y,Z)$ is an arbitrary function of $X$, $Y$ and $Z$ which are
defined by $X \equiv R$, $Y \equiv R_{ab}R^{ab}$ and $Z \equiv R_{abcd}R^{abcd}$, and $\chi $ is an
arbitrary constant which can be determined from the appropriate
Newtonian limit. The action is obtained, as usual, by integrating this
density, together with that of the matter fields, over all space. The addition
of supplementary terms to the density (\ref{density}) in order to cancel
total divergences which can be transformed to integrals on the boundary can
be problematic (see e.g. \cite{Mad88}) and so, for simplicity, they will all
be assumed to vanish.

The variation of the action derived from integrating the density
(\ref{density}) over all space is
\begin{align}
\nonumber
\delta I &= \chi^{-1} \int d \Omega \sqrt{-g} \Bigl[ \frac{1}{2} f g^{a b}
  \delta g_{a b} + f_X \delta X + f_Y \delta Y + f_Z \delta Z \Bigr]\\ 
\nonumber
&= \chi^{-1} \int d \Omega \sqrt{-g} \Bigl[ \frac{1}{2} f g^{a b}
  \delta g_{a b} - f_X (R^{a b} \delta g_{a b} - g^{a b} \delta R_{a
  b})\\
\nonumber
& \qquad \qquad \qquad \qquad \qquad \quad - 2 f_Y (R^{c (a} {R^{b)}}_{c} \delta g_{a b} - R^{a b} \delta
  R_{a b})\\
\label{var}
& \qquad \qquad \qquad \qquad \qquad  \quad - 2 f_Z ({R_{c d e}}^{(b} R^{a) e d c} \delta g_{a b} -
  {R_a}^{b c d} \delta {R^a}_{b c d}) \Bigr]
\end{align}
where use has again been made of (\ref{vars}).  Using the
relations (\ref{Rab}) and
\begin{equation*}
\delta {R^a}_{b c d} = \frac{1}{2} g^{a e} (\delta g_{e d ; b c} + \delta
g_{e b ; d c} - \delta g_{d b ; e c} - \delta g_{e c ; b d} - \delta
g_{e b ; c d} +\delta g_{c b ; e d}).
\end{equation*}
We can then write
\begin{align*}
f_X g^{a b} \delta R_{a b} &\simeq -f_{X ; c d} (g^{a b} g^{c d} - g^{a
  c} g^{b d}) \delta g_{a b}\\
2 f_Y R^{a b} \delta R_{a b} &\simeq- \Bigl[ \square (f_Y R^{a b})+(f_Y
  R^{c d})_{; c d} g ^{a b} - 2 {{(f_Y R^{c (a})_;}^{b)}}_c \Bigr]
  \delta g_{a b}\\
2 f_Z {R_a}^{b c d} \delta {R^a}_{b c d} &\simeq 4 (f_Z R^{c (a b)
  d})_{; c d} \delta g_{a b}.
\end{align*}
Substituting these expressions back into (\ref{var}) then gives
\begin{equation*}
\delta I = - \chi^{-1} \int d \Omega \sqrt{-g} P^{a b}\delta g_{a b}
\end{equation*}
where
\begin{multline}  \label{P}
P^{a b} \equiv -\frac{1}{2} f g^{a b} + f_X R^{a b}+2 f_Y R^{c (a} {R^{b)}}_{c}+2
f_Z R^{e d c (a} {R^{b)}}_{c d e} +f_{X; c d}(g^{a b} g^{c d}-g^{a c} g^{b d})
\\
+\square (f_Y R^{a b}) + g^{a b} (f_Y R^{c d})_{;c d}-2 (f_Y R^{c (a})_{;\;
\; c}^{\; b)}-4 (f_Z R^{d (a b) c})_{;c d}.
\end{multline}
The notation $f_N$ denotes partial differentiation of $f$ with respect
to $N$.  Looking for a stationary point of the action by setting the first
variation to zero gives the field equations 
\begin{equation}  \label{fequations}
P_{a b}=\frac{\chi}{2} T_{a b} - g_{a b} \Lambda
\end{equation}
when matter fields and a cosmological constant are included.  Here,
$\Lambda$ is the cosmological constant (defined independent of $%
f(X,Y,Z)$) and $T^{a b}$ is the energy-momentum tensor of the matter.
These field equations are generally of fourth-order, with the exception of
the cases in which the function $f$ is linear in the second derivatives of
the metric \cite{DeW65}.

We will often be considering gravitational theories derived from the choice
\begin{equation}
\label{action}
f=f(R)=R^{1+\delta}
\end{equation}
where $\delta \neq 0$ is a real number. The limit $%
\delta \rightarrow 0$ gives us the familiar Einstein--Hilbert Lagrangian of
GR. We take the quantity $%
R^{\delta }$ to be the positive real root of $R$ throughout this
work.  This choice of $f$ gives
\begin{multline}
P_{ab}=\delta (1-\delta ^{2})R^{\delta }\frac{R_{,a}R_{,b}}{R^{2}}-\delta (1+\delta
)R^{\delta }\frac{R_{;ab}}{R}+(1+\delta )R^{\delta }R_{ab}-\frac{1}{2}%
g_{ab}RR^{\delta }  \label{field} \\
-g_{ab}\delta (1-\delta ^{2})R^{\delta }\frac{R_{,c}R_{,}^{\ c}}{R^{2}}%
+\delta (1+\delta )g_{ab}R^{\delta }\frac{\Box R}{R}.
\end{multline}
These equations have the useful property that in vacuum they are conformally
equivalent to Einstein's equations with a scalar field in an
exponential potential, as shown below.

\subsubsection{The Newtonian Limit}

By comparing the geodesic equation to Newton's gravitational force law it
can be seen that, as usual, 
\begin{equation}
\Gamma _{00}^{\mu }=\Phi _{,\mu }  \label{ChrN}
\end{equation}%
where $\Phi $ is the Newtonian gravitational potential. All the other
Christoffel symbols have $\Gamma _{\;bc}^{a}=0$, to the required order of
accuracy.

We now seek an approximation to the field equations (\ref{field}) that is of
the form of Poisson's equation; this will allow us to fix the constant $\chi 
$. Constructing the components of the Riemann tensor from (\ref{ChrN}) we
obtain the standard results 
\begin{equation}
R_{\;0\nu 0}^{\mu }=\frac{\partial ^{2}\Phi }{\partial x^{\mu }\partial
x^{\nu }}\qquad \text{and}\qquad R_{00}=\nabla ^{2}\Phi .  \label{R00}
\end{equation}%
The $00$ component of the field equations (\ref{field}) can now be written 
\begin{equation}
(1+\delta )R_{00}-\frac{1}{2}g_{00}R=\frac{\chi }{2}\frac{T_{00}}{R^{\delta }%
}  \label{approx}
\end{equation}%
where terms containing derivatives of $R$ have been discarded as they will
contain third and fourth derivatives of $\Phi $, which will have no
counterparts in Poisson's equation. Subtracting the trace of equation (\ref%
{approx}) gives 
\begin{equation}
(1+\delta )R_{00}=\frac{\chi }{2R^{\delta }}\left( T_{00}-\frac{1}{%
2(1-\delta )}g_{00}T\right)  \label{approx2}
\end{equation}%
where $T$ is the trace of the stress--energy tensor. Assuming a
perfect--fluid form for $T$ we should have, to first--order, 
\begin{equation}
T_{00}\simeq \rho \qquad \text{and}\qquad T\simeq 3p-\rho \simeq -\rho .
\label{T00}
\end{equation}%
Substituting (\ref{T00}) and (\ref{R00}) into (\ref{approx2}) gives 
\begin{equation*}
\nabla ^{2}\Phi \simeq \frac{\chi (1-2\delta )}{4(1-\delta ^{2})}\frac{\rho 
}{R^{\delta }}.
\end{equation*}%
Comparison of this expression with Poisson's equation allows one to read off 
\begin{equation}
\chi =16\pi G\frac{(1-\delta ^{2})}{(1-2\delta )}R_{0}^{\delta }  \label{chi}
\end{equation}%
where $R_{0}$ is the value of the Ricci tensor at the time $G$ is measured.
It can be seen that the Newtonian limit of the field equations (\ref{field})
does not reduce to the usual relation $\nabla ^{2}\Phi \propto \rho $, but
instead contains an extra factor of $R^{\delta }$. This can be interpreted
as being the space--time dependence of Newton's constant, in this theory.
Such a dependence should be expected as the Lagrangian (\ref{action}) can be
shown to be equivalent to a scalar--tensor theory, after an appropriate
Legendre transformation\footnote{This equivalence to
scalar-tensor theories should not be taken to imply that bounds on the
Brans-Dicke parameter $\omega$ are immediately applicable to this
theory.  It can be shown that a potential for the scalar-field can
have a non-trivial effect on the resulting phenomenology of the theory
\cite{Olmo}.  Furthermore, the form of the perturbation to general
relativity that we are considering does not allow an expansion of the
corresponding scalar field of the form $\phi_0+\phi_1$ where $\phi_0$ is constant
and $\vert \phi_1 \vert << \vert \phi_0 \vert$, so that any
constraints obtained in a weak-field expansion of this sort cannot be
applied to this situation.} (see e.g, \cite{Mag94}).  This type of Newtonian
gravity theory admits a range of simple exact solutions in the case where
the effective value of $G$ is a power-law in time \cite{newt}.

\section{Conformal Transformations}
\label{confsec}

A conformal transformation of the metric $g_{a b}$ into $\bar{g}_{a b}$
can be written
\begin{equation}
\label{ctran}
g_{a b} = e^{2 \Gamma (x)} \bar{g}_{a b}
\end{equation}
where $\Gamma (x)$ is an arbitrary function of the coordinates $x^a$.
The line-element is then transformed according to
\begin{equation*}
ds^2 = e^{2 \Gamma (x)} d \bar{s}^2
\end{equation*}
and the square root of the determinant of the metric as
\begin{equation*}
\sqrt{-g}= e^{4 \Gamma} \sqrt{-\bar{g}},
\end{equation*}
in four dimensions.  Conformal transformations of this kind are local,
isotropic transformations of the standard of `size', as defined by the
line-element.  These transformations are known as conformal as they
leave the angle between any two vectors in the space-time unaltered.  

After performing such a transformation we use the term
`conformal frame' to distinguish the new, rescaled metric from the
original.  Among the infinite possible conformal frames there are
two which are most commonly used and have specific interpretations:
the Jordan frame and the Einstein frame.  The Jordan frame is the one
in which the energy-momentum tensor is covariantly conserved and in
which test-particles follow geodesics of the space-time metric.  The
Brans-Dicke theory \cite{Bra61}, for instance, is most usually
formulated in the Jordan frame.  The Einstein frame is the conformal
frame in which the field equations of the theory take the form of the
Einstein equations (unlike the Jordan frame, the Einstein frame can only be
defined for some theories).  In the Einstein frame the field equations
are second order but the energy-momentum tensor of the matter fields
is not always covariantly conserved and test-particles do not necessarily follow
geodesics of the space-time metric.  The Einstein frame, therefore, is
particularly useful for finding vacuum solutions, but less useful for
finding solutions with matter fields present.

Under the transformation (\ref{ctran}) it can be shown that the
Ricci tensor and scalar transform as (see e.g. \cite{Yan})
\begin{align}
\label{cRab}
R_{a b} &= \bar{R}_{a b} - 2 \Gamma_{; \bar{a b}} +2 \Gamma_{,a} \Gamma_{,b}
-2 \bar{g}_{a b} \bar{g}^{c d} \Gamma_{,c} \Gamma_{,d}-\bar{g}_{a b}
\bar{\square} \Gamma \\
\label{cR}
e^{2 \Gamma} R &= \bar{R} -6 \bar{g}^{a b} \Gamma_{,a} \Gamma_{,b}-6
\bar{\square} \Gamma
\end{align}
and the d'Alembertian transforms as
\begin{equation}
\label{cbox}
e^{2 \Gamma} \square \phi = \bar{\square} \phi +2 \bar{g}^{a b}
\Gamma_{, a} \phi_{, b}
\end{equation}
where overbars over operators or indices denote that they are defined
using the metric $\bar{g}_{a b}$.  We will now use these
transformations to show how the scalar-tensor and some fourth-order
theories can be transformed from the Jordan frames to the Einstein
frame.

\subsection{Scalar-Tensor Theories}

All of the scalar-tensor theories defined by the Lagrangian (\ref{ST2}), in
the Jordan frame, can be transformed to an equivalent in the Einstein
frame.  To see this we will consider the various terms in the
generating Lagrangian (\ref{ST2}) seperately.  Firstly, we consider
the term containing the Ricci scalar
\begin{equation*}
\mathcal{L}_1 = \frac{1}{16 \pi} \sqrt{-g} \phi R
\end{equation*}
which under the conformal transformation (\ref{ctran}) becomes
\begin{equation}
\label{L1}
\mathcal{L}_1 = \frac{1}{16 \pi} \sqrt{-\bar{g}} \phi e^{2 \Gamma}
(\bar{R}-6 \bar{g}^{a b} \Gamma_{,a} \Gamma_{,b}-6 \bar{\square}
\Gamma).
\end{equation}
The non-minimal coupling to the Ricci scalar can now be removed by
making the choice of conformal factor
\begin{equation}
\label{conformal}
e^{2 \Gamma} = \phi^{-1} \qquad \text{such that} \qquad g_{a b}=
\frac{\bar{g}_{a b}}{\phi}.
\end{equation}
This choice of $\Gamma$ defines the conformal transformation between
the Jordan and Einstein frames in the scalar-tensor theories.  Making
this choice of $\Gamma$ it can be seen that the last term on the
right-hand side of (\ref{L1}) is a total divergence, which can be
transformed to an integral over a surface at infinity and therefore ignored.

Considering now the second term in (\ref{ST2}) we get
\begin{equation*}
\mathcal{L}_2 = -\frac{1}{16 \pi} \sqrt{-g} \frac{\omega}{\phi} g^{a b} \phi_{,
  a} \phi_{, b} = -\frac{1}{4 \pi} \sqrt{-\bar{g}} \omega \bar{g}^{a b}
  \Gamma_{, a} \Gamma{, b}
\end{equation*}
under the transformation (\ref{ctran}) and using the definition of
$\Gamma$ above.  The third term then gives
\begin{equation*}
\mathcal{L}_3 = -\frac{1}{8 \pi} \sqrt{-g} \Lambda = -\frac{1}{8 \pi}
\sqrt{-\bar{g}} e^{4 \Gamma} \Lambda
\end{equation*}
so that the whole Lagrangian can be written
\begin{equation}
\label{ST3}
\mathcal{L} = \frac{1}{16 \pi} \sqrt{-\bar{g}} \left( \bar{R}- 2 (3+2
\omega) \bar{g}^{a b} \Gamma_{, a} \Gamma_{, b}- 2 e^{4 \Gamma}
\Lambda \right) +\mathcal{L}_m(\Psi, e^{2 \Gamma} \bar{g}_{a b}).
\end{equation}
Now, by making the definitions
\begin{equation}
\label{STdef}
\sqrt{\frac{4 \pi}{(3+2 \omega)}}= \frac{\partial \Gamma}{\partial \psi} \qquad
\text{and} \qquad 8 \pi V (\psi)= e^{4 \Gamma} \Lambda
\end{equation}
for the scalar $\psi$ and the function $V(\psi)$, we can write the
transformed Lagrangian (\ref{ST3}) as
\begin{equation}
\label{ST4}
\mathcal{L} = \frac{1}{16 \pi} \sqrt{-\bar{g}} \bar{R} +\sqrt{-\bar{g}} \left(
  -\frac{1}{2} \bar{g}^{a b} \psi_{, a} \psi_{, b}-V(\psi) \right)+
  \mathcal{L}_m(\Psi, e^{2 \Gamma} \bar{g}_{a b} ).
\end{equation}
In the absense of any matter fields the scalar-tensor theories can now
clearly be seen to be conformally related to Einstein's theory in the
presence of a scalar field in a potential (that potential
disappearing in the absense of $\Lambda$).  

In the Brans-Dicke theory
\cite{Bra61} the coupling constant $\omega$ is a constant quantity and
the scalar fields $\phi$ and $\psi$ are related by
\begin{equation*}
\ln \phi = \sqrt{\frac{16 \pi}{(3+2 \omega)}} \psi.
\end{equation*}
For more general theories with $\omega=\omega(\phi)$ the definition of
$\psi$, (\ref{STdef}), must be integrated to obtain a relation between
$\phi$ and $\psi$.

By extremising the action (\ref{ST4}) with respect to
$\bar{g}_{a b}$ we get the Einstein frame field equations
\begin{equation*}
\bar{R}_{a b}=8\pi(\bar{T}_{a b}-\frac{1}{2}\bar{g}_{a
  b}\bar{T}+\psi_{, a}\psi_{, b})
\end{equation*}
and by extremizing with respect to $\psi$ we get the Einstein frame propagation equation
\begin{equation*}
\bar{\square} \psi=-\sqrt{4\pi}\alpha \bar{T},
\end{equation*}
where $\alpha^{-2} \equiv 3+2 \omega$ and we have defined the energy-momentum tensor $\bar{T}_{a b}$ with
respect to $\bar{g}_{a b}$ so that
$\bar{T}^{a b}=e^{6 \Gamma} T^{a b}$.  Correspondingly, whilst the Jordan frame energy-momentum tensor is always
covariantly conserved, ${T^{a b}}_{; b}=0$, its counterpart in the
Einstein frame is not, ${\bar{T}^{a b}}_{; b}=\sqrt{4 \pi} \alpha \bar{T}
{\psi_,}^{a}$.  The interaction between the scalar
field $\psi$ and the matter fields described by $\bar{T}_{a b}$ can
now be seen explicitly.

\subsection{Fourth-Order Theories}

We will now show the conformal transformation between the Jordan
frame and the Einstein frame for theories derivable
from an abritrary function of the Ricci scalar \cite{cot, maeda}.  Extremising the action
\begin{equation}
\label{fr}
S=\int d \Omega \sqrt{-g} f(R)
\end{equation}
with respect to the metric $g_{a b}$ gives the field equations,
according to (\ref{P}),
\begin{equation}
\label{R16}
f' R_{a b} - \frac{1}{2} f g_{a b} - f'_{;a b}+g_{a b} \square f'=
\frac{\chi}{2} T_{ab} -g_{ab} \Lambda
\end{equation}
where primes denote differentiation with respect to $R$.  Under the
conformal transformation
\begin{equation*}
\bar{g}_{a b} = f' g_{a b},
\end{equation*}
and making the definition
\begin{equation*}
\phi \equiv \sqrt{\frac{3}{\chi}} \ln f',
\end{equation*}
the field equations (\ref{R16}) then become
\begin{equation}
\bar{R}_{a b}- \frac{1}{2} \bar{g}_{a b} \bar{R} = \frac{\chi}{2}
\left( \phi_{,a} \phi_{,b}-\frac{1}{2} \bar{g}_{a b} \bar{g}^{c d}
\phi_{,c} \phi_{,d} - \bar{g}_{a b} V \right) +\frac{\chi}{2} \frac{T_{ab}}{f'}
\end{equation}
where
\begin{equation*}
V = \frac{(R f'-f)}{2 f'^2} + \frac{2 \Lambda}{\chi f'}.
\end{equation*}
It can now be seen that, in the absense of any matter fields, theories
derived from an action of the form (\ref{fr}) are conformally
equivalent to GR in the presence of a scalar field in a potential.

We will now consider the more specific case of theories derived from
the Lagrangian (\ref{action}).  In this case the relevant conformal
transformation can be written as $\bar{g}_{ab}=\Omega
_{0}R^{\delta } g_{ab}$, where $\Omega_0$ is a constant.  The scalar
field can then be defined as
\begin{equation*}
\phi \equiv \sqrt{\frac{3}{16\pi G}}\ln R^{\delta },
\end{equation*}
and the field equations, in the absense of matter fields and any
cosmological constant, can be rewritten as 
\begin{equation}
\bar{G}_{ab}=8\pi G\left( \phi _{,a}\phi _{,b}-\frac{1}{2}\bar{g}_{ab}(\bar{g%
}^{cd}\phi _{,c}\phi _{,d}+2 V(\phi ))\right)  \label{conformalfield}
\end{equation}
and 
\begin{equation*}
\bar{\square} \phi =\frac{dV}{d\phi },
\end{equation*}
where $V(\phi )$ is given by 
\begin{equation}
V(\phi )=\frac{\delta \ \text{sign}(R)}{16\pi G(1+\delta )\Omega _{0}}\exp
\left\{ {\sqrt{\frac{16\pi G}{3}}\frac{(1-\delta )}{\delta }\phi }\right\} .
\label{pot1}
\end{equation}
The magnitude of the quantity $\Omega _{0}$ is not physically important and
simply corresponds to the rescaling of the metric by a constant quantity,
which can be absorbed by an appropriate rescaling of units. It is, however,
important to ensure that $\Omega _{0}>0$ in order to maintain the +2
signature of the metric.

\clearemptydoublepage
\chapter{Homogeneous and Isotropic Scalar-Tensor Cosmologies}
\label{Cosmology}

\bigskip

We will consider in this section the homogeneous and isotropic cosmological
solutions to the scalar-tensor fields equations (\ref{STfields}) and
(\ref{STfields2}).  These are space-times with a six-dimensional group
of motions, $G_6$, acting transitively on a three-dimensional
space-like subspace.

\section{Friedmann-Robertson-Walker Cosmologies}

Spatially homogeneous and isotropic space-times are described by the
Friedmann-Robertson-Walker (FRW) line-element
\begin{equation*}
ds^{2}=-dt^{2}+a^{2}(t)\left( \frac{dr^{2}}{1-kr^{2}}+r^{2}(d\theta ^{2}+\sin
^{2}\theta d\phi ^{2})\right) ,
\end{equation*}%
where $a(t)$ is the scale factor and $k$ is the spatial curvature. Using
the FRW metric in the scalar-tensor field equations (\ref{STfields}) and (\ref{STfields2}), and
setting $\Lambda=0$, gives
\begin{equation}
2\dot{H}+3H^{2}+\frac{\omega }{2}\frac{\dot{\phi}^{2}}{\phi ^{2}}+2H\frac{%
\dot{\phi}}{\phi }+\frac{\ddot{\phi}}{\phi }=-\frac{8\pi }{\phi }p-\frac{k}{%
a^{2}},  \label{Nariai1}
\end{equation}
\begin{equation}
\frac{\ddot{\phi}}{\phi }=\frac{8\pi }{\phi }\frac{(\rho -3p)}{(2\omega +3)}%
-3H\frac{\dot{\phi}}{\phi }-\frac{\dot{\omega}}{(2\omega +3)}\frac{\dot{\phi}%
}{\phi },  \label{Nariai2}
\end{equation}%
and 
\begin{equation}
\frac{8\pi }{3\phi }\rho =H^{2}+H\frac{\dot{\phi}}{\phi }-\frac{\omega }{6}%
\frac{\dot{\phi}^{2}}{\phi ^{2}}+\frac{k}{a^{2}},  \label{Friedmann}
\end{equation}%
where $H\equiv \dot{a}/a$ is the Hubble rate, an overdot denotes
differentiation with respect to comoving proper time, $t$, $\rho $ is the
matter density, and $p$ is the pressure. Each non-interacting fluid source $%
p(\rho )$ separately satisfies a conservation equation:
\begin{equation}
\dot{\rho}+3H(\rho +p)=0.  \label{fluid}
\end{equation}
Substituting equations (\ref{Nariai2}) and (\ref{Friedmann}) into equation (\ref{Nariai1}) gives
\begin{equation}
\dot{H}+H^{2}-H\frac{\dot{\phi}}{\phi }+\frac{\omega }{3}\frac{\dot{\phi}^{2}%
}{\phi ^{2}}
=-\frac{8\pi }{3\phi }\frac{(3p\omega +3\rho +\rho \omega )}{%
(2\omega +3)}+\frac{1}{2}\frac{\dot{\omega}}{(2\omega +3)}\frac{\dot{\phi}}{%
\phi }.  \label{acceleration}
\end{equation}
It is the solutions of these equations that concern us in this chapter.

\section{Brans-Dicke Theory}

For the Brans-Dicke theory ($\omega=$constant) there are a variety of
known exact homogeneous and isotropic solutions (see e.g. \cite{Bar92, Gur73, Nar68, Tup}).  We
will present the solutions here that are of particular significance,
and that will be of use later on.

In these theories $\omega (\phi
)\equiv \omega $ is a constant. The three essential field equations
for the evolution of the scalar field $\phi (t)$ and the expansion scale
factor $a(t)$ in a Brans-Dicke universe are (\ref{Nariai2}), (\ref{Friedmann}), and (%
\ref{fluid}).  Now $\omega $ is a constant parameter and the theory reduces to
GR in the limit $\omega \rightarrow \infty $ where $\phi
=G^{-1}\rightarrow $ constant. The general solutions to the
FRW Brans-Dicke theory field equations are fully understood. The
vacuum solutions are the $t\rightarrow 0$ attractors for the
perfect-fluid solutions and the general solutions with equation
of state
\begin{equation}
p=(\gamma -1)\rho  \label{pee}
\end{equation}%
and $k=0$ can all be found. At early times they approach the vacuum
solutions but at late time they approach particular power-law exact
solutions \cite{Nar68}:
\begin{equation}
a(t)=t^{[2+2\omega (2-\gamma )]/[4+3\omega \gamma (2-\gamma )]}  \label{bds1}
\end{equation}
\begin{equation}
\phi (t)=\phi _{0}t^{[2(4-3\gamma )/[4+3\omega \gamma (2-\gamma )]}
\label{bds2}
\end{equation}

These particular exact power-law solutions for $a(t)$ and $\phi (t)$ are
`Machian' in the sense that the cosmological evolution is driven by the
matter content rather than by the kinetic energy of the free $\phi $ field.

The power-law solutions (\ref{bds1}) and (\ref{bds2}) are valid for
any fluid with equation of state (\ref{pee}).  We will now consider
seperately the solutions for universes dominated by radiation, matter
and vacuum energy.  The general solutions for $k=0$ are known for
any fluid that satisfies (\ref{pee}) \cite{Gur73}.  Here we will
only present the general solution for a radiation-dominated universe
as the late-time power-law attractors for matter and vacuum dominated
eras are expected to be sufficient for any realistic universe (see below).

\subsection{Radiation Domination}

We consider first the case of a flat radiation--dominated universe.
Assuming the equation of state $p=\frac{1}{3} \rho$ and defining the
conformal time variable, $\eta$, by $a d\eta \equiv dt$, the equations
(\ref{Nariai1}), (\ref{Nariai2}) and (\ref{Friedmann}) integrate to give
\cite{Gur73}
\begin{align}
\label{eta2}
\frac{a'(\eta)}{a(\eta)}&=\frac{\eta +\eta_2}{\eta^2+(2 \eta_2 +3
  \eta_1)\eta +\eta_2^2 + 3 \eta_1 \eta_2 -\frac{3}{2}\omega
  \eta_1^2}\\
\frac{\phi'(\eta)}{\phi(\eta)}&=\frac{3 \eta_1}{\eta^2+(2 \eta_2 +3
  \eta_1)\eta +\eta_2^2 + 3 \eta_1 \eta_2 -\frac{3}{2}\omega
  \eta_1^2}
\label{eta1}
\end{align}
where $\eta_1$ and $\eta_2$ are integration constants.

For $\omega > -3/2$ the solutions to these equations are
\begin{align}
a(\eta)&=a_1 (\eta +\eta_{+})^{\frac{1}{2}+
  \frac{1}{2\sqrt{1+\frac{2}{3} \omega}}} (\eta
  +\eta_{-})^{\frac{1}{2} -
  \frac{1}{2\sqrt{1+\frac{2}{3} \omega}}} \label{aw>}\\
\phi(\eta)&=\phi_1 (\eta +\eta_{+})^{-
  \frac{1}{2\sqrt{1+\frac{2}{3} \omega}}} (\eta +\eta_{-})^{+
  \frac{1}{2\sqrt{1+\frac{2}{3} \omega}}} \label{phiw>}
\end{align}
where $\eta_{\pm}=\eta_2+\frac{3}{2}\eta_1 \pm \frac{3}{2} \eta_1
\sqrt{1+\frac{2}{3}\omega}$, $a_1$ and $\phi_1$ are integration constants, and
$8 \pi \rho_{r0}/3 \phi_1 a_1^2=1$ (subscript 0 indicates a quantity
measured at the present day and we rescale so that $a_0=1$, throughout).
\begin{figure}
\epsfig{file=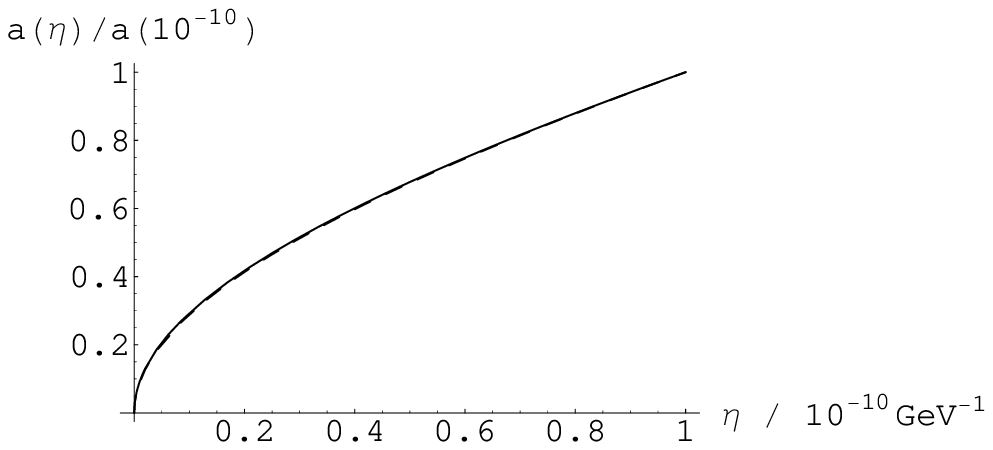,height=9cm}
\epsfig{file=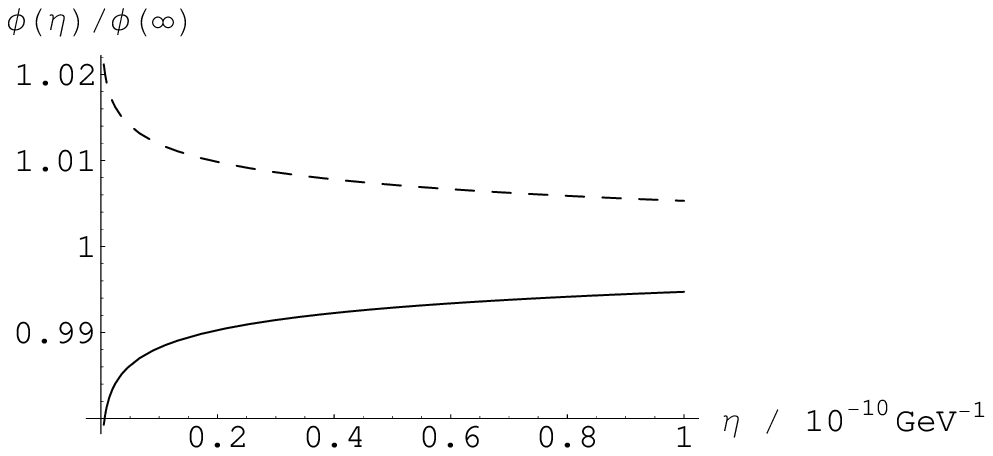,height=9cm}
\caption{{\protect {\textit{The evolution of $a$ and $\phi$ as functions of $\eta$
	for $\omega > -\frac{3}{2}$ with $\vert \eta_1 \vert=10^{-12} GeV^{-1}$,
	$\omega=40 000$, and $\eta_2$ set so that $a(0)=0$.  The solid
	line corresponds to $\eta_1 >0$ and the dashed line to  $\eta_1 <0$.}}}}
\label{raddom}
\end{figure}

For $\omega < -3/2$ we find
\begin{align}
a(\eta)&=a_1
\sqrt{(\eta+\eta_{-})^2+\eta_{+}^2} \exp \left(
  \frac{-1}{\sqrt{\frac{2}{3}\vert\omega\vert -1}} \tan^{-1}
  \frac{\eta+\eta_{-}}{\eta_{+}} \right), \label{aw<}\\
\phi(\eta)&=\phi_1
\left(\frac{2}{\sqrt{\frac{2}{3}\vert\omega\vert -1}} \tan^{-1}
  \frac{\eta+\eta_{-}}{\eta_{+}} \right), \label{phiw<}
\end{align}
where $\eta_{+}= \frac{3}{2} \eta_1
\sqrt{\frac{2}{3}\vert\omega\vert-1}$, $\eta_{-}=\eta_2+\frac{3}{2}
\eta_1$, $8 \pi \rho_{r0}/3 \phi_1 a_1^2=1$ and constants have been
absorbed into $a_1$ and $\phi_1$.
\begin{figure}
\epsfig{file=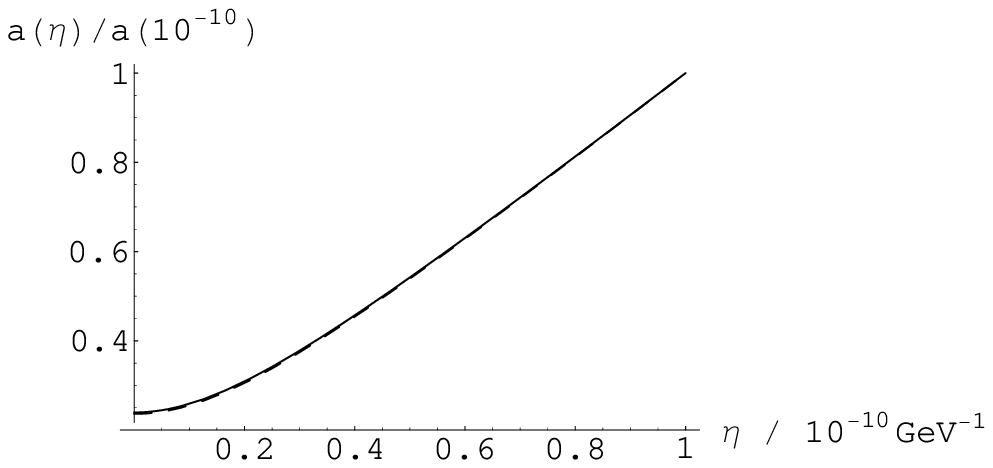,height=9cm}
\epsfig{file=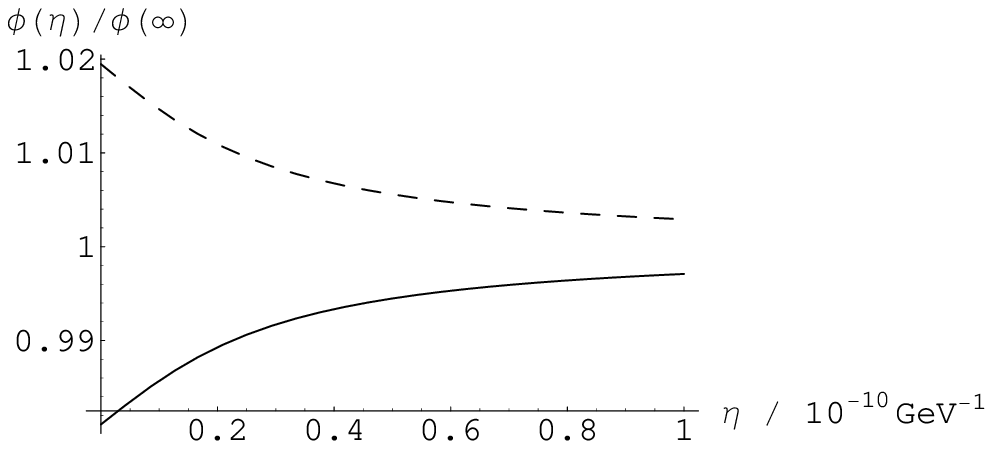,height=9cm}
\caption{\textit{The evolution of $a$ and $\phi$ as functions of $\eta$
	for $\omega < -\frac{3}{2}$ with $\vert \eta_1 \vert=10^{-13} GeV^{-1}$,
	$\omega=40 000$, and $\eta_2$ set so that $a'(0)=0$.  The solid
	line corresponds to $\eta_1 >0$ and the dashed line to  $\eta_1 <0$.}}
\label{raddom2}
\end{figure}

In these solutions $\eta_1$ determines the evolution of the scalar
field during radiation domination, and is a physically interesting
quantity; $\eta_2$ sets the origin of the conformal time coordinate.
The evolution of $a$ and $\phi$, for $\omega > - 3/2$ and $\omega < -3/2$, is
shown in figures \ref{raddom} and \ref{raddom2}, respectively.  A
value of $\omega =40000$ is chosen, in agreement with recent solar
system observations \cite{Bert}.

For $\omega >-3/2$ ($\omega <-3/2$) we see that the scale factor here
undergoes an initial period of rapid (slow)
expansion and at late times is attracted towards the solution $a(\eta)
\propto \eta$, or $a(t) \propto t^{\frac{1}{2}}$.  Similarly, $\phi$ can be seen to be
changing rapidly at early times and slowly at late times.  We
attribute these two different behaviours, at early and late times, to
periods of free scalar--field domination and radiation domination,
respectively.  In fact, setting $\eta_1=0$ in (\ref{eta2}) and
(\ref{eta1}), we remove the scalar--dominated period and gain the
power--law exact solutions $a(\eta) \propto \eta$ and
$\phi=$ constant.  These are the ``Machian'' solutions usually considered in the
literature.  Allowing $\eta_1$ to be nonzero we will have a
nonconstant $\phi$, and hence $G$, during radiation domination.  If
$\rho_{r0}=0$ is chosen then these solutions become vacuum ones that
are driven by the $\phi$ field.  For $k=0$ these solutions do not have
GR counterparts. 

We see from (\ref{phiw<}), and figure \ref{raddom2},
that for $\omega<-3/2$ the initial singularity is avoided; for a more
detailed discussion of this effect, see \cite{Bar04b}.

\subsection{Matter Domination}

When considering a matter--dominated universe we could proceed as above and
determine a set of general solutions to (\ref{Nariai1}), (\ref{Nariai2}) and (\ref{Friedmann})
that at early times are described by free scalar--field domination and at
late times by matter domination (see \cite{Gur73}), but for our
purposes this is unnecessary.  For a realistic universe we require a period of
radiation domination during which primordial nucleosynthesis can
occur.  If the scalar--field--dominated period of the Universe's history were to
impinge upon the usual matter--dominated period then we would effectively
lose the radiation--dominated era.  For this reason it is sufficient to
ignore the free scalar component of the general solution and consider
only the Machian component.  This is equivalent to imposing the
condition $\dot{\phi} a^3 \rightarrow 0$ as $a \rightarrow 0$.  With
this additional constraint the solutions to (\ref{Nariai1}), (\ref{Nariai2}) and
(\ref{Friedmann}), for $k=0$ and $p=0$, are given by \cite{Nar68}
\begin{equation}
a(t) = a_* t^{\frac{2+2\omega}{4+3\omega}} \quad \text{and} \quad
\phi(t) =\phi_* t^{\frac{2}{4+3\omega}}
\label{matBD}
\end{equation}
where 
\begin{equation*}
\frac{8 \pi}{3 \phi_*} \frac{\rho_{m0}}{a_*^3} = \frac{2 (3+2
  \omega)}{3 (4+3 \omega)}  
\end{equation*}
and $a_*$ and $\phi_*$ are constants. These solutions can be seen to approach their GR counterparts as $\omega
\rightarrow \infty$.

\subsection{Vacuum Energy Domination}

Similarly, for a vacuum ($p=-\rho$) dominated period of expansion we can
impose the condition $\dot{\phi} a^3 \rightarrow 0$ as $a
\rightarrow 0$ to get the power law exact solutions \cite{Nar68} 
\begin{equation}
\label{vacBD}
a(t) =a_{\dagger} t^{\frac{1}{2}+\omega} \quad \text{and} \quad \phi(t) =\phi_{\dagger} t^2
\end{equation}
where 
\begin{equation*}
\frac{8 \pi}{3 \phi_{\dagger}} \rho_{\Lambda 0} = \frac{1}{12}
(3+2\omega) (5+6\omega)
\end{equation*}
and $a_{\dagger}$ and $\phi_{\dagger}$ are constants.  Note that this vacuum stress does not
produce a de Sitter metric, as in GR.

\subsection{Vacuum Solutions}

In the Brans-Dicke theory it is possible to have spatially
flat and positively curved exact vacuum solutions, unlike in the
general relativistic cosmologies.

\subsubsection{Spatially flat solution}

Assuming solutions of the form $\phi \propto t^{x}$ and $a\propto t^{y}$
and setting $a(0)=0$ gives on substitution into (\ref{Nariai1}),
(\ref{Nariai2}) and (\ref{Friedmann}) the $k=0$ Brans-Dicke vacuum solutions \cite{Tup}
\begin{equation}
\label{phi_b(t)}
a(t)=t^{\frac{1}{3}(1+2(1-\sqrt{3(3+2\omega )})^{-1})}
\qquad \text{and} \qquad
\phi (t)=\phi _{b0}\left( \frac{t}{t_{0}}\right) ^{-2(1-\sqrt{3(3+2\omega )})^{-1}}.
\end{equation}

\subsubsection{Closed solution}

For the closed region we now follow the method given in  \cite{Bar92}
to find expressions for $a(\tau )$ and $\phi (\tau )$%
. We start by introducing conformal time, $\eta $, defined by $a d\eta =d\tau 
$; then equation (\ref{Nariai2}) becomes
\begin{equation*}
\phi ,_{\eta \eta }+\frac{2}{a}a,_{\eta }\phi ,_{\eta }=0.
\end{equation*}
This integrates directly to yield
\begin{equation}
\phi ,_{\eta }a^{2}=\sqrt{3}A(2\omega +3)^{-1/2}  \label{phi,n}
\end{equation}%
where $A$ is a constant. We now introduce the variable $y=\phi _{p}a^{2}$ to
write (\ref{Friedmann}) as
\begin{equation}
y,_\eta^2 = -4ky^2 + \frac{1}{3}\phi ,_\eta^2 a^4 (2 \omega +3).
\label{y,n}
\end{equation}
Equations (\ref{phi,n}) and (\ref{y,n}) then give
\begin{equation}
\frac{\phi ,_\eta}{\phi} = \sqrt{3} A y^{-1} (2 \omega +3)^{-1/2}
\qquad \text{and} \qquad
y,_\eta^2 = -4ky^2 + A^2.  \label{y,n2}
\end{equation}
The solutions of equations (\ref{y,n2}), when $k > 0$, are given by
\begin{equation}
y(\eta) = \frac{A}{2\sqrt{k}} \sin(2\sqrt{k}(\eta+B),
\end{equation}
and
\begin{equation}
\phi (\eta )=C\tan ^{\sqrt{\frac{3}{(2\omega +3)}}}(\sqrt{k}(\eta +B))
\label{phi_p(n)}
\end{equation}%
where $B$ and $C$ are arbitrary constants. We now fix the conformal time
origin by setting $B=0$, so that $y=\phi _{p}a^{2}$ gives
\begin{equation}
a(\eta) \propto \frac{\sin^{1/2}(2\sqrt{k}\eta)}{\tan^{\sqrt{\frac{3}{%
4(2\omega +3)}}}(\sqrt{k}\eta)}.  \label{S(n)}
\end{equation}

\section{More General Theories}

In order to evaluate more general scalar-tensor theories it is
convenient to work in the Einstein frame.  In this frame, substituting the FRW
line-element into the field equations gives
\begin{equation}
\label{EFriedmann1and2}
\left( \frac{\dot{\bar{a}}}{\bar{a}}\right)^2 =
\frac{8\pi}{3}(\bar{\rho}+\frac{1}{2}\dot{\psi}^2)-\frac{k}{\bar{a}^2},
\end{equation}
\begin{equation}
\label{25}
\frac{\ddot{\bar{a}}}{\bar{a}} =
-\frac{4\pi}{3}(\bar{\rho}+3\bar{p}+2\dot{\psi}^2)
\end{equation}
and
\begin{equation}
\ddot{\psi}+3\frac{\dot{\bar{a}}}{\bar{a}}\dot{\psi} = -\sqrt{4\pi}\alpha(\bar{\rho}-3\bar{p})
\label{EFriedmann3}
\end{equation}
where over--dots here denote differentiation with respect to $\bar{t}$
and $\bar{a}(\bar{t})=e^{-\Gamma} a(t)$ is the scale--factor in the Einstein
frame.  Defining $N= \ln (\bar{a}/\bar{a}_0)$, Damour and Nordtvedt \cite{Dam93} write
(\ref{EFriedmann1and2}), (\ref{25}) and (\ref{EFriedmann3}) as
\begin{equation}
\label{master}
\frac{2(1-\epsilon)}{(3-4\pi\psi'^2)}\psi''+(2-\gamma-\frac{4}{3}\epsilon)\psi'=-(4-3\gamma)\frac{\alpha}{\sqrt{4\pi}},
\end{equation}
where $\epsilon=3k/8\pi\bar{\rho}\bar{a}^2$, $\bar{p}=(\gamma-1) \bar{\rho}$ and
$'$ denotes differentiation with respect to $N$.

Now we consider a coupling parameter of the form
\begin{equation*}
\Gamma(\psi) =\Gamma(\psi_{\infty})+ \alpha_0 (\psi(\bar{t})-\psi_{\infty})+  \frac{1}{2}\beta(\psi(\bar{t})-\psi_{\infty})^2
\end{equation*}
so that
\begin{equation}
\label{alpha}
\alpha(\bar{t})=\frac{\alpha_0}{\sqrt{4\pi}} +\frac{\beta}{\sqrt{4\pi}}(\psi(\bar{t})-\psi_{\infty}).
\end{equation}
This is of the same form as chosen by Santiago, Kalligas and Wagoner
\cite{San97} and Damour and Pichon \cite{Dam}.  Santiago,
Kalligas and Wagoner arrive at this form of $\alpha$ by assuming
the evolution of the Universe has been close to the GR solutions
throughout the period from primordial nucleosynthesis to the present.  They
therefore consider themselves justified in
performing a Taylor expansion about the asymptotic value of $\psi$ and
discarding terms of second order or higher.  Damour and Pichon consider
$\Gamma$ to be a potential down which $\psi$ runs.  They assume that a particle near
a minimum of a potential experiences a generically parabolic form for
that potential. This leads them to consider a quadratic form for $\Gamma$
which gives, on differentiation, $\alpha$ as above.  These two lines
of reasoning are, of course, equivalent as the parabolic form of a
particle near its minimum of potential can be found using a Taylor
series.  Theories with this form of $\alpha$ belong to the attractor
class which approach GR at late times, if we impose the additional
condition $\alpha_0=0$.

\subsection{Radiation Domination}

For the case of a radiation--dominated flat universe, the general
solutions of (\ref{EFriedmann1and2}), (\ref{25}) and
(\ref{EFriedmann3}) with the
choice (\ref{alpha}), are, for $\omega > -3/2$:
\begin{align*}
\bar{a}^2(\eta) &= \frac{8 \pi \bar{\rho}_{r0}}{3} (\eta-\eta_2)
(\eta-\eta_2+\vert \eta_1 \vert)\\
\psi-\psi_1 &= \sqrt{\frac{3}{16 \pi}} \frac{\eta_1}{\vert \eta_1 \vert} \ln
  \left( \frac{\eta -\eta_2}{\eta -\eta_2+\vert
    \eta_1 \vert} \right),
\end{align*}
and, for $\omega < -3/2$,
\begin{align*}
\bar{a}^2(\eta) &= \frac{2 \pi \bar{\rho}_{r0}}{3} \vert \eta_1 \vert^2 +
\frac{8 \pi \bar{\rho}_{r0}}{3} (\eta-\eta_2)^2\\
\psi-\psi_1 &= -\sqrt{\frac{3 \pi}{16}} \frac{\eta_1}{\vert \eta_1 \vert} +
\sqrt{\frac{3}{4 \pi}} \frac{\eta_1}{\vert \eta_1 \vert} \tan^{-1}
\left( \frac{2(\eta -\eta_2)}{\vert \eta_1 \vert} \right).
\end{align*}
Here, $\eta_1$ determines the evolution of the scalar
field, $\psi_1$ is the value it approaches asymptotically and $\eta_2$ sets the origin of the conformal time
coordinate defined by $\bar{a}(\eta) d \eta=d \bar{t}$.
For $\omega > -3/2$, $\eta_1 \in \mathcal{R}$ such that $\psi \in
\mathcal{R}$ whilst for $\omega < -3/2$, $\eta_1 \in \mathcal{I}$ such
that $\psi \in \mathcal{I}$ and $\omega \in \mathcal{R}$.

The corresponding solutions in the Jordan frame are
\begin{equation}
\phi(\eta) = \exp \left( -\beta (\psi(\eta)-\psi_{\infty})^2 \right)
\label{STrad}
\end{equation}
and
\begin{equation}
a^2(\eta) = \frac{\bar{a}^2(\eta)}{\phi}. \label{STrad2}
\end{equation}
They exhibit the same features as their Brans-Dicke
counterparts: at early times there is a period of free--scalar--field
domination, and at late times they approach $a \propto t^{\frac{1}{2}}$ and $\phi
=$constant.

\subsection{Matter Domination}

Solutions during the matter--dominated era are difficult to find because
the energy--momentum tensor for the matter field is not
conserved.  However, it is possible to obtain an evolution
equation for $\psi$.  Using (\ref{master}), we get 
\begin{equation}
\label{matter}
\frac{2(1-\epsilon)}{(3-4\pi\psi'^2)}\psi''+(1-\frac{4}{3}\epsilon)\psi'=
-\frac{\beta}{4\pi}(\psi-\psi_{\infty}).
\end{equation}
For a flat universe $\epsilon=0$ and we can solve
(\ref{matter}) by making the simplifying assumption
$4 \pi \psi'^2\ll 3$.  This gives the solution
\begin{equation}
\label{solution}
\psi(N)-\psi_{\infty}=
A e^{-\frac{3}{4}(1+\sqrt{1-\frac{2\beta}{3\pi}})N}
+B e^{-\frac{3}{4}(1-\sqrt{1-\frac{2\beta}{3\pi}})N}
\end{equation}
where $A$ and $B$ are constants of integration.

\subsection{Vacuum Domination}

Similarly, for the case of a vacuum--dominated universe the evolution
of $\psi$ can be approximated using (\ref{master}) to obtain
\begin{equation}
\label{solution2}
\psi(N)-\psi_{\infty}=
Ce^{-\frac{3}{2}(1+\sqrt{1-\frac{2\beta}{3\pi}})N}
+De^{-\frac{3}{2}(1-\sqrt{1-\frac{2\beta}{3\pi}})N}
\end{equation}
where C and D are constants of integration.

\clearemptydoublepage
\chapter{Homogeneous and Isotropic Fourth-Order Cosmologies}
\label{Cosmology2}

\bigskip

This chapter is based on the the work of Clifton and Barrow \cite{Cli}.

\section{Freidmann-Robertson-Walker Cosmologies}

In this chapter we will again be concerned with the idealised homogeneous and
isotropic space-times described by the FRW metric:
\begin{equation}
ds^{2}=-dt^{2}+a^{2}(t)\left( \frac{dr^{2}}{(1-k r^{2})}+r^{2}d\theta
^{2}+r^{2}\sin ^{2}\theta d\phi ^{2}\right) .  \label{FRW}
\end{equation}%
Substituting this metric ansatz into the $\mathcal{L}=R^{1+\delta}$ field equations (\ref{field}), and
assuming the Universe to be filled with a perfect fluid of pressure $p$ and
density $\rho $, gives the generalised version of the Friedmann equations 
\begin{align}
(1-\delta )R^{1+\delta }+3\delta (1+\delta )R^{\delta }\left( \frac{\ddot{R}%
}{R}+3\frac{\dot{a}}{a}\frac{\dot{R}}{R}\right) -3\delta (1-\delta
^{2})R^{\delta }\frac{\dot{R}^{2}}{R^{2}}& =\frac{\chi }{2}(\rho -3p)
\label{Friedman1} \\
-3\frac{\ddot{a}}{a}(1+\delta )R^{\delta }+\frac{R^{1+\delta }}{2}+3\delta
(1+\delta )\frac{\dot{a}}{a}\frac{\dot{R}}{R}R^{\delta }& =\frac{\chi }{2}%
\rho
\label{Friedman2}
\end{align}%
where, as usual, 
\begin{equation}
R=6\frac{\ddot{a}}{a}+6\frac{\dot{a}^{2}}{a^{2}}+6\frac{k}{a^{2}}.
\label{R2}
\end{equation}%
It can be seen that in the limit $\delta \rightarrow 0$ these equations
reduce to the standard Friedmann equations of GR. A study of
the vacuum solutions to these equations for all $k$ has been made by
Schmidt (see the review \cite{schmidt}) and a qualitative study of the
perfect-fluid evolution for all $k$ has been made by Carloni et al 
\cite{Car04}. Various conclusions are also immediate from the general
analysis of $f(R)$ Lagrangians made in reference \cite{BO} by specialising them to
the case $f=R^{1+\delta }$. In what follows we shall be interested in
extracting the physically relevant aspects of the general evolution so that
observational bounds can be placed on the allowed values of $\delta $.

Assuming a perfect-fluid equation of state of the form $p=(\gamma-1) \rho $
gives the usual conservation equation $\rho \propto a^{-3 \gamma}$.
Substituting this into equations (\ref{Friedman1}) and (\ref{Friedman2}),
with $k =0$, gives the power-law exact Friedmann solution for $\gamma
\neq 0$ 
\begin{equation}
a(t)=t^{\frac{2(1+\delta )}{3 \gamma}}  \label{power}
\end{equation}%
where 
\begin{equation}
(1-2\delta )(2-3\delta \gamma-2\delta ^{2}(1+3\gamma ))=12\pi
G(1-\delta ) \gamma^{2}\rho _{c}  \label{rhoc}
\end{equation}%
and $\rho _{c}$ is the critical density of the universe.

Alternatively, if $\gamma =0$, there exists the de Sitter solution 
\begin{equation*}
a(t)=e^{nt}
\end{equation*}%
where

\begin{equation*}
3(1-2\delta )n^{2}=8\pi G(1-\delta )\rho _{c}.
\end{equation*}

The critical density (\ref{rhoc}) is shown graphically, in figure \ref%
{density8}, in terms of the density parameter $\Omega _{0}=\frac{8\pi G\rho
_{c}}{3H_{0}^{2}}$ as a function of $\delta $ for pressureless dust ($\gamma
=1$) and black-body radiation ($\gamma =4/3$). It can be seen from the graph
that the density of matter required for a flat universe is dramatically
reduced for positive $\delta $, or large negative $\delta $. In order for
the critical density to correspond to a positive matter density we require $%
\delta $ to lie in the range 
\begin{equation}
-\frac{\sqrt{73+66\omega +9\omega ^{2}}+3\gamma}{4(1+3\gamma )}<\delta <%
\frac{\sqrt{73+66\omega +9\omega ^{2}}-3\gamma}{4(1+3\gamma )}.
\label{range}
\end{equation}

\begin{figure}
\epsfig{file=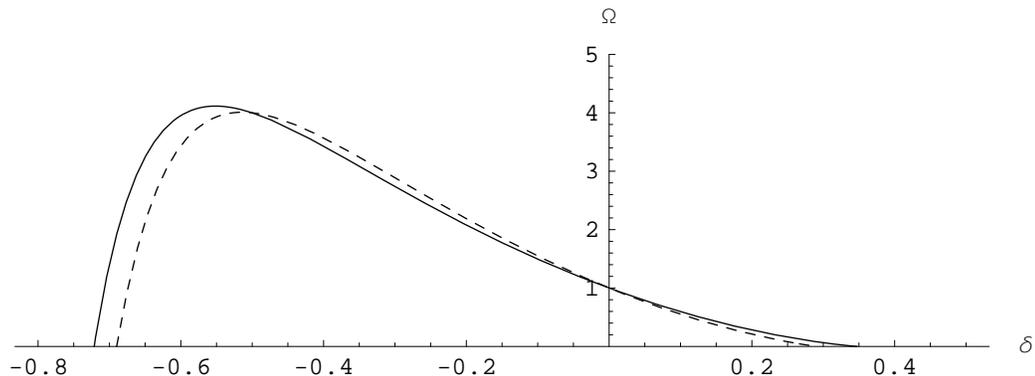,height=5cm}
\caption{{\protect{\textit{The density parameter, $\Omega_0$, as a function of $\protect%
\delta$. Solid line corresponds to pressureless dust and dashed line to
black-body radiation.}}}}
\label{density8}
\end{figure}

\section{The Dynamical Systems Approach}

The system of field equations (\ref{Friedman1}) and (\ref{Friedman2}) have been
studied previously using a dynamical systems approach by Carloni, Dunsby,
Capozziello and Troisi for general $k$ \cite{Car04}. We elaborate on
their work by studying in detail the spatially flat, $k =0$, subspace
of solutions. This allows us to draw conclusions about the asymptotic
solutions of (\ref{Friedman1}) and (\ref{Friedman2}) when $k =0$ and so
investigate the stability of the power--law exact solution (\ref{power}) and
the extent to which it can be considered an attractor solution. By
restricting to $k =0$ we avoid `instabilities' associated with the
curvature which are already present in general relativistic cosmologies.

In performing this analysis we choose to work in the conformal time
coordinate 
\begin{equation}  \label{tau8}
d \tau \equiv \sqrt{\frac{8 \pi \rho}{3 R^{\delta}}} dt.
\end{equation}

Making the definitions 
\begin{equation*}
x \equiv \frac{R^{\prime}}{R} \qquad \text{and} \qquad y \equiv \frac{%
a^{\prime}}{a},
\end{equation*}
where a prime indicates differentiation with respect to $\tau$, the field
equations (\ref{Friedman1}) and (\ref{Friedman2}) can be written as the
autonomous set of first order equations 
\begin{align}  \label{phase1}
x^{\prime}&= \frac{2- \delta (3 \gamma -2)}{\delta^2 (1+\delta)}- \frac{%
\delta x^2}{2}-\frac{(4-\delta (3 \gamma -2)) x y}{2 \delta}- \frac{2
(1-\delta) y^2}{\delta^2} \\
y^{\prime}&= -\frac{1}{\delta}+\frac{1}{2} (2+3 \delta) x y+\frac{(2+\delta
(3 \gamma -2)) y^2}{2 \delta}.
\label{phase2}
\end{align}

These coordinate definitions are closely related to those chosen by Holden
and Wands \cite{Hol98} for their phase-plane analysis of Brans-Dicke
cosmologies and allow us to proceed in a similar fashion.

\subsection{Locating the Critical Points}

The critical points at finite distances in the system of equations (\ref%
{phase1}) and (\ref{phase2}) are located at 
\begin{equation}  \label{critical1}
x_{1,2} = \pm \frac{4-3 \gamma}{\delta \sqrt{(1+ \delta) (5-3 \gamma)}}
\qquad \text{and} \qquad y_{1,2} = \pm \frac{1}{\sqrt{(1+\delta) (5-3 \gamma)%
}}
\end{equation}
and at 
\begin{equation}  \label{critical2}
x_{3,4} = \mp \frac{3 \sqrt{2 \gamma}}{\sqrt{(1+\delta) (2-3 \delta
\gamma-\delta^2 (2+6 \gamma))}} 
\end{equation}
and 
\begin{equation}
y_{3,4} = \pm 
\frac{\sqrt{2 (1+\delta)}}{\sqrt{2-3 \delta \gamma-\delta^2 (2+6 \gamma)}%
}.
\end{equation}
The critical points $(x_1, y_1)$ and $(x_2, y_2)$ do not exist for $\gamma=5/3$.

The exact form of $a(t)$ at these critical points, and the stability of
these solutions, can be easily deduced. At the critical point $(x_{i},y_{i})$
the forms of $a(\tau )$ and $R(\tau )$ are given by 
\begin{equation}
a(\tau )=a_{0}e^{y_{i}\tau }\qquad \text{and}\qquad R(\tau
)=R_{0}e^{x_{i}\tau },  \label{atau}
\end{equation}%
where $a_{0}$ and $R_{0}$ are constants of integration. In terms of $\tau $
the perfect-fluid conservation equation can be integrated to give 
\begin{equation*}
\rho =\rho _{0}e^{-3 \gamma y_{i}\tau },
\end{equation*}%
where $\rho _{0}$ is another positive constant. Substituting into the
definition of $\tau $ now gives 
\begin{equation*}
d\tau \propto e^{-\frac{3}{2} \gamma y_{i}\tau -\frac{\delta }{2}%
x_{i}\tau }dt
\end{equation*}%
or, integrating, 
\begin{equation}
t-t_{0}\propto \frac{1}{\frac{3}{2} \gamma y_{i}+\frac{\delta }{2}x_{i}}%
e^{\frac{3}{2} \gamma y_{i}\tau +\frac{\delta }{2}x_{i}\tau }.  \label{t}
\end{equation}

It can now be seen that if $3 \gamma y_i+\delta x_i >0$ then $t
\rightarrow \infty$ as $\tau \rightarrow \infty$ and $t \rightarrow t_0$ as $%
\tau \rightarrow -\infty$. Conversely, if $3 \gamma y_i+\delta x_i <0$
then $t \rightarrow t_0$ as $\tau \rightarrow \infty$ and $t \rightarrow
-\infty$ as $\tau \rightarrow -\infty$.

In terms of $t$ time the solutions corresponding to the critical points at
finite distances can now be written as 
\begin{equation*}
a(t)\propto (t-t_{0})^{\frac{2y_{i}}{3 \gamma y_{i}+\delta x_{i}}}\qquad 
\text{and}\qquad R(t)\propto (t-t_{0})^{\frac{2x_{i}}{3 \gamma y_{i}+\delta x_{i}}}.
\end{equation*}%
The critical points 1 and 2 can now been seen to correspond to $a\propto t^{%
\frac{1}{2}}$ and the points 3 and 4 correspond to (\ref{power}).

In order to analyse the behaviour of the solutions as they approach infinity
it is convenient to transform to the polar coordinates 
\begin{align*}
x &= \bar{r} \cos \phi \\
y &= \bar{r} \sin \phi.
\end{align*}
The infinite phase plane can then be compacted into a finite size by
introducing the coordinate 
\begin{equation*}
r = \frac{\bar{r}}{1+\bar{r}}.
\end{equation*}
The equations (\ref{phase1}) and (\ref{phase2}) then become 
\begin{multline}  \label{phase3}
r^{\prime}=\frac{-1}{4 \delta^2 (1+\delta)} \Biggl( 4 (1-2 r) (\delta
(1+\delta) \sin \phi-(2-\delta (3 \gamma -2)) \cos \phi) \\
- r^2 ((6+8 \delta+3 \delta^2+\delta^3-12 \delta \gamma) \cos \phi
+(1+\delta) (2-2 \delta-\delta^2-2 \delta^3) \cos 3 \phi \\
- 2 \delta (3-\delta (3 \gamma -2)+3 \cos 2 \phi) \sin \phi) \Biggr)
\end{multline}
and 
\begin{multline}  \label{phase4}
\phi^{\prime}=\frac{-1}{2 \delta^2 (1+\delta) (1-r) r} \Biggl( (2 \delta
(1+\delta) \cos \phi+2(2-\delta (3 \gamma -2)) \sin \phi) (1-2 r) \\
-\Bigl( \delta (1+\delta) \cos \phi (1-3 \cos 2 \phi)-4 \sin \phi+4
(1-\delta)^2 \sin^3 \phi \\
+2 \delta (3 \gamma -2+\delta (1+\delta) (1+2 \delta) \cos^2 \phi) \sin \phi
\Bigr) r^2 \Biggr).
\end{multline}
In the limit $r\rightarrow 1$ ($\bar{r} \rightarrow \infty$) it can be seen
that critical points at infinity satisfy 
\begin{equation*}
\sin \phi_i (\delta \cos \phi_i+\sin \phi_i) (\delta (1+2 \delta) \cos
\phi_i+2 (1-\delta) \sin \phi_i)=0
\end{equation*}
and so are located at 
\begin{align}  \label{5,6}
\phi_{5,(6)} &=0\qquad(+\pi) \\
 \label{7,8}
\phi_{7,(8)} &=\tan^{-1} (-\delta)\qquad (+\pi) \\
\phi_{9,(10)} &=\tan^{-1} \left( -\frac{\delta (1+2 \delta)}{2 (1-\delta)}%
\right) \qquad(+\pi).
 \label{9,10}
\end{align}
The form of $a(t)$ can now be calculated for each of these critical points
by proceeding as Holden and Wands \cite{Hol98}. Firstly, as $r \rightarrow 1$
equation (\ref{phase3}) approaches 
\begin{align*}
r^{\prime}&\rightarrow \frac{1}{4 \delta^2} \Biggl( \delta (1+2 \delta
(3 \gamma -2)) \sin \phi_i-3 \delta \sin 3 \phi_i \\
&\qquad -(2-\delta (2+\delta))\cos \phi_i+(2-\delta (2+\delta+2 \delta^2))
\cos 3 \phi_i\Biggr) \\
&\equiv f(\phi_i)
\end{align*}
which allows the integral 
\begin{equation*}
r-1= f(\phi_i) (\tau-\tau_0)
\end{equation*}
where the constant of integration, $\tau_0$, has been set so that $r
\rightarrow 1$ as $\tau \rightarrow \tau_0$. Now the definition of $x$
allows us to write 
\begin{equation*}
\frac{R^{\prime}}{R}=\frac{r}{(1-r)} \cos \phi_i=-\frac{f(\phi_i)
(\tau-\tau_0)+1}{f(\phi_i) (\tau-\tau_0)} \cos \phi_i \rightarrow -\frac{%
\cos \phi_i}{f(\phi_i) (\tau-\tau_0)}
\end{equation*}
as $\tau \rightarrow \tau_0$. Integrating this it can be seen that 
\begin{equation*}
R \propto \vert \tau-\tau_0 \vert^{-\frac{\cos \phi_i}{f(\phi_i)}} \qquad 
\text{as} \qquad r \rightarrow 1.
\end{equation*}
Similarly, 
\begin{equation*}
a \propto \vert \tau-\tau_0 \vert^{-\frac{\sin \phi_i}{f(\phi_i)}} \qquad 
\text{as} \qquad r \rightarrow 1.
\end{equation*}

The definition of $\tau$ (\ref{tau8}) now gives 
\begin{equation*}
d \tau \propto \vert \tau-\tau_0 \vert^{\frac{3\gamma}{2} \frac{%
\sin\phi_i}{f(\phi_i)}+\frac{\delta}{2} \frac{\cos\phi_i}{f(\phi_i)}} dt
\end{equation*}
which integrates to 
\begin{equation}  \label{tinfty}
t-t_0 \propto - \frac{f(\phi_1)}{F(\phi_i)} \vert \tau-\tau_0 \vert^{-\frac{%
F(\phi_i)}{f(\phi_i)}}
\end{equation}
where 
\begin{equation*}
F(\phi_i)=\frac{3 \gamma \sin \phi_i +\delta \cos \phi_i-2 f(\phi_i)}{2}.
\end{equation*}
The location of critical points at infinity can now be written in terms of $%
t $ as the power--law solutions 
\begin{equation}  \label{inftysol}
R(t) \propto (t-t_0)^{\frac{\cos\phi_i}{F(\phi_i)}} \qquad \text{and} \qquad
a(t) \propto (t-t_0)^{\frac{\sin\phi_i}{F(\phi_i)}}.
\end{equation}

Direct substitution of the critical points (\ref{5,6}), (\ref{7,8}) and (\ref%
{9,10}) into (\ref{inftysol}) gives 
\begin{align*}
a_{5,6}(t)& \rightarrow  \text{constant} \\
a_{7,8}(t)& \rightarrow \sqrt{t-t_{0}} \\
a_{9,10}(t)& \rightarrow (t-t_{0})^{\frac{\delta (1+2\delta )}{(1-\delta )}}
\end{align*}%
as $r\rightarrow 1$. Moreover, it can be seen from (\ref{tinfty}) that as $%
r\rightarrow 1$ and $\tau \rightarrow \tau _{0}$ so $t\rightarrow t_{0}$ as
long as $F(\phi _{i})/f(\phi _{i})<0$, as is the case for the stationary
points considered here (as long as the value of $\delta $ lies within the
range given by (\ref{range})).

The exact forms of $a(t)$ at all the critical points are summarised in the
table \ref{points1}.

\begin{table}[ht]
\par
\begin{center}
\begin{tabular}{|c|c|}
\hline
\textbf{Critical point} & \textbf{a(t)} \\ \hline
1, 2, 7 and 8 & $t^{\frac{1}{2}}$ \\ 
3 and 4 & $t^{\frac{2 (1+\delta)}{3 \gamma}}$ \\ 
5 and 6 & constant \\ 
9 and 10 & $t^{\frac{\delta (1+2 \delta)}{(1-\delta)}}$\\
\hline
\end{tabular}
\end{center}
\caption{The form of $a(t)$ at critical points of the phase plane.}
\label{points1}
\end{table}

\subsection{Stability of the Critical Points}

The stability of the critical points at finite distances can be established
by perturbing $x$ and $y$ as 
\begin{equation}
x(r)=x_{i}+u(r)\qquad \text{and}\qquad y(r)=y_{i}+v(r)  \label{lin}
\end{equation}%
and checking the sign of the eigenvalues, $\lambda _{i}$, of the linearised
equations 
\begin{equation*}
u^{\prime }=\lambda _{i}u\qquad \text{and}\qquad v^{\prime }=\lambda _{i}v.
\end{equation*}

Substituting (\ref{lin}) into equations (\ref{phase1}) and (\ref{phase2})
and linearising in $u$ and $v$ gives 
\begin{align*}
u^{\prime }& =-\left( \delta x_{i}+\frac{(4-\delta (3\gamma -2))}{2\delta }%
y_{i}\right) u-\left( \frac{(4-\delta (3\gamma -2))}{2\delta }x_{i}+4\frac{%
(1+\delta )}{\delta ^{2}}y_{i}\right) v \\
v^{\prime }& =\frac{(2+3\delta )}{2}y_{0}u+\left( \frac{(2+3\delta )}{2}%
x_{i}+\frac{(2+\delta (3\gamma -2 ))}{\delta }y_{0}\right) v.
\end{align*}%
The eigenvalues $\lambda _{i}$ are therefore the roots of the quadratic
equation 
\begin{equation*}
\lambda _{i}^{2}+B\lambda _{i}+C=0
\end{equation*}%
where 
\begin{align*}
B& =-\frac{1}{2}(2+\delta )x_{i}-\frac{3}{2}(3\gamma -2)y_{i} \\
C& =-\frac{\delta }{2}(2+3\delta )x_{i}^{2}-(2+\delta (1+3\delta
))x_{i}y_{i}+\frac{1}{2\delta }(8-8\delta -6\gamma (1+3 \delta )+9\delta
\gamma^{2})y_{i}^{2}.
\end{align*}%
If $B>0$ and $C>0$ then both values of $\lambda _{i}$ are negative, and we
have a stable critical point. If $B<0$ and $C>0$ both values of $\lambda
_{i} $ are positive, and the critical point is unstable to perturbations. $%
C<0$ gives a saddle-point.

For points $1$ (upper branch) and $2$ (lower branch) this gives 
\begin{equation*}
B= \mp \frac{(4+\delta (3 \gamma -1)-3 \gamma)}{\delta \sqrt{(1+\delta) (5-3
\gamma)}} \qquad \text{and} \qquad C= - \frac{(4+4 \delta-3 \gamma)}{\delta
(1+\delta)}
\end{equation*}
and for points $3$ (upper branch) and $4$ (lower branch) 
\begin{equation*}
B = \pm \frac{3 (2+2 \delta -\gamma (1+2 \delta))}{\sqrt{2 (1+\delta) (2-3 \delta
\gamma-2 \delta^2 (1+3 \gamma))}} \qquad \text{and} \qquad C= \frac{(4+4
\delta -3 \gamma)}{\delta (1+\delta)}.
\end{equation*}

The stability of the critical points at finite distances for a universe
filled with pressureless dust, for various different values of 
$\delta $, are given in table \ref{points2}, and for
black-body radiation are given in table \ref{points3}.
\begin{table}[ht]
\par
\begin{center}
\begin{tabular}{|c|cc|ccc|}
\hline
\textbf{Critical} & \multirow{2}{*}{\textbf{B}} & \multirow{2}{*}{\textbf{C}} & \multirow{2}{*}{$-\frac{%
\sqrt{73}+3}{16} <\delta <-\frac{1}{4}$} & \multirow{2}{*}{$-\frac{1}{4} <\delta<0$} & 
\multirow{2}{*}{$0<\delta <\frac{\sqrt{73}-3}{16} $} \\ \textbf{point} & & & & & \\ \hline
1 & $-\frac{(1+2 \delta)}{\delta \sqrt{2 (1+\delta)}}$ & $-\frac{(1+4 \delta)%
}{\delta (1+\delta)}$ & Saddle & Stable & Saddle \\ 
2 & $\frac{(1+2 \delta)}{\delta \sqrt{2 (1+\delta)}}$ & $-\frac{(1+4 \delta)%
}{\delta (1+\delta)}$ & Saddle & Unstable & Saddle \\ 
3 & $\frac{3}{\sqrt{2 (1+\delta) (2-3 \delta-8 \delta^2)}}$ & $\frac{(1+4
\delta)}{\delta (1+\delta)}$ & Stable & Saddle & Stable \\ 
4 & $-\frac{3}{\sqrt{2 (1+\delta) (2-3 \delta-8 \delta^2)}}$ & $\frac{(1+4
\delta)}{\delta (1+\delta)}$ & Unstable & Saddle & Unstable\\
\hline
\end{tabular}
\end{center}
\caption{The stability of critical points at finite distances for a
  universe filled with pressureless dust.}
\label{points2}
\end{table}

\begin{table}[ht]
\par
\begin{center}
\begin{tabular}{|c|cc|c|}
\hline
\label{w=1/3} \textbf{Critical point} & \textbf{B} & \textbf{C} & $-\frac{%
\sqrt{6}+1}{5} <\delta < \frac{\sqrt{6}-1}{5}$ \\ \hline
1 & $-\frac{3}{\sqrt{1+\delta}}$ & $-\frac{4}{(1+\delta)}$ & Saddle \\ 
2 & $\frac{3}{\sqrt{1+\delta}}$ & $-\frac{4}{(1+\delta)}$ & Saddle \\ 
3 & $\frac{(1-\delta)}{\sqrt{(1+\delta) (1-2\delta-5 \delta^2)}}$ & $\frac{4%
}{(1+\delta)}$ & Stable \\ 
4 & $-\frac{(1-\delta)}{\sqrt{(1+\delta) (1-2\delta-5 \delta^2)}}$ & $\frac{4%
}{(1+\delta)}$ & Unstable\\
\hline
\end{tabular}
\end{center}
\caption{The stability of critical points at finite distances for a
  universe filled with black-body radiation}
\label{points3}
\end{table}

Values of $\delta<-\frac{\sqrt{18+38\gamma+9\gamma^2}+3\gamma}{4 (1+3
\gamma)}$ and $\delta>\frac{\sqrt{18+38\gamma+9\gamma^2}-3\gamma}{4 (1+3
\gamma)}$ have not been considered here as they lead to negative values of $%
\rho_0$ for the solution (\ref{power}).  (The reader may note the
difference here between the range of $\delta$ for which point 3 is a
stable attractor compared with the analysis of Carloni et. al.).

Point 3 lies in the $y>0$ region and so corresponds to the expanding
power--law solution (\ref{power}). It can be seen from the table above that
this solution is stable for certain ranges of $\delta $ and a saddle-point
for others. In contrast, point 4, the contracting power--law solution (\ref%
{power}), is unstable or a saddle-point. The nature of the stability of
these points and the trajectories which are attracted towards them will be
explained further in the next subsection.

A similar analysis can be performed for the critical points at infinity.
This time only the variable $\phi $ will be perturbed as 
\begin{equation}
\phi (t)=\phi _{i}+q(t).  \label{phipert}
\end{equation}%
The conditions for stability of the critical points are now that $r^{\prime
}>0$ and the eigenvalue $\mu $ of the linearised equation $q^{\prime }=\mu q$
satisfies $\mu <0$ in the limit $r\rightarrow 1$. If both of these
conditions are satisfied then the point is a stable attractor; if only one
is satisfied the point is a saddle; and if neither are satisfied then
the point is repulsive.

Substituting (\ref{phipert}) into (\ref{phase4}) and linearising in $q(t)$
gives, in the limit $r\rightarrow 1$, 
\begin{align*}
q^{\prime }& =\frac{1}{4\delta ^{2}(1-r)}\Big( (6(1-\delta )+\delta
^{2}(1+2\delta ))\cos \phi _{i}\\
& \qquad \qquad \qquad \qquad -3(2(1-\delta )-\delta ^{2}(1+2\delta ))\cos
3\phi _{i}-3\delta \sin \phi _{i}+9\delta \sin 3\phi \Big) q \\
& \equiv \mu q.
\end{align*}%
The sign of $r^{\prime }$ in the limit $r\rightarrow 1$ can be read off from
(\ref{phase3}). The stability properties of each of the stationary points at
infinity can now be summarised in the table \ref{points4}, where
$N_1=\frac{\sqrt{18+38\gamma+9\gamma^2}+3\gamma}{4 (1+3 \gamma)}$
and $N_2=\frac{\sqrt{18+38\gamma+9\gamma^2}-3\gamma}{4 (1+3 \gamma)}$.
\begin{table}[ht]
\par
\begin{center}
\begin{tabular}{|c|cccc|}
\hline
\label{w=0infty} \textbf{Critical point} & $-N_1 <\delta <-\frac{1}{2}$ & $-%
\frac{1}{2}<\delta<0$ & $0<\delta<\frac{1}{4}$ & $\frac{1}{4}<\delta <N_2 $
\\ \hline
5 & Stable & Saddle & Unstable & Unstable \\ 
6 & Unstable & Saddle & Stable & Stable \\ 
7 & Unstable & Unstable & Saddle & Stable \\ 
8 & Stable & Stable & Saddle & Unstable \\ 
9 & Saddle & Stable & Stable & Saddle \\ 
10 & Saddle & Unstable & Unstable & Saddle\\
\hline
\end{tabular}
\end{center}
\caption{The stability of critical points at infinity.}
\label{points4}
\end{table}

\subsection{Illustration of the Phase Plane}

\begin{figure}
\epsfig{file=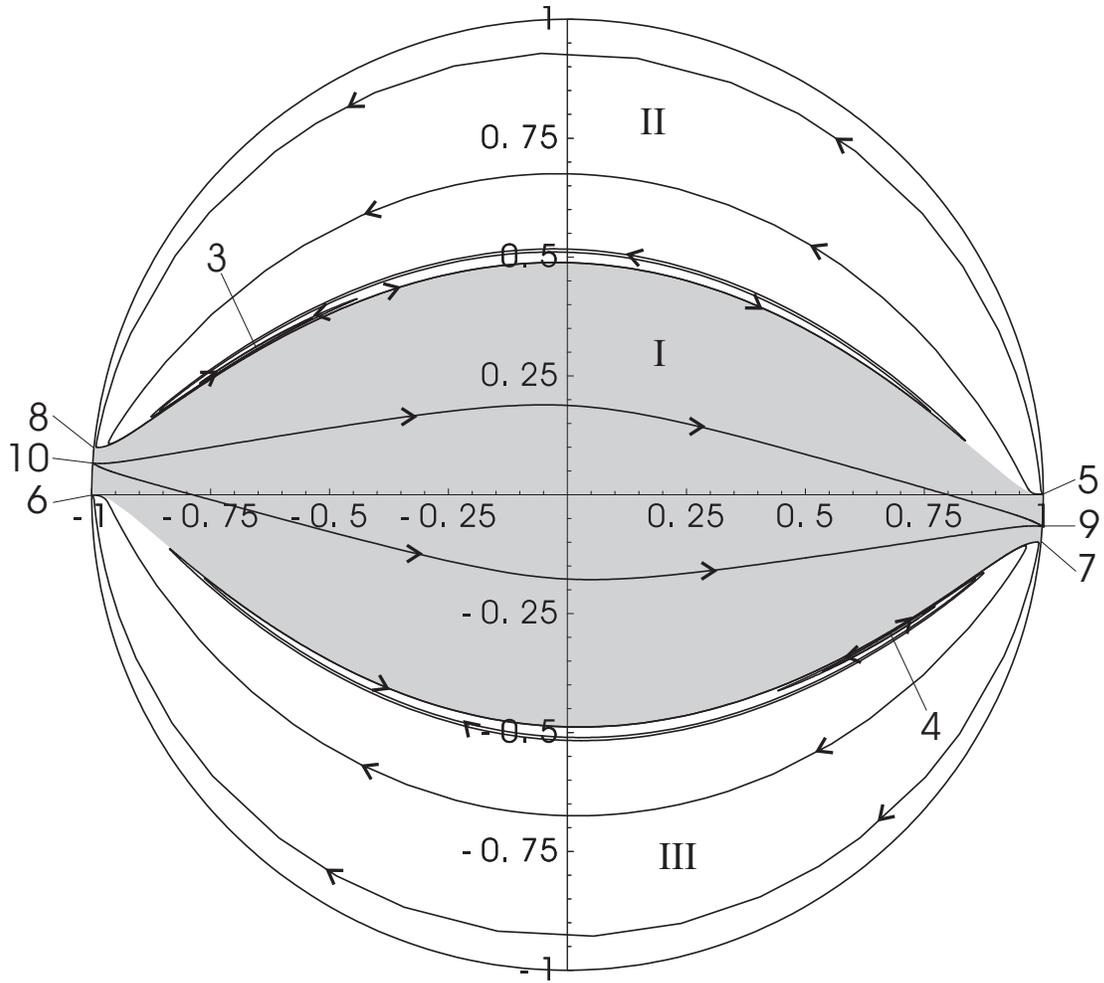,height=13cm}
\caption{\textit{Phase plane of cosmological solutions for $\protect\gamma %
=1 $ and $\protect\delta =0.1$.  $r$ is used as the radial coordinate and
$\phi$ is measured anti-clockwise from $y=0$, $x>0$.}}
\label{d0.1}
\end{figure}

Some representative illustrations of the phase plane are now presented.
Firstly, the compactified phase plane for a universe filled with
pressureless dust, $\gamma =1$, and a value of $\delta =0.1$ is shown in
figure \ref{d0.1}. 

Figure \ref{d0.1} is seen to be split into three separate regions labelled
I, II and III. The boundaries between these regions are the sub-manifolds $%
R=0$. As pointed out by Carloni et. al. the plane $R=0$ is an invariant
sub-manifold of the phase space through which trajectories cannot pass.

The equation for $R$ in a FRW universe, (\ref{R2}), can now be rewritten as 
\begin{equation}
R=\frac{16\pi \rho }{\delta R^{\delta }}((1+\delta )y^{2}+\delta (1+\delta
)xy-1).  \label{xy}
\end{equation}%
This shows that the boundary $R=0$ is given in terms of $x$ and $y$ by $%
(1+\delta )y^{2}+\delta (1+\delta )xy-1=0$ and that in region I the sign of $%
R$ must be opposite to the sign of $\delta $ in order to have a positive $%
\rho $. Similarly, in regions II and III, $R$ must have the same sign as $%
\delta $ in order to ensure a positive $\rho $.

It can be seen that regions II and III are symmetric under a rotation of $%
\pi $ and a reversal of the direction of the trajectories. As region II is
exclusively in the semi--circle $y\geqslant 0$ all trajectories confined to
this region correspond to eternally expanding (or expanding and
asymptotically static) universes. Similarly, region III is confined to the
semi--circle $y \leqslant 0$ and so all trajectories confined to this region
correspond to eternally contracting (or contracting and asymptotically
static) universes. Region I, however, spans the $y=0$ plane and so can have
trajectories which correspond to universes with both expanding and
contracting phases. In fact, it can be seen from figure \ref{d0.1} that, for 
$\delta=0.1$ all trajectories in region I are initially expanding and
eventually contracting.

It can be seen from figure \ref{d0.1} that in region I the only stable
attractors are, at early times, the expanding point 10 and at late times the
contracting point 9. (By `attractors at early times' we mean the critical
points which are approached if the trajectories are followed backwards in
time). Both of these points correspond to the solution

\begin{equation*}
a\propto t^{\delta \frac{(1+2\delta )}{(1-\delta )}}
\end{equation*}
which describes a slow evolution independent of the matter content of the
universe. Notably, region I only has stable attractor points, at both early
and late times, which are reached in a finite time as
$t\rightarrow$constant.  In
region II the only stable attractors can be seen to be the static point 5 at
some early finite time, $t_{0}$, and the expanding matter-driven expansion
described by point 3 as $t\rightarrow \infty $. Conversely, in region III
the only stable attractors are the contracting point 4 for $t\rightarrow
-\infty $ and the static point 6 for $t\rightarrow t_{0}$.

\begin{figure}
\epsfig{file=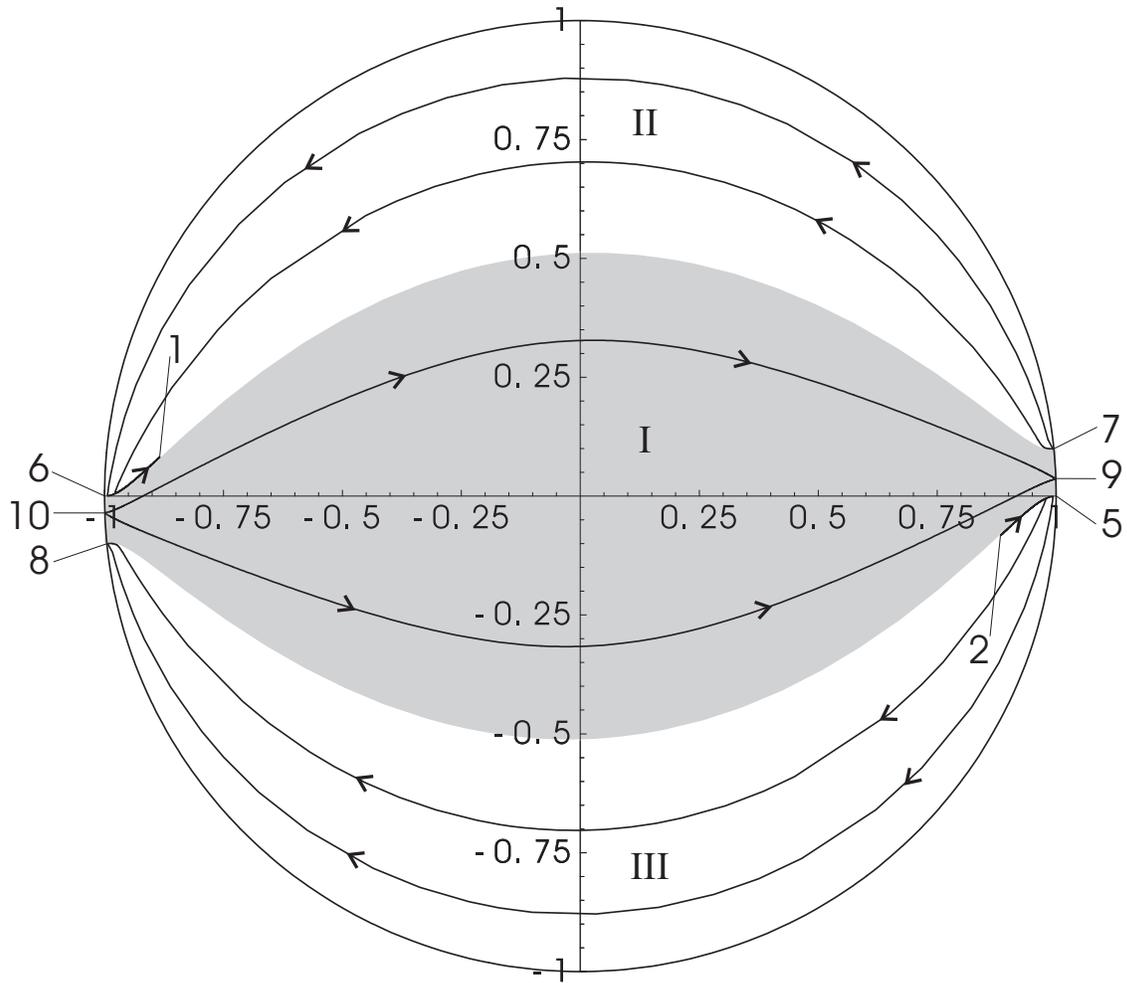,height=13cm}
\caption{\textit{Phase plane of cosmological solutions for $\protect\gamma %
=1 $ and $\protect\delta =-0.1$.  $r$ is used as the radial coordinate and
$\phi$ is measured anti-clockwise from $y=0$, $x>0$.}}
\label{d-0.1}
\end{figure}

Figure \ref{d-0.1} shows the compactified phase plane for a universe
containing pressureless dust and having $\delta =-0.1$. 
Figure \ref{d-0.1} is split into three separate regions in a similar way to
figure \ref{d0.1}, with the boundary between the regions again corresponding
to $R=0$ and is given in terms of $x$ and $y$ by (\ref{xy}). Regions II and
III again correspond to expanding and contacting solutions, respectively.
Region I, still has point 10 as the early-time attractor and point 9 as the
late-time attractor, but now has all trajectories initially contracting and
eventually expanding.  Stable attractors in Region I are still reached
in a finite time. Region II now has point
7 as an early-time stable attractor solution and point 1 as a late-time
stable attractor solution, corresponding to $a\rightarrow t^{\frac{1}{2}}$
as $t\rightarrow \infty $. Point 3, which was the stable attractor at late
times when $\delta =0.1$, is now no longer located in Region II and can
instead be located in region I where it is now a saddle-point in the phase
plane. Interestingly, the value of $\delta $ for which point 3 ceases to
behave as a stable attractor ($\delta =0$) is exactly the same value of $%
\delta $ at which the point moves from region II into region I; so as long
as point 3 can be located in region I, it is the late-time stable attractor
solution and as soon as it moves into region I it becomes a saddle-point. At
this same value of $\delta ,$ point 1 ceases to be a saddle-point and
becomes the late-time stable attractor for region II, so that region II
always has a stable late-time attractor as $t\rightarrow \infty $. Region
III behaves in a similar way to the description given for region II above,
under a rotation of $\pi $ and with the directions of the trajectories
reversed.

Phase planes diagrams for $\gamma =1$ with values of $\delta $ other than $%
0.1$ and $-0.1$, but still within the range

\begin{equation*}
-\frac{\sqrt{18+38\gamma +9\gamma ^{2}}+3\gamma}{4(1+3\gamma )}<\delta <%
\frac{\sqrt{18+38\gamma +9\gamma ^{2}}-3\gamma}{4(1+3\gamma )},
\end{equation*}%
look qualitatively similar to those above with some of the attractor
properties of the critical points being exchanged as they pass each
other.  In particular, for $\delta <-\frac{1}{4}$ point 3 returns to region II and once
again becomes the stable late-time attractor for trajectories in that
region. The points that are the stable attractors for any particular value
of $\delta $ can be read off from the tables in the previous section.

Universes filled with perfect fluid black-body radiation also retain
qualitatively similar phase-plane diagrams to the ones above; with the
notable difference that the point 3 is always located in region II and is
always the late-time stable attractor of that region. This can be seen
directly from the Ricci scalar for the solution (\ref{power}) which is given
by 
\begin{equation*}
R=\frac{3\delta (1+\delta )}{t^{2}}
\end{equation*}%
and can be seen to have the same sign as $\delta $, for $\delta >-1$, and so
is always found in region II.

For a spatially-flat, expanding FRW universe containing black-body radiation
we therefore have that (\ref{power}) is the generic attractor as $%
t\rightarrow \infty $. Similarly, for a spatially-flat, expanding,
matter-dominated FRW universe (\ref{power}) is the attractor solution as $%
t\rightarrow \infty $; except when $-\frac{1}{4}<\delta <0$, in which case
it is point 1 ($a\propto t^{\frac{1}{2}}$).

If we require a stable period of matter domination, during which $a(t)\sim
t^{\frac{2}{3}}$, we therefore have the theoretical constraint $\delta >0$
(or $\delta <-\frac{1}{4}$). Such a period is necessary for structure to
form through gravitational collapse in the post-recombination era of the
Universe's expansion.

The effect of a non--zero curvature, $k \neq 0$, on the cosmological
dynamics is similar to the general relativistic case. The role of negative
curvature ($k =-1$) can be deduced by noting that its effect is similar
to that of a fluid with $\gamma =2/3$. The solution (\ref{power}) is
unstable to any perturbation away from flatness and will diverge away
from $k =0$ as $t\rightarrow \infty $. This is usually referred to as the
`flatness problem' and can be seen to exist in this theory from the analysis
of Carloni et. al. \cite{Car04}.

\clearemptydoublepage
\chapter{Other Homogeneous Fourth-Order Cosmologies}
\label{Cosmology3}

\bigskip

We will consider in this chapter the conditions required for gravity theories that are derived
from Lagrangians that are functions of the scalar curvature and Ricci and
Riemann curvature invariants to possess solutions which are homogeneous
space-times other than the standard FRW solutions.  We will firstly
consider space-times of the G\"{o}del  \cite{God49, heck, tsag},
Einstein static \cite{ellis}, and de Sitter \cite{HE}
types.  The G\"{o}del solution admits a $G_5$ group of motions, the
Einstein static solution has a $G_7$ and the de Sitter solution has
the maximal $G_{10}$.  In the G\"{o}del case we determine the
conditions for the existence or non-existence of closed time-like
curves.  These three space times share the common property that the three
curvature invariants $X=R$, $Y=R^{a b} R_{a b}$ and $Z=R^{a b c d}
R_{a b c d}$ are constant for each of them.
This gives the fourth-order field equations a more simple form,
allowing solutions to be found more readily.

We will also consider the vacuum Kasner solutions \cite{kas} and their fluid-filled counterparts that
form type I of the Bianchi classification of three-dimensional homogeneous
spaces.  These universes have a $G_3$ group of motions acting
transitively on a three dimensional subspace and are geometrically
special.  In vacuum (or
perfect-fluid) cases they are defined by just one (or two) free constant(s)
compared to the four (eight) that specify the most general spatially
homogeneous vacuum (or perfect fluid) solutions. However, they have proved
to provide an excellent dynamical description of the evolution of the most
general models over finite time intervals.  In GR the chaotic vacuum Mixmaster
universe of Bianchi Type IX undergoes a infinite sequence of chaotic
space-time oscillations on approach to its initial or final singularities
which is well approximated by a sequence of different Kasner epochs which
form a Poincar\'{e} return mapping for the chaotic dynamical system
\cite{bkl, misner, jb1, jb2, Che83, rend}.

This chapter is based on the work of Clifton and Barrow
\cite{Clif}, \cite{further} and Barrow and Clifton \cite{Bar06}.

\section{G\"{o}del, Einstein and de-Sitter Universes}

These three homogeneous space-times, and the investigations of their stability, have played a
central role in our understanding of the dynamics of GR and
in the possible astrophysical consequences of general relativistic
cosmologies. Higher-order modifications of GR are of
importance in assessing the corrections that might be introduced in
high-curvature environments and can also be of use in
explaining the late-time acceleration of the Universe. Furthermore,
investigations of these theories allows for an evaluation of the special
nature of GR itself.  Here we study the conditions under which the G\"{o}del,
Einstein static, and de Sitter universes exist in a wide class of
non-linear gravity theories.

Considerable simplification of the fourth-order field equations (\ref{P}) occurs if
the three curvature scalars $X$, $Y$ and $Z$ are constant. In this
case the fourth-order field equations, (\ref{P}), reduce to 
\begin{equation}
P^{ab}=-\frac{1}{2}%
fg^{ab}+R^{ab}(f_{X}-2Rf_{Z})-2R^{acdb}R_{cd}(f_{Y}+4f_{Z})+\frac{1}{2}%
g^{ab}(X^{2}-4Y+Z)f_{Z}  \label{P2}
\end{equation}%
where use has been made of the identities \cite{DeW65} 
\begin{align*}
{R^{abcd}}_{;bc}& =-\square R^{ad}+{{{R^{ac}}_{;}}^{d}}_{c} \\
{{{R^{ac}}_{;}}^{d}}_{c}& =\frac{1}{2}R_{;}^{\;ad}+R^{abed}R_{be}+R_{%
\;c}^{a}R^{dc} \\
{R^{ab}}_{;ab}& =\frac{1}{2}\square R
\end{align*}
and
\begin{equation*}
2{R^{a}}_{cde}R^{bcde}+2g^{ab}Y+2RR^{ab} =\frac{1}{2}%
g^{ab}Z-4R^{acdb}R_{cd}+4{R^{a}}_{c}R^{bc}+\frac{1}{2}g^{ab}R^{2}.
\end{equation*}
Equation (\ref{P2}) is only of second order and is therefore a significant
simplification of the original system of equations.  We will
investigate the G\"{o}del, Einstein static and de Sitter universes 
solutions of these field equations.

\subsection{G\"{o}del Universes}

\label{godeluniverses}

The G\"{o}del universe \cite{God49, heck, tsag} is a rotating homogeneous space-time
given by the line-element 
\begin{align}
ds^{2}& =-(dt+C(r)d\psi )^{2}+D^{2}(r)d\psi ^{2}+dr^{2}+dz^{2}  \label{godel}
\\
& =-dt^{2}-2C(r)dtd\psi +G(r)d\psi ^{2}+dr^{2}+dz^{2} \nonumber
\end{align}
where 
\begin{align*}
C(r)& =\frac{4\Omega }{m^{2}}\sinh ^{2}\left( \frac{mr}{2}\right) \\
D(r)& =\frac{1}{m}\sinh (mr) \\
G(r)& =D^{2}(r)-C^{2}(r) \\
& =\frac{4}{m^{2}}\sinh ^{2}\left( \frac{mr}{2}\right) \left[ 1+\left( 1- 
\frac{4\Omega ^{2}}{m^{2}}\right) \sinh ^{2}\left( \frac{mr}{2}\right) %
\right]
\end{align*}
and $m$ and $\Omega $ are constants, and $\Omega$ controls the rate of
rotation. The existence of G\"{o}del universes in
one particular $f(X,Y)$ theory has been previously studied by Accioly \cite%
{Acc87} where a solution was found in vacuum. We extend this analysis to the
more general class of theories above and to universes filled with matter
fields.

The G\"{o}del universe is of particular theoretical interest as it allows
the possibility of closed time-like curves, and hence time-travel \cite{tsag,
bt2}. The condition required to avoid the existence of closed time-like
curves is \cite{reb, bd} 
\begin{equation*}
G(r)>0\qquad \text{for}\qquad r>0
\end{equation*}%
or 
\begin{equation}
m^{2}\geqslant 4\Omega ^{2}.  \label{ineq}
\end{equation}%
By investigating the existence of G\"{o}del universes in this general class
of gravity theories we will also be able to determine those theories in
which the condition (\ref{ineq}) is satisfied. Consequently, we
will be able to determine those theories in which time travel is a
theoretical possibility.

It is convenient to work in the non-holonomic basis defined by 
\begin{align*}
{e^{(0)}}_0 &= {e^{(2)}}_2 = {e^{(3)}}_3 = 1 \\
{e^{(1)}}_1 &= D(r) \\
{e^{(0)}}_1 &= C(r).
\end{align*}
The G\"{o}del line-element (\ref{godel}) then becomes 
\begin{equation*}
ds^2=-(\theta^{(0)})^2+(\theta^{(1)})^2+(\theta^{(2)})^2+(\theta^{(3)})^2
\end{equation*}
where the one-forms $\theta^{(A)}$ are given by $\theta^{(A)}=e^{(A)}_{\quad
a}dx^a$ (capital Latin letters denote tetrad indices and lower case Latin
letters denote space-time indices). The inverses of $e^{(A)}_{\quad a}$ can
be calculated from the relations $e^{(A)}_{\quad a} e^{a}_{\; (B)}={%
\delta^{(A)}}_{(B)}$ and the non-zero elements of the Riemann tensor in this
basis are then given by 
\begin{align*}
R_{(0)(1)(0)(1)} &= R_{(0)(2)(0)(2)} = \Omega^2 \\
R_{(1)(2)(1)(2)} &= 3 \Omega^2 - m^2.
\end{align*}

The perfect-fluid energy-momentum tensor is defined in the usual way with
respect to the comoving 4-velocity $U^{a}=(-1,0,0,0)$ and its covariant
counterpart $U_{a}=(1,0,C(r),0)$ such that its non-zero components in the
non-holonomic basis are given by 
\begin{equation*}
T_{(0)(0)}=\rho \qquad \text{and}\qquad T_{(1)(1)}=T_{(2)(2)}=T_{(3)(3)}=p.
\end{equation*}

The field equations, (\ref{fequations}), for this space-time can then be
manipulated into the form 
\begin{align}
&\Lambda -\frac{\chi }{2}p =\frac{1}{2}f  \label{1t} \\
&0 =(2\Omega ^{2}-m^{2})(f_{X}-4[\Omega
^{2}-m^{2}]f_{Z})+2(f_{Y}+4f_{Z})(8\Omega ^{4}-5\Omega ^{2}m^{2}+m^{4})
\label{1a} \\
&\frac{\chi }{2}(\rho +p) =2\Omega ^{2}(f_{X}-4[\Omega
^{2}-m^{2}]f_{Z})+4(f_{Y}+4f_{Z})\Omega ^{2}(2\Omega ^{2}-m^{2}).  \label{1b}
\end{align}%
Solving the field equations has now been reduced to solving these three
algebraic relations for some specified $f(X,Y,Z)$.

\subsubsection{$\boldsymbol{1 \; \; \; \; f=f(X)}$}

For $\rho +p\neq 0$ we see from (\ref{1b}) that $f_{X}\neq 0$. From equation
(\ref{1a}) it can then be seen that $m^{2}=2\Omega ^{2}$, as in general
relativity. Therefore for any theory of the type $f=f(X)$ the inequality (%
\ref{ineq}) is not satisfied and closed time-like curves exist, when $\rho
+p\neq 0$.

It now remains to investigate the case $\rho +p=0$. It can immediately be
seen from (\ref{1b}) that we must have $f_{X}=0$ in order for a solution to
exist. This sets the relation between $m$ and $\Omega $ and automatically
satisfies equation (\ref{1a}). The required value of $\Lambda $ can then be
read off from equation (\ref{1t}). The condition $f_{X}=0$ will now be
investigated for a variety of specific theories.

\subsubsection{$\boldsymbol{1a \; \; \; \; f=f(X)=X+\alpha X^2}$}

Theories of this kind have been much studied \cite{BO, ker} as they have a
number of interesting properties, not least of which is that they display
divergences which are normalisable at the one loop level \cite{Ste77}.

In these theories, the condition $f_{X}=0$ is equivalent to 
\begin{equation*}
m^{2}=\Omega ^{2}+\frac{1}{4\alpha }.
\end{equation*}%
The condition under which this theory then satisfies the inequality (\ref%
{ineq}), and hence does not admit closed time-like curves, is $\Omega^2
\leqslant (12 \alpha)^{-1}$. Therefore, for any given theory of this kind,
defined only by a choice of the constant $\alpha $, there is a range of
values of $\Omega $ for which closed time-like curves do not exist, when $%
\alpha>0$ and $\rho +p=0$. However, when $\alpha<0$ this condition is never
satisfied and closed time-like curves exist for all values of $\Omega$.

\subsubsection{$\boldsymbol{1b \; \; \; \; f=f(X)=X+\frac{\protect\alpha^2}{X}}$}

Theories of this type have generated considerable interest as they introduce
cosmological effects at late times, when $R$ is small, which may be able to
mimic the effects of dark energy on the Hubble flow \cite{Car05,Noj03, Nav}.
The square in the factor $\alpha ^{2}$ is introduced here as these theories
require a positive value for this coefficient in order for the field
equations to have a solution.

When $f_{X}=0$ we have the two possible relations 
\begin{equation*}
m^{2}=\Omega ^{2}\pm \frac{\alpha }{2}
\end{equation*}%
where $\alpha $ is the positive real root of $\alpha ^{2}$. The upper branch
of this solution then allows the condition (\ref{ineq}) to be satisfied if $%
6\Omega ^{2}\leqslant \alpha $; or, for the lower branch, if $6\Omega
^{2}\leqslant -\alpha $. For any particular value of $\alpha >0$, the upper
branch always admits a range of $\Omega $ for which closed time-like curves
do not exist. For the lower branch, however, the inequality (\ref{ineq}) is
never satisfied and closed time-like curves are permitted for any value of $%
\Omega $.

\subsubsection{$\boldsymbol{1c \; \; \; \; f=f(X)=\vert X \vert^{1+\protect\delta}}$}

This scale-invariant class of theories is of interest as its particularly
simple form allows a number of physically relevant exact solutions to be
found \cite{Cli05,Bar06,schm,Car04}. In order for solutions of the G\"{o}del
type to exist in these theories we must impose upon $\delta $ the constraint 
$\delta \geqslant 0$.

The condition $f_{X}=0$ now gives 
\begin{equation*}
m^{2}=\Omega ^{2}.
\end{equation*}%
Evidently, the condition (\ref{ineq}) is never satisfied in this case: G%
\"{o}del solutions always exist and closed time-like curves are permitted
for any value of $\Omega $.

\subsubsection{$\boldsymbol{2 \; \; \; \; f=f(X,Y,Z)}$}

In this general case, equations (\ref{1a}) and (\ref{1b}) can be manipulated
into the form 
\begin{align}
f_{X}-4(\Omega ^{2}-m^{2})f_{Z}& =\frac{(8\Omega ^{4}-5\Omega
^{2}m^{2}+m^{4})}{4\Omega ^{2}(4\Omega ^{2}-m^{2})}\chi (\rho +p)  \label{4t}
\\
f_{Y}+4f_{Z}& =-\frac{(2\Omega ^{2}-m^{2})}{8\Omega ^{2}(4\Omega ^{2}-m^{2})}%
\chi (\rho +p)  \label{5}
\end{align}%
when $m^{2}\neq 4\Omega ^{2}$. The special case $m^{2}=4\Omega ^{2}$ gives 
\begin{align}
\rho +p& =0 \\
f_{X} &= 4 \Omega^2 (f_Y+f_Z).
\end{align}

First we consider $\rho +p>0$. It is clear that $m^{2}=4\Omega ^{2}$ is not
a solution in this case. Equations (\ref{4t}) and (\ref{5}) show that in
order to satisfy the inequality (\ref{ineq}), and avoid the existence of
closed time-like curves, the following two conditions must be satisfied
simultaneously 
\begin{align}
f_{X}+4(m^{2}-\Omega ^{2})f_{Z}& <0  \label{G2} \\
f_{Y}+4f_{Z}& <0.  \label{G3}
\end{align}%
For $\rho +p<0$ the inequalities must be reversed in these two equations. It
is now clear that if it is possible to construct a theory which has a
solution of the G\"{o}del type without closed time-like curves for a
non-zero $\rho +p$ of a given sign, then this theory cannot ensure the
non-existence of closed time-like curves for the opposite sign of $\rho +p$.

It remains to investigate the case $\rho+p=0$. In order to have a solution
in this case we require that the two equations 
\begin{align}
f_X-4 (\Omega^2-m^2) f_Z &= 0 \\
f_Y+4 f_Z &= 0
\end{align}
are simultaneously satisfied, for $m^2 \neq 4 \Omega^2$, or 
\begin{equation}  \label{G4}
f_X = 4 \Omega^2 (f_Y+f_Z)
\end{equation}
for $m^2 = 4 \Omega^2$. Any theory which satisfies (\ref{G4}), therefore,
does not allow the existence of closed time-like curves when $\rho+p=0$.

We now illustrate these considerations by example.

\subsubsection{$\boldsymbol{2a \; \; \; \; f=X+ \protect\alpha X^2 +\protect\beta Y +\protect\gamma Z}$}

Using the identity \cite{DeW65} 
\begin{equation*}
\frac{\delta }{\delta g_{ab}}\left( \sqrt{g}(X^{2}-4Y+Z)\right) =\text{pure
divergence}
\end{equation*}%
it is possible to rewrite $f$ as $f=X+\hat{\alpha}X^{2}+\hat{\beta}Y$ where $%
\hat{\alpha}=\alpha -\gamma $ and $\hat{\beta}=\beta +4\gamma $. This is the
theory considered by Accioly \cite{Acc87}.

For the case $\rho +p>0,$ the inequalities (\ref{G2}) and (\ref{G3}) then
become 
\begin{equation}
\hat{\beta}<0\qquad \text{and}\qquad \hat{\alpha}>\frac{1}{2|R|}.
\label{ineq2}
\end{equation}%
The first of these inequalities can be satisfied trivially. The second can
be only be satisfied for all $\Omega $ if $|R|$ has some non-zero minimum
value. From the field equations (\ref{1a}) and (\ref{1b}) we can obtain 
\begin{equation*}
(2\Omega ^{2}-m^{2})(1+4\hat{\alpha}[\Omega ^{2}-m^{2}])=-2\hat{\beta}%
(8\Omega ^{4}-5\Omega ^{2}m^{2}+m^{4})
\end{equation*}%
which gives two solutions for $m$ as a function of $\Omega $, $\hat{\alpha}$
and $\hat{\beta}$. Substituting either of these values of $m$ into $%
R=2(\Omega ^{2}-m^{2})$ then gives that $R\rightarrow 0$ as $\Omega
^{2}\rightarrow \frac{1}{8 \vert \hat{\beta} \vert}$. It can now be seen that the second
equality in (\ref{ineq2}) cannot be satisfied for all $\Omega $ if $\hat{%
\alpha}$ is finite. It is, therefore, not possible to construct a theory of
this type which excludes the possibility of closed time-like curves for all $%
\Omega $, when $\rho +p\neq 0$.

For $\rho+p<0$ the inequalities in (\ref{ineq2}) must be reversed. In this
case $\hat{\beta}>0$ and $\hat{\alpha}>0$ allows a range of $\Omega$ in
which closed time-like curves are not permitted and $\hat{\beta}>0$ and $%
\hat{\alpha}<0$ does not allow closed time-like curves for any values of $%
\Omega$.

The case $\rho +p=0$ was studied by Accioly \cite{Acc87}, who found
that equation (\ref{G4}) is satisfied if 
\begin{equation*}
4\Omega ^{2}=m^{2}=\frac{1}{(3\hat{\alpha}+\hat{\beta})}.
\end{equation*}%
Hence closed time-like curves do not exist in G\"{o}del universes for these
theories, when $\rho +p=0$.

\subsubsection{$\boldsymbol{2b \; \; \; \; f=\protect\alpha X+\protect\beta Y}$}

This class of theories is introduced as an example which excludes the
possibility of closed time-like curves for $\rho+p>0$. The inequalities (\ref%
{G2}) and (\ref{G3}) in this case are 
\begin{equation*}
\alpha <0 \qquad \text{and} \qquad \beta <0,
\end{equation*}
which can be trivially satisfied. For $\rho+p=0$ the only non-trivial
solution is given by $m^2 = 4 \Omega^2$, the value of $\Omega$ then being
given in terms of $\alpha$ and $\beta$ by (\ref{G4}).

This example shows explicitly that it is possible to construct a theory in
which closed time-like curves do not occur in G\"{o}del universes when $\rho
+p\geqslant 0$ (though we do not consider it as physically viable as $\alpha 
$ is required to have the `wrong' sign \cite{BO, Ruz}).

Table \ref{table} summarises the results found in this section.
\begin{table}[ht]
\par
\begin{center}
\begin{tabular}{|c|c|c|c|c|}
\hline
\multirow{2}{*}{$f(X,Y,Z)$} & Additional & \multicolumn{3}{|c|}{Closed
time-like curves exist when} \\ \cline{3-5}
& conditions & $\; \rho+p > 0 \;$ & $\; \rho+p < 0 \;$ & $\; \rho+p = 0 \;$
\\ \hline
$X+ \alpha X^2$ & $\alpha >0$ & $\checkmark$ & $\checkmark$ & $\checkmark$/$%
\times$ \\ 
& $\alpha <0$ & $\checkmark$ & $\checkmark$ & $\checkmark$ \\ 
&  &  &  &  \\ 
$X+\frac{\alpha^2}{X} $ & $+$ve branch & $\checkmark$ & $\checkmark$ & $%
\checkmark$/$\times$ \\ 
& $-$ve branch & $\checkmark$ & $\checkmark$ & $\checkmark$ \\ 
&  &  &  &  \\ 
$\vert X \vert^{1+\delta}$ & $\delta >0$ & $\checkmark$ & $\checkmark$ & $%
\checkmark$ \\ 
&  &  &  &  \\ 
$\; X+\alpha X^2 + \beta Y + \gamma Z \;$ & $\; \alpha-\gamma >0$, $\; \beta
+ 4 \gamma<0 \;$ & $\checkmark$/$\times$ & $\checkmark$ & $\times$ \\ 
& $\alpha-\gamma <0$, $\; \beta + 4 \gamma<0$ & $\checkmark$ & $\checkmark$
& $\times$ \\ 
& $\alpha-\gamma >0$, $\; \beta + 4 \gamma>0$ & $\checkmark$ & $\checkmark$/$%
\times$ & $\times$ \\ 
& $\alpha-\gamma <0$, $\; \beta + 4 \gamma>0$ & $\checkmark$ & $\times$ & $%
\times$ \\ 
&  &  &  &  \\ 
$\alpha X + \beta Y$ & $\alpha >0$, $\; \beta <0$ & $\checkmark$ & $%
\checkmark$ & $\times$ \\ 
& $\alpha >0$, $\; \beta >0$ & $\checkmark$ & $\times$ & $\times$ \\ 
& $\alpha <0$, $\; \beta <0$ & $\times$ & $\checkmark$ & $\times$ \\ 
& $\alpha <0$, $\; \beta >0$ & $\checkmark$ & $\checkmark$ & $\times$ \\ 
\hline
\end{tabular}%
\end{center}
\caption{A summary of the conditions under which closed time-like curves can
exist in G\"{o}del universes, for various different gravitational theories,
defined by $f(X,Y,Z)$: $\checkmark $ denotes their existence for all values
of $\Omega $ and $\times $ denotes that they are not allowed for any value
of $\Omega $. The symbol $\checkmark $/$\times $ means that closed time-like
curves are allowed to exist for some restricted range of $\Omega $ only; the
ranges are given in the main text.}
\label{table}
\end{table}

\subsection{Einstein Static Universes}
\label{einsteinstatic}

The Einstein static universe is a homogeneous and isotropic space-time with
line-element 
\begin{equation*}
ds^{2}=-dt^{2}+\frac{dr^{2}}{(1-k r^{2})}+r^{2}(d\theta ^{2}+\sin
^{2}\theta d\phi ^{2}).
\end{equation*}%
Here, $k$ parametrizes the curvature of the space-like slices
orthogonal to $t$ and the scale-factor has been rescaled to 1. For a
universe containing pressureless dust the field equations (\ref{fequations})
can now be written as 
\begin{align} 
\chi \rho & =4 k f_{X}+16 k^{2} (f_{Y}+f_{Z})
\label{static1} \\
 \label{static2}
\Lambda & =\frac{1}{2}f-2 k f_{X}-8 k^{2} (f_{Y}+f_{Z}).
\end{align}%
It can now be seen immediately that solutions exist for an Einstein static
universe for any $f(X,Y,Z)$ that is differentiable in all its arguments. The
corresponding values of $\rho $ and $\Lambda $ are simply read off from
equations (\ref{static1}) and (\ref{static2}). It remains to be studied
under what circumstances these solutions are stable. The investigation by
Barrow, Ellis, Maartens, and Tsagas \cite{ellis} shows that this is an issue
that depends upon the material content and the equation of state of matter
in a delicate fashion. In GR, there is first-order stability
against density perturbations when the sound speed exceeds a critical value (%
$1/\sqrt{5}$ of the speed of light) because the Jeans length exceeds the
size of the universe \cite{gibb, ellis}. However, there is instability
against homogeneous gravitational-wave modes of Bianchi IX type \cite{ellis}%
. In general, we expect a universe with compact space sections and
Killing vectors to display linearisation instability \cite{BT}.

\subsection{de Sitter Universes}
\label{desitter}

The line element for the maximally symmetric de Sitter vacuum universe can
be written as 
\begin{equation*}
ds^{2}=-dt^{2}+e^{2\sqrt{\frac{\Lambda }{3}}t}(dr^{2}+r^{2}(d\theta
^{2}+\sin ^{2}\theta d\phi ^{2}))
\end{equation*}%
where $\Lambda $ is the cosmological constant in GR. For de Sitter space-time all
components of the Riemann and Ricci tensors can be written in terms of the
Ricci scalar using the equations 
\begin{align*}
R_{abcd}& =-\frac{1}{12}R(g_{ad}g_{bc}-g_{ac}g_{bd}) \\
R_{ab}& =\frac{1}{4}Rg_{ab}
\end{align*}%
where the Ricci scalar is $R=4\Lambda $. The field equations (\ref%
{fequations}) can now be reduced to the single equation 
\begin{equation}
\frac{1}{2}f-\Lambda =\Lambda f_{X}+2\Lambda ^{2}f_{Y}+\frac{4}{3}\Lambda
^{2}f_{Z}.
\end{equation}%
This equation must be satisfied by $f$ if the de Sitter universe is to be a
solution in any particular gravitational theory. This problem reduces to
that studied by Barrow and Ottewill \cite{BO} in the case $f=f(X)$. This
result establishes the situations where inflation of de Sitter sort can
arise from higher-order corrections to the gravitational Lagrangian. In the
case where $f=f(X)$ alone it is appreciated that the resulting theory is
conformally equivalent to general relativity plus a scalar field \cite{cot,
maeda} with an asymmetric exponential potential and so either de Sitter or
power-law inflation is possible. Our results establish when de Sitter
inflation is possible in situations where the other invariants, $Y$ and $Z$
contribute to the Lagrangian, and the conformal equivalence with GR is
broken. The stability of the de Sitter solutions in these
theories will be studied elsewhere.

\section{Anisotropic Cosmologies}

The way in which the Kasner metric has played a central role in the
elucidation of the existence and structure of anisotropic cosmological
models and their singularities in GR makes it the obvious
starting point for an extension of that understanding to cosmological
solutions of higher-order gravity theories.  We will determine the
conditions for the existence of Kasner-like solutions and find their exact forms.
In some cases these solutions are required to be isotropic and correspond to
exact FRW vacuum solutions with zero spatial
curvature. By studying gravity theories whose Lagrangians are derived from
arbitrary powers of the curvature invariants we are able to find simple
exact solutions. Past studies have usually focussed on the addition of
higher-order curvature terms to the Einstein-Hilbert Lagrangian. This
results in enormous algebraic complexity and exact solutions cannot be
found. The results presented here provide a tractable route into
understanding the behaviours of anisotropic cosmological models in
situations where the higher-order curvature corrections are expected to
dominate. The solutions we present are vacuum solutions but we provide a
simple analysis which determines when the introduction of perfect fluids
with non-comoving velocities has a negligible effect on the cosmological
evolution at early times.

\subsection{Field Equations}

We will be looking for spatially homogeneous, vacuum solutions of Bianchi
type I, described by the line--element 
\begin{equation}
ds^{2}=-dt^{2}+\sum_{i=1}^{3}t^{2p_{i}}dx_{i}^{2}  \label{kasner}
\end{equation}%
where $p_{1}$, $p_{2}$ and $p_{3}$ are constants to be determined. For the
special case $p_{1}=p_{2}=p_{3}$ these solutions correspond to spatially
flat FRW metrics, which will be found to exist for various higher-order
theories both in vacuum (where none exist in GR) and in the
presence of a perfect fluid.

We can determine the number of independently arbitrary functions of three
space variables that will characterise the \textit{general} vacuum solution
of the field equations on a Cauchy surface of constant time in these
higher-order theories. The field equations are in general fourth order in
time; so if we choose a synchronous reference system then we need 6
functions each for the symmetric $3\times 3$ tensors $g_{\alpha \beta }$, $%
\dot{g}_{\alpha \beta }$, $\dddot{g}_{\alpha \beta }$ and $\ddddot{g}%
_{\alpha \beta }$. This gives 24 functions, but they may be reduced to 20 by
using the 4 constraint equations, and finally again to 16 by using the 4
coordinate covariances. If a perfect fluid were included as a matter source
the final number would rise to 20 due to the inclusion of the density and 3
non-comoving velocity components in the initial data count. We note that in
GR the general vacuum solution is prescribed by 3 arbitrary
functions of 3 space variables, and that in GR the Kasner vacuum
solution is prescribed by one free constant, as the three $p_{i}$ satisfy
two algebraic constraints.

\subsection{Exact Solutions}

For a realistic theory we should expect the dominant term of the analytic
function $f(X,Y,Z)$ to be of the Einstein-Hilbert form in the Newtonian
limit. However, there is no reason to expect such a term to dominate in the
high curvature limit.  In fact, this is the limit in which quantum
corrections should become dominant. We therefore allow the dominant term in
a power series expansion of $f(X,Y,Z)$ to be of the form $R^{n}$, $%
(R_{ab}R^{ab})^{n}$ or $(R_{abcd}R^{abcd})^{n}$ on approach to the
singularity, where $n$ is a constant. These three different cases will be
investigated separately below.

\subsubsection{$\boldsymbol{f=R^n}$}

Substituting $f=f(X)=R^{n}$ into the fourth-order field equations (\ref{P})
together with the line-element (\ref{kasner}) gives the two independent
algebraic constraints, 
\[
(2n^{2}-4n+3)P+(n-2)Q-3(n-1)(2n-1)=0 
\]
and 
\[
2(n^{2}-1)P+(n-2)(P^{2}-Q)-3(n-1)^{2}(2n-1)=0, 
\]
where we have defined 
\begin{equation}  \label{Xn1}
P \equiv \sum_{i} p_{i} \qquad \text{and} \qquad Q \equiv \sum_{i}
p_{i}^{2}.
\end{equation}
The constraint equations have two classes of solution. The first is given by 
\begin{eqnarray*}
&&P=\frac{3(n-1)(2n-1)}{(2-n)} \\
&&Q=\frac{P^{2}}{3}
\end{eqnarray*}%
which is only solved by the isotropic solution $p_{1}=p_{2}=p_{3}=P/3$. This
is the zero-curvature isotropic vacuum universe found by Bleyer and Schmidt 
\cite{schmidt, schm}. It can be seen that these isotropic cosmologies
are valid for all $n\neq 2$ and correspond to expanding universes for $n<1/2$
or $1<n<2$ and to collapsing universes for $1/2<n<1$ or $2<n$.

The second class of solutions to (\ref{Xn1}) is given by 
\begin{eqnarray}
&&P=2n-1  \label{Xnkasner} \\
&&Q=(2n-3)(1-2n).  \nonumber
\end{eqnarray}%
This class represents a generalisation of the Kasner metric of
GR. It can be shown that it only exists for $n$ in the
range $1/2\leqslant n\leqslant 5/4$ and that the values of the constants $%
p_{i}$ must then lie within the ranges 
\begin{eqnarray*}
2n-1-2\sqrt{(2n-1)(5-4n)} &\leqslant &3p_{1}\leqslant 2n-1-\sqrt{(2n-1)(5-4n)%
}, \\
2n-1-\sqrt{(2n-1)(5-4n)} &\leqslant &3p_{2}\leqslant 2n-1+\sqrt{(2n-1)(5-4n)}%
, \\
2n-1+\sqrt{(2n-1)(5-4n)} &\leqslant &3p_{3}\leqslant 2n-1+2\sqrt{(2n-1)(5-4n)%
},
\end{eqnarray*}%
where it has been assumed without loss of generality that $p_{1}\leqslant
p_{2}\leqslant p_{3}$. These ranges are shown on figure 5.1 and can be
read off by drawing a horizontal line of constant $n$ on the plot. The four
points at which the horizontal line crosses the two closed curves gives the
four boundary values for the allowed ranges of the constants $p_{i}$. It can
be seen that the points at which the curves cross the abscissa, which
corresponds to $n=1$, give the boundary values $-1/3\leqslant p_{1}\leqslant
0\leqslant p_{2}\leqslant 2/3\leqslant p_{3}\leqslant 1$, in agreement with
the Kasner solution of GR. For $n>1/2$ these solutions
correspond to expanding universes with a curvature singularity at $t=0$. For 
$n=1/2$ the only solution is Minkowski space. 
\begin{figure}[ht]
\label{figure51}
\centerline{\epsfig{file=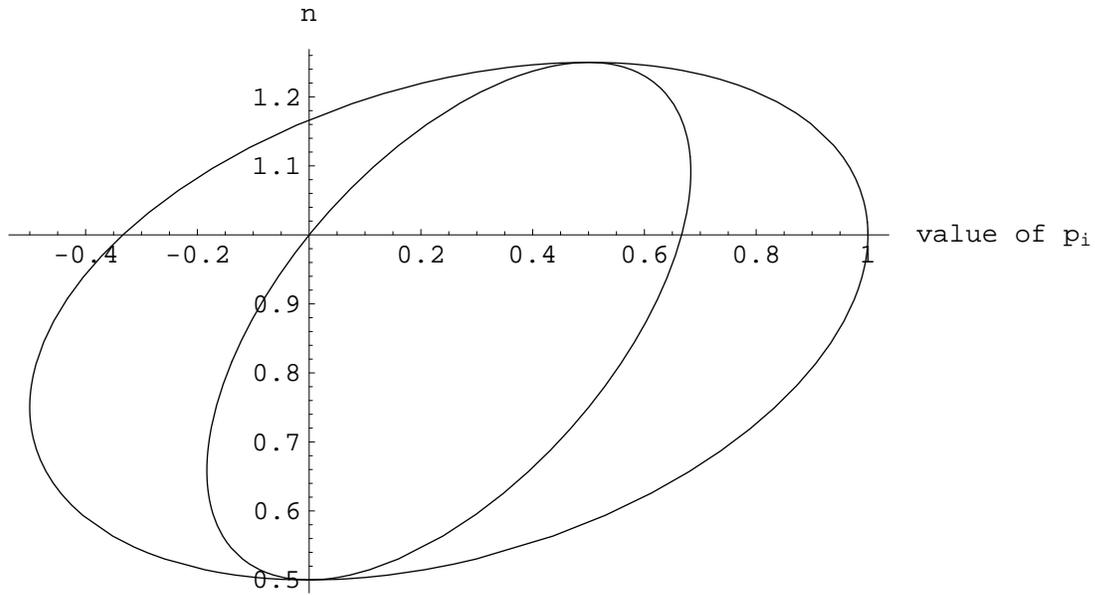,height=9cm}}
\caption{\textit{For the solution (\protect\ref{Xnkasner}), the intervals in
which the Kasner indices $p_{i}$ are allowed to lie can be read off from
this graph. For any value of $n$ in the range $\frac{1}{2}\leqslant
n\leqslant \frac{5}{4}$ a horizontal line is drawn; the boundaries of the
intervals in which the $p_{i}$ lie are then given by the four points at
which the horizontal line crosses the two closed curves. For $n=1$ these
boundaries can be seen to be $-\frac{1}{3}$, $0$, $\frac{2}{3}$ and $1$, as
expected for the Kasner solution of general relativity.}}
\end{figure}

For a universe filled with a perfect fluid having an equation of state $%
p=(\gamma -1)\rho $, $\gamma $ constant, relating the fluid pressure $p$ to
its density $\rho ,$ the field equations (\ref{P}) have the
isotropic solution 
\begin{equation}
p_{1}=p_{2}=p_{3}=\frac{2n}{3\gamma },
\end{equation}%
for $\gamma \neq 0$. This reduces to the spatially-flat FRW solution of
GR in the limit $n\rightarrow 1$. For $n>0,$ these isotropic
cosmologies are expanding, and for $n<0$ they are contracting, with $n=0$
giving Minkowski space. The stability and observational consequences of
cosmologies of this type were investigated in \cite{Cli05, Car04}, where
primordial nucleosynthesis and the microwave background were used to impose
observational constraints on the admissible values of $n$.

\subsubsection{$\boldsymbol{f=(R_{ab} R^{a b})^n}$}

Substituting $f=f(Y)=(R_{ab}R^{ab})^{n}$ into the fourth-order field
equations (\ref{P}) along with the metric ansatz (\ref{kasner}) gives the two
independent equations, 
\begin{multline}
Y^{n-1} \Bigl( (P^{2}+Q-4PQ+P^{2}Q+Q^{2})-  \label{Yn1} \\
\;2(3P^{2}+P^{3}+3Q-9PQ+2Q^{2})n+2(4P^{2}+4Q-8PQ)\Bigr)=0  
\end{multline}%
and 
\begin{multline}
Y^{n-1}\Bigl( (24P-2P^{2}-2P^{3}-30Q+10PQ)n+(8P^{2}-32P+24Q)n^{2}
\label{Yn2} \\
\;\;\;+(P^{2}-4P+2P^{3}+9Q-10PQ-P^{2}Q+3Q^{2})\Bigr)=0,
\end{multline}%
where $P$ and $Q$ are defined as before. Equations (\ref{Yn1}) and (\ref{Yn2}%
) have five classes of solutions.

The first of these classes is given by 
\[
P=Q=0.
\]%
This is only satisfied for $p_{1}=p_{2}=p_{3}=0$, which is
Minkowski space. It can be seen from (\ref{Yn1}) and (\ref{Yn2}) that as
this solution corresponds to $Y=0$ it only exists for $n\geqslant 0$ (for $%
n<0$ the premultiplicative factor in (\ref{Yn1}) and (\ref{Yn2}) causes the
left-hand side of those equations to diverge).

The second class of solutions is given by 
\[
P=Q=1.
\]%
This is just the Kasner solution of GR, for which the values
of the constants $p_{i}$ are constrained to lie within the ranges $%
-1/3\leqslant p_{1}\leqslant 0\leqslant p_{2}\leqslant 2/3\leqslant
p_{3}\leqslant 1$. Again, this solution corresponds to $Y=0$ and so is only
valid for $n\geqslant 0$.

The third class of solutions is given by 
\begin{eqnarray*}
&&P=\frac{3(1-3n+4n^{2})\pm \sqrt{3(-1+10n-5n^{2}-40n^{3}+48n^{4})}}{2(1-n)}
\\
&&Q=\frac{P^{2}}{3}.
\end{eqnarray*}%
The only solution belonging to this class corresponds to an isotropic and
spatially flat vacuum FRW cosmology. The values of the constants $p_{i}$ are
all equal to $P/3$ in this case and the solution is valid for all $n\neq 1$.

The fourth class of solutions is given by 
\begin{eqnarray}
&&P=(1-2n)^{2}  \label{Ynkasner} \\
&&Q=1-8n+16n^{2}-8n^{3}.  \nonumber
\end{eqnarray}%
The solutions belonging to this class are anisotropic cosmologies which
reduce to the standard Kasner form, $P=Q=1$, in the limit $n\rightarrow 1$.
For this class of solutions, the values of the constants $p_{i}$ are in
general constrained to lie within the ranges 
\begin{eqnarray*}
(1-2n)^{2}-2A &\leqslant &3p_{1}\leqslant (1-2n)^{2}-A \\
(1-2n)^{2}-A &\leqslant &3p_{2}\leqslant (1-2n)^{2}+A \\
(1-2n)^{2}+A &\leqslant &3p_{3}\leqslant (1-2n)^{2}+2A
\end{eqnarray*}%
where $A=\sqrt{(1-2n)(1-6n+4n^{3})}$ and the $p_{i}$ have again been ordered
so that $p_{1}\leqslant p_{2}\leqslant p_{3}$. These boundaries are shown in
figure 5.2 which can be read in the same way as figure 5.1, by
taking a horizontal line of constant $n$ and noting the four points at which
that line intersects the curves. These intercepts determine the allowed
intervals for the values of the $p_{i}$. 
\begin{figure}[ht]
\label{5.2} \centerline{\epsfig{file=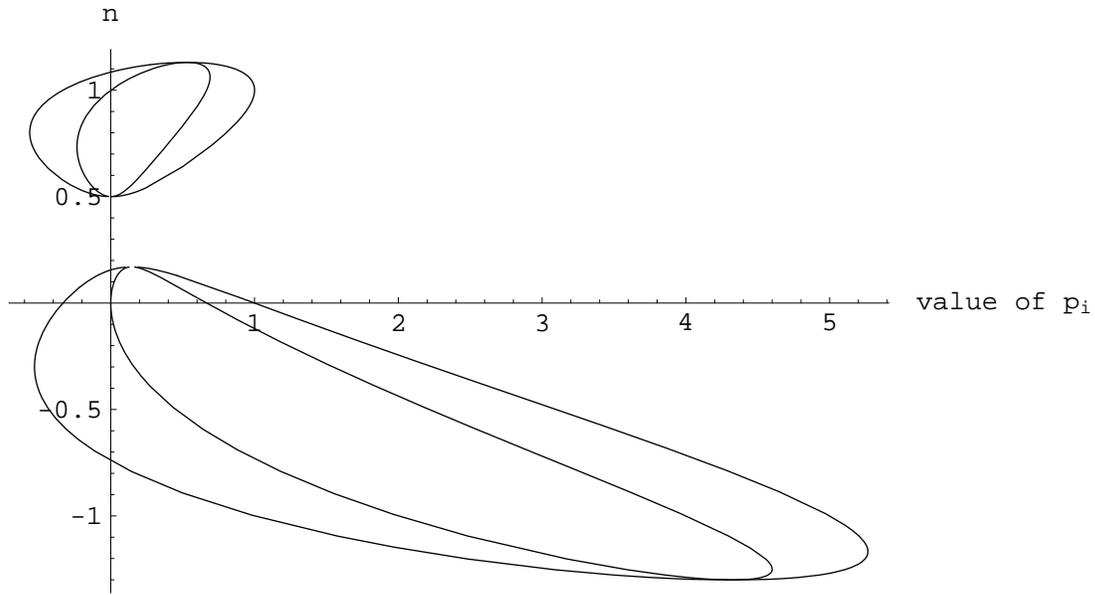,height=9cm}}
\caption{\textit{For the solution (\protect\ref{Ynkasner}) the intervals in
which the Kasner indices $p_{i}$ are constrained to lie can be read off this
graph, as with figure \protect5.1. A horizontal line of constant $n$ is
drawn; the boundaries of the intervals in which the $p_{i}$ lie are then
given by the four points at which the horizontal line crosses the curves.
For $n=1$ these boundaries can be seen to be $-\frac{1}{3}$, $0$, $\frac{2}{3%
}$ and $1$, as expected for the Kasner solution of general relativity.}}
\end{figure}

For real-valued $p_{i}$, the value of $n$ must lie either in the range $%
n_{1}\leqslant n\leqslant n_{2}$ or in the range $1/2\leqslant n\leqslant
n_{3}$, where $n_{1}$, $n_{2}$ and $n_{3}$ are the roots of the cubic
polynomial $1-6n+4n^{3}=0$ and are chosen such that $n_{1}<n_{2}<n_{3}$.
These generalisations of the Kasner metric always correspond to expanding
cosmologies, independent of the value of $n$.

The fifth class of solutions to (\ref{Yn1}) and (\ref{Yn2}) is given by 
\begin{eqnarray*}
&&P=4n-1 \\
&&Q=-3+12n-8n^{2}\pm 2(1-2n)\sqrt{2(1-4n+2n^{2})}.
\end{eqnarray*}%
This class describes complex-valued $p_{i}$ for all values of $n$ (except
for $n=1/4$, for which this class of solutions reduces to the first class)
and is therefore of limited interest.

The isotropic solution for a universe filled with a fluid with equation of
state $p=(\gamma -1)\rho $ is given by the choice 
\[
p_{1}=p_{2}=p_{3}=\frac{4n}{3\gamma }.
\]%
This solution reduces to the spatially-flat FRW cosmology of GR in the
limit $n\rightarrow 1/2$ and corresponds to an expanding
universe for $n>0$ and to a contracting universe for $n<0$.

\subsubsection{$\boldsymbol{f=(R_{a b c d} R^{a b c d})^n}$}

Substituting $f=f(Z)=(R_{abcd}R^{abcd})^{n}$ and (\ref{kasner}) into the
field equations (\ref{P}) gives the two independent equations 
\begin{multline}
Z^{n-1}\Biggl( \Bigl(\frac{P^{2}Q}{2}-\frac{P^{4}}{12}-Q-\frac{3Q^{2}}{4}%
+2S-\frac{2PS}{3}\Bigr) +8(S-Q)n^{2} \label{Zn1} \\
+\Bigl(\frac{P^{4}}{3}+6Q+2PQ-2P^{2}Q+3Q^{2}-10S+\frac{2PS}{3}\Bigr)n \Biggr)=0
\end{multline}%
and 
\begin{multline}
Z^{n-1}\Biggl( \Bigl(\frac{P^{4}}{4}-P-\frac{3P^{2}}{2}-\frac{P^{3}}{2}+%
\frac{13Q}{2}+\frac{7PQ}{2}-\frac{3P^{2}Q}{2}+\frac{9Q^{2}}{4}-8S\Bigr)
\label{Zn2} \\
+(6P+4P^{2}-16Q-2PQ+8S)n +8(Q-P)n^{2}\Biggr)=0
\end{multline}%
where $P$ and $Q$ are defined as before and we have also now defined

\[
S \equiv \sum_{i}p_{i}^{3}. 
\]
Equations (\ref{Zn1}) and (\ref{Zn2}) have four different classes of
solution.

The first class of solutions is given by 
\[
P=Q=S=0. 
\]%
The only solution that belongs to this class is Minkowski space. As $%
R_{abcd}R^{abcd}=0$ for Minkowski space it can only be a solution for $%
n\geqslant 0$, due to the premultiplier in equations (\ref{Zn1}) and (\ref%
{Zn2}).

The second class of solutions is given by 
\[
P=Q=S=1. 
\]%
The only solution that belongs to this class corresponds to $p_{1}=p_{2}=0$
and $p_{3}=1$. Making the coordinate transformations $\bar{z}=t\sinh z$ and $%
\bar{t}=t\cosh z$ \cite{landau} allows the line-element (\ref{kasner}) to
then be written in the form 
\[
ds^{2}=-d\bar{t}^{2}+dx^{2}+dy^{2}+d\bar{z}^{2} 
\]%
which is clearly Minkowski space again. This is only a solution for $%
n\geqslant 0$, as with the first class of solutions.

The third class of solutions to (\ref{Zn1}) and (\ref{Zn2}) is given by 
\begin{eqnarray*}
&&P=\frac{3(1-2n+4n^{2}\pm \sqrt{-1+10n-16n^{2}+16n^{4}}}{2(1-n)}, \\
&&Q=\frac{P^{2}}{3}, \\
&&S=\frac{P^{3}}{9},
\end{eqnarray*}%
which only has the isotropic and spatially flat vacuum FRW cosmology as a
solution, where $p_{1}=p_{2}=p_{3}=P/3$. This is a solution for all $n\neq 1$%
.

The fourth and last class of solutions is given by 
\begin{eqnarray*}
&&P=4n-1 \\
&&Q=\frac{1}{3}\{16n^{2}-8n-1\pm 4\sqrt{2[n(1-2n)+S(1-n)]}\}.
\end{eqnarray*}%
This class of solutions corresponds to anisotropic metrics with $Z=0$ and is
unusual in that it cannot be expressed in the form $P=$ constant and $Q=$
constant. This feature means that the standard picture of a plane
intersecting an ellipsoid is no longer valid for this class. Solutions
in this class do not appear to have any range of $n$ for which the constants 
$p_{i}$ take real values, and so are of limited physical interest.

The isotropic solution for a universe filled with a fluid with equation of
state $p=(\gamma -1)\rho $ is given by 
\[
p_{1}=p_{2}=p_{3}=\frac{4n}{3\gamma }.
\]%
As before, this reduces to the spatially flat FRW solution of GR in
the limit $n\rightarrow 1/2$ and corresponds to an expanding
universe for $n>0$ and to a contracting universe for $n<0$.

\subsection{Investigation of the Effects of Matter}

We have identified above two generalisations of the Kasner metric of
GR which are solutions to the scale-invariant theories of gravity we
are investigating. The first of these is a solution to the theory defined by 
$\mathcal{L}=R^{n}$; the second is a solution to the theory $\mathcal{L}=(R_{ab}R^{ab})^{n}$.
These are the first exact anisotropic solutions to be found for higher-order
gravity theories. In this section we will investigate some of the properties
of these cosmologies.

The behaviour of these solutions is of particular interest when considering
the Bianchi type VIII or type IX `mixmaster' cosmologies. The field
equations for these cosmologies can be cast into the form of the equations
of motion of a particle moving inside an exponentially steep triangular
potential well with open channels in the corners \cite{misner}. The three
steep-sided walls are created by the 3-curvature anisotropies. In the region
where the potential is negligible (far from the walls) the behaviour of the
solutions approaches that of the Kasner metric. As the exponentially steep
potential wall is approached the universe `particle' is reflected and
re-enters a Kasner--like regime with the Kasner indices $p_{i}$
systematically changed to some new values by the rule of reflection from the
potential wall. In general relativistic cosmologies of Bianchi types VIII
and IX this process is repeated an infinite number of times as the
singularity is approached \cite{bkl, misner, jb1, jb2, Che83, rend} so long
as matter obeys an equation of state with $p<\rho $. After reflection from
the potential wall the Kasner index that was previously negative is permuted
to a positive value and the lowest-valued positive Kasner index is permuted
to a negative value. This is repeated \textit{ad infinitum} in the general
relativistic cosmology as one of the Kasner indices must be kept negative
while the other two are positive. In the generalisations of the Kasner
metric presented above this is no longer the case; for some solutions it is
possible for all of the Kasner indices to be positive. If the Kasner indices
are being permuted in a chaotic fashion for long enough then, eventually,
they will end up in such a configuration. Once this occurs the oscillations
will end, as all spatial dimensions will be contracting, and the singularity
will be reached monotonically without further oscillation of the scale factors 
\cite{BK, Bar78}.

From figure 5.1 it can be seen that for solution (\ref{Xnkasner}) all
Kasner indices can be made positive for values of $1<n<5/4$ in the $%
\mathcal{L}=R^{n}$ theories. Similarly, from figure 5.2 it can be seen
that for the solution (\ref{Ynkasner}), with $n_{1}<n<n_{2}$ or $1<n<n_{3},$
all the Kasner indices can be made positive in the theory $\mathcal{L}%
=(R_{ab}R^{ab})^{n}$ (where the $n_{i}$ are defined as before). For both of
these solutions, with these ranges of $n$, there is therefore no infinite
series of chaotic mixmaster oscillations as the singularity is approached.
However, the solutions to both of these theories are still expected to
exhibit an infinite number of chaotic oscillations if $1/2<n<1$, as at least
one of the Kasner indices must be kept negative whilst another is kept
positive. A separate detailed analysis is require to determine where there
are differences between this behaviour and the chaotic oscillations
found in GR.

It now remains to show that the vacuum solutions found above are the
asymptotic attractor solutions in the presence of non-comoving matter
motions as $t\rightarrow 0$. This analysis will follow closely that of \cite%
{Bar06} and \cite{landau}. We aim to show that the fluid stresses diverge
more slowly than the vacuum terms as $t\rightarrow 0$ and so produce
negligible metric perturbations to an anisotropic Kasner like universe. If this
is the case, then the vacuum solutions above can indeed be considered as the
asymptotic attractor solutions even in the presence of matter. Matter will
just be carried along by the expansion and behave like a test fluid.

We now consider a perfect fluid with equation of state $p=(\gamma -1)\rho $
where $1\leqslant \gamma <2$ and the non-zero 4-velocity components,
$U^{i}$, are normalised so that $U^{i}U_{i}=-1$. The conservation equations ${T^{ab}}%
_{;b}=0$ on the metric background (\ref{kasner}) can then be written in the
form \cite{landau,Bar78,jb2} 
\begin{eqnarray*}
&&\frac{\partial }{\partial x^{i}}(t^{p}\rho ^{\frac{1}{\gamma }}U^{i})=0 \\
&&\gamma \rho U^{k}(U_{i,k}-\frac{1}{2}U^{l}g_{kl,i})=-\frac{1}{3}\rho
_{,i}-(\gamma -1)U_{i}U^{k}\rho _{,k}.
\end{eqnarray*}%
Neglecting spatial derivatives with respect to time derivatives, these
equations integrate to 
\begin{eqnarray*}
t^{p}U_{0}\rho ^{\frac{1}{\gamma }} &=& \text{constant} \\
U_{\alpha }\rho ^{\frac{(\gamma -1)}{\gamma }} &=& \text{constant}.
\end{eqnarray*}%
The neglecting of spatial derivatives means that these equations are valid
on scales larger than the particle horizon as in the velocity-dominated
approximation (although we have not so far restricted the fluid motions to
be non-relativistic) \cite{landau, els, tomita}.

From the second of the integrals above, it can be seen that all the
covariant components of the spatial 3-velocity, $U_{\alpha }$, are
approximately equal. This is not true of the contravariant components as the
Kasner indices, $p_{i}$, in the metric elements used to raise indices are
not equal in these solutions. The contravariant component that diverges the
fastest, and dominates the others in the $t\rightarrow 0$
velocity-dominated limit, is
therefore $U^{3}=U_{3}t^{2p_{3}}$, as $p_{3}$ is the largest of the $p_{i}$.
If the 4-velocity is normalised, so that $U_{a}U^{a}=-1$, and the
contravariant 3-velocity component $U^{3}$ diverges the fastest as $%
t\rightarrow 0,$ then we must have $U_{0}U^{0}\sim
U_{3}U^{3}=(U_{3})^{2}t^{-2p_{3}}$ in that limit. The integrated
conservation equations above can now be solved approximately in this limit
to give 
\begin{eqnarray*}
&&\rho \sim t^{-\gamma (p_{1}+p_{2})/(2-\gamma )} \\
&&U_{\alpha }\sim t^{(p_{1}+p_{2})(\gamma -1)/(2-\gamma )}
\end{eqnarray*}%
as $t\rightarrow 0$.

It is now possible to calculate the leading-order contributions to the
energy-momentum tensor, ${T^{a}}_b=(\rho +p)U^{a}U_{b}+p {\delta^{a}}_b$:
\begin{eqnarray*}
T_{0}^{0} \sim& \rho U^{0}U_{0} \sim& t^{-P-p_{3}}\\
T_{1}^{1} \sim& \rho \qquad     \sim& t^{-\gamma (P-p_{3})/(2-\gamma )}\\
T_{2}^{2} \sim& \rho U^{2}U_{2} \sim& t^{-2p_{2}-P+p_{3}}\\
T_{3}^{3} \sim& \rho U^{3}U_{3} \sim& t^{-P-p_{3}}
\end{eqnarray*}
for $\gamma<4/3$.  The component which diverges the fastest here as $t\rightarrow 0$ is $%
T_{3}^{3}\sim t^{-P-p_{3}}$, for general $\gamma $. For the case $\gamma =%
\frac{4}{3}$ where the 4-velocity of the fluid is comoving, $U_{i}=\delta
_{i}^{0}$, all the components of $T_{b}^{a}$ diverge as $\rho \sim
t^{-2P+2p_{3}}$.

For the generalisation of the Kasner metric in the $\mathcal{L}=R^{n}$
theory, (\ref{Xnkasner}), the vacuum terms diverge as $t^{-2n}$. We
therefore require that $2n>P+p_{3}$, or $p_{3}<1$, in order for the vacuum
terms to dominate over the matter terms for a non-comoving perfect fluid,
with general $\gamma $, in the limit $t\rightarrow 0$. For the $\gamma =%
\frac{4}{3},$ comoving perfect fluid ($U_{i}=\delta _{i}^{0}$), the
condition for vacuum domination is $2n>2P-2p_{3}$, or $p_{3}>n-1$. Both of
these conditions are ensured by the boundary values on $p_{3}$ given above.
In the velocity-dominated limit, the vacuum solutions given here are
therefore the appropriate asymptotic solutions on approach to the
singularity.

For the generalised Kasner metric in the $\mathcal{L}=(R_{ab}R^{ab})^{n}$
theory, (\ref{Ynkasner}), the vacuum terms diverge as $t^{-4n}$. For vacuum
domination as $t\rightarrow 0$ in this solution we therefore require $%
4n>P+p_{3}$, or $p_{3}<-1+8n-4n^{2}$, for the general non-comoving $\gamma $
fluid; and $4n>2P-2p_{3}$, or $p_{3}>1-6n+4n^{2}$, for the $\gamma =4/3$
comoving fluid. From figure 5.2 it can be seen that solutions of this
class can be in one of two regions, $n_{1}\leqslant n\leqslant n_{2}$ or $%
1/2\leqslant n\leqslant n_{3}$. The validity of the vacuum solution as $%
t\rightarrow 0$ is different in these two different regions. For the region $%
1/2\leqslant n\leqslant n_{3}$ the boundary conditions on the index $p_{3}$
mean that the conditions above are automatically satisfied. For the region $%
n_{1}\leqslant n\leqslant n_{2}$ these inequalities are not always
satisfied. For general $\gamma ,$ the condition for vacuum domination is
only met if $n$ lies in the narrow range $1/6<n<n_{2}$; for any other value
of $n$ in this region the index $p_{3}$ can be such that the fluid
diverges faster than the vacuum terms. Similarly, for the $\gamma =4/3$
comoving fluid the condition for vacuum domination is only satisfied if $n$
lies in the narrow range $n_{4}<n<n_{2}$, where $n_{4}$ is the real root of
the cubic

\[
4n^{3}-12n^{2}+10n-1=0. 
\]

To understand the evolution of an anisotropic solution of the form (\ref%
{kasner}) in the theory $\mathcal{L}=(R_{ab}R^{ab})^{n}$, where $%
n_{1}\leqslant n\leqslant n_{2}$, it is therefore necessary to take into
account the relativistic motions of any fluid that is present (except in the
narrow ranges of $n$ identified above). This range of $n$ can, however, be
regarded as not belonging to physically interesting theories on other
grounds. In order to agree with weak-field experiments it is necessary for a
gravitational theory to contain a term which approximates the
Einstein-Hilbert action in the weak-field limit (see ref. \cite{Cli05}). Any
theory containing such a limit must therefore have a term in its Lagrangian
that diverges as $t^{-2}$. In considering the behaviour of alternative
theories close to the singularity it is necessary for the extra terms in the
Lagrangian to diverge faster than this if they are to be influential as $%
t\rightarrow 0$. For the theory $\mathcal{L}=(R_{ab}R^{ab})^{n}$ this
requires $n>1/2$. For $n<1/2$ such a term will diverge slower than the
Einstein-Hilbert term which will then dominate and display the standard
Kasner behaviour of GR. For the anisotropic solution (\ref%
{Ynkasner}) we are therefore only interested in the region lying in the
range $1/2\leqslant n\leqslant n_{3}$, which has vacuum terms that dominate
the fluid stresses in the velocity-dominated approximation.

It remains to investigate the effects of fluids with equations of
state, $\gamma >4/3$. For sufficiently stiff fluids the velocity-dominated
approximation is not valid and $U^{\alpha }U_{\alpha }\rightarrow 0$ as $%
t\rightarrow 0$ due to the very high inertia of the fluid producing a slow
down under contraction \cite{jb5}. In such a limit $U_{0}\rightarrow 1$
and the conservation equations can be solved to give 
\begin{eqnarray*}
&&\rho \sim t^{-\gamma P} \\
&&U_{\alpha }\sim t^{(\gamma -1)P}.
\end{eqnarray*}%
In this approximation we can write $U_{\alpha }U^{\alpha }\sim
(U_{3})^{2}t^{-2p_{3}}\sim t^{2(\gamma -1)P-2p_{3}}$. It can now be seen
that this behaviour occurs when 
\begin{equation}
p_{3_{\ }}\leqslant (\gamma -1)P.  \label{stiff}
\end{equation}%
In this limit it is only required that $\rho $ diverges more slowly than the
vacuum terms, as $\rho U_{\alpha }U^{\alpha }<<\rho $ when $t\rightarrow 0$.

For the solution (\ref{Xnkasner}) to the theory $\mathcal{L}=R^n$ it can be
seen from the condition (\ref{stiff}) that it is necessary for $\gamma
\geqslant 4/3$ in order for the velocity-dominated approximation to break
down (this is derived using the upper limit on $n$ and the lower limit on $%
p_3$). The condition that the fluid stresses diverge more slowly than the
vacuum terms is now 
\begin{equation}
\gamma< \frac{2 n}{ (2 n-1)}
\end{equation}
where it has been assumed $n>1/2$. A sufficient condition to satisfy this
for all allowed values of $n$ is $\gamma < 5/3$. For fluids with a stiffer
equation of state, $5/3 \leqslant \gamma < 2$, the vacuum terms dominate
over the fluid motions provided that $n<1$. For $n<1$, however, the
Einstein-Hilbert term will be the leading one in the gravitational action,
as discussed above.

Similarly, for the solution (\ref{Ynkasner}) to the theory $\mathcal{L}%
=(R_{ab}R^{ab})^{n}$ it can be seen from the condition (\ref{stiff}) that it
is again necessary to have $\gamma \geqslant 4/3$ in order for the
velocity-dominated approximation to break down. In the region $1/2\leqslant
n\leqslant n_{3}$ the condition for the vacuum solution to be unperturbed by
the stiff fluid is always satisfied for any fluid with equation of state $%
\gamma <2$. However, in the region $n_{1}<n<n_{2}$ the condition for the
vacuum to diverge faster than the energy density of the stiff fluid is never
satisfied for a fluid which satisfies the necessary condition for the
velocity-dominated approximation to break down (except in the narrow range $%
(7-\sqrt{33})/8<n<n_{2}$ where there are some values of $\gamma $ for which
the vacuum terms can diverge fastest).

\subsection{Discussion}

We have investigated some anisotropic cosmological solutions to higher-order
Lagrangian theories of gravity. Whist the standard general relativistic
theory appears to be consistent
with all weak-field tests, there is less reason to think that it should be
valid in high curvature regimes, such as in the vicinity of a possible
initial cosmological curvature singularity. In fact, it is in the
high-curvature limit that quantum effects should become important and we
should expect to see deviations from the standard theory. Without knowing
the exact form of such deviations, we have approached this problem by
considering a general class of theories that can be derived from an
arbitrary analytic function of the three curvature invariants $R$, $R_{ab}R^{ab}
$ and $R_{abcd}R^{abcd}$. Expanding this function as a power series in these
variables we then expect the dominant term in the Lagrangian to be of the
form $R^{n}$, $(R_{ab}R^{ab})^{n}$ or $(R_{abcd}R^{abcd})^{n}$ as the
singularity is approached. We have found all of the solutions to these
theories that can be expressed in terms of the Bianchi type I line--element (%
\ref{kasner}) in vacuum. These solutions provide simple testing grounds for
the exploration of quantum cosmological effects in higher-order gravity
theory. We have also found the homogeneous, isotropic and spatially flat
solutions to these theories in the presence of a perfect fluid.

Exact vacuum anisotropic solutions of the form (\ref{kasner}) were found for
the theories $\mathcal{L}=R^{n}$ and $\mathcal{L}=(R_{ab}R^{ab})^{n}$,
whilst it was shown that no such solutions exist for theories defined by $%
\mathcal{L}=(R_{abcd}R^{abcd})^{n}$. The properties of these solutions, with
respect to their relation to the more general Bianchi type VIII and IX
cosmological behaviour, has been investigated. We have argued that for all
of the physically relevant new solutions of these theories, the Universe
will not experience an infinite number of mixmaster oscillations as the
singularity is approached. The extent to which these vacuum solutions can be
considered as realistic in the presence of a perfect fluid has also been
investigated. It has been shown in the velocity-dominated approximation that
all the anisotropic vacuum solutions found for plausible theories are valid
in the limit $t\rightarrow 0$. The case of stiff fluids that do not satisfy
the velocity-dominated approximation as $t\rightarrow 0$ has also been
investigated. For the solutions to the theories $\mathcal{L}=R^{n}$ it was
found that for a fluid with equation of state $\gamma <5/3$ the vacuum
solutions are good approximations in the vicinity of the singularity. For
the theories $\mathcal{L}=(R_{ab}R^{ab})^{n}$ it was found that the vacuum
solutions are good approximations for all $\gamma <2$.

\clearemptydoublepage
\chapter{Static Solutions}
\label{Static}

\bigskip

In this chapter the vacuum static and spherically symmetric solutions to the
Brans-Dicke and $R^n$ theories will be presented.  For these theories
there is no equivalent of `Birkhoff's theorem' (in fact, in the next
chapter we will present solutions which explicitly violate Birkhoff's
theorem).  The condition of being static is therefore added as an
extra condition, in contrast to the usual treatment in GR.

This chapter is based on the work of Clifton and Barrow \cite{Cli}.

\section{Scalar-Tensor Theories}

The static and spherically symmetric solutions of the Brans-Dicke
theory were found long ago by Brans himself \cite{Bra62}.  These solutions have
been investigated and rediscovered in many papers since, and so
we will only present a very brief summary of the situation here, for
completeness of this study.  In isotropic coordinates
\begin{equation*}
ds^2 = -e^{2 \alpha} dt^2+e^{2 \beta} (dr^2+r^2 d\Omega^2)
\end{equation*}
where $\alpha=\alpha(r)$ and $\beta=\beta(r)$, there are four solutions of
the Brans-Dicke field equations (\ref{STfields}) and (\ref{STfields2})
which are given by \cite{Bra62}
\begin{flalign*}
\textbf{I} & & e^{\alpha} &= e^{\alpha_0}
\left[\frac{1-\frac{B}{r}}{1+\frac{B}{r}}\right]^k  & \\
& & e^{\beta} &= e^{\beta_0} \left(1-\frac{B}{r} \right)^2
\left[\frac{1-\frac{B}{r}}{1+\frac{B}{r}}\right]^{\frac{(k-1)
    (k+2)}{k}} & \\
& & \phi &= \phi_0
\left[\frac{1-\frac{B}{r}}{1+\frac{B}{r}}\right]^{\frac{(1-k^2)}{k}} &
\\
\textbf{II} & & \alpha &= \alpha_0 + \frac{2}{\Lambda} \tan^{-1} \left(
\frac{r}{B} \right) & \\
& & \beta &= \beta_0 - \frac{2 (C+1)}{\Lambda}  \tan^{-1} \left(
\frac{r}{B} \right) - \ln \left[\frac{r^2}{(r^2+B^2)} \right] & \\
& & \phi &= \phi_0 e^{\frac{2 C}{\Lambda} \tan^{-1} \left( \frac{r}{B}
  \right)} & \\
\textbf{III} & & \alpha &= \alpha_0 -\frac{r}{B} & \\
& & \beta &= \beta_0 -2 \ln \left( \frac{r}{B} \right) + \frac{(C+1)
  r}{B} & \\
& & \phi &= \phi_0 e^{-\frac{C r}{B}} & \\
\textbf{IV} & & \alpha &= \alpha_0 -\frac{1}{B r} & \\
& & \beta &= \beta_0 + \frac{(C+1)}{B r} & \\
& & \phi &= \phi_0 e^{-\frac{C}{ Br}} & 
\end{flalign*}
where
\begin{align*}
k^2 &= \frac{(4+2 \omega)}{(3+2 \omega)} \\
\Lambda^2 &= C \left(1-\frac{\omega C}{2} \right) - (C+1)^2 >0 \\
C &= \frac{-1 \pm \sqrt{-2 \omega -3}}{(\omega+2)}
\end{align*}
and $B$, $\alpha_0$, $\beta_0$ and $\phi_0$ are constants.  Whilst
solution $I$ is valid for all values of $\omega$, solutions $II$,
$III$ and $IV$ are only valid for $\omega< -3/2$.  It is probably for this
reason that it is most often solution $I$ that is used in the literature
for analyses of the static and spherically symmetric situation.
Solution $I$ is conformally related to the minimally coupled massless
scalar field solution of Buchdahl \cite{buch59}.

It can be seen that these solutions are not all independent of each
other.  By a transformation of the form
\begin{equation*}
r \rightarrow \frac{1}{r},
\end{equation*}
and some redefinition of constants, solution $II$ can be
transformed into the $\omega <-3/2$ range of solution $I$ \cite{Bhad} and
solution $III$ can be transformed into solution $IV$
\cite{Bhad2}.  It was further shown in \cite{Bhad} that the independent
solutions $I$ and $IV$ are both conformally related to the general
solution of the static and spherically symmetric case in the Einstein
conformal frame, given by Wyman \cite{Wy}.

\section{Fourth-Order Theories}

In the absence of any matter the $\mathcal{L}=R^{1+\delta}$ field equations (\ref{field}) can be
written as 
\begin{equation}
R_{ab}=\delta \left( \frac{R_{;}^{\ cd}}{R}-(1-\delta )\frac{R_{,}^{\
c}R_{,}^{\ d}}{R^{2}}\right) \left( g_{ac}g_{bd}+\frac{1}{2}\frac{(1+2\delta
)}{(1-\delta )}g_{ab}g_{cd}\right) .  \label{staticfield}
\end{equation}
We find that an exact static spherically symmetric solution of these field
equations is given in Schwarzschild coordinates by the line-element 
\begin{equation}
ds^{2}=-A(r)dt^{2}+\frac{dr^{2}}{B(r)}+r^{2}(d\theta ^{2}+\sin ^{2}\theta
d\phi ^{2})  \label{Chan}
\end{equation}%
where 
\begin{align*}
A(r)& =r^{2\delta \frac{(1+2\delta )}{(1-\delta )}}+\frac{C}{r^{\frac{%
(1-4\delta )}{(1-\delta )}}} \\
B(r)& =\frac{(1-\delta )^{2}}{(1-2\delta +4\delta ^{2})(1-2\delta (1+\delta
))}\left( 1+\frac{C}{r^{\frac{(1-2\delta +4\delta ^{2})}{(1-\delta )}}}%
\right)
\end{align*}%
and $C=$ constant. This solution is conformally related to the $Q=0$ limit
of the one found by Chan, Horne and Mann for a static spherically--symmetric
space-time containing a scalar-field in a Liouville potential
\cite{Cha95}.  It reduces to Schwarzschild in the limit of GR: $\delta
\rightarrow 0$.

In order to evaluate whether or not this solution is physically relevant we
will proceed as follows. A dynamical systems approach will be used to
establish the asymptotic attractor solutions of the field equations (\ref%
{staticfield}). The field equations will then be perturbed around these
asymptotic attractor solutions and solved to first order in the perturbed
quantities. This linearised solution will then be treated as the physically
relevant static and spherically-symmetric weak-field limit of the field
equations (\ref{staticfield}) and compared with the exact solution (\ref%
{Chan}).

\subsection{Dynamical System}

The dynamical systems approach has already been applied to a situation of
this kind by Mignemi and Wiltshire \cite{Mig89}.  We present a brief summary
of the relevant points of their work which is relevent to this study,
in the above notation.

Taking the value of sign$(R)$ from (\ref{Chan}) as sign$(-\delta (1+\delta
)/(1-2\delta (1+\delta )))$ and making a suitable choice of $\Omega _{0}$
allows the scalar-field potential (\ref{pot1}) to be written as 
\begin{equation}
V(\phi )=-\frac{3\delta ^{2}}{8\pi G(1-2\delta (1+\delta ))}\exp \left( {%
\sqrt{\frac{16\pi G}{3}}\frac{(1-\delta )}{\delta }\phi }\right) .
\label{pot}
\end{equation}

In four dimensions Mignemi and Wiltshire's choice of line--element
corresponds to 
\begin{equation}
d\bar{s}^{2}=e^{2U(\xi )}\left( -dt^{2}+\bar{r}^{4}(\xi )d\xi ^{2}\right) +%
\bar{r}^{2}(\xi )(d\theta ^{2}+\sin ^{2}\theta d\phi ^{2})  \label{metric}
\end{equation}%
which, after some manipulation, gives the $\mathcal{L}=R^{1+\delta}$
Einstein frame field equations (\ref{conformalfield}) as 
\begin{align}
\zeta ^{\prime \prime }& =-\frac{2c_{1}^{2}(1-\delta )^{2}+6\delta ^{2}\eta
^{\prime }{}^{2}-24\delta ^{2}\eta ^{\prime }\zeta ^{\prime }-2(1-2\delta
-8\delta ^{2})\zeta ^{\prime }{}^{2}}{1-2\delta +4\delta ^{2}}-e^{2\zeta }
\label{wilt1} \\
\eta ^{\prime \prime }& =\frac{(1-2\delta -8\delta ^{2})(c_{1}^{2}(1-\delta
)^{2}+3\delta ^{2}\eta ^{\prime }{}^{2}-12\delta ^{2}\eta ^{\prime }\zeta
^{\prime }-(1-2\delta -8\delta ^{2})\zeta ^{\prime }{}^{2})}{3\delta
^{2}(1-2\delta +4\delta ^{2})} \nonumber \\ & \qquad
\qquad+\frac{(1-2\delta (1+\delta ))}{3\delta^{2}}%
e^{2\zeta }  \label{wilt2}
\end{align}%
and 
\begin{multline}
e^{2\eta }=-\frac{(1-2\delta (1+\delta ))}{3\delta ^{2}(1-2\delta +4\delta
^{2})}\left( c_{1}^{2}(1-\delta )^{2}+3\delta ^{2}\eta ^{\prime
}{}^{2}-12\delta ^{2}\eta ^{\prime }\zeta ^{\prime } \right.\\\left. -(1-2\delta -8\delta
^{2})\zeta ^{\prime }{}^{2}+(1-2\delta +4\delta ^{2})e^{2\zeta } \right)
\label{wilt3}
\end{multline}%
where 
\begin{align*}
\zeta (\xi )& =U(\xi )+\log \bar{r}(\xi ) \\
\eta (\xi )& =-\frac{(1-2\delta (1+\delta ))}{3\delta ^{2}}U(\xi )+2\log 
\bar{r}(\xi )-\frac{(1-\delta )^{2}}{3\delta ^{2}}c_{1}\xi +\text{constant.}
\end{align*}%
Primes denote differentiation with respect to $\xi $ and $c_{1}$ is a
constant of integration.

\begin{figure}[tbp]
\epsfig{file=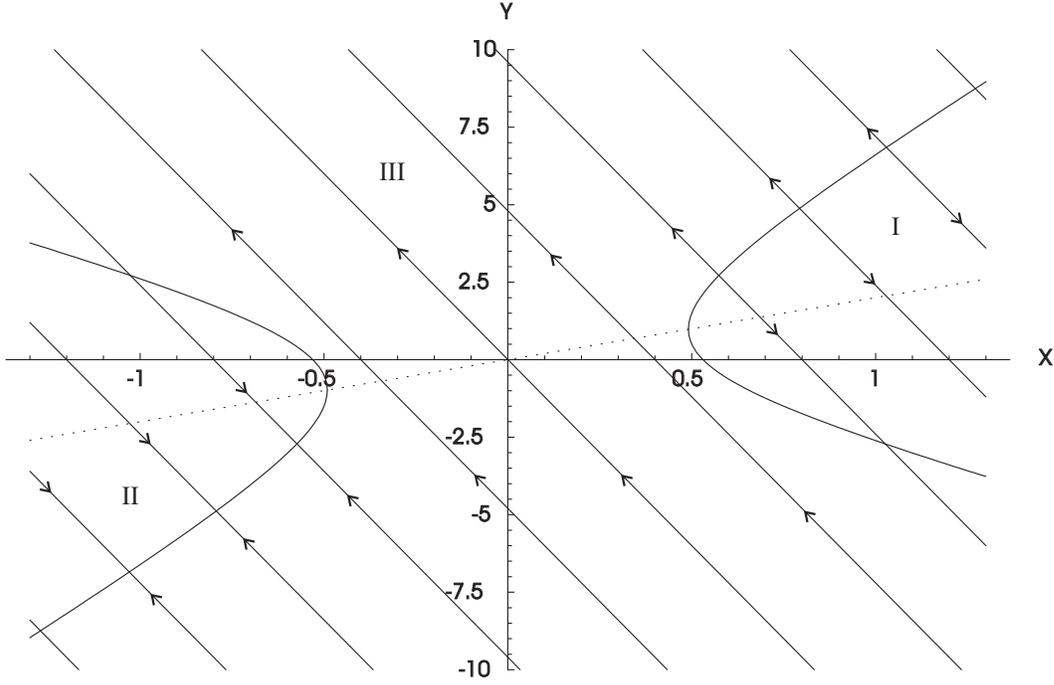,height=9cm}
\caption{\textit{The $Z=0$ plane of the phase space defined by $X$,$Y$ and $%
Z $ for $\protect\delta =0.1$ and $c_{1}=0.5$. The bold lines show the
critical points in this plane and the diagonal lines show the unphysical
trajectories confined to this plane. The dotted line is $Y=2X$ and separates
the critical points where $\protect\xi \rightarrow \infty $ from the points
where $\protect\xi \rightarrow -\infty $ }}
\label{finitephase}
\end{figure}

The variables $X$, $Y$ and $Z$ are here defined to be 
\begin{equation*}
X=\zeta ^{\prime }\qquad Y=\eta ^{\prime }\qquad Z=e^{\zeta }
\end{equation*}%
(not to be confused with the definitions of the scalar curvature
invariants used in other chapters).  The field equations (\ref{wilt1})
and (\ref{wilt2}) can then be written in terms of these new variables as
the following set of first-order autonomous differential equations 
\begin{align}
X^{\prime }& =-\frac{2c_{1}^{2}(1-\delta )^{2}+6\delta ^{2}Y^{2}-24\delta
^{2}XY-2(1-2\delta -8\delta ^{2})X^{2}}{1-2\delta +4\delta ^{2}}-Z^{2}
\label{X'}
\\
\label{Y'}
Y^{\prime }& =\frac{(1-2\delta -8\delta ^{2})(c_{1}^{2}(1-\delta
)^{2}+3\delta ^{2}Y^{2}-12\delta ^{2}XY-(1-2\delta -8\delta ^{2})X^{2})}{%
3\delta ^{2}(1-2\delta +4\delta ^{2})} \nonumber \\ &\qquad \qquad
+\frac{(1-2\delta (1+\delta ))}{%
3\delta ^{2}}Z^{2} \\
Z^{\prime }& =XZ.  \label{Z'}
\end{align}%
(The reader should note the different definition of $Z$ here to that of
Mignemi and Wiltshire). As identified by Mignemi and Wiltshire, the only
critical points at finite values of $X$, $Y$ and $Z$ are in the plane $Z=0$
along the curve defined by 
\begin{equation*}
c_{1}^{2}(1-\delta )^{2}+3\delta ^{2}Y^{2}-12\delta ^{2}XY-(1-2\delta
-8\delta ^{2})X^{2}=0.
\end{equation*}%
These curves are shown as bold lines in figure \ref{finitephase}, together
with some sample trajectories from equations (\ref{X'}) and (\ref{Y'}). From
the definition above we see that the condition $Z=0$ is equivalent to $\bar{r%
}e^{U}=0$. Whilst we do not consider trajectories confined to this plane to
be physically relevant, we do consider the plot to be instructive as it gives
a picture of the behaviour of trajectories close to this surface and
displays the attractive or repulsive behaviour of the critical points, which
can be the end points for trajectories which could be considered as
physically meaningful. 
The dotted line in figure \ref{finitephase} corresponds to the line $Y=2X$
and separates two different types of critical points. The critical points
with $Y>2X$ can be seen to be repulsive to the trajectories in the $Z=0$ plane
and correspond to the limit $\xi \rightarrow -\infty $. Conversely, the
points with $Y<2X$ are attractive and correspond to the limit $\xi
\rightarrow \infty $. As $Z=\bar{r}e^{U},$ all critical points of this type
in the $Z=0$ plane correspond either to naked singularities, $\bar{r}%
\rightarrow 0$, or regular horizons, $\bar{r}\rightarrow $constant.

The two bold lines in figure \ref{finitephase} are the points at which the
surface defined by 
\begin{equation*}
c_1^2 (1-\delta)^2+3 \delta^2 Y^2 -12 \delta^2 X Y -(1-2 \delta-8\delta^2)
X^2 +(1-2 \delta+4\delta^2) Z^2 =0
\end{equation*}
crosses the $Z=0$ plane. This surface splits the phase space into three
separate regions between which trajectories cannot move. These regions are
labelled $I$, $II$ and $III$ in figure \ref{finitephase}. It can be seen
from (\ref{wilt3}) that trajectories are confined to either regions $I$ or $%
II$ for the potential defined by (\ref{pot}). If we had chosen the opposite
value of sign$(R)$ in (\ref{pot1}) then trajectories would be confined to
region $III$. We will show, however, that region $III$ does not contain
solutions with asymptotic regions in which $\bar{r} \rightarrow \infty$ and
so is of limited interest for our purposes.

\begin{figure}[tbp]
\epsfig{file=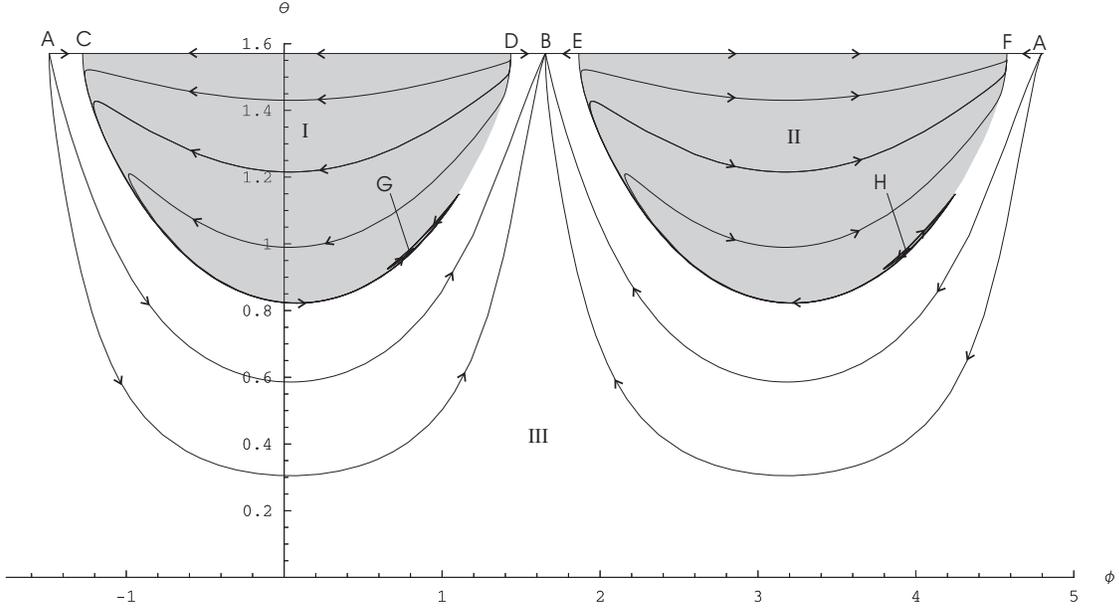,height=8cm}
\caption{\textit{The surface at infinity of the phase space defined by $X$,$%
Y $ and $Z$ for $\protect\delta =0.1$. The shaded areas show where regions $I
$ and $II$. Region $III$ is unshaded.}}
\label{infinityphase}
\end{figure}

In order to find the remaining critical points it is necessary to analyse
the sphere at infinity. This can be done by making the transformation 
\begin{equation*}
X=\rho \sin \theta \cos \phi \qquad Y=\rho \sin \theta \sin \phi \qquad
Z=\rho \cos \theta
\end{equation*}%
and taking the limit $\rho \rightarrow \infty $. The set of equations (\ref%
{X'}), (\ref{Y'}) and (\ref{Z'}) then give 
\begin{multline*}
\frac{d\theta }{d\tau }\rightarrow -\frac{\cos \theta }{24\delta
^{2}(1-2\delta +4\delta )}(6\delta ^{2}\cos \phi (3-3\delta (2-9\delta
)+(1-\delta (2+11\delta ))\cos 2\theta ) \\
-(3-3\delta (4-\delta (15-22\delta -32\delta ^{2}))\\+(5-\delta (20-\delta
(3+34\delta +32\delta ^{2})))\cos 2\theta )\sin \phi \\
-2(18\delta ^{2}(1-\delta (2+7\delta ))\cos 3\phi \\-(1-\delta (4+\delta
(9-26\delta +32\delta ^{2})))\sin 3\phi )\sin ^{2}\theta )
\end{multline*}%
and 
\begin{multline*}
\frac{d\phi }{d\tau }\rightarrow -\frac{1}{24\delta ^{2}(1-2\delta +4\delta
^{2})}(6\delta ^{2}(1-\delta (2+41\delta )-5(1-\delta (2+5\delta ))\cos
2\theta )\text{cosec}\theta \sin \phi \\
+2((1-\delta (4+\delta (9-26\delta +32\delta ^{2})))\cos 3\phi \\ +18\delta
^{2}(1-\delta (2+7\delta ))\sin 3\phi )\sin \theta \\
-2\cos \phi (4(1-2\delta (2-\delta (3-2\delta -4\delta ^{2})))\text{cosec}%
\theta \\-(7-\delta (28+\delta (15-2\delta (43+128\delta ))))\sin \theta ))
\end{multline*}%
where $d\tau =\rho d\xi $. These equations can be used to plot the positions
of critical points and trajectories on the sphere at infinity. The result of
this is shown in figure \ref{infinityphase}. Once again, these trajectories
do not correspond to physical solutions in the phase space but are
illustrative of trajectories at large distances and help to show the
attractive or repulsive nature of the critical points. 
The surface at infinity has eight critical points, labelled $A$-$H$ in
figure \ref{infinityphase}. Points $A$ and $B$ are the end-points of the
trajectory that goes through the origin in figure \ref{finitephase} and are
located at 
\begin{equation*}
\theta =\frac{\pi }{2},\qquad \text{and}\qquad \phi _{1,(2)}=\cos
^{-1}\left( \frac{-6\delta ^{2}}{\sqrt{1-4\delta -12\delta ^{2}+32\delta
^{3}+100\delta ^{4}}}\right) (+\pi )
\end{equation*}%
or, in terms of the original functions in the metric (\ref{metric}), 
\begin{equation*}
\bar{r}\rightarrow (\xi -\xi _{1})^{\frac{3\delta ^{2}}{(1-2\delta -8\delta
^{2})}}\qquad \text{and}\qquad e^{U}\rightarrow (\xi -\xi _{1})^{\frac{%
3\delta ^{2}}{(1-2\delta -8\delta ^{2})}},
\end{equation*}%
where $\xi _{1}$ is a constant of integration. The points $A$ and $B$
therefore both correspond to $\xi \rightarrow \xi _{1}$ and hence to $\bar{r}%
\rightarrow 0$.

Points $C$, $D$, $E$ and $F$ are the four end points of the two curves in
figure \ref{finitephase} and therefore correspond to $\xi \rightarrow \infty$
or $-\infty$ and $\bar{r} \rightarrow 0$ or constant.

The remaining points, $G$ and $H$, are located at 
\begin{equation*}
\phi _{1,(2)}=\frac{\pi }{4}(+\pi )\qquad \text{and}\qquad \theta =\frac{1}{2%
}\cos ^{-1}\left( -\frac{1-2\delta +10\delta ^{2}}{3-6\delta +6\delta ^{2}}%
\right)
\end{equation*}%
or 
\begin{equation*}
\bar{r}^{\frac{(1-2\delta +4\delta ^{2})}{(1-\delta )^{2}}}\rightarrow \pm 
\sqrt{\frac{(1-2\delta -2\delta ^{2})}{(1-2\delta +4\delta ^{2})}}\frac{1}{%
(\xi -\xi _{2})}
\end{equation*}
and
\begin{equation}
e^{\frac{(1-2\delta +4\delta ^{2})U}{%
3\delta ^{2}}}\rightarrow \pm \sqrt{\frac{(1-2\delta -2\delta ^{2})}{%
(1-2\delta +4\delta ^{2})}}\frac{1}{(\xi -\xi _{2})},  \label{GH}
\end{equation}%
where $\xi _{2}$ is another integration constant.  The positive branch corresponds
to point $H$ and the negative branch to point $G$. These points are,
therefore, the asymptotic limit of the exact solution (\ref{Chan}) and
correspond to $\xi \rightarrow \xi _{2}$ and hence $\bar{r}\rightarrow
\infty $.

Whilst it may initially appear that trajectories are repelled from the point 
$H$, this is only the case in terms of the coordinate $\xi $. In terms of
the more physically relevant quantity $\bar{r},$ the point $H$ is an
attractor. This can be seen from the first equation in (\ref{GH}). Taking
the positive branch here it can be seen that $\bar{r}$ increases as $\xi $
decreases. So, in terms of $\bar{r}$ the points $G$ and $H$ are both
attractors, as $\bar{r}\rightarrow \infty $.

We can now see that in region $I$ all trajectories appear to start at
critical points corresponding to either $\bar{r} \rightarrow 0$ or constant
and end at point $G$ where $\bar{r} \rightarrow \infty$. Region $II$ appears
to share the same features as region $I$ with all trajectories starting
at either $\bar{r} \rightarrow 0$ or constant and ending at $H$ where $\bar{r%
} \rightarrow \infty$. Region $III$ has no critical points corresponding to $%
\bar{r} \rightarrow \infty$ and so all trajectories both begin and end on
points corresponding to $\bar{r} \rightarrow 0$ or constant.

Therefore solutions with an asymptotic region in which $\bar{r}%
\rightarrow \infty $ only exist in regions $I$ and $II$ where the potential can
be described by equation (\ref{pot}). Furthermore, all trajectories in
regions $I$ and $II$ are attracted to the solution 
\begin{equation}
ds^{2}=-\bar{r}^{\frac{6\delta ^{2}}{(1-\delta )^{2}}}dt^{2}+\frac{%
(1-2\delta +4\delta ^{2})(1-2\delta -2\delta ^{2})}{(1-\delta )^{4}}d\bar{r}%
^{2}+\bar{r}^{2}(d\theta ^{2}+\sin ^{2}\theta d\phi ^{2}),
\label{asymptotic}
\end{equation}%
which is the asymptotic behaviour of the solution found by Chan, Horne and
Mann \cite{Cha95}. We therefore conclude that all solutions with an
asymptotic region in which $\bar{r}\rightarrow \infty $ are attracted
towards the solution (\ref{asymptotic}) as $\bar{r}\rightarrow \infty $.

Rescaling the metric back to the original conformal frame, we therefore
conclude that the generic asymptotic attractor solution to the field
equations, (\ref{staticfield}), is 
\begin{equation}
ds^{2}=-r^{2\delta \frac{(1+2\delta )}{(1-\delta )}}dt^{2}+\frac{(1-2\delta
+4\delta ^{2})(1-2\delta -2\delta ^{2})}{(1-\delta )^{2}}dr^{2}+r^{2}(d%
\theta ^{2}+\sin ^{2}\theta d\phi ^{2})  \label{asymptotic2}
\end{equation}%
as $r\rightarrow \infty $, which reduces to Minkowski space in the $\delta
\rightarrow 0$ limit of GR.

\subsection{Linearised Solution}

We now proceed to find the general solution, to first order in
perturbations, around the background described by (\ref{asymptotic2}).
Writing the perturbed line-element as 
\begin{multline}
ds^{2}=-r^{2\delta \frac{(1+2\delta )}{(1-\delta )}}(1+V(r))dt^{2}\\+\frac{%
(1-2\delta +4\delta ^{2})(1-2\delta -2\delta ^{2})}{(1-\delta )^{2}}%
(1+W(r))dr^{2}+r^{2}(d\theta ^{2}+\sin ^{2}\theta d\phi ^{2})
\label{asymptotic3}
\end{multline}%
and making no assumptions about the order of $R$, the field equations (\ref%
{staticfield}) become, up to first order in $V$ and $W$, 
\begin{multline}
\frac{\delta (1+2\delta )(1+2\delta ^{2})}{(1-\delta )^{2}r^{2}}+\frac{%
(1+2\delta ^{2})}{(1-\delta )}\frac{V^{\prime }}{r}-\frac{\delta (1-2\delta )%
}{2(1-\delta )}\frac{W^{\prime }}{r}+\frac{V^{\prime \prime }}{2}  \label{rr}
\\
=\frac{\delta (1+2\delta )}{2}\frac{R^{\prime }{}^{2}}{R^{2}}-\frac{\delta
(1+2\delta )}{2(1-\delta )}\frac{R^{\prime \prime }}{R}-\frac{3\delta }{%
4(1-\delta )}\frac{R^{\prime }}{R}V^{\prime }\\-\frac{\delta (1+2\delta
)(2+\delta )}{2(1-\delta )^{2}r}\frac{R^{\prime }}{R}+\frac{\delta
(1+2\delta )}{4(1-\delta )}\frac{R^{\prime }}{R}W^{\prime },
\end{multline}%
\begin{multline}
\frac{\delta (1+2\delta )(1-2\delta -2\delta ^{2})}{(1-\delta )^{2}r^{2}}-%
\frac{\delta (1+2\delta )}{(1-\delta )}\frac{V^{\prime }}{r}+\frac{(2-\delta
+2\delta ^{2})}{2(1-\delta )}\frac{W^{\prime }}{r}-\frac{V^{\prime \prime }}{%
2}  \label{tt} \\
=-\frac{3\delta }{2}\frac{R^{\prime }{}^{2}}{R^{2}}+\frac{3\delta }{%
2(1-\delta )}\frac{R^{\prime \prime }}{R}+\frac{\delta (1+2\delta )(2-\delta
+2\delta ^{2})}{2(1-\delta )^{2}r}\frac{R^{\prime }}{R}\\+\frac{\delta
(1+2\delta )}{4(1-\delta )}\frac{R^{\prime }}{R}V^{\prime }-\frac{3\delta }{%
4(1-\delta )}\frac{R^{\prime }}{R}W^{\prime }
\end{multline}%
and 
\begin{multline}
-\frac{2\delta (3-4\delta +2\delta ^{2}+8\delta ^{3})}{(1-\delta )^{2}r^{2}}+%
\frac{2(1-2\delta +4\delta ^{2})(1-2\delta -2\delta ^{2})}{(1-\delta
)^{2}r^{2}}W-\frac{V^{\prime }}{r}+\frac{W^{\prime }}{r}  \label{thth} \\
=-\delta (1+2\delta )\frac{R^{\prime }{}^{2}}{R^{2}}+\frac{\delta (1+2\delta
)}{(1-\delta )}\frac{R^{\prime \prime }}{R}+\frac{\delta (4-\delta +2\delta
^{2}+4\delta ^{3})}{(1-\delta )^{2}r}\frac{R^{\prime }}{R}\\+\frac{\delta
(1+2\delta )}{2(1-\delta )}\frac{R^{\prime }}{R}V^{\prime }-\frac{\delta
(1+2\delta )}{2(1-\delta )}\frac{R^{\prime }}{R}W^{\prime }.
\end{multline}

Expanding $R$ to first order in $V$ and $W$ gives 
\begin{equation}  \label{R}
R=-\frac{6 \delta (1+\delta)}{(1-2\delta-2 \delta^2)} \frac{1}{r^2}+R_1
\end{equation}
where 
\begin{multline}  \label{R1}
R_1= \frac{2 (1+\delta+\delta^2)}{(1-2\delta-2\delta^2)} \frac{W}{r^2} -%
\frac{2 (1-\delta) (1+2 \delta^2)}{(1- 2\delta-2\delta^2)(1-2 \delta+4
\delta^2)} \frac{V^{\prime}}{r} \\
+\frac{(1-\delta)(2-\delta+2 \delta^2)}{(1- 2\delta-2\delta^2)(1-2 \delta+4
\delta^2)} \frac{W^{\prime}}{r} \\ - \frac{(1-\delta)^2}{(1-
2\delta-2\delta^2)(1-2 \delta+4 \delta^2)} V^{\prime\prime}.
\end{multline}

Substituting (\ref{R}) into the field equations (\ref{rr}), (\ref{tt}) and (%
\ref{thth}) and eliminating $R_1$ using (\ref{R1}) leaves 
\begin{multline*}
\frac{(1+\delta+\delta^2)(5-12\delta+12 \delta^2+4 \delta^3)}{3 (1-\delta)^2
(1+\delta)} \frac{W}{r} +\frac{(16-47\delta+76 \delta^2-34 \delta^3-16
\delta^4+32 \delta^5)}{6 (1-\delta^2) (1-2\delta+4 \delta^2)} W^{\prime} \\
-\frac{(1+\delta +7 \delta^2-19 \delta^3+44 \delta^4+20 \delta^5)}{3
(1-\delta^2) (1-2\delta+4 \delta^2)} \frac{Y}{r} -\frac{(8-15 \delta+18
\delta^2 +16 \delta^3)}{6 (1+\delta) (1-2\delta+4\delta^2)} Y^{\prime} \\
= -\frac{(1-2\delta-2\delta^2)(5-12\delta+12\delta^2+4\delta^3)}{12
(1-\delta^2) (1+\delta)} \frac{\psi}{r} -\frac{(1-2 \delta-2\delta^2)}{4
(1-\delta^2)} \psi^{\prime},
\end{multline*}
\begin{multline*}
-\frac{(1+2 \delta) (1+\delta+\delta^2) (3-4 \delta+4 \delta^2)}{3
(1-\delta)^2 (1+\delta)} \frac{W}{r} -\frac{(1+2\delta)(2-\delta+2
\delta^2)(3 -4\delta+4\delta^2)}{6 (1-\delta^2) (1-2\delta+4\delta^2)}
W^{\prime} \\
+\frac{(3-2\delta +17\delta^2 -4 \delta^3+40 \delta^2)}{3 (1-\delta^2) (1-2
\delta+4 \delta^2)} \frac{Y}{r}+ \frac{(6-\delta+2 \delta^2+20 \delta^3)}{ 6
(1+\delta) (1-2\delta+4\delta^2)} Y^{\prime} \\
= \frac{(1+2 \delta) (1-2\delta-2 \delta^2) (3-4\delta+4 \delta^2)}{12
(1-\delta)^2 (1+\delta)} \frac{\psi}{r} + \frac{(1+2 \delta)(1-2
\delta-2\delta^2)}{12 (1-\delta^2)} \psi^{\prime}.
\end{multline*}
and 
\begin{multline*}
-\frac{2 (8-8 \delta+3 \delta^2+10 \delta^3-28 \delta^4-12 \delta^5)}{3
(1-\delta^2) (1+\delta) (1-2\delta-2 \delta^2)} \frac{W}{r} - \frac{(13-22
\delta+12 \delta^2+26 \delta^3-56 \delta^4)}{3 (1-\delta^2) (1-2 \delta-2
\delta^2)(1-2 \delta +4 \delta^2)} W^{\prime} \\
+\frac{2 (4+9 \delta^2+8 \delta^3-12\delta^4)}{3 (1+\delta) (1-2
\delta-2\delta^2)(1-2\delta+4 \delta^2)} \frac{Y}{r} + \frac{(5 -4\delta-4
\delta^2+12 \delta^3)}{3 (1+\delta) (1-2 \delta-2\delta^2)(1-2\delta+4
\delta^2)} Y^{\prime} \\
= \frac{(5-4 \delta-4 \delta^2+12 \delta^3)}{6 (1-\delta)^2 (1+\delta)} 
\frac{\psi}{r} +\frac{(1+2 \delta)}{6 (1-\delta^2)} \psi^{\prime}
\end{multline*}
where $Y= r V^{\prime}$ and $\psi=r^3 R_1^{\prime}$, subject to the
constraint (\ref{R1}).

For $-\frac{(7+3 \sqrt{21})}{20}<\delta<-\frac{(7-3\sqrt{21})}{20}$ the
general solution to this first order set of coupled equations is given, in
terms of $V$ and $W$, by 
\begin{align}  \label{V}
V(r) &= c_1 V_1(r)+c_2 V_2(r)+ c_3 V_3(r) +\text{constant} \\
W(r) &= -c_1 V_1(r)+c_2 W_2(r)+c_3 W_3(r)
\end{align}
where 
\begin{align*}
V_1 &=-r^{-\frac{(1-2 \delta+4 \delta^2)}{(1-\delta)}} \\
V_2 &= \frac{ (1+2 \delta) r^{-\frac{(1-2 \delta+4 \delta^2)}{2 (1-\delta)}}%
}{2 (2-3 \delta+12 \delta^2+16 \delta^3)} \left( (1+2 \delta)^2 \sin (A \log
r) + 2 A (1-\delta) \cos(A \log r) \right) \\
W_2 &= r^{-\frac{(1-2 \delta+4 \delta^2)}{2 (1-\delta)}} \sin (A \log r) \\
V_3 &= \frac{ (1+2 \delta) r^{-\frac{(1-2 \delta+4 \delta^2)}{2 (1-\delta)}}%
}{2 (2-3 \delta+12 \delta^2+16 \delta^3)} \left( (1+2 \delta)^2 \cos (A \log
r) - 2 A (1-\delta) \sin(A \log r) \right) \\
W_3 &= r^{-\frac{(1-2 \delta+4 \delta^2)}{2 (1-\delta)}} \cos (A \log r)
\end{align*}
and 
\begin{equation*}
A= -\frac{\sqrt{7-28 \delta+36 \delta^2-16 \delta^3-80 \delta^4}}{2
(1-\delta)}
\end{equation*}
which is a real number for the any $\delta$ in the range specified
above.  The extra constant in (\ref{V}) is from the integration of $Y$ and can be
trivially absorbed into the definition of the time coordinate. The above
solution satisfies the constraint (\ref{R1}) without imposing any conditions
upon the arbitrary constants $c_1$, $c_2$ and $c_3$.

It can be seen by direct comparison that the constant $c_1$ is linearly
related to the constant $C$ in (\ref{Chan}) by a factor that is a function
of $\delta$ only. The constants $c_2$ and $c_3$ correspond to two new
oscillating modes.

\clearemptydoublepage
\chapter{Inhomogeneous Cosmological Solutions}
\label{Inhomogeneous}

\bigskip

We will present in this chapter inhomogeneous cosmological solutions
to the scalar-tensor and fourth-order field equations.  Considerations
will be limited to space-times admitting spherically symmetric three
dimensional space-like hypersurfaces.  The existence of vacuum
solutions with this symmetry shows explicitly that Birkhoff's theorem
cannot be formulated in these theories.

In the scalar-tensor case we will also present a study of how the
spherical collapse model works in these theories.  Closed universes
are matched to flat universes at a boundary and allowed to evolve
seperately.  Following the evolution of the scalar field in each
region then allows for consideration of how spatial inhomogeneity in
Newton's gravitational constant $G$ can arise in a realistic universe.

This chapter is based on the work by Clifton, Mota and Barrow
\cite{Cli05} and Clifton \cite{Clifton}.

\section{Scalar-Tensor Theories}

Extensive studies have been made of cosmological solutions of scalar-tensor
gravity theories \cite{Fuji}, although they are limited in two respects.
First, they usually focus on the simplest case of isotropic expansion with zero
spatial curvature, where simple exact solutions exist. Second, they are
almost exclusively concerned with spatially homogeneous cosmologies. The latter
restriction means that the value of $G$ and its rate of change in time, $%
\dot{G},$ are required to be the same everywhere in the universe.  We
want to understand how $G$ and $\dot{G}$ are expected to vary in space in a
realistic inhomogeneous universe. Since, even on the scale of a typical
galaxy, the amplitude of visible density inhomogeneities are of order
$10^{6}$, we need to go beyond linear perturbation theory for such an
analysis.

Previous studies on inhomogenieties in physical constants have
focussed on the formation of black holes (gravitational memory) and
perturbative expansions (see
\cite{mem,mem1,mem2,mem3,mem4,mem5,mem6}).  In this work we confront this problem by tracing the evolution
of $G$, first in exact inhomogeneous solutions and then in a simple, but not
unrealistic, inhomogeneous universe in which a zero-curvature
Brans-Dicke FRW background universe is populated by spherical
overdensities which are modelled by positive curvature Brans-Dicke FRW
universes in the dust-dominated era of the Universe's history. This will
enable us to track the different evolution followed by $G(t)$ in the
background universe and in the overdense regions, which eventually separate
off from the background universe and start to contract to high-density
separate closed universes. This process can produce significant differences
between $G$ and $\dot{G}$ in the background and in the overdensities.
Eventually, the collapse of the spherical overdensities will be stopped by
pressure and a complicated sequence of dissipative and relaxation processes
will lead to virialisation. This state will provide the gravitational environment out of
which which stars and planetary systems like our own will form, directly
reflecting the local value of $G(t)$ inherited from their virialised
protogalaxy or its parent protocluster. The simple model we use for
inhomogeneities in density and in $G$ has many obvious limitations, notably
in its neglect of pressure, deviations from spherical symmetry, accretion,
and interactions between inhomogeneities. Nonetheless, we expect that it
will be indicative of the importance of taking spatial inhomogeneity into
account.  It provides the first step in a clear path
towards improved realism in the modelling of inhomogeneities that mirrors
the route followed in standard cosmological studies of galaxy formation with
constant $G$.

\subsection{Exact Solutions}

Starting with a solution of Einstein's field equations with a scalar field
we can apply a conformal transformation to arrive at a solution of the
Brans-Dicke field equations in a vacuum. A spherically symmetric exact solution for the
collapse of a minimally-coupled scalar field, $\psi $, in GR is known
and is given by \cite{Hussain}
\begin{equation}
ds^{2}=-(qt+b)(f^{2}(r)dt^{2}-f^{-2}(r)dr^{2}) + R^{2}(r,t)(d\theta ^{2}+\sin
^{2}\theta d\phi ^{2}),  \label{GRsolution}
\end{equation}%
where 
\begin{align*}
f^{2}(r) &=(1-\frac{2c}{r})^{\alpha }, \\
R^{2}(r,t) &=(qt+b)r^{2}(1-\frac{2c}{r})^{1-\alpha },\\
\alpha &=\pm \frac{\sqrt{3}}{2}. 
\end{align*}
The evolution of
the minimally coupled scalar field, $\psi $, in the Einstein frame, is given
by
\begin{equation}
\psi (r,t)=\pm \frac{1}{4\sqrt{\pi }}ln\left[ d\left( 1-\frac{2c}{r}\right)
^{\frac{\alpha }{\sqrt{3}}}(qt+b)^{\sqrt{3}}\right] .  \label{psi}
\end{equation}%
Here, $q$, $b$, $c$ and $d$ are constants. Now under the
transformation to the Jordan frame, (\ref{conformal}), where $\phi
=\exp \left[ \psi \sqrt{\frac{8\pi }{\omega +\frac{3}{2}}} \right]$, we obtain
\begin{multline}
d\bar{s}^{2}=-\frac{B(t)^{1-\sqrt{3}/\beta }}{d^{1/\beta }A(r)^{\alpha /\sqrt{%
3}\beta }}\left[ A(r)^{\alpha }dt^{2}-A(r)^{-\alpha }dr^{2}\right] \\ + \frac{%
A(r)^{1-\alpha \frac{1+\sqrt{3}\beta }{\sqrt{3}\beta }}B(t)^{1-\sqrt{3}%
/\beta }r^{2}}{d^{1/\beta }}(d\theta ^{2}+sin^{2}\theta d\phi ^{2}) \label{JBDsolution1}
\end{multline}%
and
\begin{equation}
\phi (r,t)=\left[ dA(r)^{\alpha /\sqrt{3}}B(t)^{\sqrt{3}}\right] ^{1/\beta }
\label{phi1}
\end{equation}%
where $A(r)=1-\frac{2c}{r}$, $B(t)=qt+b$ and $\beta =\pm \sqrt{2\omega +3}$.
We now assume that $q\neq 0$ (i.e. the metric is not static) and define the
new time coordinate $\bar{t}=(qt+b)^{\frac{3}{2}-\frac{\sqrt{3}}{2\beta }}$.

In the limit that $c\rightarrow 0$ the $r$-dependence of the metric is
removed and the space becomes homogeneous. In this case we expect (\ref%
{JBDsolution1}) to reduce to the FRW Brans-Dicke metric given in the last
section. We see from the form of (\ref{JBDsolution1}) that the metric should
reduce to that of a flat FRW Brans-Dicke universe. Insisting on this limit
requires us to set $\beta =\sqrt{2\omega +3}$, $q=\frac{2\beta }{3\beta -%
\sqrt{3}}$ and $d=1$. This leaves the metric:
\begin{multline}
d\bar{s}^{2}=-A(r)^{\alpha (1-\frac{1}{\sqrt{3}\beta })}d\bar{t}%
^{2}\\+A(r)^{-\alpha (1+\frac{1}{\sqrt{3}\beta })}\bar{t}^{\frac{2(\beta -%
\sqrt{3})}{3\beta -\sqrt{3}}} \left[ dr^{2}+A(r)r^{2}(d\theta ^{2}+\sin^{2}\theta
d\phi ^{2})\right] .  \label{JBDsolution2}
\end{multline}%
Rewriting (\ref{phi1}) with these coordinates and constants gives
\begin{equation}
\phi(r,t)=\left( 1-\frac{2c}{r} \right)^{\pm\frac{1}{2\beta}}\bar{t}^{2/(%
\sqrt{3}\beta -1)}.  \label{phi2}
\end{equation}
A comparison of (\ref{JBDsolution2}) with the FRW solution (\ref{phi_b(t)}) shows
that (\ref{JBDsolution2}) does indeed reduce to a flat vacuum FRW metric in
the limit $c\rightarrow 0$ (an inhomogeneous universe requires $c\neq 0$).
The metric (\ref{JBDsolution2}) is asymptotically flat and has singularities
at $\bar{t}=0$ and $r=2c$; the coordinates $r$ and $\bar{t}$ therefore cover
the ranges $0\leq \bar{t}<\infty $ and $2\leq \frac{r}{c}<\infty $.

\begin{figure}[ht]
\epsfig{file=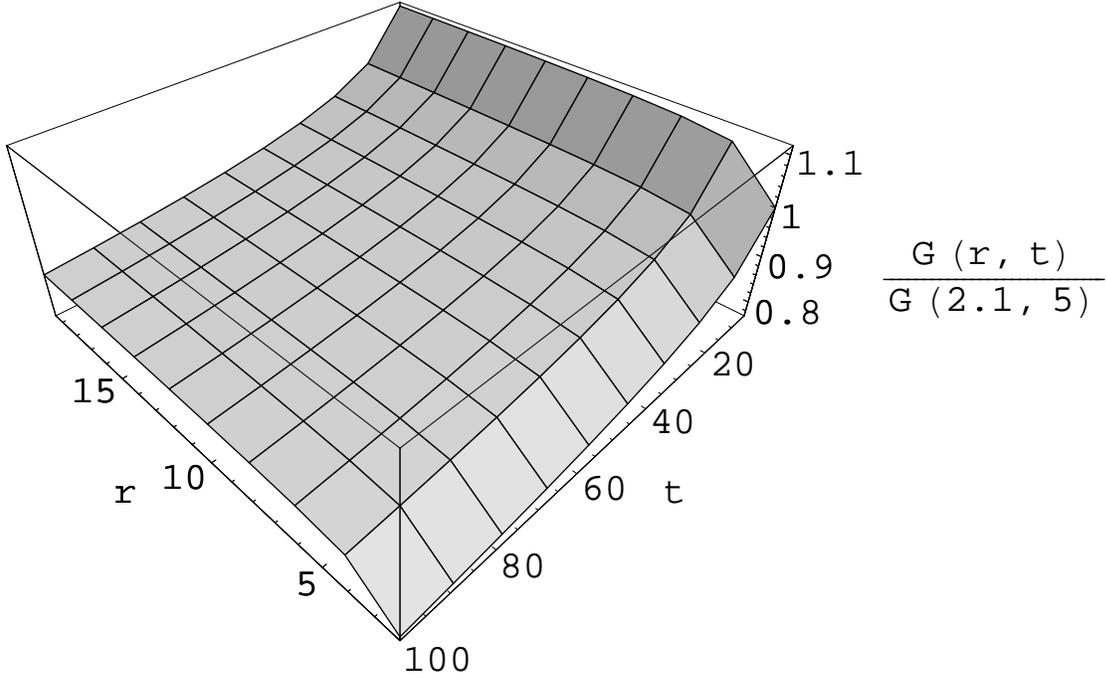,height=12cm}
\caption{{\protect {\textit{This graph illustrates the possible space and time
variations that can arise in G in inhomogeneous solutions to the Brans-Dicke
field equations with $\omega=100$, normalised at $r=2.1$ and $t=5$.}}}}
\label{GTB}
\end{figure}

The equations (\ref{Gc}) and (\ref{phi2}) can now be used to construct
a plot of $G(r,t)$; this is done in figure \ref{GTB} which
was constructed by choosing the minus sign in (\ref{phi2}). From the form of $%
G(r,t) $ we see that this choice corresponds to an overdensity in the
mass distribution (identified by comparison with the inhomogeneous Brans-Dicke
solution with matter, found below). In this figure $\omega $ was set equal
to $100$ and $c$ to $1$. This plot shows how $G$ can vary in
space and time in an inhomogeneous universe which consists of a static
Schwarzschild-like mass sitting at $r=0$ in an expanding universe. As
$r\rightarrow \infty $ it can be seen that this solution approaches
the behaviour of a Brans-Dicke FRW universe.

\subsubsection{An Inhomogeneous Brans-Dicke Solution With Matter}

\label{vacsol}

We now seek a solution of the Brans-Dicke field equations (\ref{STfields}) with the form
\begin{equation}  \label{metric5}
ds^2=-e^{\nu}dt^2+e^{\mu}a^2(dr^2+r^2d\theta^2+r^2 \sin^2\theta d\phi^2)
\end{equation}
where $e^{\nu }=e^{\nu (r)}$, $e^{\mu }=e^{\mu (r)}$ and $a=a(t)$.
Substituting (\ref{metric5}) into (\ref{STfields}) gives, for a
perfect fluid
\begin{multline}  \label{T11}
\frac{8\pi}{\phi}p = -\left[ 2\frac{\ddot{a}}{a}+\left(\frac{\dot{a}}{a}%
\right)^2 +\frac{\omega}{2} \left( \frac{\dot{\phi}}{\phi} \right)^2 +2\frac{%
\dot{a}}{a}\frac{\dot{\phi}}{\phi} + \frac{\ddot{\phi}}{\phi} \right]
e^{-\nu} \\
+\left[\frac{\mu^{{\prime}^2}}{4} +\frac{\mu^{\prime}\nu^{\prime}}{r} + \frac{%
\mu^{\prime}\nu^{\prime}}{2} -\frac{\omega}{2}\left(\frac{\phi^{\prime}}{\phi%
}\right)^2 +\frac{\phi^{\prime}}{\phi} \left(\mu^{\prime}+\frac{\nu^{\prime}%
}{2} +\frac{2}{r}\right)\right] e^{-\mu}a^{-2},
\end{multline}
\begin{multline}  \label{T22}
\frac{8\pi}{\phi}p=-\left[2\frac{\ddot{a}}{a}+\left(\frac{\dot{a}}{a}
\right)^2 +\frac{\omega}{2} \left(\frac{\dot{\phi}}{\phi} \right)^2+2\frac{%
\dot{a}}{a} \frac{\dot{\phi}}{\phi} + \frac{\ddot{\phi}}{\phi} \right]
e^{-\nu} \\
+\left[ \frac{\mu^{\prime\prime}\nu^{\prime\prime}}{2} + \frac{\nu^{{\prime}^2}%
}{4} + \frac{\mu^{\prime}+\nu^{\prime}}{2r} +\frac{\omega}{2} \left(\frac{%
\phi^{\prime}}{\phi} \right)^2 + \frac{\phi^{\prime\prime}}{\phi} + \left(%
\frac{\nu^{\prime}}{2} +\frac{1}{r}\right) \frac{\phi^{\prime}}{\phi} \right]
e^{-\mu}a^{-2},
\end{multline}
and
\begin{multline}  \label{T44}
-\frac{8\pi}{\phi} \rho = -\left[3\left(\frac{\dot{a}}{a}\right)^2 - \frac{%
\omega}{2} \left(\frac{\dot{\phi}}{\phi}\right)^2+3\frac{\dot{a}}{a}\frac{%
\dot{\phi}}{\phi}\right] e^{-\nu} \\
+\left[ \mu^{\prime\prime}+\frac{\mu^{{\prime}^2}}{4} + \frac{2\mu^{\prime}}{r}
+\frac{\omega}{2} \left(\frac{\phi^{\prime}}{\phi} \right)^2 +\frac{%
\phi^{\prime\prime}}{\phi} +\left(\frac{\nu^{\prime}}{2} +\frac{1}{r}\right)%
\frac{\phi^{\prime}}{\phi}\right] e^{-\mu} a^{-2}
\end{multline}
as the $T^{11}$, $T^{22}$ and $T^{00}$ equations, respectively. The
propagation equation (\ref{STfields2}) now becomes
\begin{equation}  \label{wave}
\frac{8\pi(\rho-3p)}{(2\omega+3)\phi} = \left[\frac{\ddot{\phi}}{\phi} + 3%
\frac{\dot{a}}{a}\frac{\dot{\phi}}{\phi}\right] e^{-\nu} -\left[\frac{%
\phi^{\prime\prime}}{\phi} + \frac{(\mu^{\prime}+\nu^{\prime})}{2} \frac{%
\phi^{\prime}}{\phi} + \frac{2}{r} \frac{\phi^{\prime}}{\phi}\right]%
e^{-\mu}a^{-2}
\end{equation}
and the only other non-trivial field equation is that for $T^{10}$:
\begin{equation}  \label{T14}
\nu^{\prime}\frac{\dot{a}}{a} - \omega\frac{\dot{\phi}}{\phi}\frac{%
\phi^{\prime}}{\phi} - \frac{\dot{\phi}^{\prime}}{\phi} +\frac{\dot{a}}{a}%
\frac{\phi^{\prime}}{\phi} + \frac{\nu^{\prime}}{2} \frac{\dot{\phi}}{\phi}%
=\frac{8 \pi}{\phi} (\rho+p) v e^{\frac{\nu}{2}}
\end{equation}
where $v$ is the three velocity of the perfect fluid.  We now assume
$\phi$ is of the form $\phi(r,t)=\phi(r)\phi(t)$ and look for
solutions to the set of equations
\begin{equation}  \label{S11}
\frac{\mu^{{\prime}^2}}{4} +\frac{\mu^{\prime}\nu^{\prime}}{r} + \frac{%
\mu^{\prime}\nu^{\prime}}{2} =\frac{\omega(r)}{2}\left(\frac{\phi^{\prime}(r)%
}{\phi(r)}\right)^2 -\frac{\phi^{\prime}(r)}{\phi(r)} \left(\mu^{\prime}+%
\frac{\nu^{\prime}}{2} +\frac{2}{r}\right),
\end{equation}
\begin{equation}  \label{S22}
\frac{\mu^{\prime\prime}\nu^{\prime\prime}}{2} + \frac{\nu^{{\prime}^2}}{4} + 
\frac{\mu^{\prime}+\nu^{\prime}}{2r} =-\frac{\omega}{2} \left(\frac{%
\phi^{\prime}(r)}{\phi(r)} \right)^2 - \frac{\phi^{\prime\prime}(r)}{\phi(r)}
- \left(\frac{\nu^{\prime}}{2} +\frac{1}{r}\right) \frac{\phi^{\prime}(r)}{%
\phi(r)},
\end{equation}
\begin{equation}  \label{S44}
\mu^{\prime\prime}+\frac{\mu^{{\prime}^2}}{4} + \frac{2\mu^{\prime}}{r} =-%
\frac{\omega}{2} \left(\frac{\phi^{\prime}(r)}{\phi(r)} \right)^2 -\frac{%
\phi^{\prime\prime}(r)}{\phi(r)} -\left(\frac{\nu^{\prime}}{2} +\frac{1}{r}%
\right)\frac{\phi^{\prime}(r)}{\phi(r)},
\end{equation}
and
\begin{equation}  \label{Swave}
\frac{\phi^{\prime\prime}(r)}{\phi(r)} + \frac{(\mu^{\prime}+\nu^{\prime})}{2%
} \frac{\phi^{\prime}(r)}{\phi(r)} + \frac{2}{r} \frac{\phi^{\prime}(r)}{%
\phi(r)}=0.
\end{equation}
Such solutions are given by \cite{Nariai68} as
\begin{equation}
e^{\nu }=\left( \frac{1-\frac{c}{2kr}}{1+\frac{c}{2kr}}\right) ^{2k},
\label{ev2}
\end{equation}
\begin{equation}
e^{\mu }=\left( 1+\frac{c}{2kr}\right) ^{4}\left( \frac{1-\frac{c}{2kr}}{1+%
\frac{c}{2kr}}\right) ^{2(k-1)(k+2)/k},  \label{eu2}
\end{equation}
and
\begin{equation}
\phi (r)=\phi _{0}\left( \frac{1-\frac{c}{2kr}}{1+\frac{c}{2kr}}\right)
^{-2(k^{2}-1)/k}  \label{phir2}
\end{equation}
where $k=\sqrt{\frac{4+2\omega}{3+2\omega}}$. For $e^{\nu}$ and $e^{\mu}$ of
this form, equations (\ref{T11}), (\ref{T44}) and (\ref{wave}) become
\begin{equation}  \label{t11}
\frac{8\pi(\rho e^{\nu}-3p e^{\nu})}{(2\omega+3)\phi(t)} = \frac{\ddot{%
\phi}(t)}{\phi(t)} + 3\frac{\dot{a}}{a}\frac{\dot{\phi}(t)}{\phi(t)},
\end{equation}
\begin{equation}  \label{t44}
\frac{8\pi}{\phi(t)} \rho e^{\nu} = 3\left(\frac{\dot{a}}{a}\right)^2 - 
\frac{\omega}{2} \left(\frac{\dot{\phi}(t)}{\phi(t)}\right)^2+3\frac{\dot{a}%
}{a}\frac{\dot{\phi}(t)}{\phi(t)},
\end{equation}
and
\begin{equation}  \label{wave2}
-\frac{8\pi}{3\phi(t)}\frac{(3\omega pe^{\nu} + 3\rho e^{\nu} + \omega\rho
e^{\nu})}{(2\omega+3)} =\frac{\ddot{a}}{a} -\frac{\dot{a}}{a}\frac{\dot{%
\phi}(t)}{\phi(t)} + \frac{\omega}{3}\left(\frac{\dot{\phi}(t)}{\phi(t)}%
\right)^2.
\end{equation}
Here (\ref{wave2}) was obtained by substituting (\ref{T11}) and (\ref{T44})
into (\ref{wave}) and discarding the terms involving $r$ derivatives, as
these now sum to zero. We see that (\ref{t11}), (\ref{t44}) and (\ref{wave2}%
) are simply (\ref{Nariai1}), (\ref{Nariai2}) and (\ref{Friedmann}) with $%
k=0 $, $p_{FRW}=pe^{\nu }$ and $\rho _{FRW}=\rho e^{\nu }$, where
subscript FRW denotes a quantity derived from the field equations using the FRW
metric. We also have, from ${T^{a b}}_{; b}=0$, that
\begin{equation}  \label{fluid2}
\frac{d}{dt}(\rho e^{\nu}) + 3H(\rho e^{\nu} +p e^{\nu}) = 0.
\end{equation}
Looking for solutions of the form $a\propto t^{x}$ and $\phi (t)\propto t^{y}
$ gives, on substitution into (\ref{t11}), the relation $y=2-3x\gamma $
(assuming an equation of state for the Universe of the form $p=(\gamma
-1)\rho $). Using (\ref{t11}), (\ref{t44}), and this relation then gives the
solutions
\begin{equation}  \label{a2}
a(t)=a_0\left(\frac{t}{t_0}\right)^{\frac{2\omega(2-\gamma)+2}{%
3\omega\gamma(2-\gamma)+4}}
\end{equation}
and
\begin{equation}  \label{phi(t)2}
\phi(t)=\phi_0\left(\frac{t}{t_0}\right)^{\frac{2(4-3\gamma)}{%
3\omega\gamma(2-\gamma)+4}}.
\end{equation}
These are exactly the same as would be expected for the scale factor and the
Brans-Dicke field in a flat FRW universe \cite{Nar68}. The form of $\rho$ is then given
by (\ref{fluid2}) and (\ref{ev2}) as
\begin{equation}
\rho (r,t)=\rho _{0}\left( \frac{a_{0}}{a(t)}\right) ^{3\gamma }\left( \frac{%
1-\frac{c}{2kr}}{1+\frac{c}{2kr}}\right) ^{-2k},  \label{rho(rt)2}
\end{equation}
and $\phi(r,t)$ is given as
\begin{equation}
\phi (r,t)=\phi _{0}\left( \frac{t}{t_{0}}\right) ^{\frac{2(4-3\gamma )}{%
3\omega \gamma (2-\gamma )+4}}\left( \frac{1-\frac{c}{2kr}}{1+\frac{c}{2kr}}%
\right) ^{-2(k^{2}-1)/k}.  \label{phi(rt)2}
\end{equation}

\begin{figure}[ht]
\epsfig{file=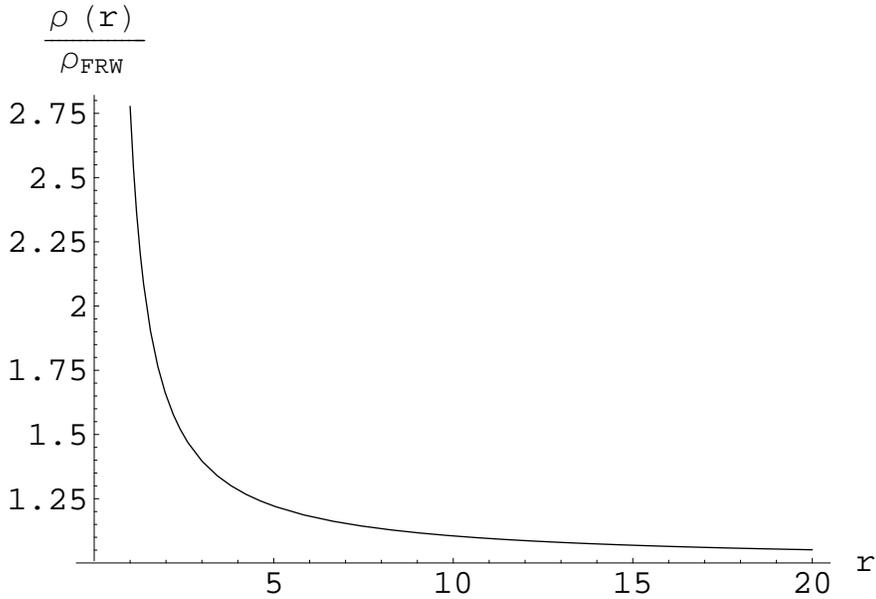,height=8cm}
\caption{{\protect {\textit{Distribution of $\protect\rho $ as a
function of $r$, with $\protect\omega =100$, from equation (\protect\ref{rho(rt)2}%
) and }}}${\protect {\mathit{c=0.5}}}$.}
\label{rho5}
\end{figure}

This separable solution
displays the same time dependence as the power-law FRW Brans-Dicke
universes, (\ref{bds1})-(\ref{bds2}), but with an additional inhomogeneous $%
r$-dependence created by the matter source at $r=0$. Such a distribution of
matter in space is illustrated by figure \ref{rho5}. Here we have chosen, for
illustrative purposes, $\gamma =1$, $\omega =100$ and a background value set
by the choice $\rho =\rho _{FRW}$ ($\rho _{FRW}$ being the matter density that
would be expected in the corresponding homogeneous universe). The temporal
evolution of $\rho $ is the same as the FRW case.

\begin{figure}[ht]
\epsfig{file=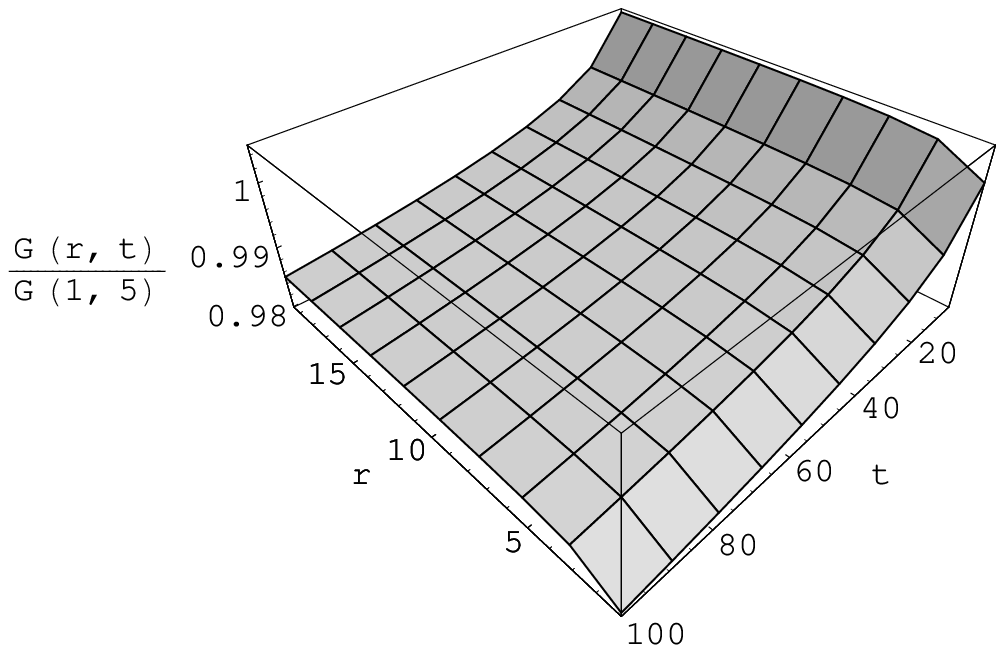,height=12cm}
\caption{{\protect {\textit{Evolution of $G(r,t)$ in space and time in
an inhomogeneous matter dominated Universe, with $\protect\omega =100$, from
equation (\protect\ref{G(rt)}) and }}}$c=0.5.$}
\label{GTB3}
\end{figure}

We see from figure \ref{rho5} that the matter density is isotropic and
asymptotically constant as $r\rightarrow \infty $ with a sharp power-law
peak near the origin. Now (\ref{phi(rt)2}) gives us
\begin{equation}
G(r,t)=G_{0}\left( \frac{1-\frac{c}{2kr}}{1+\frac{c}{2kr}}\right)
^{2(k^{2}-1)/k}t^{-\frac{2(4-3\gamma )}{3\omega \gamma (2-\gamma )+4}}.
\label{G(rt)}
\end{equation}
Equation (\ref{G(rt)}) is used, with the values $\omega =100$, $\gamma =1$
and $c$ $=0.5$ to create figure \ref{GTB3}, which shows the space-time
evolution of $G(r,t)$.

These results show how $G(r,t)$ can vary in space and time in an
asymptotically-flat universe with a peak of matter at the origin. Observers
located near the mass concentration will determine different values of $G$
locally, although they will find the same values of $\dot{G}/G$ everywhere
because of the separable nature of the $G(r,t)$ evolution in
equation (\ref{rho(rt)2}). This was also the case for solution (\ref{phi2}) given in
subsection \ref{vacsol}. In the next section we shall consider a more
realistic model in which both $G$ and $\dot{G}/G$ are different from place
to place. Plots like figure \ref{GTB3} can be generated for universes
dominated by other types of cosmological fluid and with different rates of
density fall off with $r$.

\subsection{Spherical Collapse Model}

\subsubsection{Matching Two Vacuum FRW Brans-Dicke Universes}

\label{vaccollapse}

We will now consider a simple model of a spherically symmetric cosmological
inhomogeneity produced by matching together flat and positively curved
vacuum FRW Brans-Dicke universes. This is a well studied technique, first introduced
by Lema\^{\i}tre, for studying the non-linear evolution of overdensities in
general relativistic FRW universes. The overdense region is modelled as a
closed universe that at first expands more slowly than the background,
before reaching an expansion \ maximum and collapsing back to high density,
whilst the background continues to expand. In this section we consider
vacuum universes only, with $\rho =p=0$.

For flat vacuum FRW universes the Friedmann equation (\ref{Friedmann}) gives

\begin{equation}
\left( \frac{\dot{a}}{a}\right) ^{2}+\frac{\dot{a}}{a}\frac{\dot{\phi _{b}}}{%
\phi _{b}}=\frac{\omega }{6}\left( \frac{\dot{\phi _{b}}}{\phi _{b}}\right)
^{2}  \label{backFried}
\end{equation}%
where $\dot{}=\frac{d}{dt}$, $\phi _{b}=\phi _{b}(t)$ is the Brans-Dicke scalar
field and $a$ is the scale factor in the flat background. For a positively curved ($k=+1$) region the
scale factor is taken to be $S(\tau)$, which satisfies the Friedmann
equation for the closed vacuum Brans-Dicke universe:

\begin{equation}
\left( \frac{S^{\prime }}{S}\right) ^{2}+\frac{S^{\prime }}{S}\frac{\phi
_{p}^{\prime }}{\phi _{p}}=\frac{\omega }{6}\left( \frac{\phi _{p}^{\prime }%
}{\phi _{p}}\right) ^{2}-\frac{k}{S^{2}},  \label{pertFried}
\end{equation}%
where $^{\prime }=\frac{d}{d\tau }$, $\phi _{p}=\phi _{p}(\tau )$ and $\tau $
and $k$ are the proper time and curvature of this perturbed region.

In matching these two regions at $t=t_0=\tau_0$ we must satisfy the boundary
conditions

\begin{equation*}
S(\tau _{0}) =a(t_{0}), \qquad \phi_{p}(\tau_0) =\phi _{b}(t_0)
\end{equation*}
and
\begin{equation*}
\left( \frac{dS}{d\tau }\right) _{0} =\left( \frac{da}{dt}\right) _{0},
\qquad \left( \frac{d\phi _{p}}{d\tau }\right) _{0} =\left(
\frac{d\phi _{b}}{dt}\right)_{0}.
\end{equation*}

\subsubsection{From $\protect\eta$ to $t$}

The function $\tau (\eta )$ is now obtained by integrating $Sd\eta =d\tau $
and $\tau (\eta )$ can then be used to obtain $S(\tau )$. We now require a
relation between $t$ and $\tau $.  For this we proceed as in
reference \cite{Barrow and Kunze} and use the equation of relativistic hydrostatic
equilibrium \cite{Harrison7073,Harrison70731,landau}
\begin{equation}
\frac{\partial \Phi }{\partial r}=-\frac{\partial p/\partial r}{p+\rho }
\label{hydroeq}
\end{equation}%
where $\Phi $ is the Newtonian gravitational potential and $r$ is radial
distance. Equation (\ref{hydroeq}) is derived under the assumptions that the
configuration is static and the gravitational field is weak, so $\Phi $
completely determines the metric; then (\ref{hydroeq}) is given by the
conservation of energy-momentum for a perfect fluid. We now use $d\tau
=e^{\Phi }dt$, and for a scalar field we have an effective equation of
state with $p=\rho$ and $\rho =\frac{\omega }{\phi }\dot{\phi}^{2}$. Combining these
results gives
\begin{equation}
\frac{d\tau}{dt} = \frac {\dot{\phi_b}}{\phi_p^{\prime}} \frac{\phi_p^{1/2}}{%
\phi_b^{1/2}}.  \label{dT/dt}
\end{equation}
Now $\dot{\phi}_{b}\propto a^{-3}$ and $\phi _{p}^{\prime }\propto S^{-3}$,
and so with the solutions (\ref{phi_p(n)}) and (\ref{phi_b(t)}) this gives
\begin{equation}
\frac{d\tau }{dt}=\frac{\sin ^{1/2}(2\sqrt{k}\eta )}{\sin ^{1/2}(2\sqrt{k}%
\eta _{0})}\frac{S^{2}}{S_{0}^{2}}\frac{a^{\sqrt{1+\frac{2}{3}\omega }-4}}{%
a_{0}^{\sqrt{1+\frac{2}{3}\omega }-4}}  \label{dTdt2}
\end{equation}
Equation (\ref{dTdt2}) and the relation $Sd\eta =d\tau $ allow us to obtain $S(t)$
from (\ref{S(n)}). This is done numerically. Now fixing the constants of
proportionality together with $k$ in equations (\ref{S(n)}), (\ref{phi_p(n)}) and
(\ref{phi_b(t)}), in order to satisfy the boundary conditions, we find
equations for the evolution of the scale factors and scalar fields in the
flat background and perturbed region. These are matched at a boundary, at
time $t_{0}=\tau _{0}=\eta _{0}$.

\subsubsection{Results}

\begin{center}
\begin{figure}[tbp]
\epsfig{file=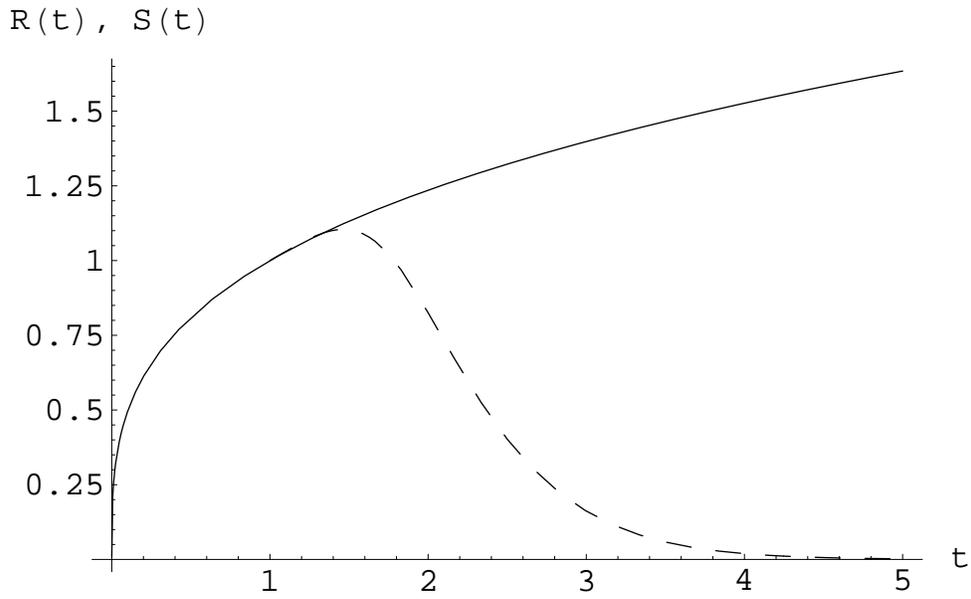,height=8cm}
\caption{{\protect {\textit{Evolution of the scale factor $S$ in the
perturbed overdense region (dashed line) and in the background $a$(solid
line) with respect to the comoving proper time in the flat background.}}}}
\label{RandS}
\end{figure}
\end{center}

Figure \ref{RandS} shows the evolution of $a(t)$ and $S(t)$ when the region
described by $S(t)$ becomes positively curved at initial time $t_{0}=1$. In
figure \ref{RandS} we choose $\omega =100$, for illustrative purposes, and $%
a_{0}=S_{0}=1$ so that the boundary condition for the matching of the first
derivatives of the scale factors is given by $\left( \frac{dS}{d\tau }%
\right) _{0}=\left( \frac{dS}{d\eta }\right) _{0}=\left( \frac{da}{dt}%
\right) _{0}.$

\begin{figure}[tbp]
\epsfig{file=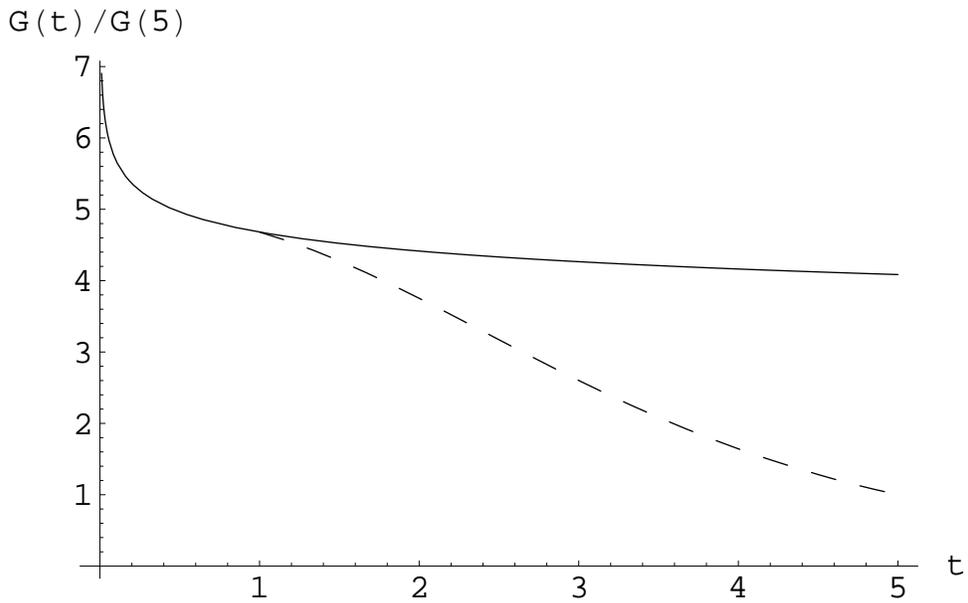,height=8cm}
\caption{{\protect {\textit{Evolution of $G(t)$ in the overdense
perturbed overdense region of positive curvature (dashed line) and in the
spatially flat background universe (solid line).}}}}
\label{FRWG}
\end{figure}

We can now express $G=G(t)$ in the regions of different curvature using
equation (\ref{Gc}) and the flat and closed FRW solutions (\ref{phi_p(n)}) and (\ref{phi_b(t)}), along with the
appropriate coordinate transformations. This gives figure \ref{FRWG}.
It is clearly seen that the evolution of $G(t)$ is quite different in the
two regions, as expected. The collapsing overdensity evolves faster than the
background and possesses a smaller value of $G$ but a larger value of $|\dot{%
G}/G|\ $at all times after the matching. 

\subsection{A More Refined Spherical Collapse Model}
\label{Results}

We now generalise the spherical collapse model described in the last section
to the more realistic case of a flat universe containing
matter and a cosmological constant
(see. e.g. refs. \cite{Padmanabhan}, \cite{peacock} or \cite{lahav}).
As before, we match a flat Brans-Dicke FRW
background to a spherically symmetric overdensity at an appropriate boundary
and allow the two regions to evolve separately.

\subsubsection{The background universe}
\label{background}

Again, we consider a flat $(k=0)$, homogeneous and isotropic background
universe. Since we are interested in the matter-dominated epoch, when
structure formation starts, we can assume that our universe contains only
matter and a vacuum energy contribution so that $\rho =\rho _{m}+\rho
_{\Lambda }$ and $p=p_{\Lambda }=-\rho _{\Lambda }$ give the total density
and pressure, respectively. So, for the flat background, the Friedmann
equation (\ref{Friedmann}) gives for a general $\omega (\phi )$ theory,
\begin{equation}
\left( \frac{\dot{a}}{a}\right) ^{2}+\frac{\dot{a}}{a}\frac{\dot{\phi}}{\phi 
}=\frac{\omega }{6}\frac{\dot{\phi}^{2}}{\phi ^{2}}+\frac{8\pi }{3\phi }%
(\rho _{m}+\rho _{\Lambda })  \label{fried}
\end{equation}%
and the scalar field equation (\ref{Nariai2}) gives
\begin{equation}
\ddot{\phi}+3\frac{\dot{a}}{a}\dot{\phi}=\frac{8\pi }{(2\omega +3)}(\rho
_{m}+4\rho _{\Lambda })-\frac{\dot{\omega}\dot{\phi}}{(2\omega +3)}.
\label{phidot}
\end{equation}
Here, $\rho _{m}\propto a(t)^{-3}$ and $\rho _{\Lambda }=\text{constant}$.
These equations govern the evolution of $\phi (t)$ and $a(t)$ in the flat
expanding cosmological background. The contribution of the vacuum energy
stress ($p=-\rho $) to the Friedmann equation (\ref{fried}) in Brans-Dicke cosmology
differs from that in GR because of the presence of the
variable $\phi $ field: it is not the same as the addition of a cosmological
constant term to the right-hand side of (\ref{fried}). However, with this proviso, we
shall continue to refer to $\Lambda $CDM models in Brans-Dicke theories in
the following sections.

\subsubsection{The overdensity}
\label{perturbation}

Again, we consider a spherical overdense region of radius $S$ and model the
interior space-time as a closed FRW Brans-Dicke universe, ignoring
any anisotropic effects of gravitational instability or collapse. As usual,
we assume there is no shell-crossing; this implies mass conservation inside
the overdensity \cite{Padmanabhan}.  The evolution equations can now be written in a form that ignores the
spatial dependence of the fields.  Now $\rho =\rho _{CDM}+\rho _{\Lambda }$
and $p=p_{\Lambda }=-\rho _{\Lambda }$, where `CDM' corresponds to cold dark
matter, so that the acceleration equation (\ref{acceleration}) gives
\begin{multline}
\ddot{S}-\dot{S}\frac{\dot{\phi _{c}}}{\phi _{c}}=-S\left( \frac{\omega _{c}
}{3}\frac{\dot{\phi _{c}}^{2}}{\phi _{c}^{2}}-\frac{1}{2}\frac{\dot{\omega}_{c}}{(2\omega _{c}+3)}\frac{\dot{\phi}
_{c}}{\phi _{c}}  \right.\\\left.  + \frac{8\pi }{3\phi _{c}}\frac{
(\rho _{CDM}(3+\omega _{c})+\rho _{\Lambda }(3-2\omega _{c}))}{(2\omega
_{c}+3)} \right) \label{rc},
\end{multline}%
while the scalar field equation (\ref{Nariai2}) reduces to
\begin{equation}
\ddot{\phi _{c}}+3\frac{\dot{S}}{S}\dot{\phi _{c}}=\frac{8\pi }{(2\omega
_{c}+3)}(\rho _{CDM}+4\rho _{\Lambda })-\frac{\dot{\omega _{c}}\dot{\phi _{c}%
}}{(2\omega _{c}+3)}  \label{phic}
\end{equation}%
where $S=S(t)$ is the scale factor and $\phi _{c}=\phi _{c}(t)$ is the Brans-Dicke
scalar field in the collapsing region of positive curvature where $\rho
_{CDM}\propto S(t)^{-3}$ and $\rho _{\Lambda }=\text{constant}$. These
equations give the evolution of $\phi _{c}(t)$ and $S(t)$.

We have assumed that the equation of motion of the field inside the cluster
overdensity is described by the local space-time geometry. This means that
the field follows the dark-matter collapse from the beginning of the
cluster's formation. We do not consider this to be fully realistic since
there is expected to be an outflow of energy associated with $\phi $ from
the overdensity to the background universe, as first noticed by Mota
and van de Bruck (\cite{carsten}).  The details of this outflow of
energy and its effect on the collapse can only be determined by a fully
relativistic hydrodynamical calculation, which is beyond the scope of this
study. Nevertheless, at late times during the collapse of the dark matter
(and especially when the density contrast in the dark matter is very large)
the field should no longer feel the effects of the expanding background and
will decouple from it. We are also neglecting the effects of deviations from
spherical symmetry, which grow during the collapse in the absence of
pressure, along with rotation, gravitational tidal interactions between
different overdensities, and all forms of non-linear hydrodynamical
complexity.

\subsubsection{Evolution of the overdensity}

Consider a spherical perturbation in the dark matter fluid with a spatially
constant internal density. Initially, this perturbation is assumed to have a
density amplitude $\delta _{i}>0$ where $|\delta _{i}|\ll 1$. The initial
density of dark matter inside the overdensity is therefore $\rho _{CDM}=\rho
_{m}(1+\delta _{i})$.

Four characteristic phases of the overdensity's evolution can be identified:
\begin{itemize}
\item \textit{Expansion}: we employ the initial boundary condition $\phi
_{c}=\phi $ and assume that at early times the overdensity expands along
with the background.

\item \textit{Turnaround}: for a sufficiently large $\delta _{i},$ gravity
prevents the overdensity from expanding forever; the spherical overdensity
breaks away from the general expansion and reaches a maximum radius.
Turnaround is defined as the time when $S=S_{max}$, $\dot{S}=0$ and $\ddot{S}%
<0$.

\item \textit{Collapse}: the overdensity subsequently collapses ($\dot{S}<0$%
). If pressure and dissipative physics are ignored the overdensity would
collapse to a singularity where the density of matter would tend to
infinity. In reality this singularity does not occur; instead, the kinetic
energy of collapse is transformed into random motions.

\item \textit{Virialisation}: dynamical equilibrium is reached and the
system becomes stationary with a fixed radius and constant energy density.
\end{itemize}

We require our spherical overdensity to evolve from the linear perturbation
regime at high redshift until it becomes non-linear, collapses, and
virialises.

\subsubsection{Virialisation}

In scalar-tensor theories we expect that the gravitational potential will
not be of the standard local $r^{-1}$ form. This requires reconsideration of
the virial condition. According to the virial theorem, equilibrium will be
reached when \cite{Goldstein}
\begin{equation}
T=\frac{1}{2}S\frac{\partial U}{\partial S}  \label{virial}
\end{equation}%
where $T$ is the average total kinetic energy, $U$ is the average total
potential energy and $S$ here denotes the radius of the spherical
overdensity.

The potential energy for a given component $x$ can be calculated from its
general form in a spherical region \cite{landau}
\begin{equation*}
U_{x}=2\pi \int_{0}^{S}\rho _{tot}\Phi _{x}r^{2}dr,
\end{equation*}%
where $\rho _{tot}$ is the total energy density and $\Phi _{x}$ is the
gravitational potential due to the density component $\rho _{x}$.

The gravitational potential $\Phi _{x}$ can be obtained from the weak-field
limit of the field equations (\ref{STfields}). This results in a
Poisson equation where the terms associated to the scalar field can be
absorbed into the definition of the Newtonian constant as \cite{will}
\begin{equation}  \label{Gc}
G_c=\frac{4+2\omega_c(\phi_c)}{3+2\omega_c(\phi_c)}\frac{1}{\phi_c}.
\end{equation}

This results in the usual form for the Newtonian potential
\begin{equation*}
\Phi _{x}\left( s\right) =-2\pi G_{c}\rho _{x}(3\gamma _{x}-2)\left( S^{2}-%
\frac{r^{2}}{3}\right)
\end{equation*}%
where $G_{c}$ is given by equation (\ref{Gc}) and $\gamma _{x}-1$ is $%
p_x/\rho_x$ for the fluid component with density $\rho _{x}$ and pressure $%
p_x$ (appearing due to the relativistic correction to Poisson's equation: $%
\Delta \Phi =4\pi G\left( \rho +3p\right) $).

In $\Lambda $CDM models of structure formation it is entirely plausible to
set $\gamma _{x}=1$ as the energy density of the cosmological constant is
negligible on the virialised scales we are considering \cite{wang,carsten}. The
potential energy associated with a given component $x$ inside the
overdensity is now given by
\begin{equation}
U_{x}=-\frac{16\pi ^{2}}{15}G_{c}\rho _{tot}\rho _{x}S^{5}.
\end{equation}
Therefore, the virial theorem will be satisfied when
\begin{equation*}
T_{vir}=\frac{1}{2}S_{vir}\left( \frac{\partial U}{\partial S}\right) _{vir},
\end{equation*}
where
\begin{equation}
\frac{\partial U}{\partial S}=-\frac{16\pi ^{2}}{15}\left[ \frac{\partial
G_{c}}{\partial S}\rho _{tot}\rho _{x}S^{5}+ G_{c}\frac{\partial \rho _{tot}}{%
\partial S}\rho _{x}S^{5}+G_{c}\rho _{tot}\frac{\partial \rho _{x}}{\partial
S}S^{5}+G_{c}\rho _{tot}\rho _{x}5S^{4}\right] \label{du/dr}
\end{equation}
and we have used $U_{tot}=U_{CDM}+U_{\Lambda }+U_{\phi _{c}}$, $\rho
_{tot}=\rho _{CDM}+\rho _{\Lambda }+\rho _{\phi _{c}}$, $\rho _{\phi
_{c}}=\omega _{c}\dot{\phi}^{2}/\phi $ and
\begin{equation*}
\frac{\partial G_{c}}{\partial S}=\frac{\dot{G}_{c}}{\dot{S}}=-G_{c}\ \dot{%
\phi}_{c}\frac{3+2\omega _{c}}{4+2\omega _{c}}\left( G_{c}+\frac{2\omega
_{c}^{^{\prime }}(\phi _{c})}{(3+2\omega _{c})^{2}}\right) .
\end{equation*}
The other components of equation (\ref{du/dr}) are obtained from equations (\ref{rc})
and (\ref{phic}).

Using equation (\ref{virial}), together with energy conservation at turnaround
and virialisation, we obtain an equilibrium condition in terms of potential
energies only

\begin{equation}
\frac{1}{2}S_{vir}\left( \frac{\partial U}{\partial S}\right)
_{z_{v}}+U_{tot}(z_{v})=U_{tot}(z_{ta}),  \label{virialcond}
\end{equation}%
where $z_{v}$ is the redshift at virialisation and $z_{ta}$ is the redshift
of the over-density at its turnaround radius. The behaviour of $G$ during
the evolution of an overdensity can now be obtained by numerically evolving
the background equations (\ref{fried}) and (\ref{phidot}) and the
overdensity equations (\ref{rc}) and (\ref{phic}) until the virial
condition (\ref{virialcond}) holds.

We point out here an inconsistency when one makes use of equation
(\ref{virialcond}) together with the assumption that energy is conserved.
This inconsistency is removed by assuming a negligible outflow of $\phi $
from the overdensity, in which case we regain energy conservation within the
system and so retain self-consistency.

\subsubsection{Overdensities vs Background}

\subsubsection{Brans-Dicke theory}
\label{bransd}

\begin{figure*}
\begin{center}
\subfigure[$\omega=40000$]{\epsfig{figure=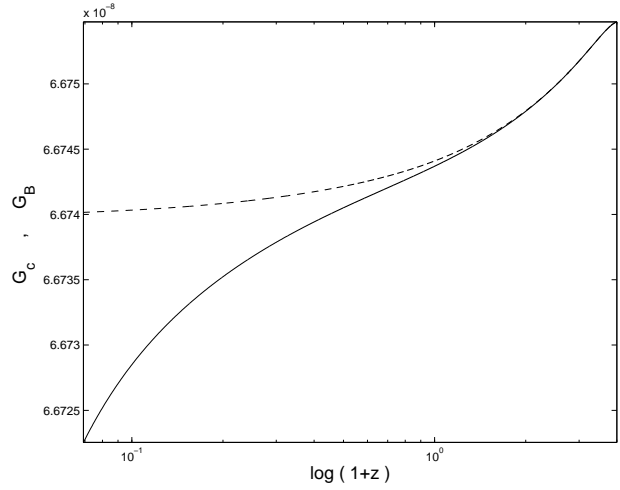, height=6.5cm}}
\subfigure[$\omega=500$]{\epsfig{figure=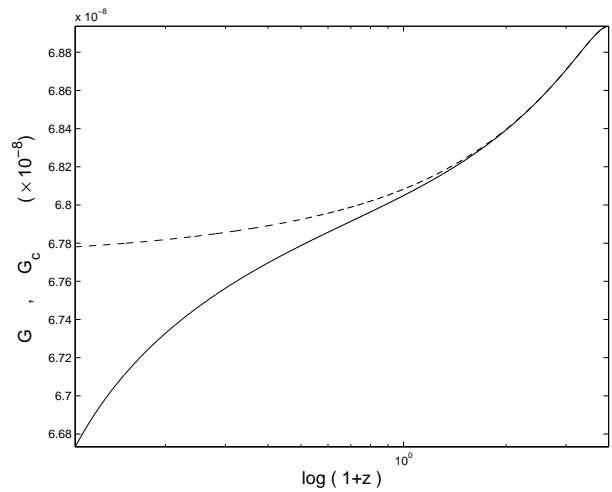, height=6.5cm}}
\end{center}
\caption{{\protect {\textit{Plots of $G$ against $\ln (1+z)$ in the
background (dashed-line) and in an overdensity (solid-line), for different
values of $\protect\omega .$ Initial conditions are chosen in both cases so
as to give $G=G_{0}$, the present value of the Newton constant, at
virialisation.  We note that increasing $\omega$ decreases the
difference in $G$ between the overdensity and the background.}}}}
\label{JBDplot}
\end{figure*}

The Brans-Dicke coupling parameter $\omega $ is constant and constrained by
a variety of local gravitational tests (see \cite{Uzan} for a review). The
strongest constraint to date is derived from observations of the Shapiro
time delay of signals from the Cassini space craft as it passes behind the
Sun. These considerations led Bertotti, Iess and Tortora \cite{Bert},
after a complicated data analysis process, to claim that $\omega $ must have
a value greater than $40000$, to $2\sigma$. This limit on $\omega $ must
be satisfied at all times in all parts of the universe, and leads to the
conclusion that Brans-Dicke theory must be phenomenologically very similar
to GR throughout most of the history of the universe.
However, we do still expect a cosmological evolution of the Brans-Dicke
field $\phi $ which determines the value of Newton's $G$; and we expect this
evolution to be different in regions that collapse to form the structure probed
by Cassini compared to that in the idealised expanding cosmological
background, as described above. Hence we expect the measurable value of $G$
to be different in these two distinct regions with different
histories.

The plots in figure \ref{JBDplot} were constructed using the representative
values $\omega =40000$ and $\omega =500$ and the boundary condition $\phi
_{c0}\simeq G_{0}^{-1}$, so that the value of $G$ measured inside the
overdensity at present is equal to the value of Newton's $G$, as measured
locally. The evolution of the background was determined by matching $\phi
_{c}$ to $\phi $ at the time when the overdensity decouples from the
background, $t_{i}$. We see a clear difference in the evolution of $G$ in
the two regions, as expected.  This example shows that we expect different values of $G$ and $\dot{G}/G$
inside and outside virialised overdensities. The present value of $G$ and $%
\dot{G}/G$ depends on the history of the region where it was sampled,
as well as on the Brans-Dicke coupling parameter, $\omega$.  

It can
be seen from the plots in figure \ref{JBDplot} that increasing
$\omega$ has the effect of decreasing the difference in $G$
between the background universe and the overdensity.  The size of this
inhomogeneity is found to be of order $1/\omega$ and, correspondingly,
reduces to zero as $\omega \rightarrow \infty$.  This is an important
consistency check for the methods used as we expect Brans-Dicke
theory to reduce to GR, with a constant $G$, in this limit.

\subsubsection{Scalar-tensor theory with $2\protect\omega +3=2A\left\vert 1-%
\frac{\protect\phi }{\protect\phi _{\infty }}\right\vert ^{-p}$}
\label{theory1}

\begin{figure*}
\begin{center}
\subfigure[$2\protect\omega (\protect\phi )+3$ for p=2 and A=1, 2
and 5 (solid, dashed and dotted lines, respectively).]{\epsfig{figure=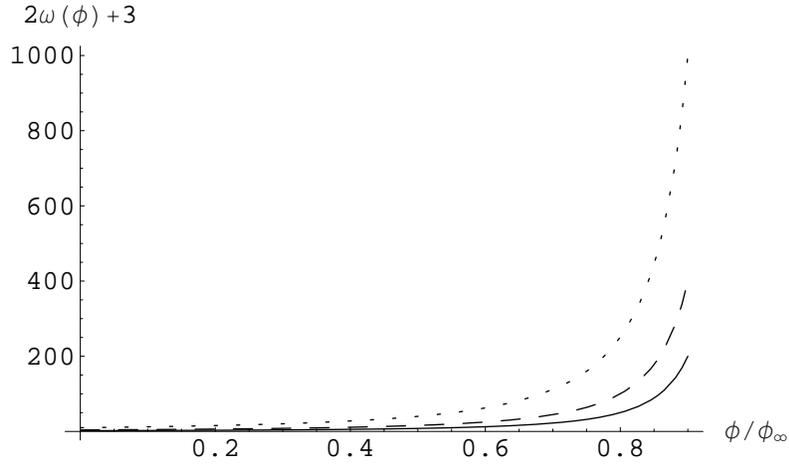, height=6.5cm}}
\subfigure[ $2\protect%
\omega (\protect\phi )+3$ for A=1 and p=1.5, 2 and 2.5 (solid, dashed and
dotted lines, respectively).]{\epsfig{figure=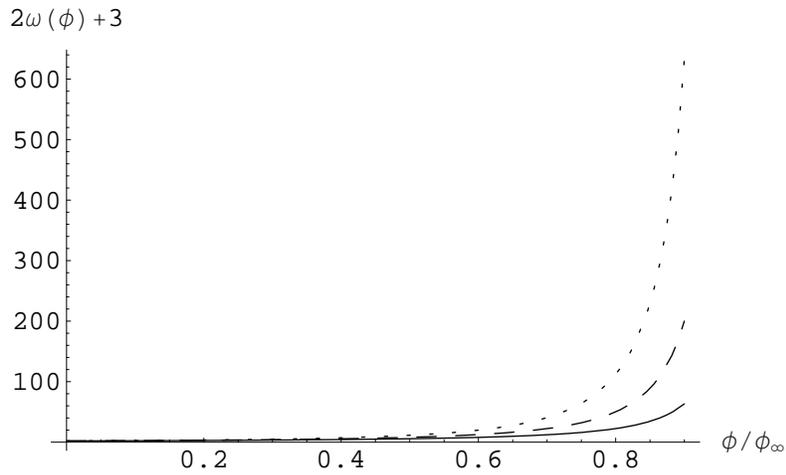, height=6.5cm}}
\end{center}
\caption{In these gravity theories there is fast
approach to general relativity at late times when $\protect\phi \rightarrow 
\protect\phi _{\infty }$, but significantly different behaviour at early
times.}
\label{w(phi)}
\end{figure*}

Next, we consider a scalar-tensor theory with a variable $\omega (\phi )$.
We investigate the class of theories defined by the choice of coupling
function 
\begin{equation*}
2\omega (\phi )+3=2A\left\vert 1-\frac{\phi }{\phi _{\infty }}\right\vert
^{-p},
\end{equation*}
where $A$, $\phi _{\infty }$, and $p$ are positive definite constants. We
refer to this as Theory 1. Such a choice of coupling was considered by
Barrow and Parsons and was solved exactly for the case of a flat FRW
universe containing a perfect fluid \cite{Barrow and Parsons}.

Setting the constants as $2A=\left( \phi _{\infty }/\beta \right) ^{2}$ and $%
p=2$ gives us the scalar-tensor theory considered by Damour and
Pichon \cite{Dam} and by Santiago, Kalligas and Wagoner \cite{San97}. This
choice of $\omega (\phi )$ corresponds to setting $\ln A(\phi )=\ln A(\phi
_{0})+\frac{1}{2}\beta (\phi _{\infty }-\phi )^{2}$, where $A^{2}(\phi )$ is
the conformal factor $1/\phi$ from equation (\ref{conformal}). Damour and
Nordvedt \cite{Dam93} consider this function as a potential and therefore justify
its choice in relation to the generic parabolic form near a potential
minimum. Expecting the function to be close to zero (i.e. GR), Santiago,
Kalligas and Wagoner justify its expression as a perturbative expansion.
This choice of $\omega (\phi )$ with $p>1/2$ corresponds to a general
two-parameter class of scalar-tensor theories that are close to GR and
will be drawn ever closer to it with $\omega \rightarrow \infty $ and $%
\omega ^{\prime }/\omega ^{3}\rightarrow 0$ as the universe expands and $%
\phi \rightarrow \phi _{\infty }$. We therefore consider it as a
representative example of a wide family of plausible varying-$G$ theories
that generalise Brans-Dicke theory.

The evolution of this form of $\omega (\phi )$ is shown graphically in
figure \ref{w(phi)} for different values of $A$ and $p$. Clearly the
evolution of $\omega (\phi )$ is sensitive to both $A$ and $p$ and so the
choice of these parameters is important for the form of the underlying
theory. For illustrative purposes we choose here the values $p=1.5,2$
and $5$ and $A=1,2$ and $5$.

\begin{figure*}
\epsfig{file=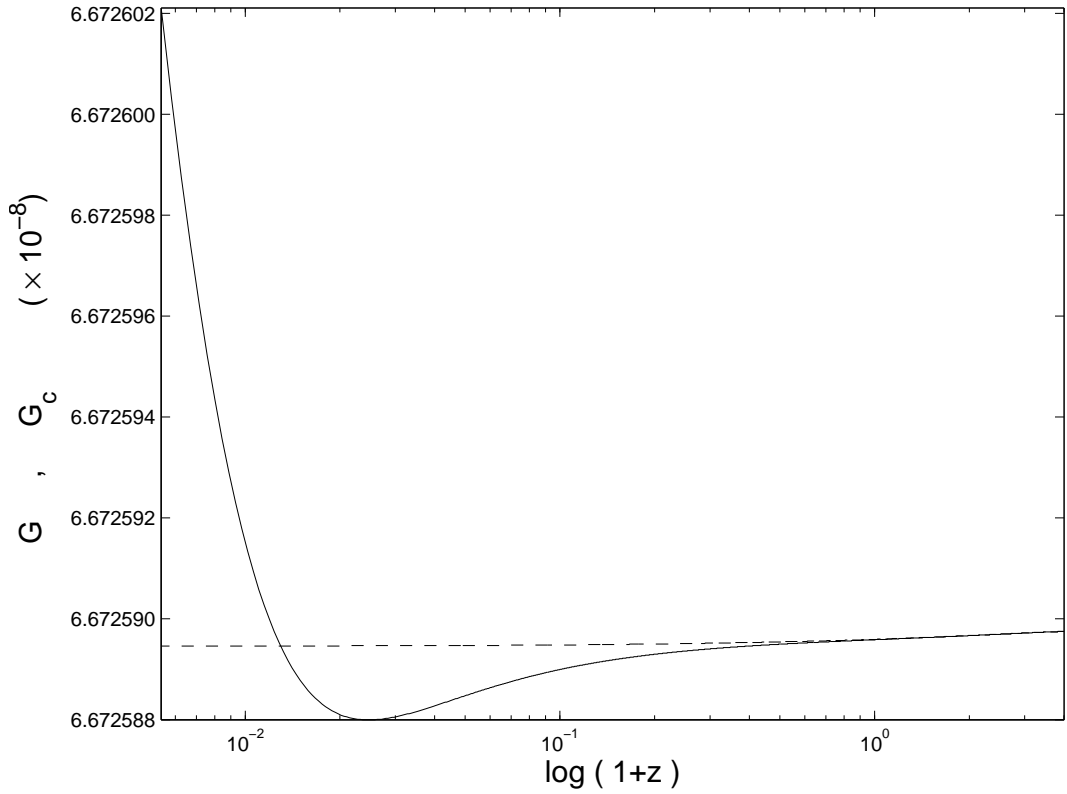,height=11cm}
\caption{{\protect {\textit{{$2\protect\omega +3=2A\left\vert 1-\frac{%
\protect\phi }{\protect\phi _{\infty }}\right\vert ^{-2}$ A graph of $\ $the
variation $G$ against $\ln (1+z)$ for the background universe (dashed-line)
and inside an overdensity (solid-line) which give $G=G_{0},$the presently
observed terrestrial value, at virialisation.}}}}}
\label{w(phi)plot}
\end{figure*}

In a similar way to the Brans-Dicke case we now create an evolution of $\phi_c$
that virialises at $z=0$ to give the value $G_{c0}=6.673\times
10^{-11}$, as observed experimentally.  The corresponding evolution
for $\phi_b$ is calculated as before by matching it to the value of
$\phi_c$ at the time the overdensity decouples from the background and
begins to collapse.  In creating these plots we have used the
conservative parameter values $p=2$, $\omega_{c0} = 1.2 \times10^5$ and 
$A= 6\times10^{-7}$ which are consistent with observation and allow
structure formation to occur in a similar way to GR.
The results of this are plotted in figure \ref{w(phi)plot}.

Again, we note the different evolution of $G(t)$ in the two regions, and the
difference in the asymptotic values of $G$. We note
that experimental measurements of $G$ on Earth have a significant
uncertainty with the 1998 CODATA value carrying an uncertainty 12 times 
\textit{greater} than the standard value adopted in 1987. The 1998 value is
given as \cite{codata, sch}

\begin{equation*}
G_{1998}=6.673\pm 0.010\times 10^{-8}cm^{3}gm^{-1}s^{-2},
\end{equation*}%
while the 2002 CODATA pre-publication announcement reverts to the earlier
higher accuracy consensus with \cite{codata2002}

\begin{equation*}
G_{2002}=6.6742\pm 0.0010\times 10^{-8}cm^{3}gm^{-1}s^{-2}\ 
\end{equation*}

We could re-run the above analysis with different values of $A$ and $p$,
but expect that the results would look qualitatively similar. From figure %
\ref{w(phi)} we see that increasing (decreasing) the values of $A$ and $p$
will increase (decrease) the value of $\omega (\phi )$ for a given $\phi $,
thereby making the theory more (less) like GR. We therefore expect an
analysis with a higher (lower) value of $A$ and/or $p$ to look very similar
to the analysis presented above with a less (more) rapid evolution of $G(t)$%
. For the sake of brevity we omit such an analysis here. \emph{\ }

\subsubsection{Space and Time variations of $G$}

We now calculate how time and space variations of $G$ evolve with redshift
and depend on the dark matter density contrast, $\Delta _{c}$. In order to do this,
we make the definitions
\begin{equation*}
\frac{\Delta G}{G}(t)\equiv \frac{G(t)-G_{0}}{G_{0}},\qquad \frac{\delta G}{G%
}(t)\equiv \frac{G_{c}(t)-G_{b}(t)}{G_{b}(t)}
\end{equation*}
and
\begin{equation*}
\Delta_{c}\equiv \frac{\rho _{CDM}(z_{v})}{\rho _{m}(z_{v})},
\end{equation*}
where $G_{c}$ and $G_{b}$ correspond to $G$ as measured in the overdensity
and in the background universe, respectively.

\begin{figure}
\epsfig{file=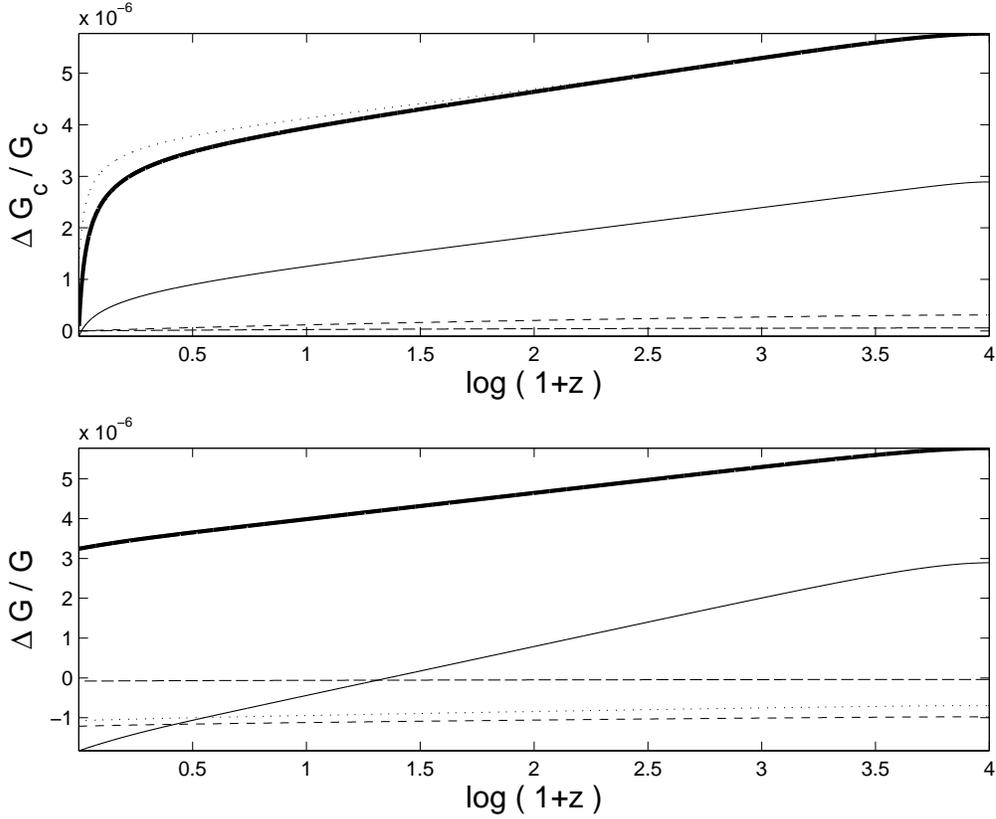,height=11cm}
\caption{{\protect {\textit{Evolution of $\Delta G/G$ as a function of 
$log(1+z)$ for overdensities which virialise at $z=0$ in a $\Lambda $CDM
model. Upper plot: evolution inside the overdensity. Lower plot: evolution
in the background universe. Thick solid line $2\protect\omega +3=4
\times 10^6$,
thin solid line $2\protect\omega +3=B^{2}\protect\phi $,
dashed-line $2\protect\omega +3=2A|1-\frac{\protect\phi }{\protect\phi _{0}}%
|^{-2}$, dash-dotted line $2\protect\omega +3=C|\ln (\frac{\protect\phi }{%
\protect\phi _{0}})|^{-4}$, dotted line $2\protect\omega +3=D|1-(\frac{%
\protect\phi }{\protect\phi _{0}})^{2}|^{-1}$ . Each model is normalised in
order to have $G_{0}=G_{c}(z=0)$ inside the overdensities.}}}}
\label{DeltaG}
\end{figure}

\begin{figure}
\epsfig{file=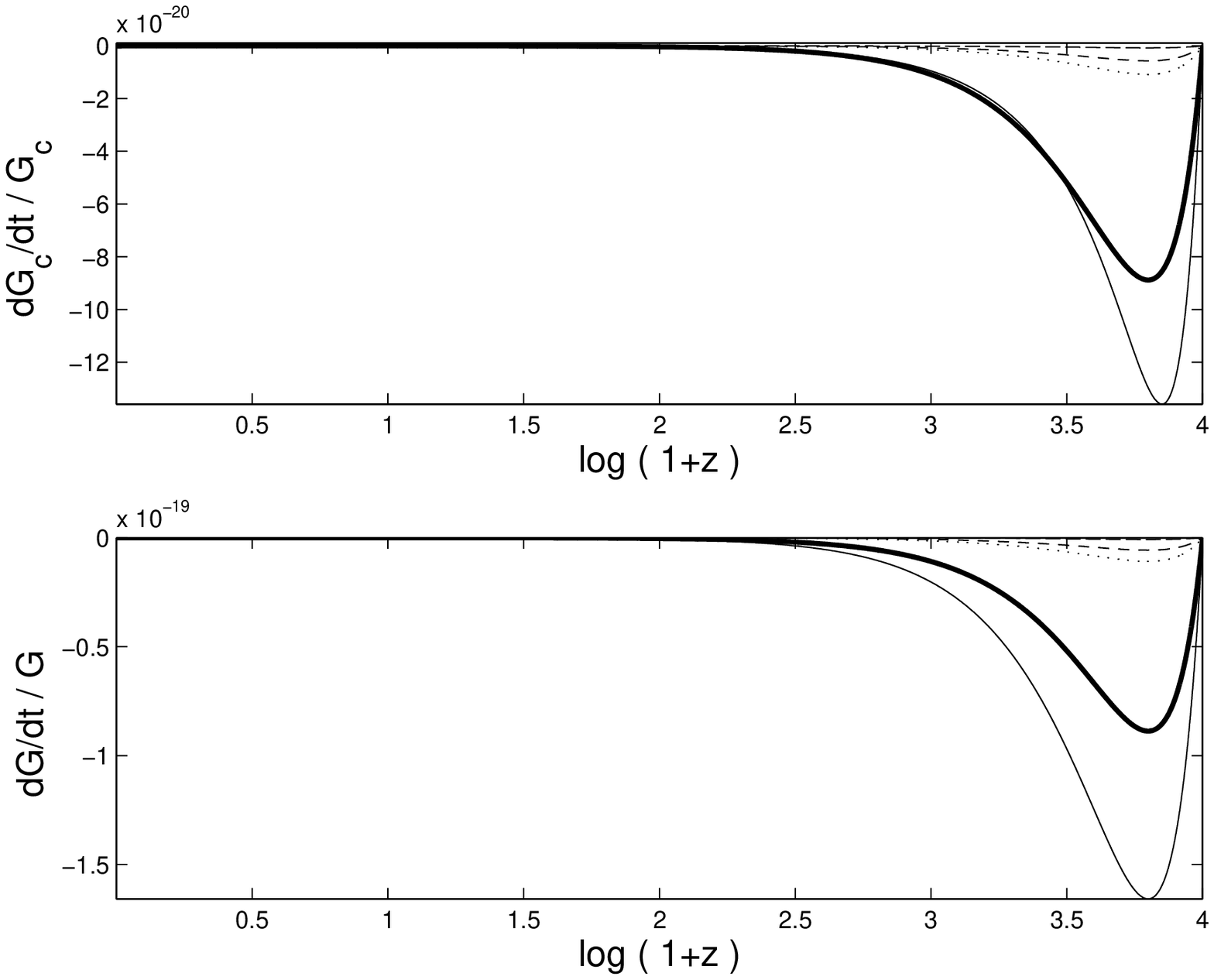,height=11cm}
\caption{{\protect {\textit{Evolution of $\dot{G}/G$ as a function of $%
log(1+z)$ for overdensities which virialise at $z=0$ in a $\Lambda $CDM
model. Upper plot: evolution inside the overdensity. Lower plot: evolution
in the background universe. Thick solid line $2\protect\omega +3=4 \times 10^6$,
thin solid line $2\protect\omega +3=B^{2}\protect\phi $,
dashed-line $2\protect\omega +3=2A|1-\frac{\protect\phi }{\protect\phi _{0}}%
|^{-2}$, dash-dotted line $2\protect\omega +3=C|\ln (\frac{\protect\phi }{%
\protect\phi _{0}})|^{-4}$, dotted line $2\protect\omega +3=D|1-(\frac{%
\protect\phi }{\protect\phi _{0}})^{2}|^{-1}$ . Each model is normalised in
order to have $G_{0}=G_{c}(z=0)$ inside the overdensities.}}}}
\label{dotG}
\end{figure}

\begin{figure}
\epsfig{file=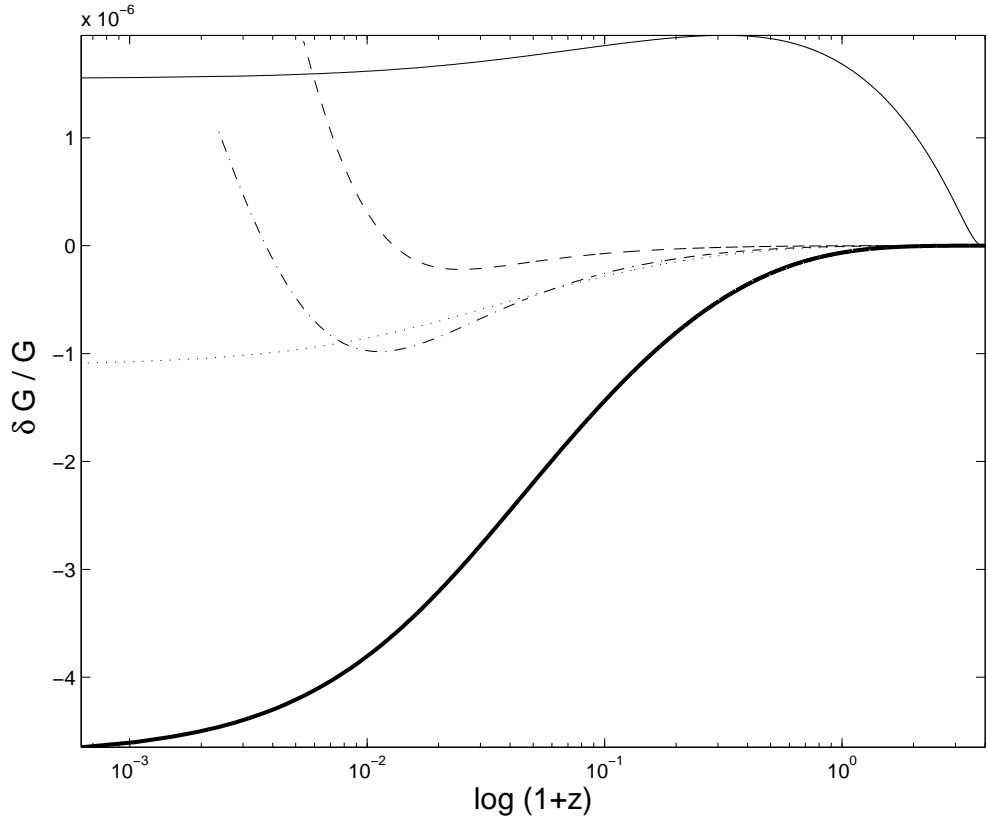,height=11cm}
\caption{{\protect {\textit{Evolution of $\protect\delta G/G$ as a
function of $log(1+z)$ for overdensities which virialise at $z=0$ in a $%
\Lambda $CDM model. Thick solid line $2\protect\omega +3=4 \times 10^6$, thin
solid line $2\protect\omega +3=B^{2}\protect\phi ^{2(p+1)}$, dashed-line $2%
\protect\omega +3=2A|1-\frac{\protect\phi }{\protect\phi _{0}}|^{-2}$,
dash-dotted line $2\protect\omega +3=C|\ln (\frac{\protect\phi }{\protect%
\phi _{0}})|^{-4}$, dotted line $2\protect\omega +3=D|1-(\frac{\protect\phi 
}{\protect\phi _{0}})^{2}|^{-1}$ . Each model is normalised in order to have 
$G_{0}=G_{c}(z=0)$ inside the overdensities.}}}}
\label{deltaG}
\end{figure}

The results of our numerical calculations, for a cluster which virialises at 
$z_{v}=0$, are presented in figures \ref{DeltaG}, \ref{dotG} and \ref{deltaG}%
, respectively. These plots display the evolution of $\dot{G}/G,\Delta G/G,$
and $\delta G/G$ with redshift for Brans-Dicke theory, the theory of subsection \ref{theory1}, and some other
choices of $\omega (\phi )$ that are specified in the captions.  The
parameters used in generating these plots are $B=0.4$, $C=10^{-16}$,
$D=80$, $A=6 \times 10^{-7}$, $p=2$ and $\omega= 4 \times 10^6$.  These values
were chosen so as to agree with observation and so that structure
formation is not significantly different from that which occurs in GR. 
\begin{figure}
\epsfig{file=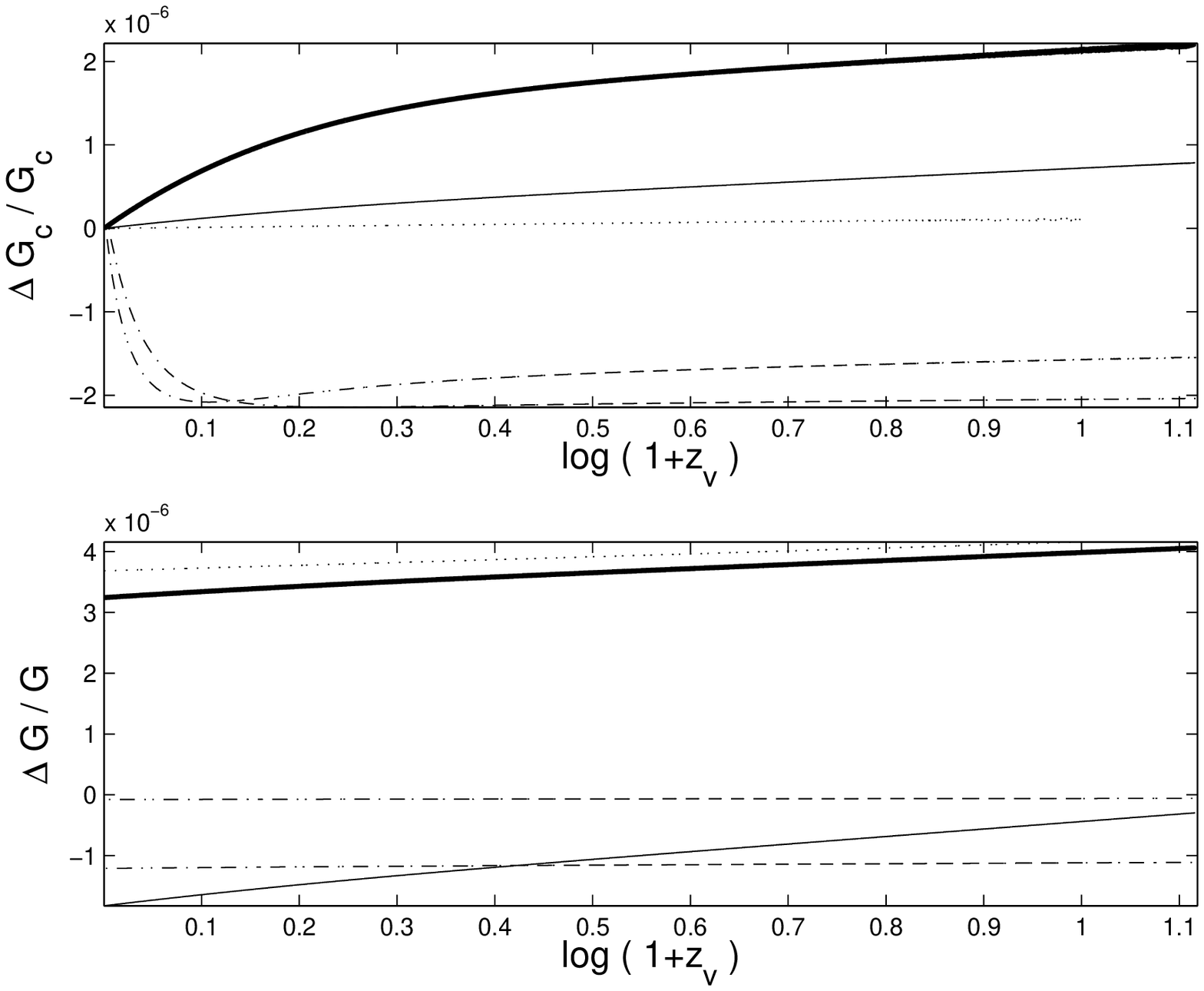,height=11cm}
\caption{{\protect {\textit{Values of $\Delta G/G$ as a function of $%
log(1+z_{v})$ at virialisation. Upper plot: evolution inside the
overdensity. Lower plot: evolution in the background universe. Thick solid
line $2\protect\omega +3=4 \times 10^6$; thin solid line $2\protect\omega +3=B^{2}%
\protect\phi $; dashed-line $2\protect\omega +3=2A|1-\frac{\protect%
\phi }{\protect\phi _{0}}|^{-2}$; dash-dotted line $2\protect\omega +3=C|\ln
(\frac{\protect\phi }{\protect\phi _{0}})|^{-4}$; dotted line $2\protect%
\omega +3=D|1-(\frac{\protect\phi }{\protect\phi _{0}})^{2}|^{-1}$ . Each
model is normalised to have $G_{0}=G_{c}(z=0)$ inside the overdensities.}}}}
\label{DeltaGv}
\end{figure}

It is clear from the plots that different scalar-tensor theories lead to
different variations of $G$. The predictions of these models can be quite
diverse. While some models produce higher values of $G$ inside the
overdensity, others produce a lower one. A feature common to all models is
that the value of $G$ and $\dot{G}/G$ inside an overdensity is different
from $G$ and $\dot{G}/G$ in the background universe. The reason for these
differences is that in the non-linear regime, when the overdensity
decouples from the background expansion at turnaround, the field $\phi $
that drives variations in the Newtonian gravitational `constant' stops
feeling the background expansion. After
turnaround, the field inside the overdensity, $\phi _{c}$, deviates from the
field, $\phi $, in the background universe, leading to spatial variations in 
$G$. In reality, such spatial inhomogeneities in the value of $G$ are small: 
$\delta G/G\approx 10^{-6}$, figure \ref{deltaG}. The time variations of $G$
are even smaller than the spatial inhomogeneities but with a marked
difference between the inside and outside rates of change. We find $\dot{G}%
_{c}/G_{c}\leq 10^{-20}s^{-1}$ inside the clusters and $\dot{G}/G\leq
10^{-19}s^{-1}$ in the background, figure \ref{dotG}.

\begin{figure}
\epsfig{file=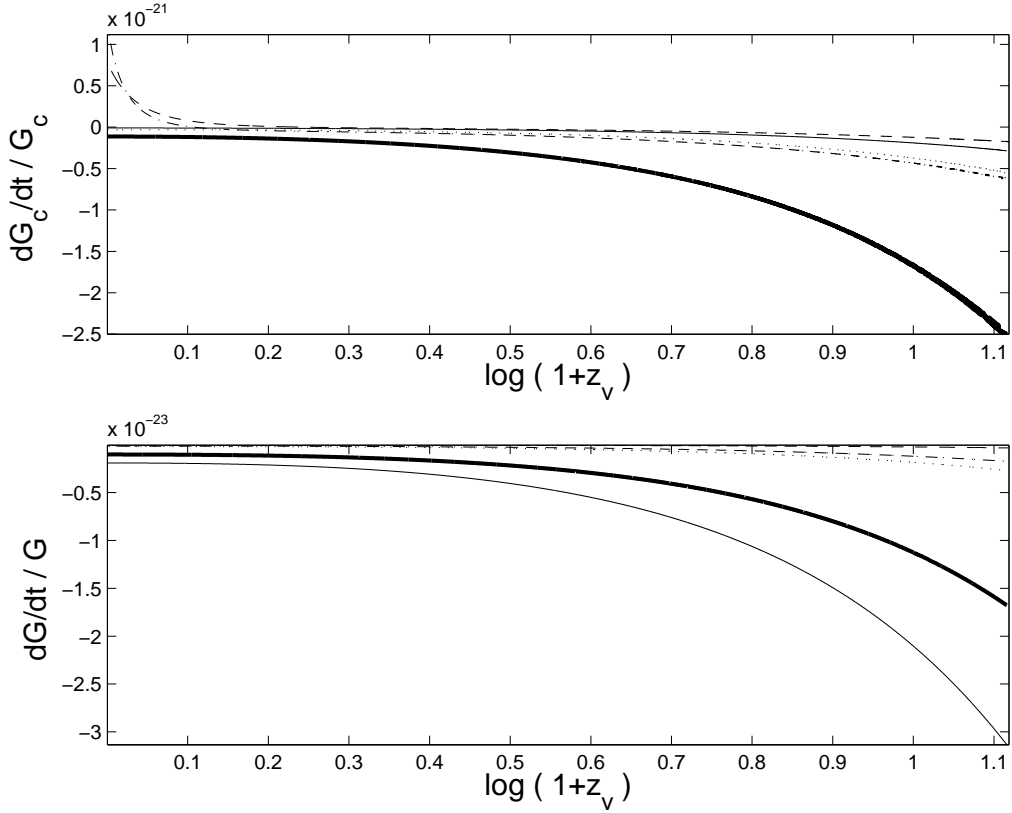,height=11cm}
\caption{{\protect {\textit{Values of $\dot{G}/G$ as a function of $%
log(1+z_{v})$ at virialisation. Upper plot: evolution inside the
overdensity. Lower plot: evolution in the background universe. Thick solid
line $2\protect\omega +3=4 \times 10^6$; thin solid line $2\protect\omega +3=B^{2}%
\protect\phi$; dashed-line $2\protect\omega +3=2A|1-\frac{\protect%
\phi }{\protect\phi _{0}}|^{-2}$; dash-dotted line $2\protect\omega +3=C|\ln
(\frac{\protect\phi }{\protect\phi _{0}})|^{-4}$; dotted line $2\protect%
\omega +3=D|1-(\frac{\protect\phi }{\protect\phi _{0}})^{2}|^{-1}$. Each
model is normalised to have $G_{0}=G_{c}(z=0)$ inside the overdensities.}}}}
\label{dotGv}
\end{figure}

\begin{figure}
\epsfig{file=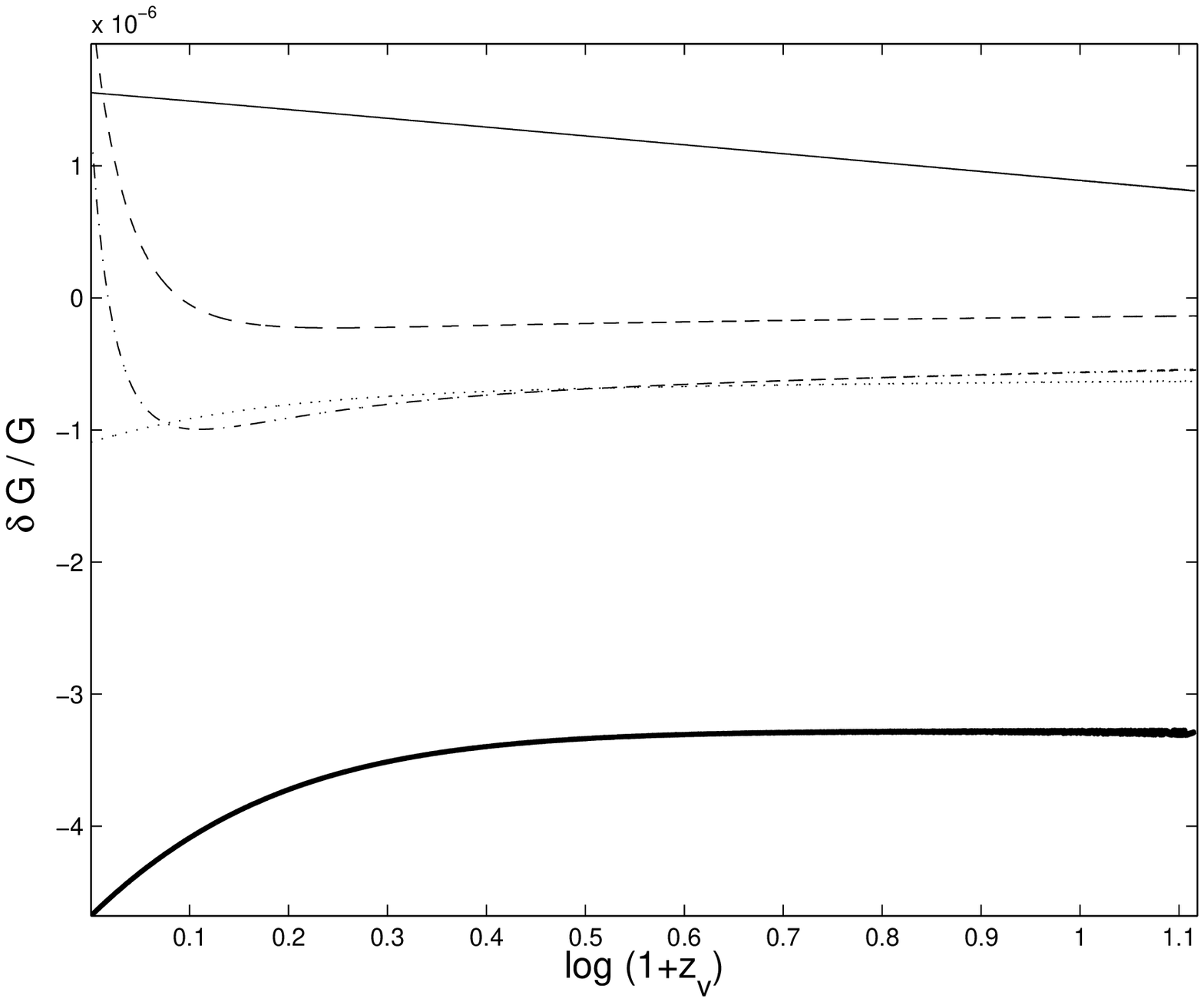,height=11cm}
\caption{{\protect {\textit{Values of $\protect\delta G/G$ as a
function of $log(1+z_{v})$ at virialisation. Thick solid line $2\protect%
\omega +3=4 \times 10^6$; thin solid line $2\protect\omega
+3=B^{2}\protect\phi %
$; dashed-line $2\protect\omega +3=2A|1-\frac{\protect\phi }{%
\protect\phi _{0}}|^{-2}$; dash-dotted line $2\protect\omega +3=C|\ln (\frac{%
\protect\phi }{\protect\phi _{0}})|^{-4}$; dotted line $2\protect\omega %
+3=D|1-(\frac{\protect\phi }{\protect\phi _{0}})^{2}|^{-1}$. Each model is
normalised to have $G_{0}=G_{c}(z=0)$ inside the overdensities.}}}}
\label{deltaGv_1}
\end{figure}

Figures \ref{DeltaGv}, \ref{dotGv} and \ref{deltaGv_1} represent the values of 
$\Delta G/G$, $\delta G/G$ and $\dot{G}/G$ at the redshift of virialisation (%
$z_{v}$). Once again the differences between the different scalar-tensor
theories and between $G_{c}(z_{v})$ and $G(z_{v})$ are evident. 

\begin{figure}
\epsfig{file=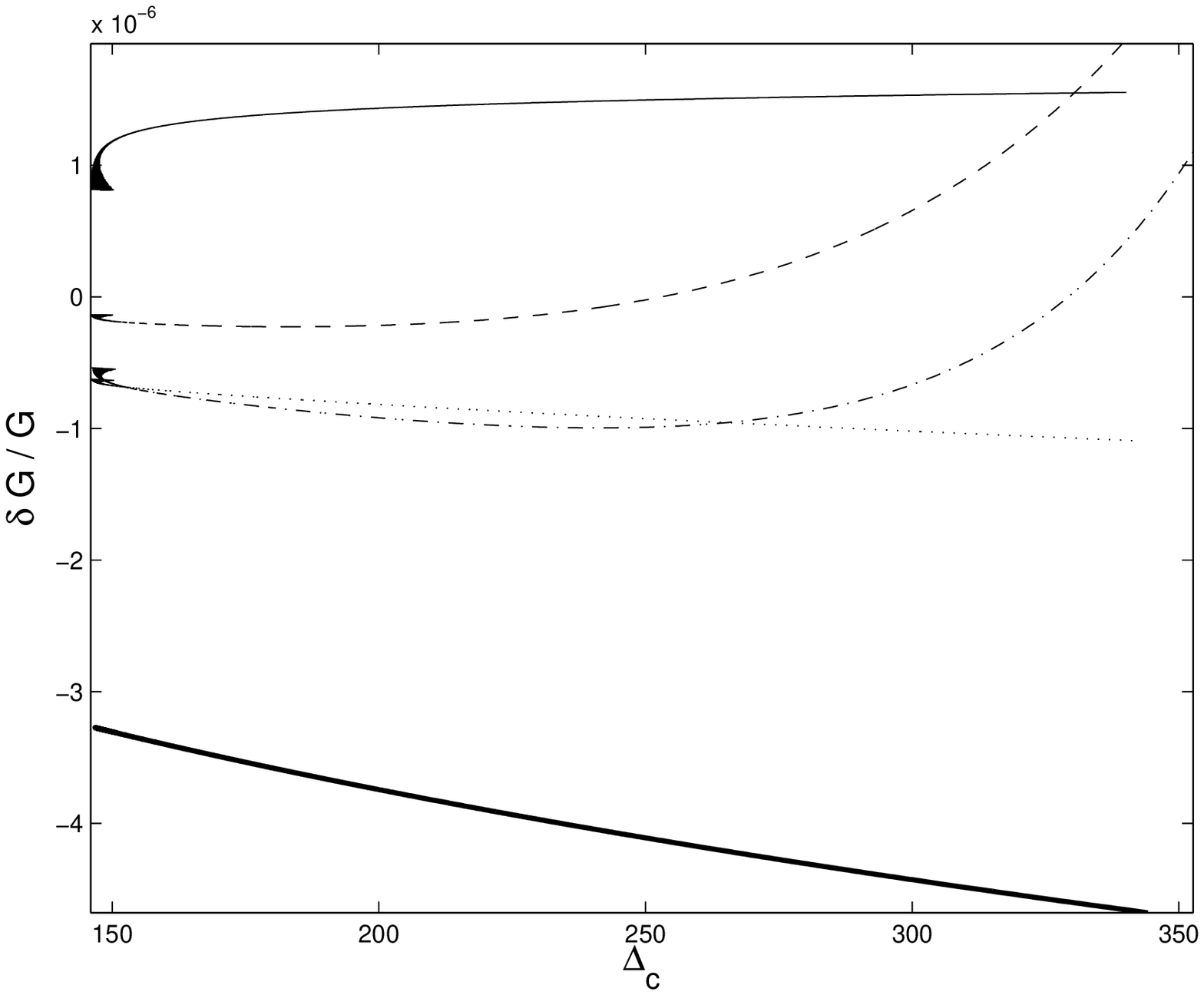,height=11cm}
\caption{{\protect {\textit{Values of $\protect\delta G/G$ as a
function of $\Delta _{c}$ for a $\Lambda $CDM model. Thick-solid line $2%
\protect\omega +3=4 \times 10^6$; solid line $2\protect\omega +3=B^{2}\protect%
\phi $; dashed-line $2\protect\omega +3=2A|1-\frac{\protect\phi }{%
\protect\phi _{0}}|^{-2}$; dash-dotted line $2\protect\omega +3=C|\ln (\frac{%
\protect\phi }{\protect\phi _{0}})|^{-4}$; dotted line $2\protect\omega %
+3=D|1-(\frac{\protect\phi }{\protect\phi _{0}})^{2}|^{-1}$. Each model is
normalised to have $G_{0}=G_{c}(z=0)$ inside the overdensities.}}}}
\label{deltaGrhov}
\end{figure}

Figure \ref{deltaGrhov} shows how the dark matter contrast, $\Delta _{c}\equiv \rho
_{CDM}(z_{v})/\rho _{m}(z_{v})$, affects the difference between the value of 
$G$ inside an overdensity and in the cosmological background. (Recall that,
in the Einstein-de Sitter model $\Delta _{\mathrm{c}}\equiv \rho
_{CDM}(z_{v})/\rho _{m}(z_{v})\approx 147$ at virialisation, and $\Delta _{%
\mathrm{c}}\equiv \rho _{CDM}(z_{v})/\rho _{m}(z_{c})\approx 187$ at
collapse). It is interesting to see that different scalar-tensor theories
produce a different dependence. 

From the numerical simulations, we see that a variation of the Newtonian
gravitational constant, \textit{of the order presented here}, does not
affect the predictions of the structure formation models. For instance, the
virialisation radius and the density contrast $\Delta _{c}$ of the
virialised clusters are very similar to the $\Lambda $CDM and standard CDM
models with constant $G$ . Nevertheless, one should point out that different
scalar-tensor theories lead to slightly different results. Besides that,
there is also a dependence on the initial conditions, as with many other
cosmological scalar fields (see e.g. \cite{carsten,mota1,mota2}). Different initial
conditions will lead to slightly different allowed values of $\omega $ and
to different cosmological behaviours of $G$ and $G_{c}$.

\subsection{Discussion}

We have studied the inhomogeneous cosmological evolution of the Newtonian
gravitational `constant' $G$ within the framework of relativistic
scalar-tensor theories of gravity, of which Brans-Dicke theory is the
simplest and best known case. We began by first exploiting the conformal
equivalence between these theories and GR to transform an
existing solution of Einstein's field equations to a new exact spherically
symmetric inhomogeneous vacuum cosmological solution of the Brans-Dicke
field equations. We then present a second spherically symmetric
perfect-fluid solution to the Brans-Dicke field equations which corresponds
to an overdense inhomogeneity in an asymptotically homogenous  and isotropic
expanding cosmological background. These exact solutions have simple enough
form to allow the $G(r,t)$ evolution to be investigated directly and can be
used to model the presence of a Schwarzschild-like mass in an expanding
Brans-Dicke universe that approaches an idealised FRW universe
asymptotically as $r\rightarrow \infty $. It is noted that far from the mass 
$G$ varies very slowly with $r$ and that as $r\rightarrow \infty $ the
variation of $G$ with $r$ is removed altogether and only the cosmological
evolution of $G$ with time is relevant. Close to the mass, the $r$ variation
of $G$ becomes more significant and we see that $G$ $\rightarrow 0$ at small 
$r$. We also note that in the limit $\omega \rightarrow \infty $ all space
and time variation in $G$ is removed from these solutions and GR is recovered.

Next, we increased the complexity of the model by matching together two
vacuum FRW Brans-Dicke universes of zero and positive curvature on a
spacelike time slice. This results in a simple model for a spherically
symmetric perturbation in the density and curvature that is followed into
the non-linear regime. The different evolutions of $G$ in the two regions
were determined and a comparison was made of the different evolutions of $G$
on spacelike slices of differing time. We see from this toy model that the
value of $G$ does indeed have a different value in regions that have
decoupled from the expanding cosmological background and collapsed, compared
to its value in the background itself. We were then able to repeat this
construction for a more realistic matching of two FRW $\Lambda $CDM
universes of different curvature in an arbitrary scalar-tensor gravity
theory. We followed the development of spherical overdensities through their
expansion, separation from the background, turnaround, and subsequent
collapse. Applying a simple approximation to virialise the collapsing
overdensities, we were then able to study the differences in $G(x^{\mu},t)$
between the non-expanding overdensity and the expanding background universe.
We highlighted as a special case the simple Brans-Dicke theory but also
presented results for other scalar-tensor gravity theories, specified by
their defining coupling function $\omega (\phi )$.  Although each theory
predicts a different detailed cosmological evolution of $G$, a feature
common to all of them is the difference between the value and time
evolution of $G$ inside bound overdensities and in the background universe.
These differences will produce spatial inhomogeneities in $G$ with a value
which depends on the scalar-tensor theory used. While some models produce
higher values of $G$ inside the overdensity, others produce lower ones. In
spite of these differences, such spatial inhomogeneities are small: $\delta
G/G\approx 10^{-6}$. The differences in the time variations of $G$ were
found, with typically $\dot{G}_{c}/G_{c}\leq 10^{-20}s^{-1}$ inside the
clusters and $\dot{G}/G\leq 10^{-19}s^{-1}$ in the background universe in
the range of theories studied. Such variations in $G$ do not significantly
alter the virialisation radius and the density contrast $\Delta _{c}$ of the
virialised clusters from those in standard $\Lambda $CDM and CDM models with
constant $G$. Nevertheless, different scalar-tensor theories lead to
slightly different results. There is also a dependence on the initial
conditions. Different initial conditions will lead to a different value of $%
\omega $ and to different cosmological behaviours of $G$ and $G_{c}$. Taken
as a whole, these analyses show that local observational limits on varying $G$
made within our solar system or Galaxy must be used with caution when
placing constraints upon the allowed cosmological variation of $G$ on
extragalactic scales or in the early universe. The universe is not spatially
homogeneous and, in cosmological models where it can vary, nor is $G$.

\section{Fourth-Order Theories}

It has been known for some time that Birkhoff's theorem is not ensured
in the vacuum solutions of generalised fourth-order theories of
gravity, as they are in GR (see e.g. \cite{Pech, Ste77}).  We will now
present an exact solution that shows this behaviour explicitly; a
non-static, vacuum solution admitting spherically symmetric
three-dimensional space-like hypersurfaces.  A perturbation expansion will be
performed about the background ($r \rightarrow \infty$ limit) of the
exact solution.

\subsection{Exact Solution}

A solution to the spherically symmetric vacuum field equations of $R^n$ theory
(\ref{staticfield}) is given by
\begin{equation}
ds^2= -A_2(r) dt^2 +a^2(t) B_2(r) (dr^2 +r^2 d \Omega^2) \label{Fon}
\end{equation}
where
\begin{align*}
A_2(r)& = \left(
\frac{1-\frac{C_2}{r}}{1+\frac{C_2}{r}}\right)^{\frac{2}{q}}& 
a(t)& = t^{\delta \frac{(1+2 \delta)}{(1-\delta)}}\\
B_2(r)& = \left(1+\frac{C_2}{r}\right)^4 A(r)^{q+2 \delta-1}&
q^2& = 1-2 \delta +4\delta^2
\end{align*}
and $C_2$ is a constant.  This solution is conformally related to one
found by Fonarev \cite{Fonarev} and reduces to the Schwarzschild
solution in the limit $\delta \rightarrow 0$.  It shows explicitly the lack of validity
of Birkhoff's theorem in these theories.  It becomes $r$-independent in the
limit $r \rightarrow \infty$, but still displays strong curvature in
this limit as the metric reduces to the spatially flat vacuum
FRW cosmology found in \cite{Cli}.

We will now continue to find the general form of spherically symmetric
perturbations to the background of the above solution.  It will be seen that there exist extra
modes which are not excited in the exact solutions, but that the modes
corresponding to the linearised exact solutions above are the ones most
interesting for performing gravitational experiments in this
space-time.

\subsection{Linear perturbation analysis}

Perturbative analyses in the literature are often performed about
Minkowski space or de Sitter space.  This is
perfectly acceptable practise in GR and fourth-order theories in which
an Einstein-Hilbert term dominates in the low curvature regime.  In
other fourth-order theories, of the type considered here, in which the
Einstein-Hilbert term does not dominate the low curvature regime then
there is good reason to consider perturbing about other backgrounds.
We have shown explicitly, with exact solutions, the existence of other
spherically symmetric space-times.  We will now proceed to perform a
linear perturbation analysis about the background of solution
(\ref{Fon}).  The general solution to first order in perturbations
will be found.

Writing the perturbed line-element as
\begin{equation}
\label{lin5}
ds^2=-(1+P(r)) dt^2+b^2(t) (1+Q(r)) (dr^2+r^2 d\Omega^2)
\end{equation}
allows the vacuum field equations to be linearised in $P$ and $Q$.
These linearised field equations are
\begin{multline*}
6 \delta (1-2 \delta -2 \delta^2) (5-14 \delta -12 \delta^2) \nabla P
\\-12 (1-2 \delta -2 \delta^2) (1-6 \delta +4 \delta^2+4 \delta^3)
\nabla Q \\+(1+2 \delta) (1-\delta)^2 t^{\frac{2 (1-2 \delta-2
    \delta^2)}{(1-\delta)}} \nabla \psi =0
\end{multline*}
\begin{multline*}
(2- 17 \delta +18 \delta^2 +52 \delta^3 +8 \delta^4) P'' -2 (1
  -\delta +3 \delta^2 -34 \delta^3 -32 \delta^4) \frac{P'}{r} \\+2 (2 -8
  \delta -9 \delta^2 +16 \delta^3 +8 \delta^4) Q''+ 2 (1 -4 \delta -15
  \delta^2 +20 \delta^3 +16 \delta^4) \frac{Q'}{r} \\+3
  (1-\delta)^2 t^{\frac{2 (1-2 \delta-2 \delta^2)}{(1-\delta)}} \left(
  \frac{\psi''}{2 (1+2 \delta)} +\frac{\psi'}{3 r} \right)=0
\end{multline*}
and
\begin{equation*}
\psi'=0
\end{equation*}
where 
\begin{equation*}
\psi=\nabla P + 2 \nabla Q.
\end{equation*}
Primes here denote differentiation with respect to $r$ and $\nabla$
is the Laplacian on the three-dimensional subspace.  These equations
have the general solution
\begin{align*}
P &= -\frac{c_4}{r} +\frac{2 c_5 (1-6 \delta+4 \delta^2+4
  \delta^3)}{(5-14 \delta-12 \delta^2)} r^2+ \text{constant}\\
Q &= \frac{(1-2 \delta) c_4}{r} +\delta c_5 r^2 + \text{constant}
\end{align*}
where $c_4$ and $c_5$ are constants and the two other constant terms
in $P$ and $Q$ are independent of each other and can be absorbed into
$t$ and $s$ by redefinitions.  It can be seen that one of the
modes, $c_4$, corresponds to the linearised version of the exact
solution (\ref{Fon}).

We immediately see that the form of these linearised solutions
are quite different to those obtained by expanding around the
background of the static spherically symmetric solution (\ref{Chan}).  Whist the expansion about
(\ref{Chan}) produces damped oscillatory modes, as well as the mode
corresponding to the linearised exact solution, the expansion about
(\ref{Fon}) produces more familiar looking terms proportional to
$r^2$.  Aside from the different form of these extra modes, there is
also a noticeable difference in the terms corresponding to the
linearised exact solutions, which both go as $r^{-1}$ in the limit
$\delta \rightarrow 0$, but behave differently from each other when
$\delta \neq 0$.  This shows explicitly the differences
that can arise when linearising about different backgrounds.  Not only
are there extra modes which can take different functional forms,
but even the modes which reduce to the Schwarzschild limit as $\delta
\rightarrow 0$ are different, depending on the background that has been chosen.  

\clearemptydoublepage
\chapter{Scalar-Tensor Cosmologies with Energy Exchange}
\label{Energy exchange}

\bigskip

Non-minimally coupled scalar fields arise in a variety of different
theories, including Kaluza-Klein theory \cite{KK, KK2}, string theories
\cite{string} and brane-worlds \cite{branes, branes2}. The same mechanism that creates a
scalar field non-minimally coupled to the curvature in these theories can
also lead to a coupling between the scalar and matter fields. This coupling
manifests itself through the matter Lagrangian becoming a function of $\phi $%
. The possibility of such a coupling is usually neglected in the literature,
where the matter Lagrangian is \textit{a priori} assumed to be independent
of $\phi $. It is the possibility of a coupling between the scalar and
matter fields in scalar-tensor theories that will be the subject of this
chapter.

The introduction of a coupling between $\phi $ and matter greatly enlarges
the phenomenology of the theory. Potentially, this allows greater
variability of $G$ in the early universe whilst still satisfying the solar
system bounds on varying $G$ \cite{Bert}. We will investigate
the extent to which $G$ can vary when energy is exchanged between the $\phi $
field and ordinary matter. As well as giving a window into the
four-dimensional cosmologies associated with higher-dimensional theories, we
hope that this direction of study might be useful in understanding why
the present value of $G$ is so small compared to the proton mass scale ($%
Gm_{pr}^{2}\sim 10^{-39}$). The direct exchange of energy between $\phi $
and the matter fields offers a non-adiabatic mechanism for $G$ to `decay'
towards its present value from a potentially different initial value. There
have been a variety of studies which investigate the drain of energy from
ordered motion by entropy generation, due to bulk viscosity
\cite{bulk, bulk2, bulk3} or
direct decay \cite{Bar04, fr} or energy exchange \cite{tolman}, but few
studies of the drain of energy by non-adiabatic processes from a scalar
field that defines the strength of gravity \cite{p1,p2,p3}. This creates a range of new
behaviours in scalar-tensor cosmologies.

In considering a coupling between $\phi $ and matter we are forced to
reconsider the equivalence principle. The energy-momentum tensor of
perfect-fluid matter fields will no longer be covariantly conserved and the
trajectories of test-particles will no longer follow exact geodesics of the
metric. These violations of the experimentally well verified weak
equivalence principle exclude most possible couplings between $\phi $ and
matter \cite{will}. Such violations are not necessarily fatal though. We
show that whilst energy-momentum is not separately conserved by the matter
fields there is still an exact concept of energy-momentum conservation when
the energy density of the scalar field is included. Furthermore, the
non-geodesic motion of test particles is only problematic if the coupling
increases above experimentally acceptable levels as the Universe ages. The
theory we consider is still a geometric one and it remains true that at any
point on the space-time manifold it is possible to choose normal coordinates
so that it looks locally flat.

This chapter is based on the work of Clifton and Barrow \cite{BDexchange}.

\section{Field Equations}

The conservation of $T_{{}}^{ab}$, whilst appealing, is not absolutely
necessary in deriving a theory in which $G$ can vary. There are numerous
examples where one might expect ${T_{{}}^{ab}}_{;b}\neq 0$. For example,
when considering two fluids the energy-momentum tensor of each fluid is not
separately conserved unless the fluids are completely non-interacting \cite{tim}. It is
only required that the energy-momentum being lost by one of the fluids is
equal to the energy-momentum being gained by the other.

In what follows we will consider the scalar and matter fields as
two fluids (or more than two fluids if there is more than one matter fluid
present) and introduce a transfer of energy and momentum between them. Such
an interaction can be introduced by allowing $\mathcal{L}_{m}$ to be a
function of $\phi $ and will change the nature of the resulting FRW
cosmologies.

The field equations take their usual form in the Brans-Dicke theory
(\ref{STfields}) and the scalar-field propagation equation and matter energy-momentum
conservation equations are now
\begin{align}
\square \phi & =\frac{8\pi T}{(2\omega +3)}-\frac{16\pi \phi }{(2\omega +3)}%
\frac{\sigma ^{a}}{{\phi _{;}}^{a}}  \label{field21} \\
{T^{ab}}_{;b}& =\sigma ^{a}  \label{field3}
\end{align}%
where $T$ is the trace of the energy-momentum tensor and $\sigma ^{a}$ is an
arbitrary vector function of the space-time coordinates $x^{b}$ that
determines the rate of transfer of energy and momentum between the scalar
field $\phi $ and the ordinary matter fields. The precise form of $\sigma
^{a}$ depends on the detailed form of the interaction between the scalar and
matter fields in $\mathcal{L}_{m}$. For example, a conformal transformation
of the form $g_{ab}\rightarrow A^{2}(\phi )g_{ab}$ from a frame in which ${%
T_{\ \ ;b}^{ab}}=0$ gives 
\begin{equation*}
\sigma ^{a}=\frac{T}{A}\frac{dA}{d\phi }{\phi _{;}}^{a}.
\end{equation*}%
This particular choice of energy transfer can be interpreted as a space-time
variation of the rest masses of matter described by $\mathcal{L}_{m}$. For
the moment, we consider the case of more general interactions by leaving $%
\sigma ^{a}$ as an arbitrary function. Later, we will consider specific
forms of $\sigma ^{a}$ that allow direct integration of the field equations.

Substituting the spatially flat FRW metric into the field equations (\ref{STfields}), (\ref%
{field21}) and (\ref{field3}) gives the generalised Friedmann equations: 
\begin{align}
\left( \frac{\dot{V}}{V}\right) ^{2}& =-3\frac{\dot{V}}{V}\frac{\dot{\phi}}{%
\phi }+\frac{3\omega }{2}\left( \frac{\dot{\phi}}{\phi }\right)
^{2}+3(3+2\omega )\alpha \frac{\rho }{\phi }  \label{1} \\
\frac{(\dot{V}\phi )^{\cdot }}{\rho V}& =3\alpha ((2-\gamma )\omega +1)+3%
\frac{\alpha }{\rho }\frac{\phi }{\dot{\phi}}\sigma ^{0}  \label{2} \\
\frac{(V\dot{\phi})^{\cdot }}{\rho V}& =\alpha (4-3\gamma )-2\frac{\alpha }{%
\rho }\frac{\phi }{\dot{\phi}}\sigma ^{0}  \label{3} \\
\dot{\rho}+\gamma \frac{\dot{V}}{V}\rho & =\sigma ^{0}  \label{4}
\end{align}%
where we have defined a comoving volume $V=a^{3}$ and a constant $\alpha =%
\frac{8\pi }{(3+2\omega )}$; the energy-momentum tensor is assumed to be a
perfect barotropic fluid with density $\rho $ and pressure $p$ which are
linked by a linear equation of state $p=(\gamma -1)\rho $, and over-dots
denote differentiation with respect to the comoving proper time, $t$. It is
this set of differential equations that we need to solve in order to
determine the evolution of $a(t)$ and $G\propto \phi (t)^{-1}$ in
cosmological models of this type.

\section{Transfer of Energy and Entropy}

The conservation of energy and momentum, as well as the second law of
thermodynamics, are of basic importance to physics.  In considering an
interaction between a gravitational scalar field $\phi$ and
the matter fields $\Psi$, it is therefore necessary to investigate the extent to which
we can consider energy and momentum to be conserved and the second law
to be obeyed.

When we consider the thermodynamics of an exchange of energy between the
scalar field and matter it is useful to define an effective energy density, $%
\rho _{\phi }$, for the scalar field $\phi $. Defining 
\begin{equation}
\rho _{\phi }\equiv \frac{\dot{\phi}^{2}}{16\pi \omega \phi },  \label{r1}
\end{equation}%
the scalar-field propagation equation (\ref{2}) can then be rewritten as 
\begin{equation}
\dot{\rho}_{\phi }+2\frac{\dot{V}}{V}\rho _{\phi }=-\frac{R}{16\pi }\dot{\phi%
}-\sigma ^{0}.  \label{r2}
\end{equation}%
Comparison of this equation with (\ref{4}) shows that $\phi $ acts as a
fluid with equation of state $\gamma =2$ ($p_{\phi }=\rho _{\phi }$). The
two terms on the right hand side of this equation act as sources for the
energy density $\rho _{\phi }$. The first is the standard Brans-Dicke source
term for the scalar field and the second, $\sigma ^{0}(t)$, is new and
describes the energy exchange between $\phi $ and the matter fields. It can
be seen that the second term is exactly the opposite of the source term in
equation (\ref{1}), and it is in this sense that the total energy is
conserved in this theory.

It is also useful to consider the entropy. Contracting the divergence of the
energy-momentum tensor with the comoving four-velocity $U^{a}$ we obtain 
\begin{align*}
U^{a}\sigma _{a}& =U^{a}{{T_{a}}^{b}}_{;b} \\
& =U^{a}p_{;a}+U^{a}((\rho +p)U_{a}U^{b})_{;b} \\
& =U^{a}p_{;a}-((\rho +p)U^{b})_{:b},
\end{align*}%
where, in the last line, we have used the normalisation $U^{a}U_{a}=-1$ and $%
(U^{a}U_{a})_{;b}={U^{a}}_{;b}U_{a}+U^{a}{U_{a}}_{;b}=0$. Defining the
particle current\ by $N^{a}\equiv nU^{a}$, where $n$ is the number density
in a comoving Lorentz frame, this expression can be rewritten as 
\begin{align*}
U^{a}\sigma _{a}& =U^{a}\left[ p_{;a}-n\left( \frac{(\rho +p)}{n}\right)
{}_{;a}\right] -\frac{(\rho +p)}{n}{N^{a}}_{;a} \\
& =-nU^{a}\left[ p\left( \frac{1}{n}\right) {}_{;a}+\left( \frac{\rho }{n}%
\right) {}_{;a}\right] ,
\end{align*}%
where we have used the conservation of particle number, ${N^{a}}_{;a}=0$.
Recalling the first law of thermodynamics, 
\begin{equation*}
\Theta dS=pdV+dE=pd\left( \frac{1}{n}\right) +d\left( \frac{\rho }{n}\right)
,
\end{equation*}%
where $\Theta $ is the temperature and $S$ is the entropy, we now get 
\begin{equation*}
U^{a}\sigma _{a}=-n\Theta U^{a}S_{;a}
\end{equation*}%
or, making use of our assumption of spatial homogeneity, 
\begin{equation*}
\dot{S}=\frac{\sigma ^{0}}{n\Theta }.
\end{equation*}%
This tells us that, as energy is transferred from $\phi $ to the matter
fields, the entropy of the matter fields increases, as expected. Conversely,
the matter fields can decrease their entropy by transferring energy into $%
\phi $.

Unfortunately, there is currently no known way of defining
the entropy of a non-static gravitational field so it is not possible to perform
an explicit calculation of the entropy changes in $\phi$ and $g_{a b}$.  We can
only assume that if the Universe can be treated as a closed system, and
the exchange of energy is an equilibrium process, then the entropy
that is lost or gained by the matter through this exchange will be
gained or lost by the gravitational fields.  This direct interaction
of the matter with $\phi$ then allows an additional mechanism for
increasing or decreasing the entropy of the matter content of the Universe.

\section{General Solutions}

It is convenient to define a new time coordinate $\tau $ by

\begin{equation}
d\tau \equiv \rho Vdt  \label{tau}
\end{equation}
and to re-parametrise the arbitrary function $\sigma ^{0}$ by 
\begin{equation}
\sigma ^{0}=\rho ^{2}V\frac{\phi ^{\prime }}{\phi }\lambda ^{\prime }
\label{sig}
\end{equation}%
where a prime denotes differentiation with respect to $\tau $ and $\lambda
(\tau )$ is a new arbitrary function.  This re-parameterisation of the
interaction is chosen to
enable a direct integration of the field equations and does not imply
any loss of generality, as $\lambda$ is an arbitrary function.  The field
equations (\ref{2}) and (\ref{3}) can now be integrated to 
\begin{align}
\rho \phi VV^{\prime }& =3\alpha ((2-\gamma )\omega +1)(\tau -\tau
_{1})+3\alpha \lambda  \label{V'} \\
\rho V^{2}\phi ^{\prime }& =\alpha (4-3\gamma )\tau +\alpha \tau
_{2}-2\alpha \lambda  \label{phi'}
\end{align}%
where $\tau _{1}$ and $\tau _{2}$ are constants of integration. We have a
freedom in where we define the origin of $\tau $ and can, therefore, absorb
the constant $\tau _{1}$ into $\tau $ and the definition of $\tau _{2}$ by
the transformations $\tau \rightarrow \tau +\tau _{1}$ and $\tau
_{2}\rightarrow \tau _{2}-(4-3\gamma )\tau _{1}$. It can now be seen from (%
\ref{phi'}) that $\phi ^{\prime }$ is sourced by three terms. The first
corresponds to the source term in (\ref{field3}) and can be seen to
disappear for $\gamma =4/3$, as expected for black-body radiation. The
second term is constant and is the contribution of the free scalar to the
evolution of $\phi $; it is this term which distinguishes the general
spatially-flat Brans-Dicke FRW solutions \cite{Gur73} from the power-law
late-time attractor solutions \cite{Nar}. The third term is new and gives
the effect of the energy transfer on the evolution of $\phi $. This term is
dependent on the arbitrary function $\lambda $, which specifies the
interaction between $\phi $ and the matter fields.

The problem is now reduced to solving the coupled set of first-order
ordinary differential equations (\ref{V'}) and (\ref{phi'}) with the
constraint equation (\ref{4}). The remaining equation (\ref{1}) is rewritten
in terms of $\tau $ and $\lambda $ as 
\begin{equation}
\frac{\rho ^{\prime }}{\rho }+\gamma \frac{V^{\prime }}{V}=\lambda ^{\prime }%
\frac{\phi ^{\prime }}{\phi },  \label{rho}
\end{equation}%
and can be solved for $\rho$ once $V$ and $\phi $ have been found for some $%
\lambda $.

We can decouple the set of equations (\ref{V'}) and (\ref{phi'}) by
differentiating (\ref{phi'}) and substituting for (\ref{V'}) to get the
second-order ordinary differential equation 
\begin{align*}
& \left( (4-3\gamma )\tau +\tau _{2}-2\lambda \right) \frac{\phi ^{\prime
\prime }}{\phi }-\left( (4-3\gamma )-2\lambda ^{\prime }\right) \frac{\phi
^{\prime }}{\phi } \\
=& -\left[ ((4-3\gamma )\tau +\tau _{2}-2\lambda )\lambda ^{\prime
}+3(2-\gamma )(((2-\gamma )\omega +1)\tau +\lambda )\right] \left( \frac{%
\phi ^{\prime }}{\phi }\right) ^{2}
\end{align*}%
which can be integrated to 
\begin{equation}
\frac{\phi ^{\prime }}{\phi }=\frac{(4-3\gamma )\tau +\tau _{2}-2\lambda }{%
(A\tau ^{2}+B\tau +C)},  \label{phi'2}
\end{equation}%
where 
\begin{align*}
A& =3\gamma ^{2}\omega /2-3\gamma (1+2\omega )+(5+6\omega ) \\
B& =\tau _{2}+(4-3\gamma )\lambda \\
C& =-\lambda ^{2}+\lambda \tau _{2}+D
\end{align*}%
and $D$ is a constant of integration. The three source terms for $\phi
^{\prime }$ appear in the numerator on the right-hand side of equation (\ref%
{phi'2}). The equations (\ref{V'}) and (\ref{phi'}) can now be combined to
give $a^{\prime }/a$ in terms of $\phi ^{\prime }/\phi $ as 
\begin{equation}
\frac{a^{\prime }}{a}=\frac{((2-\gamma )\omega +1)\tau +\lambda }{(A\tau
^{2}+B\tau +C)}  \label{V'2}
\end{equation}%
where $A$, $B$, $C$ and $D$ are defined as before. The constant $D$ can be
set using the constraint equation (\ref{4}). Using (\ref{V'}) and (\ref{V'2}%
) we obtain the expression 
\begin{equation*}
\rho V^{2}\phi =\alpha (A\tau ^{2}+B\tau +C).
\end{equation*}%
This can then be substituted into (\ref{4}) which, in terms of $\tau $ and $%
a $, gives the generalised Friedmann equation: 
\begin{equation*}
3\left( \frac{a^{\prime }}{a}\right) ^{2}+3\frac{a^{\prime }}{a}\frac{\phi
^{\prime }}{\phi }-\frac{\omega }{2}\left( \frac{\phi ^{\prime }}{\phi }%
\right) ^{2}=\frac{(3+2\omega )}{(A\tau ^{2}+B\tau +C)}.
\end{equation*}%
Substituting (\ref{phi'2}) and (\ref{V'2}) into this, we find that 
\begin{equation*}
D=-\frac{\tau _{2}^{2}\omega }{2(3+2\omega )}.
\end{equation*}

\section{Particular Solutions}

If $\lambda $ is specified in terms of $\tau$, we now have a set of two
decoupled first-order ordinary differential equations for the two variables $%
a$ and $\phi $. It is the solution of these equations, for specific choices
of $\lambda (\tau )$, that we give in this section.

\subsubsection{$\boldsymbol{\protect\lambda (\protect\tau)= c_1+c_2 \protect\tau}$}

A simple form for $\lambda $ that allows direct integration of equations (%
\ref{phi'2}) and (\ref{V'2}) is the linear function $\lambda
=c_{1}+c_{2}\tau $. From equations (\ref{V'}) and (\ref{phi'}) it can be
seen that the constant $c_{1}$ can be absorbed into $\tau _{1}$ and $\tau
_{2}$ by simple redefinitions. The equations (\ref{phi'2}) and (\ref{V'2})
then become 
\begin{align*}
\frac{\phi ^{\prime }}{\phi }& =\frac{(4-3\gamma -2c_{2})\tau +\tau _{2}}{(%
\hat{A}\tau ^{2}+\hat{B}\tau +\hat{C})} \\
\frac{a^{\prime }}{a}& =\frac{((2-\gamma )\omega +1+c_{2})\tau }{(\hat{A}%
\tau ^{2}+\hat{B}\tau +\hat{C})}
\end{align*}%
where $\hat{A}=A-c_{2}^{2}$, $\hat{B}=B+c_{2}\tau _{2}$ and $\hat{C}=D$. The
solutions of these equations depend upon the sign of the discriminant 
\begin{equation}
\Delta =\hat{B}^{2}-4\hat{A}\hat{C}.  \label{del}
\end{equation}%

For the case $\Delta =0,$ there exist simple exact power-law solutions 
\begin{align}
a(\tau )& \propto \tau ^{\frac{2(2-\gamma )\omega +2+2c_{2}}{3\gamma
^{2}\omega -6\gamma (1+2\omega )+2(5+6\omega )-2c_{2}^{2}}}  \label{powera}
\\
\phi (\tau )& \propto \tau ^{\frac{2(4-3\gamma )-4c_{2}}{3\gamma ^{2}\omega
-6\gamma (1+2\omega )+2(5+6\omega )-2c_{2}^{2}}}.  \label{powerphi}
\end{align}%
Substituting these power-law solutions into (\ref{rho}), we can obtain the
corresponding power-law form 
\begin{equation*}
\rho \sim \tau ^{\frac{4c_{2}(2-3\gamma -c_{2})-6\gamma (1+(2-\gamma )\omega
)}{3\gamma ^{2}\omega -6\gamma (1+2\omega )+2(5+6\omega )-2c_{2}^{2}}}.
\end{equation*}%
The relationship between $\tau $ and the cosmological time $t$ can now be
obtained by integrating the definition $d\tau =\rho a^{3}dt$ given in eq. (%
\ref{tau}). This gives (\ref{powera}) and (\ref{powerphi}) in terms of $t$ time as 
\begin{align}
a(t)& \sim t^{\frac{2+2(2-\gamma )\omega +2c_{2}}{4+3\gamma \omega (2-\gamma
)-2c_{2}(7-6\gamma -c_{2})}}  \label{power1} \\
\phi (t)& \sim t^{\frac{2(4-3\gamma )-4c_{2}}{4+3\gamma \omega (2-\gamma
)-2c_{2}(7-6\gamma -c_{2})}}. \label{power2}
\end{align}%
The condition required for the occurrence of power-law inflation is obtained
by requiring the power of time in equation (\ref{power2}) to exceed unity (for
the case with out energy transfer see refs. \cite{maeda, stein}).  For $\omega
>-3/2$ we always have $\Delta \geqslant 0,$ and the case $\Delta >0$
possesses the exact solutions 
\begin{align}
a(\tau )& =a_{0}(\hat{A}\tau ^{2}+\hat{B}\tau +\hat{C})^{\frac{(2-\gamma
)\omega +1+c_{2}}{2\hat{A}}}\left( \frac{2\hat{A}\tau +\hat{B}+\sqrt{\Delta }%
}{2\hat{A}\tau +\hat{B}-\sqrt{\Delta }}\right) ^{\frac{\hat{B}((2-\gamma
)\omega +1+c_{2})}{2\hat{A}\sqrt{\Delta }}}  \label{a+}
\\
\phi (\tau )& =\phi _{0}(\hat{A}\tau ^{2}+\hat{B}\tau +\hat{C})^{\frac{%
4-3\gamma -2c_{2}}{2\hat{A}}}\left( \frac{2\hat{A}\tau +\hat{B}+\sqrt{\Delta 
}}{2\hat{A}\tau +\hat{B}-\sqrt{\Delta }}\right) ^{\frac{(4-3\gamma -2c_{2})%
\hat{B}-\tau _{2}\hat{A}}{2\hat{A}\sqrt{\Delta }}}.  \label{phi+}
\end{align}%
where $a_{0}$ and $\phi _{0}$ are constants of integration.  For
$\omega <-3/2$ we have $\Delta \leqslant 0$, and the case $\Delta <0$ has the exact
solutions 
\begin{multline}
a(\tau ) =a_{0}(\hat{A}\tau ^{2}+\hat{B}\tau +\hat{C})^{\frac{(2-\gamma
)\omega +1+c_{2}}{2\hat{A}}} \\ \times \exp \left[ -\frac{((2-\gamma )\omega +1+c_{2})%
\hat{B}}{\hat{A}\sqrt{-\Delta }}\tan ^{-1}\left( \frac{\hat{B}+2\hat{A}\tau 
}{\sqrt{-\Delta }}\right) \right]   \label{a-}
\end{multline}
\begin{multline}
\phi (\tau ) =\phi _{0}(\hat{A}\tau ^{2}+\hat{B}\tau +\hat{C})^{\frac{%
4-3\gamma -2c_{2}}{2\hat{A}}} \\ \times \exp \left[ \frac{2\tau _{2}\hat{A}-2(4-3\gamma
-2c_{2})\hat{B}}{\hat{A}\sqrt{-\Delta }}\tan ^{-1}\left( \frac{\hat{B}+2\hat{%
A}\tau }{\sqrt{-\Delta }}\right) \right] .  \label{phi-}
\end{multline}%
These solutions have the same functional form as those found by Gurevich,
Finkelstein and Ruban \cite{Gur73} for Brans-Dicke theory, in the
absence of energy exchange ($\lambda =$ constant), and reduce to them in the limit $%
c_{2}\rightarrow 0$. The behaviour of these solutions at early and late
times will be discussed in the next section.

\subsubsection{$\boldsymbol{\protect\lambda (\protect\tau)=c_3 \protect\tau^n}$, $\boldsymbol{n \neq 1}$}

We now consider forms of $\lambda (\tau )$ that are more general than a
simple linear function of $\tau $. Making the choice $\lambda =c_{3}\tau
^{n} $, where $n\neq 1$ and $c_{3}$ is constant, and setting the free scalar
component to zero ($\tau _{2}=0$), we find that (\ref{phi'2}) and (\ref{V'2}%
) can be integrated exactly. The form of the solutions again depends upon
the roots of the denominator. For real roots we require $\omega >-3/2$, for
which we find the solutions 
\begin{multline}
a(\tau ) =a_{0}\tau ^{\frac{2+2(2-\gamma )\omega }{\kappa }}\left[ \pm
2c_{3}\tau ^{n-1}\mp (4-3\gamma )\pm (2-\gamma )\sqrt{3(3+2\omega )}\right]
^{-\frac{3+3(2-\gamma )\omega -\sqrt{3(3+2\omega )}}{3\kappa (n-1)}}
\label{na+} \\
\times \left[ \pm 2c_{3}\tau ^{n-1}\mp
(4-3\gamma )\mp (2-\gamma )\sqrt{3(3+2\omega )}\right] ^{-\frac{3+3(2-\gamma
)\omega +\sqrt{3(3+2\omega )}}{3\kappa (n-1)}}
\end{multline}
\begin{multline}
\phi (\tau ) =\phi _{0}\tau ^{\frac{2(4-3\gamma )}{\kappa }}\left[ \pm
2c_{3}\tau ^{n-1}\mp (4-3\gamma )\pm (2-\gamma )\sqrt{3(3+2\omega )}\right]
^{-\frac{(4-3\gamma )+(2-\gamma )\sqrt{3(3+2\omega )}}{\kappa (n-1)}}
\label{nphi+} \\
\times \left[ \pm 2c_{3}\tau ^{n-1}\mp
(4-3\gamma )\mp (2-\gamma )\sqrt{3(3+2\omega )}\right] ^{-\frac{(4-3\gamma
)-(2-\gamma )\sqrt{3(3+2\omega )}}{\kappa (n-1)}} 
\end{multline}%
where $\kappa \equiv 2(5-3\gamma )+3(2-\gamma )^{2}\omega $. For a
denominator with imaginary roots we require $\omega <-3/2$, for which we
find 
\begin{multline}
a(\tau ) =a_{0}\left[ \pm 2(4-3\gamma )c_{3}\tau ^{1-n}\pm \kappa \tau
^{2(1-n)}\mp 2c_{3}^{2}\right] ^{\frac{1+(2-\gamma )\omega }{\kappa (1-n)}}
\label{na-} \\
\times \exp \left\{ \frac{2\sqrt{-3(3+2\omega )%
}}{3\kappa (1-n)}\tan ^{-1}\left( \frac{(4-3\gamma )-2c_{3}t^{n-1}}{%
(2-\gamma )\sqrt{-3(3+2\omega )}}\right) \right\}
\end{multline}
\begin{multline}
\phi (\tau ) =\phi _{0}\left[ \pm 2(4-3\gamma )c_{3}\tau ^{1-n}\pm \kappa
\tau ^{2(1-n)}\mp 2c_{3}^{2}\right] ^{\frac{(4-3\gamma )}{\kappa (1-n)}}
\label{nphi-} \\
\times \exp \left\{ -\frac{2(2-\gamma )\sqrt{%
-3(3+2\omega )}}{\kappa (1-n)}\tan ^{-1}\left( \frac{(4-3\gamma
)-2c_{3}t^{n-1}}{(2-\gamma )\sqrt{-3(3+2\omega )}}\right) \right\} 
\end{multline}
with $\kappa $ defined as above. The $\pm $ and $\mp $ signs here indicate
that there are multiple solutions that satisfy the field equations. These
signs should be chosen consistently within each set of square brackets
(solutions (\ref{na+}) and (\ref{nphi+}) can have upper or lower branches
chosen independently in each set of square brackets, as long as a consistent
branch is taken within each separate set of square brackets). The physical
branch should be chosen as the one for which the quantity in brackets
remains positive as $\tau \rightarrow \infty $, so ensuring the existence of
a positive real root in this limit,

These solutions display interesting new behaviours at both early and late
times, which will be discussed in the next section.

\section{Behaviour of Solutions}

\subsubsection{$\boldsymbol{\protect\lambda (\protect\tau)= c_1+c_2 \protect\tau}$}

At late times, as $\tau \rightarrow \infty $, the solutions (\ref{a+})-(\ref%
{phi-}) all approach the\ exact power-law solutions (\ref{powera}) and (\ref%
{powerphi}). It can be seen that these solutions reduce to the usual
spatially flat FRW Brans-Dicke power-law solutions \cite{Nar} in the limit
that the rate of energy transfer goes to zero, $c_{2}\rightarrow 0$. It can
also be seen that these solutions reduce to the spatially-flat FRW general
relativistic solutions in the limit $\omega \rightarrow \infty $,
irrespective of any (finite) amount of energy transfer.

The early-time behaviour of these solutions approaches that of the general
Brans-Dicke solutions \cite{Gur73}, without energy transfer. Generally, we
expect an early period of free-scalar-field domination except in the case $%
\tau _{2}=0$, in which case the power-law solutions (\ref{powera}) and
(\ref{powerphi}) are valid right up
to the initial singularity. For $\omega >-3/2,$ the scalar-field dominated
phase causes an early period of power-law inflation. In this case there is
always an initial singularity and the value of the scalar field diverges to
infinity or zero as it is approached, depending on the sign of $\tau _{2}$.
For $\omega <-3/2$ there is a `bounce' and the scale factor has a minimum
non-zero value. In these universes there is a phase of contraction followed
by a phase of expansion, with no singularity separating them. Solutions of
this type were the focus of \cite{Bar04} where the evolution of $\phi $
through the bounce was used to model the variation of various physical
constants in such situations. The energy exchange term does not play a
significant role at early times in these models. The asymptotic solutions as
the singularity (or bounce) is approached are the same as if the energy
exchange term had been neglected, and are given by \cite{Tup}, up to the
absorption of $c_{1}$ into $\tau _{2}$ previously described.

\subsubsection{$\boldsymbol{\protect\lambda (\protect\tau)=c_3 \protect\tau^n}$, $\boldsymbol{n \neq 1}$}

The behaviour of the solutions (\ref{na+})-(\ref{nphi-}) depends upon the
signs of $n-1$ and $c_{3},$ as well as on the sign of $\omega +3/2$. For
illustrative purposes we will consider the radiation case $\gamma =4/3$
which is appropriate for realistic universes dominated by
asymptotically-free interactions at early times.

For $n>1$ it can be seen that the late-time attractors of solutions (\ref%
{na+})-(\ref{nphi-}) are $a\rightarrow $ constant and $\phi \rightarrow $
constant as $\tau \rightarrow \infty $, for both $\omega >-3/2$ and $\omega
<-3/2$. At late times these universes are asymptotically static; the
evolution of the scale-factor ceases as $\tau \rightarrow \infty $ and both $%
\phi $ and $\rho $ become constant. Further analysis is required to
establish whether these static universes are stable or not (we expect them
to be stable as no tuning of parameters or initial conditions has been
performed to obtain these solutions).

\begin{figure}
\epsfig{file=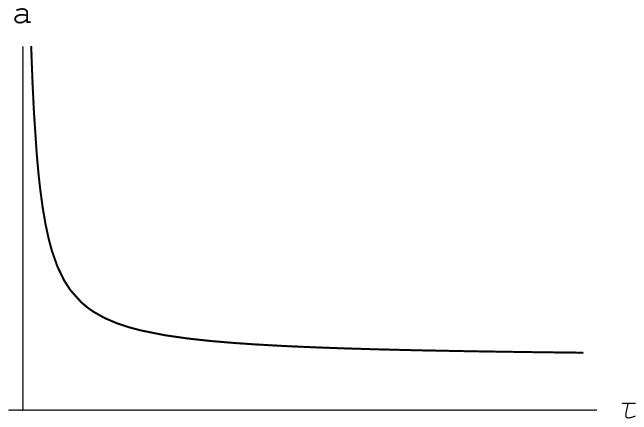,height=4.7cm}
\epsfig{file=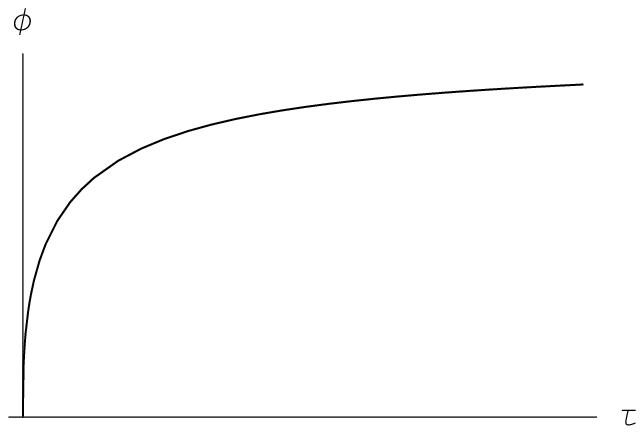,height=4.7cm}
\caption{{\protect {\textit{The evolution of $a$ and $\protect\phi $ for $n=2$, $\protect%
\omega =10$, $c_3=10$ and $\protect\gamma =4/3$.}}}}
\label{a-c}
\end{figure}

\begin{figure}
\epsfig{file=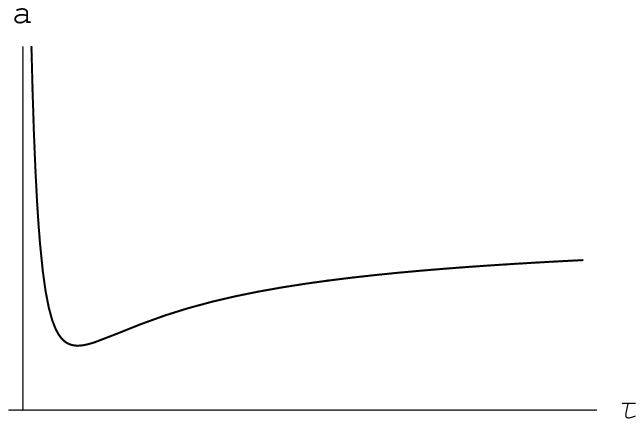,height=4.7cm}
\epsfig{file=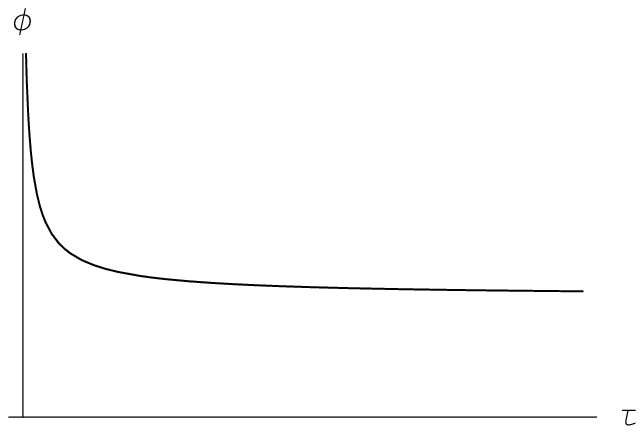,height=4.7cm}
\caption{{\protect {\textit{The evolution of $a$ and $\protect\phi $ for $n=2$, $\protect%
\omega =10$, $c_3=-10$ and $\protect\gamma =4/3$.}}}}
\label{b-d}
\end{figure}

The early-time behaviour of solutions with $n>1$ depends upon the sign of $%
c_{3}$ as well as whether $\omega $ is greater or less than $-3/2$. We will
consider first the case of $\omega >-3/2$. For $c_{3}>0$, we see that $%
a\rightarrow \infty $ as $\tau ^{n-1}\rightarrow \tau _{0}^{+};$ for $%
c_{3}<0 $ we see that $a\rightarrow \infty $ as $\tau ^{n-1}\rightarrow \tau
_{0}^{-} $ (where $\tau _{0}^{+}=((2-\gamma )\sqrt{3(3+2\omega )}+(4-3\gamma
))/2c_{3} $ and $\tau _{0}^{-}=-((2-\gamma )\sqrt{3(3+2\omega )}-(4-3\gamma
))/2c_{3}$). For $n>1$ and $\omega >-3/2$ we therefore have the generic
behaviour that $a\rightarrow \infty $ at early times and $a\rightarrow $
constant at late times. The behaviour of $a$ at intermediate times varies in
form depending on the sign of $c_{3}$, as can be seen in figures \ref{a-c} and %
\ref{b-d}. The asymptotic form of $\phi $ for $n>1$ and $\omega >-3/2$ depends
critically on the sign of $c_{3}$. For $c_{3}>0$ it can be seen that $\phi
\rightarrow 0 $ as $\tau ^{n-1}\rightarrow \tau _{0}^{+}$, whereas for $%
c_{3}<0$ it can be seen that $\phi \rightarrow \infty $ as $\tau
^{n-1}\rightarrow \tau _{0}^{-} $. The behaviour of $\phi $ in these two
cases is illustrated in figures \ref{a-c} and \ref{b-d}.

\begin{figure}
\epsfig{file=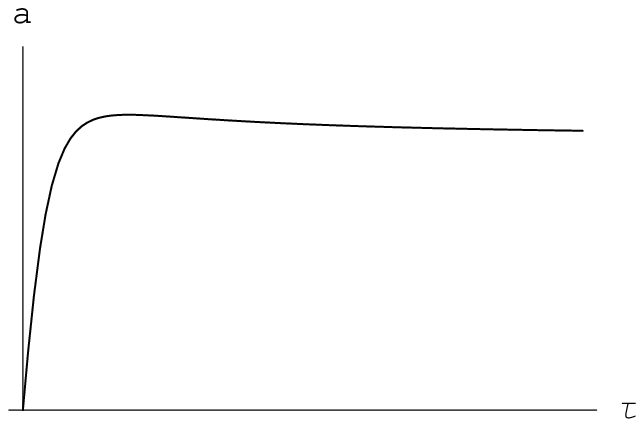,height=4.7cm}
\epsfig{file=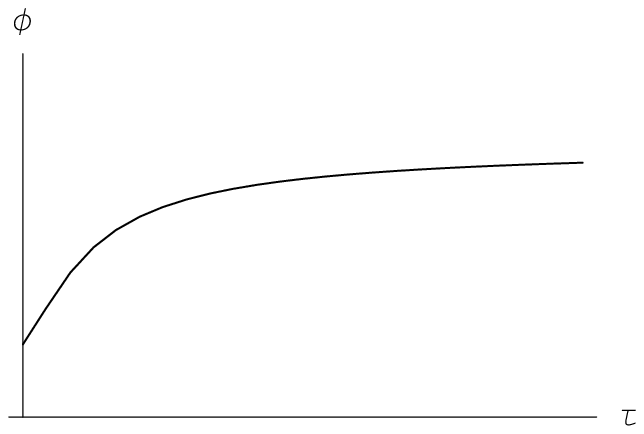,height=4.7cm}
\caption{{\protect {\textit{The evolution of $a$ and $\protect\phi $ for $n=2$, $\protect%
\omega =-10$, $c_3=10$ and $\protect\gamma =4/3$.}}}}
\label{e-g}
\end{figure}

\begin{figure}
\epsfig{file=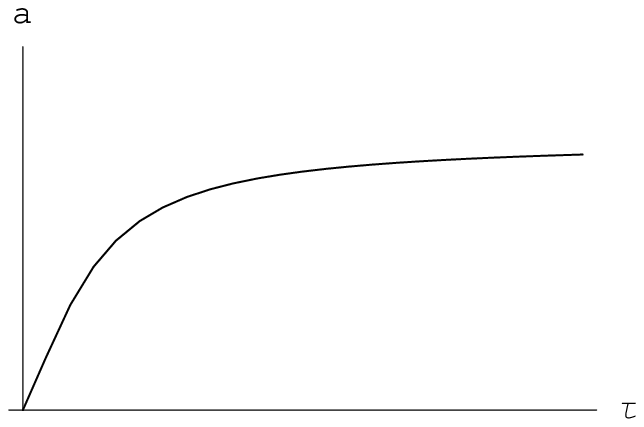,height=4.7cm}
\epsfig{file=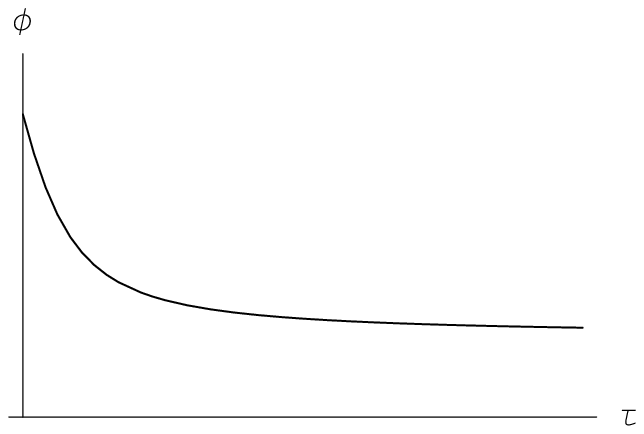,height=4.7cm}
\caption{{\protect {\textit{The evolution of $a$ and $\protect\phi $ for $n=2$, $\protect%
\omega =-10$, $c_3=-10$ and $\protect\gamma =4/3$.}}}}
\label{f-h}
\end{figure}

We now consider the early-time behaviour of solutions with $n>1$ and $\omega
<-3/2$. It can be seen that $a\rightarrow 0$ as $\tau \rightarrow 0$
irrespective of the sign of $c_{3}$, so that we find the generic behaviour $%
a\rightarrow 0$ at early times and $a\rightarrow $ constant at late times
(this is in contrast to the standard theory where an initial singularity is
avoided when $\omega <-3/2$). Again, the behaviour of $a$ at intermediate
times is dependent on the sign of $c_{3}$, as can be seen from figures \ref%
{e-g} and \ref{f-h}. As $\tau \rightarrow 0$ we see that $\phi $ has a finite
non-zero value and is either increasing or decreasing depending on the sign
of $c_{3}$. This behaviour is shown in figures \ref{e-g} and \ref{f-h}. 
\begin{figure}
\epsfig{file=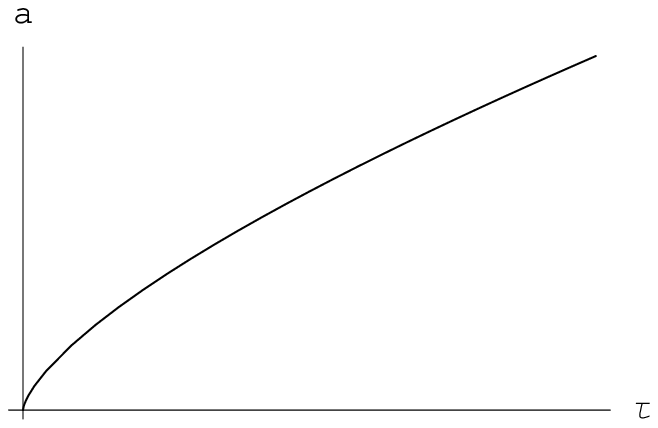,height=4.7cm}
\epsfig{file=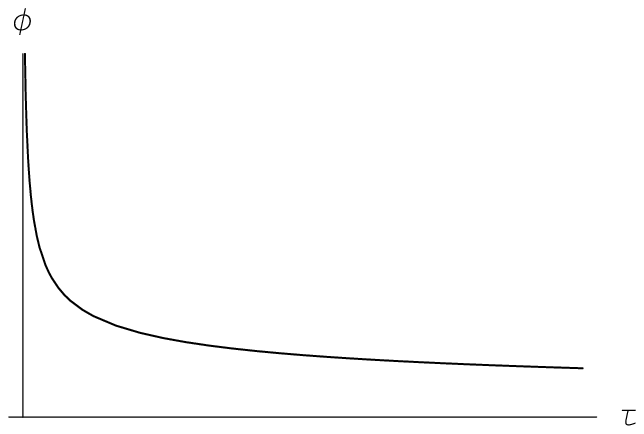,height=4.7cm}
\caption{{\protect {\textit{The evolution of $a$ and $\protect\phi $ for $n=0$, $\protect%
\omega =10$, $c_3=10$ and $\protect\gamma =4/3$.}}}}
\label{i-k}
\end{figure}
\begin{figure}
\epsfig{file=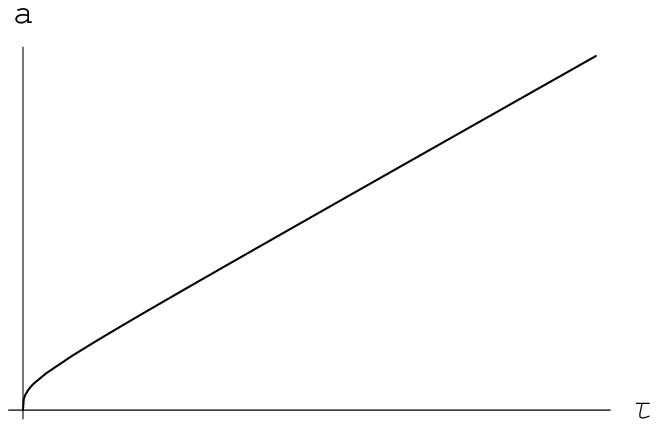,height=4.7cm}
\epsfig{file=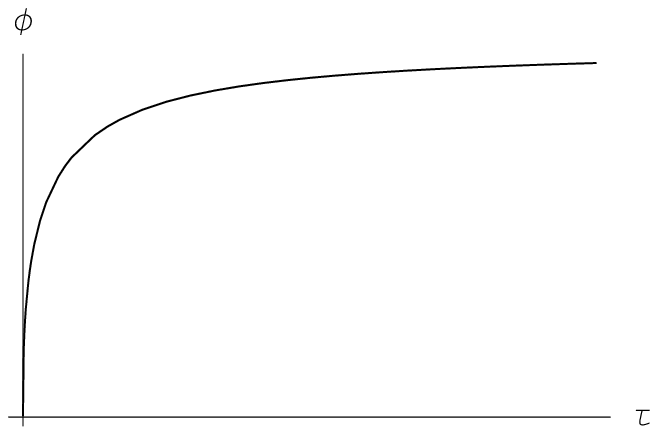,height=4.7cm}
\caption{{\protect {\textit{The evolution of $a$ and $\protect\phi $ for $n=0$, $\protect%
\omega =10$, $c_3=-10$ and $\protect\gamma =4/3$.}}}}
\label{j-l}
\end{figure}
\begin{figure}
\epsfig{file=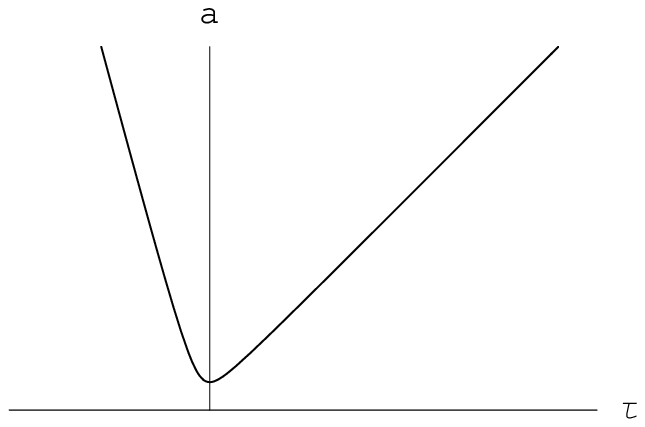,height=4.7cm}
\epsfig{file=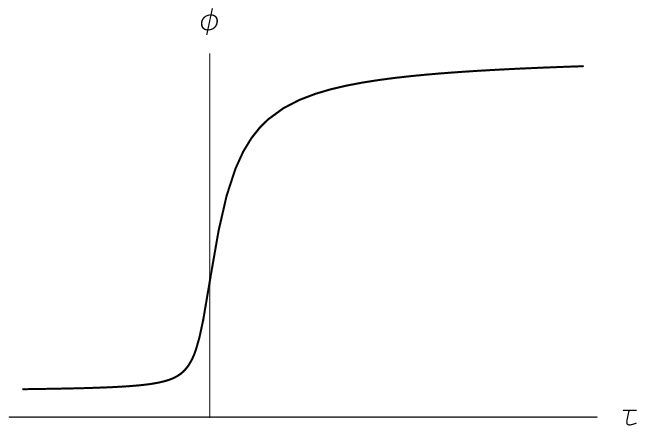,height=4.7cm}
\caption{{\protect {\textit{The evolution of $a$ and $\protect\phi $ for $n=0$, $\protect%
\omega =-10$, $c_3=10$ and $\protect\gamma =4/3$.}}}}
\label{m-o}
\end{figure}
\begin{figure}
\epsfig{file=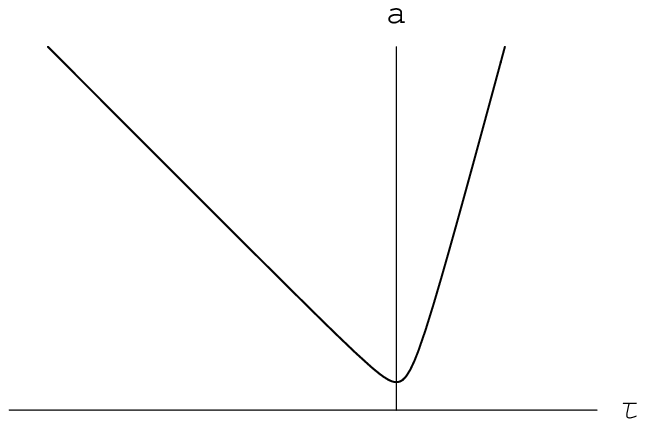,height=4.7cm}
\epsfig{file=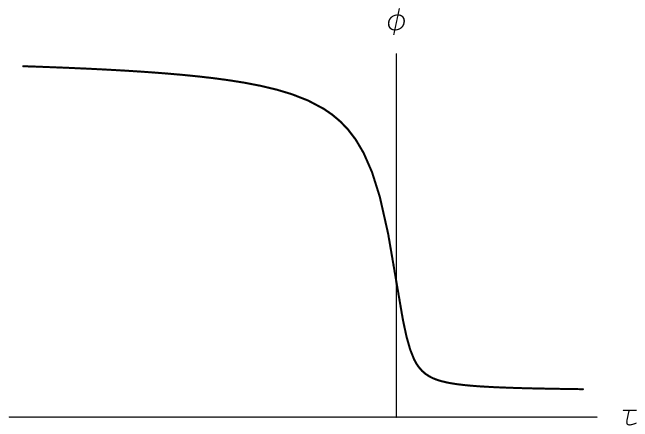,height=4.7cm}
\caption{{\protect {\textit{The evolution of $a$ and $\protect\phi $ for $n=0$, $\protect%
\omega =-10$, $c_3=-10$ and $\protect\gamma =4/3$.}}}}
\label{n-p}
\end{figure}

It remains to investigate the nature of the solutions with $n<1$. At late
times we see that 
\begin{align*}
a& \rightarrow \tau ^{\frac{2+2(2-\gamma )\omega }{2(5-3\gamma )+3(2-\gamma
)^{2}\omega }}, \\
\phi & \rightarrow \tau ^{\frac{2(4-3\gamma )}{2(5-3\gamma )+3(2-\gamma
)^{2}\omega }},
\end{align*}%
as $\tau \rightarrow \infty $, irrespective of the sign of $c_{3}$ or the
value of $\omega $. These late-time attractors are the flat FRW power-law
Brans-Dicke solutions \cite{Nar} which reduce to the standard
general relativistic solutions in the limit $\omega \rightarrow \infty $.

The early-time behaviour when $n<1$ depends on the sign of $c_{3}$ and the
sign of $\omega +3/2$. We consider first the case $\omega >-3/2$. In this
case it can be seen that $a\rightarrow 0$ as $\tau ^{n-1}\rightarrow \tau
_{0}^{+}$ or $\tau ^{n-1}\rightarrow \tau _{0}^{-}$, for $c_{3}>0$ or $%
c_{3}<0$ respectively. The behaviour of $a$ for both $c_{3}>0$ and $c_{3}<0$
is shown in figure \ref{i-k} and \ref{j-l}. The behaviour of $\phi $ at early
times depends on the sign of $c_{3}$ and goes to $\infty $ for $c_{3}>0$ or
to $0$ for $c_{3}<0$. The behaviour of $\phi $ in these cases is shown in
figures \ref{i-k} and \ref{j-l}. For $\omega <-3/2$ the scale factor $a$
contracts to a finite, but non-zero, minimum value and then expands. The
exact form of the minimum depends on the values of $n$, $\omega $ and $%
c_{3}, $ but it is interesting to note that odd values of $n$ produce
symmetric bounces and even values of $n$ produce asymmetric bounces, as
illustrated in figures \ref{m-o} and \ref{n-p}. The evolution of $\phi $ through
these bounces is smooth with a time direction prescribed by the value of $%
c_{3}$, as shown in figures \ref{m-o} and \ref{n-p} (increasing for $c_{3}>0$
and decreasing for $c_{3}<0$). The effect of changing the sign of $c_{3}$ is
seen to be a mirroring of the evolution of $a$ and $\phi $ in the $y-$axis. 

\section{Physical Consequences}

The solutions found in the previous sections are of physical interest
for a number of reasons.  The transfer of energy and momentum between a
non-minimally coupled scalar field $\phi$ and matter fields is a
prediction of a number of fundamental theories of current interest,
including string theories, Kaluza-Klein theories and brane-worlds.  The
cosmologies produced by such an interaction, therefore, should be of
direct interest in the consideration of these theories.  Furthermore,
the solutions we have found display modified behaviour at both early
and late times.  The investigation of modified theories of gravity at
early times is of particular interest as it is in the high-energy limit that
deviations from general relativity are usually expected.  Modified
behaviour at late times is also of interest as it is
at these times that we can make direct observations which can be used
to constrain deviations from the standard
general relativistic model.  We will now summarise the behaviour
of the solutions found in the previous section, highlighting
the physically significant results and constraints that can be placed
on the theory by observations.

For the case of $\lambda $ linear in $\tau $ it was shown that the late-time
attractors of the general solutions are no longer the power-law solutions of
the Brans-Dicke theory, but are given by equations (\ref{power1}) and (\ref%
{power2}).  These attractors are of special interest as they have a
simple power-law form that reduces to the general relativistic result
in the limit $\omega \rightarrow \infty$ and to the Brans-Dicke result
in the limit $c_2 \rightarrow 0$.  Observations of cosmic microwave background anisotropies and
the products of primordial nucleosynthesis will therefore be able to constraint any
potential late-time deviations of this kind, and hence the underlying
model.  The process of primordial nucleosynthesis in scalar-tensor theories has been
used by a number of authors to place constraints on the coupling
parameter $\omega (\phi )$ \cite{Cas1, Cas92, Ser92, Ser2, San97, Dam, Cli}.  In these
studies the different value of $G$ during nucleosynthesis causes the
weak interactions to freeze out at a different time and hence the
proton to neutron ratio at this time is different to the standard
case.  This modification causes different abundances of the light
elements to be produced, which can be compared with observations to
constrain the underlying theory.  Studies of this kind usually assume
$G$ to be constant during nucleosynthesis, which will not be the case
when energy is allowed to be exchanged between $\phi$ and the matter
fields.  The effects of a non-constant $G$ were studied in
\cite{Cli}.  A similar study would be required to place constraints
upon the parameters $c_2$ and $\omega$ in this theory.  The
cosmic microwave background power spectrum has also often been used to constrain
scalar-tensor theories of gravity \cite{cmb1, cmb2, Nag02, cmb4, Nag04,
  cmb6}.  In these studies the redshift of matter-radiation equality
is different from its usual general relativistic value due to
the modified late-time evolution of the Universe.  This change in the
redshift of equality is imprinted on the spectrum of perturbations
as it is only after equality that
sub-horizon scale perturbations are allowed to grow.  The main effect
is seen as a shift in the first peak of the power-spectrum, which can be
compared with observations to constrain the theory.  Again, the
late-time evolution of the Universe is modified from the usual Brans-Dicke
case by the energy exchange that we consider, so that the previous
constraints are not directly applicable.

For the case of a non-linear power-law exchange of energy, described by $%
\lambda \propto \tau ^{n}$, the late-time evolution of $a$ and $\phi $ can
be significantly modified. For $n>1,$ the solutions do not continue to
expand eternally, but are attracted towards a static state where the
time-evolutions of $a$, $\phi $ and $\rho $ cease. For $n<1$ the generic
late-time attractor is the power-law solution of a flat FRW Brans-Dicke
universe. It appears that theories of energy exchange with $n>1$ are ruled
out immediately by observations of an expanding universe whilst the case of $%
n<1$ is subject to the same late-time constraints as the standard
Brans-Dicke theory \cite{Cas1, Cas92, Ser92, Ser2, San97, Dam, Cli,cmb1,
  cmb2, Nag02, cmb4, Nag04, cmb6}.

It remains to investigate the physical consequences of the early-time
behaviour of our solutions.  For $\lambda $ linear in $\tau $ these solutions
approach those of the standard Brans-Dicke theory as either the initial
singularity or the minimum of the bounce are approached, according to the
sign of $\omega -3/2$.  The physical significance of this behaviour
has been discussed many times before, usually focusing on the
avoidance of the initial singularity and the inflation that can result
from the presence of the free component of the scalar field.

\begin{figure}
\epsfig{file=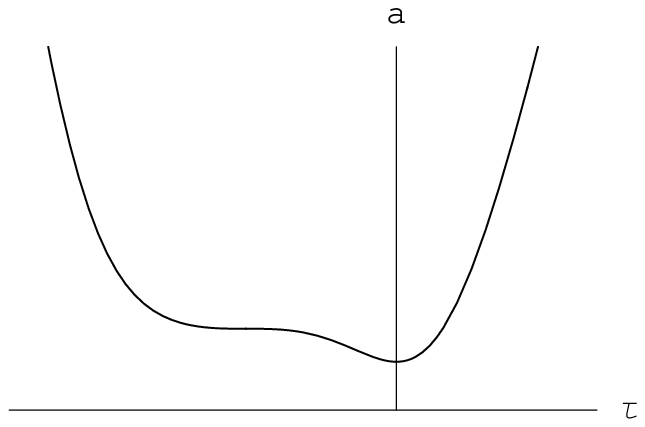,height=4.7cm}
\epsfig{file=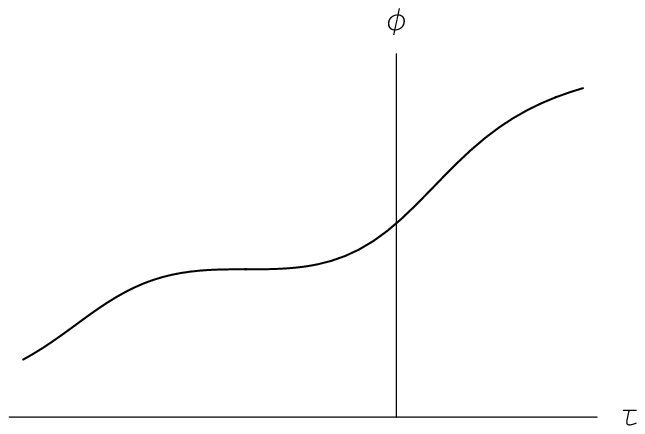,height=4.7cm}
\caption{{\protect {\textit{The evolution of $a$ and $\protect\phi $ for $n=-2$, $\protect%
\omega =-10$, $c_{3}=10$ and $\protect\gamma =4/3$.}}}}
\label{bounce}
\end{figure}

For $\lambda \propto \tau ^{n}$ the early-time
behaviour can be significantly changed from that of the standard theory. For 
$n>1$ the scale factor $a$ either approaches infinity or zero, depending on
our choice of $\omega $ and $c_{3}$, as previously described. For the more
realistic case of $n<1$ the evolution of $a$ at early times either undergoes
a period of rapid expansion or a non-singular bounce, depending on whether $%
\omega $ is greater or less than $-3/2$. This behaviour is similar to that
of the general solutions of the standard theory, but in this case the free
scalar-field-dominated epoch has not been invoked and there is more freedom
as to the exact form of the evolution. For example, with a suitable choice
of parameters it is possible to create a universe that contracts and then is
briefly static before `bouncing' and continuing on to its late-time
power-law evolution. This is shown in figure \ref{bounce} for the case $%
\omega =-10$, $n=-2$ and $c_{3}=10$. (It is interesting to note that Peter
and Pinto-Neto remark that a static period followed by a bounce could
potentially produce a scale-invariant spectrum of perturbations \cite{Peter}%
). 
For the physically reasonable models with $n\leqslant 1$ the evolution of $%
\phi $, and hence of $G$, can be significantly altered at early times from
what is generally assumed to be the case in scalar-tensor theories of
gravity. For the case $\omega >-3/2,$ the value of $G$ can be made to
diverge to infinity or to zero as the initial singularity is approached,
independent of whether or not there was an early scalar-dominated phase to
the universe's history. For the case $\omega <-3/2,$ the value of $G$
evolves smoothly through the bounce in the scale-factor, and is again
independent of whether or not there was a scalar-dominated phase. More
complicated evolutions of $\phi $ can also be constructed, as can be seen
in figure \ref{bounce}.

In performing this analysis we have allowed the coupling constant
$\omega$ to lie in the ranges $\omega > -3/2$ and $\omega <-3/2$.  It
should be noted that when $\omega <-3/2$ the scalar field $\phi$
should be considered as having negative energy.  Allowing an
interaction with matter fields may then be considered as problematic
as energy could potentially be transferred from the scalar field to
the matter fields without bound, allowing runaway solutions.

\section{Discussion}

We have considered spatially flat FRW universes in scalar-tensor theories of
gravity where energy is allowed to be exchanged between the Brans-Dicke
scalar field that determines the strength of gravity and any perfect fluid
matter fields in the space-time. We have presented a prescription for
integrating the field equations exactly for some unknown function $\lambda $
which describes the rate at which energy is exchanged. For the case of $%
\lambda $ being a linear function of $\tau $ we have found the general
solutions to the problem and for the case of $\lambda $ being a non-linear
power law function of $\tau $ we have been able to find a wide class of
exact solutions. These solutions display behaviours that can deviate
substantially from the corresponding solutions in the standard case, where
the exchange of energy is absent. Depending upon the values of the
parameters defining the theory and the exchange of energy, deviations in the
evolution of $a$ and $\phi $ can occur at both early and late times,
providing a richer phenomenology than is available in the standard
theory.  

We have found that the parameter $n$ must be bounded by the
inequality $n \leqslant 1$ if the Universe is to be expanding at late
times.  For $n=1$ we have found late-time power-law attractor
solutions which can be used to constrain the parameters $c_2$ and
$\omega$.  For $n <1$ we have seen that the late-time evolution will
be the same as in the standard Brans-Dicke case, and so is subject
to the same observational constraints as these theories.  The
parameters $c_3$ and $\tau_2$ have been shown to be influential only
in the vicinity of the initial singularity, or at the minimum of expansion in
non-singular solutions.  These parameters are therefore less
accessible to constraint by late-time observations (see, however, \cite{Cli} where
the influence on primordial nucleosynthesis is used to constrain $\tau_2$).  The parameter
$\omega$ is, as always, subject to the very tight solar system
constraint $\omega > 40 000$ to $2 \sigma$ \cite{Bert}.

These results could be of interest in
attempting to explain why $G$ is so small in the present day Universe
compared to the proton mass scale ($Gm_{pr}^{2}\sim 10^{-39}$).  In
these models the value of $G$ can decay away by a coupling between the
scalar field $\phi$ and the matter fields which allows energy to be
transferred.  The small value of $G$ is then due to the age of the
Universe.  It remains to see
whether or not the late-time modifications found above are consistent
with observations of the primordial abundance of light elements,
microwave background formation and other late-time physical
processes.  These
studies should be able to be performed in an analogous way to the ones
that already exist for the standard Brans-Dicke theory.

Using the late-time solutions that have been found it is possible to comment
on the case of FRW cosmologies with non-zero spatial curvature. At early
times it is expected that the effect of any spatial curvature on the
evolution of $a(t)$ should be negligible. From the solution (\ref{powera})
we can see that spatial curvature will dominate the late time evolution if
the condition 
\begin{equation*}
\frac{2+2(2-\gamma )\omega +2c_{2}}{4+3\gamma \omega (2-\gamma
)-2c_{2}(7-6\gamma -c_{2})}<1
\end{equation*}%
is satisfied. If this condition is not satisfied, then the power-law solution
(\ref{powera}) will be an attractor as $t\rightarrow \infty $ even in the
case of non-zero spatial curvature, offering a potential solution to the
flatness problem. This behaviour corresponds to power-law inflation
and it can be seen that the condition for equation (\ref{powera}) to dominate over the spatial curvature
at late times is, indeed, also the condition that power-law inflation should
occur.

In conclusion, we have found that a direct coupling between $\phi $ and the
matter fields in scalar-tensor cosmologies provides a richer frame-work
within which one can consider variations of $G$. We have shown that it is
possible to construct models where the late-time violations of the
equivalence principle can be made arbitrarily small (for $\lambda \propto
\tau $) or are attracted to zero (for $\lambda \propto \tau ^{n}$ where $n<1$%
). This enlarged phenomenology is of interest for the consideration of the
four-dimensional cosmologies associated with higher-dimensional theories as
well as for more general considerations of the variation of $G$ and its
late-time value. This study has been limited to scalar-tensor theories with
constant coupling parameters, to flat FRW cosmologies, and to special cases
of $\sigma ^{0}$ that allow direct integration of the field equations.
Obvious extensions exist in which these assumptions are partially or
completely relaxed.
\clearemptydoublepage
\chapter{Cosmological Constraints}
\label{Cosmological Constraints}

\bigskip

Observations of physical processes occuring in an expanding universe
can be used to constrain the underlying gravitational theory in a
number of ways.  These observations can range from galaxy surveys in
the nearby Universe to the results of processes occuring in the
very early Universe.  We will concentrate in this work on the
processes of primordial nucleosynthesis and microwave background
formation.  Both of these processes occur early in the Universe's
history and the results of both are well observed by astronomers and
astrophysicists.

This chapter is based on the work of Clifton, Barrow and Scherrer
\cite{clift} and Clifton and Barrow \cite{Cli}.

\section{Scalar-Tensor Theories and Primordial Nucleosynthesis}

Using the scalar-tensor theories we investigate the earliest well understood
physical process, primordial nucleosynthesis.  Previous studies on
this subject have been carried out by a number of
authors.  In particular, the Brans-Dicke theory has been especially
well studied in this context by, for example, Casas, Garcia-Bellido and Quiros
\cite{Cas1}, \cite{Cas92} and Serna, Dominguez-Tenreiro and Yepes
\cite{Ser92}.  The more general class of Scalar-Tensor theories has also been well
studied, most notably by Serna and Alimi \cite{Ser2}, Santiago, Kalligas and Wagoner
\cite{San97} and Damour and Pichon \cite{Dam}.  More recently, a
number of studies have been performed investigating the effect of a nonminimally
coupled quintessence field on primordial nucleosynthesis (see
e.g. \cite{Che01},\cite{Pet04}).  Although
these studies are very detailed, they all make the simplifying assumption of a
constant $G$ during the radiation-dominated phase of the Universe's
history (with the exception of \cite{Ser92}, who
numerically investigate the effect of an early scalar-dominated phase
on the Brans-Dicke theory, and \cite{Dam} who use the idea of a ``kick'' on
the scalar field during electron-positron annihilation).  We relax
this assumption and investigate the effects of entering the
radiation-dominated phase with a non-constant $G$.  The constraints we impose
upon the variation of $G$ during primordial nucleosynthesis are then used to
constrain the parameters of the theory.  In carrying out this study we
consider the more general class of scalar-tensor theories, paying
particular attention to the Brans-Dicke theory.

\subsection{Modelling the Form of $G(t)$}

If primordial nucleosynthesis were to occur during the
scalar-dominated period then the very different expansion rate would have disastrous
consequences for the light-element abundances (see e.g. \cite{Ser92}).
Therefore we limit our study to times at which the scale factor can be
approximated by a form that is close to $a(t) \propto t^{\frac{1}{2}}$, and so
primordial nucleosynthesis can safely be described as occurring during radiation
domination.  Performing a power-series expansion of the FRW
Brans-Dicke solutions
(\ref{aw>}), (\ref{phiw>}), (\ref{aw<}) and (\ref{phiw<}) in
$\eta_1/(\eta+\eta_2)$ we find
\begin{align}
a(\eta) &= a_1 (\eta+\eta_2 +3 \eta_1) +O(\eta_1^2) \label{a1}\\
\phi(\eta) &= \phi_1 \left( 1-\frac{3 \eta_1}{\eta+\eta_2} \right)
+O(\eta_1^2) \label{a22}
\end{align}
for both $\omega >-3/2$ and $\omega <-3/2$.  We can then set the
origin of the $\eta$ coordinate such that $a(0)=0$ with the choice
$\eta_2=-3 \eta_1$.  The solutions (\ref{a1}) and ({\ref{a22}) then
  become, in terms of the time coordinate $t$,
\begin{align*}
a(t) &= a_1 t^{\frac{1}{2}} +O(\eta_1^2)\\
\phi(t) &= \phi_1 \left(1+\frac{a_2}{a(t)} \right) +O(\eta_1^2),
\end{align*}
where $a_2=-3 \eta_1 a_1$ and the origin of $t$ has been chosen to coincide with the origin of
$\eta$.

Similarly, expanding the FRW $\omega (\phi)$ solutions (\ref{STrad}) and (\ref{STrad2}) in
$\eta_1/(\eta+\eta_2)$ we find that, for both $\omega >-3/2$ and $\omega <-3/2$,
\begin{align*}
a(t) &= a_1 t^{\frac{1}{2}} + O(\eta_1^2)\\
\phi(t) &= \phi_1 \left( 1+\frac{a_2}{a(t)} \right) +O(\eta_1^2)
\end{align*}
where 
\begin{align*}
a_1 &= \sqrt{8 \pi \bar{\rho}_{r0}/3} \exp
(\beta(\psi_1-\psi_{\infty})^2/2)\\
a_2 &=  \sqrt{2 \bar{\rho}_{r0}}\beta \eta_1
(\psi_1-\psi_{\infty})  \exp
(\beta(\psi_1-\psi_{\infty})^2/2)\\
\phi_1 &= \exp (-\beta
(\psi_1-\psi_{\infty})^2)
\end{align*}
and $\eta_2$ has been set so that $a(0)=0$.

In the limit $\dot{\phi} \rightarrow 0$ we can see from equations
(\ref{Nariai2}), (\ref{Friedmann}) and (\ref{acceleration}) that the standard GR Friedmann equations are
recovered, with a different value of Newton's constant given by
\begin{equation}
\label{Gt}
G(t) = \frac{1}{\phi(t)} = G_1 \frac{a(t)}{a(t)+a_2}
\end{equation}
where $G_1=1/\phi_1$.  We conclude that the solutions found above correspond to a
situation that can be described using the GR Friedmann equations with a
different, and adiabatically changing, value of $G$.  This is just the
situation considered by Bambi, Giannotti and Villante
\cite{Bam05}, but we now have an explicit form for the evolution of $G(t)$
derived from scalar-tensor gravity theory.

\begin{figure*}
\begin{center}
\subfigure[2$\sigma$ bounds for $a_2 > 0$]{\epsfig{figure=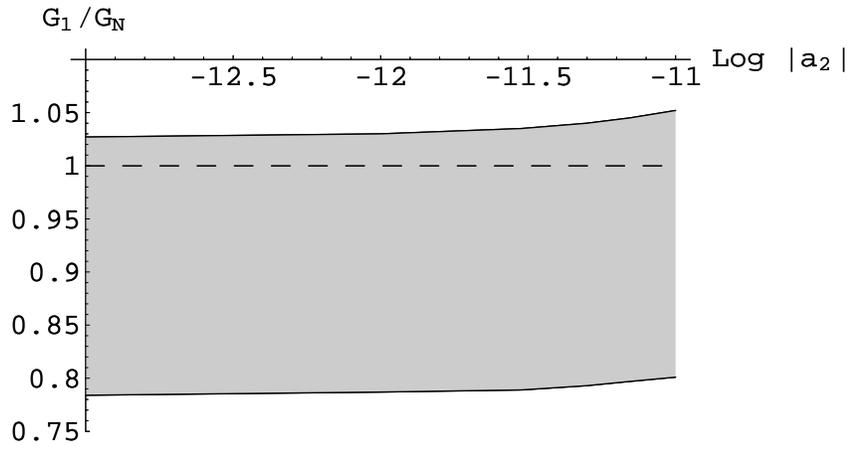, height=7cm}}
\subfigure[2$\sigma$ bounds for $a_2 < 0$]{\epsfig{figure=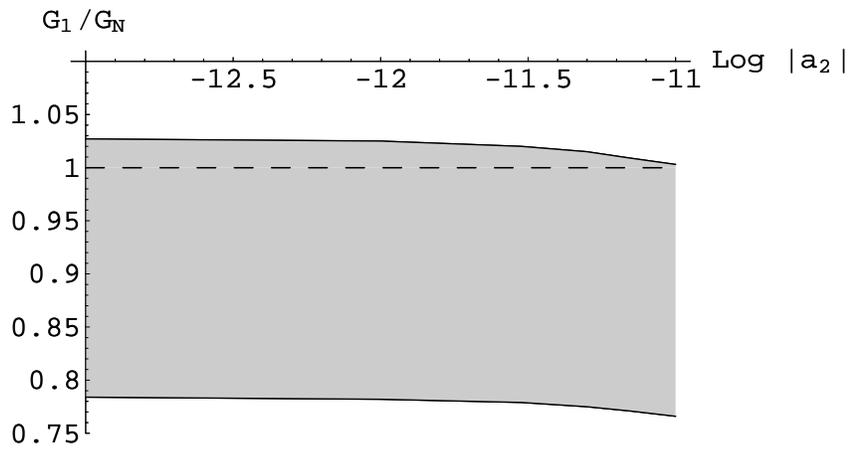, height=7cm}}
\end{center}
\caption{\textit{The upper and lower bounds on $G_1/G_N$ as a
    function of $a_2$ are shown by the solid lines.  The shaded regions correspond to the allowed
    parameter space and the dashed lines show $G_1=G_N$.}}
\label{scherrerplot}
\end{figure*}

\subsection{The Effect of a Non-Constant $G$}

The time at which weak interactions freeze out in the early Universe
is determined by equality between
the rate of the relevant weak interactions and the Hubble rate.  When the
weak interaction rate is the greater then the ratio of neutrons
to protons tracks its equilibrium value, $n/p = \exp(-(m_n-m_p)/T)$
where $m_n$ and $m_p$ are the neutron and proton masses.
If, however, the Hubble rate is greater than the weak-interaction rate
then the ratio of neutrons to
protons is effectively `frozen-in', and $\beta$ decay is the only
weak process that still operates with any efficiency.  This will be the
case until the onset of deuterium formation, at which time the
neutrons become bound and $\beta$-decay ceases.  The onset of
deuterium formation is primarily determined by the photon to
baryon ratio, $\eta_{\gamma}$, which inhibits the formation of
deuterium nuclei until the critical temperature for photodissociation is past.  As the vast
majority of neutrons end up in $^4$He the primordial abundance of this
element is influenced most significantly by the number of neutrons at
the onset of deuterium formation, which is most sensitive to the
temperature of weak-interaction freeze-out, and hence the Hubble
rate, and so $G$, at this time.  Conversely, the primordial abundances of
the other light elements are most sensitive to the temperature of
deuterium formation, and hence $\eta_{\gamma}$, when
nuclear reactions occur and the light elements form.  (See
\cite{Bam05} for a more detailed discussion of these points).

Using the simple forms of $a(t)$ and $G(t)$ derived above
we use a modified version of the Kawano code \cite{Kaw92} to
investigate the effect of this variation of $G$ on primordial nucleosynthesis directly.  We
use the deuterium abundance estimated by Kirkman et al. \cite{Kir03}
\begin{equation}
\label{Dbound}
\log \left( \frac{D}{H} \right) = -4.556 \pm 0.064 \quad \text{to} \;
1\sigma,
\end{equation}
and the $^4$He abundance estimated by Barger et al. \cite{Bar03}
\begin{equation}
\label{Hebound}
Y_P=0.238 \pm 0.005 \quad \text{to} \; 1\sigma,
\end{equation}
to create a parameter space in ($\eta_{\gamma}$, $G_1$, $a_2$).  The
three-dimensional 95$\%$ $\chi^2$ confidence region is projected into
the $G_1$, $a_2$ plane to give figure \ref{scherrerplot}, where three
species of light neutrinos and a neutron
mean lifetime of $\tau=885.7$ seconds have been assumed.

The apparent disfavoring of the value of $G_1$ much greater than $G_N$
in figure \ref{scherrerplot} is due to the observational limits we
have adopted for $Y_P$ (equation (\ref{Hebound})), which
suggest a helium abundance that is already uncomfortably low compared to the 
theoretical prediction for $Y_P$ derived from the baryon density corresponding 
to equation (\ref{Dbound}).  We expect these constraints to be
updated by future observations (see e.g. \cite{Tro04}); this
will require a corresponding update of any work, such as this,
that seeks to use these constraints to impose limits on physical
processes occuring during primordial nucleosynthesis.

These limits on the parameters $G_1$ and $a_2$ can be used to
construct plots showing the explicit evolution of $G(t)$ for various
limiting combinations of the two parameters, this is done in figure
\ref{G}.  It is interesting to note that in both of these plots the
lines corresponding to different values of $a_2$ all appear to cross at approximately the same point, $\log a \sim
-9.4$.  This confirms our earlier discussion, and the results of \cite{Bam05}, that
the $^4$He abundance is mostly only sensitive to the value of $G$ at the
time when the weak interactions freeze out.  In reality, this
freeze-out happens over a finite time interval, but from figure
\ref{G} we see that it is a good approximation to consider it happening
instantaneously - where the lines cross.  To a
reasonable accuracy one could then take $G$ throughout primordial nucleosynthesis
to be its value during the freeze-out process.  
\begin{figure*}
\begin{center}
\subfigure[$a_2 > 0$]{\epsfig{file=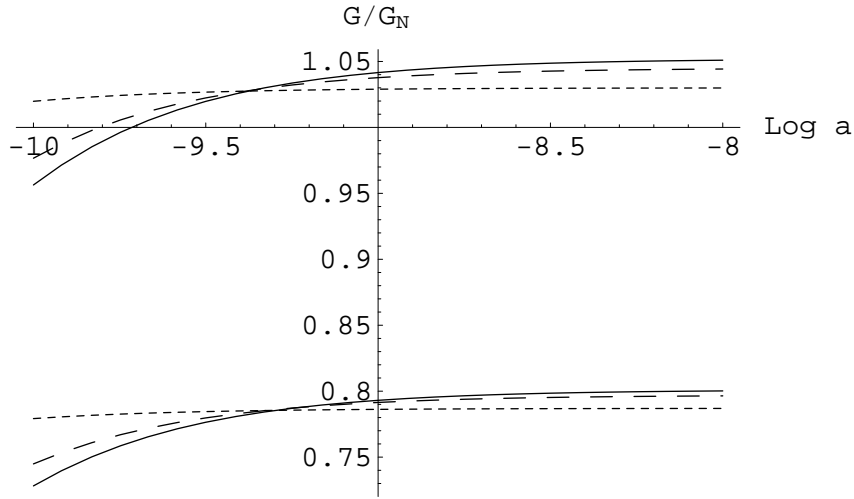,height=7cm}}
\subfigure[$a_2 < 0$]{\epsfig{file=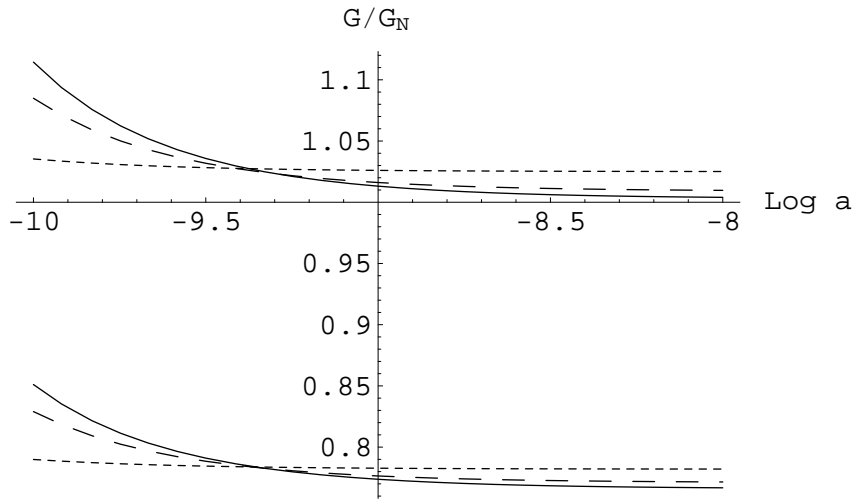,height=7cm}}
\end{center}
\caption{\textit{These plots show the explicit evolution of G for some
limiting values of $G_1$, as taken from figure \ref{scherrerplot}, for
various values of $a_2$.  Solid lines correspond to $\vert a_2 \vert =
10^{-11}$, dashed lines correspond to $\vert a_2 \vert =
10^{-11.155}$ and dotted lines correspond to $\vert a_2 \vert =
10^{-12}$.}}
\label{G}
\end{figure*}

\subsection{Constraining the Theory}

Using the results in the previous section it is possible to constrain
the underlying scalar-tensor gravitational theory.  This is done separately for the
Brans-Dicke theory and for the more general scalar-tensor theories.
For each theory we consider the case of a universe containing matter
and radiation only and then a universe containing matter, radiation and a
non-zero vacuum energy, with equation of state $p=-\rho$.

\subsubsection{Brans-Dicke theory}

\subsubsection{Universe containing matter and radiation}

Using the Brans-Dicke matter-dominated FRW solutions (\ref{matBD}) we can write the
ratio of $G$ at matter-radiation equality, $G_{eq}$, to its
present-day laboratory value, $G_N$, as
\begin{equation}
\frac{G_{eq}}{G_N}= \frac{(2 \omega+3)}{(2
  \omega+4)}\frac{\phi_N}{\phi_{eq}} = \frac{(2 \omega+3)}{(2
  \omega+4)}\left( \frac{a_0}{a_{eq}}\right)^{1/(\omega+1)} 
\equiv  \frac{(2 \omega+3)}{(2 \omega+4)}(1+z_{eq})^{1/(\omega+1)}
\label{speedup}
\end{equation}
where we have used the expression for $G$ in the weak-field limit \cite{will}:
\begin{equation}
G(\phi)=\frac{(2\omega+4)}{(2\omega+3)}\frac{1}{\phi},
\label{JG}
\end{equation}
for $G_N$, and $G(\phi)=1/\phi$ for $G_{eq}$, as in
(\ref{Gt}).  In reality, the laboratory value of $G$ today is
not equal to that in the background cosmology \cite{Cli05}, although we
take it to be so here for simplicity as this effect should be small.

We now proceed by calculating $1+z_{eq}$ in the
Brans-Dicke cosmology, following Liddle, Mazumdar and Barrow
\cite{cmb1}.  As ${T^{a b}}_{; b}=0$ in the Jordan frame we have that $\rho_r\propto
a^{-4}$ and $\rho_m\propto a^{-3}$.
For a universe containing matter and radiation only, this gives
\begin{equation*}
1+z_{eq}=\frac{a_0}{a_{eq}}=\frac{\rho_{m0}}{\rho_{r0}}\frac{\rho_{req}}{\rho_{meq}}
=\frac{\rho_{m0}}{\rho_{r0}}
\end{equation*}
as well as the usual relation $T \propto a^{-1}$, when entropy
increase is neglected.

Photons and neutrinos both contribute to the value of $\rho_{r0}$, the
present-day energy-density of radiation.  From $T_{\gamma 0}=2.728 \pm
0.004K$ \cite{Fix96}, we get $\rho_{\gamma 0}=4.66\times 10^{-34}
g \; cm^{-3}$ and using the well known result $T_{\nu 0}=
(4/11)^{1/3}T_{\gamma 0}$, and assuming three families of light neutrinos,
we also have $\rho_{\nu 0}=0.68\rho_{\gamma 0}$.  This gives the total
present-day radiation density as
$\rho_{r0}=7.84 \times 10^{-34}g \; cm^{-3}$.  Now, recalling our
assumption of spatial flatness and (\ref{Friedmann}), we can write
\begin{equation*}
\rho_{tot0}=\rho_{m0}+\rho_{r0}=\frac{3H_0^2}{8\pi G_N}\frac{(4+3\omega)(4+2\omega)}{6(1+\omega)^2}.
\end{equation*}
For $G_N=6.673\times10^{-11} \;  Nm^2kg^{-2}$, $H_0=100h \;
km s^{-1} Mpc^{-1}$, and the value of $\rho_{r0}$ above, we
have 
\begin{align*}
1+z_{eq} &=2.39 \times 10^4 h^2
\frac{(4+3\omega)(4+2\omega)}{6(1+\omega)^2} -1\\
&= 2.39 \times 10^4 h^2 \left( 1+ \frac{4}{3 \omega} \right) +O(\omega^{-2}).
\end{align*}
This correction to $1+z_{eq}$ has direct observational consequences in the
power spectrum of cosmic microwave background perturbations.  After $1+z_{eq}$ the
subhorizon scale perturbations, that were previously effectively
frozen, are allowed to grow.  Changing the value of $1+z_{eq}$
therefore causes a shift in the power-spectrum peaks, which is
potentially observable (see \cite{cmb1}, \cite{cmb2}, \cite{Nag02}
\cite{cmb4}, \cite{Nag04} and \cite{cmb6} for a more detailed discussion of the effect of a varying
$G$ on microwave background formation).  For our purposes this modified expression
for $1+z_{eq}$ can then be substituted into (\ref{speedup}) to give an equation
in terms of $\omega$, $G_{eq}$ and $h$.  Assuming a value of $h=0.7$
and that $G_{eq}=G_1$, as appears a very good approximation for the
models above, we can use our bounds on $G_1$ in terms of $a_2$ to
create an allowed parameter space in the ($\omega$,$a_2$) plane.  This
is shown in figure \ref{BDresult}.

We remind the reader that a large $\omega$ corresponds to a slowly
varying Machian component of $\phi(1+z)$, as can be seen directly from
equations (\ref{matBD}), (\ref{vacBD}) and the late-time limits of (\ref{aw>}),
(\ref{phiw>}), (\ref{aw<}) and (\ref{phiw<}).  The parameter $a_2$ determines the
evolution of the free component of $\phi$, as can be seen from the
early time limits of (\ref{aw>}), (\ref{phiw>}), (\ref{aw<}) and (\ref{phiw<}).  In
the limits $\omega \rightarrow \infty$ and $a_2 \rightarrow 0$ the
Machian and free components of $\phi(1+z)$ both become constant,
respectively, resulting in a constant $G(1+z)$.  The constraints imposed
upon $\omega$ and $a_2$ in figure \ref{BDresult} therefore correspond
to constraints upon the evolution of $G(1+z)$ in this theory, valid both during primordial
nucleosynthesis and at other cosmological epochs.

\subsubsection{Universe containing matter, radiation and a nonzero
  vacuum energy}

A more realistic constraint would involve taking into account a
late-time period of vacuum domination; so as well as $\rho_r\propto
a^{-4}$ and $\rho_m\propto a^{-3}$, we now also have $\rho_{\Lambda}
=\text{constant}$.  Hence,
\begin{align*}
\frac{G_{eq1}}{G_N} &=\frac{(2\omega+3)}{(2\omega+4)}\frac{\phi_0}{\phi_{eq1}} 
= \frac{(2\omega+3)}{(2\omega+4)} \frac{\phi_{eq2}}{\phi_{eq1}}
    \frac{\phi_{0}}{\phi_{eq2}}\\
&=\frac{(2\omega+3)}{(2\omega+4)}  \left(
    1+z_{eq1}\right)^{\frac{1}{(1+\omega)}}
    \left(1+z_{eq2} \right)^{-\frac{\omega}{(1+\omega)(1+2\omega)}}
\end{align*}  
where we have defined the redshift of matter-radiation equality, $z_{eq1}$, and
the redshift of matter-vacuum equality, $z_{eq2}$, as
\begin{equation*}
1+z_{eq1}=\frac{\rho_{m0}}{\rho_{r0}}
\quad \text{and} \quad
1+z_{eq2}=\left(\frac{\rho_{\Lambda0}}{\rho_{m0}}
\right)^{\frac{1}{3}}.  
\end{equation*}
As above, we still have $\rho_{r0}=7.84 \times 10^{-34}g \; cm^{-3}$ but
now our assumption of spatial flatness gives
\begin{equation*}
\rho_{tot0}=\rho_{\Lambda0}+\rho_{m0}+\rho_{r0}=\frac{3H_0^2}{8\pi
  G_N}\frac{(4+3\omega)(4+2\omega)}{6(1+\omega)^2} .
\end{equation*}
or, using $G_N=6.673\times10^{-11} \;  Nm^2kg^{-2}$ and $H_0=100h \;
km s^{-1} Mpc^{-1}$,
\begin{equation*}
1+z_{eq1}=\frac{\left( 2.39 \times 10^4 h^2
\frac{(4+3\omega)(4+2\omega)}{6(1+\omega)^2} -1 \right)}{\left( 1+\frac{\rho_{\Lambda 0}}{\rho_{m0}} \right)}.
\end{equation*} 
Taking the value $\rho_{\Lambda0}/\rho_{m0}=2.7$, consistent with
WMAP observations \cite{Ben03}, we get the results shown in figure \ref{BDresult}.
\begin{figure}[ht]
\begin{center}
\epsfig{file=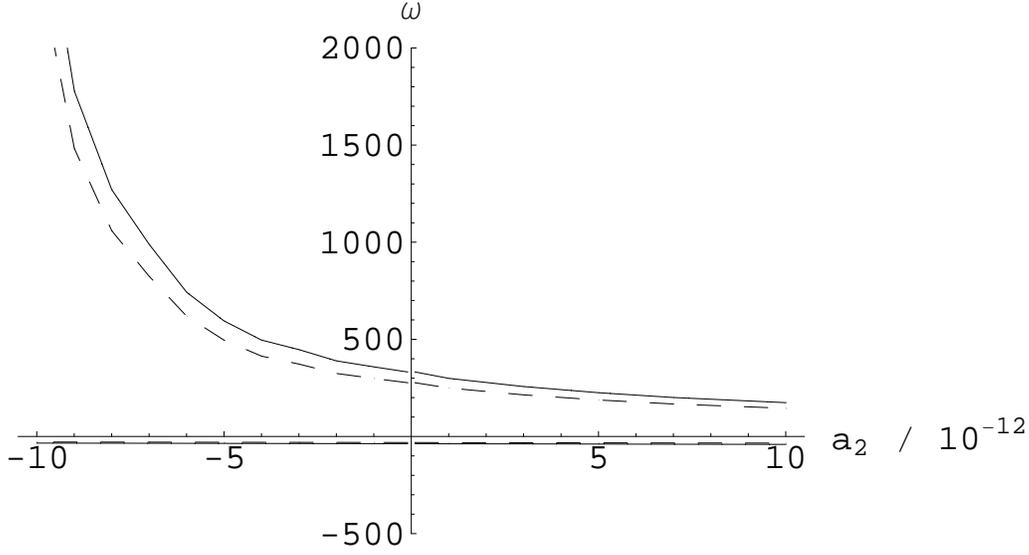,height=8.5cm}
\end{center}
\caption{\textit{The allowed parameter space in this plot is the
    region above the higher
    solid line and below the lower solid line, with $h=0.7$ to 2$\sigma$
    for a universe containing matter and radiation only.  The dashed
    lines show the corresponding result for a universe with a
    non-zero vacuum energy.}}
\label{BDresult}
\end{figure}

\subsubsection{Dynamically-coupled theories}

\subsubsection{Constraints on $\omega(\phi)$ at matter-radiation equality}

Recalling that $e^{2 \Gamma}=1/\phi$ and $\alpha^{-2}=2\omega+3$ allows us to re-write (\ref{JG}) as
\begin{equation*}
G=e^{2 \Gamma}(1+\alpha^2),
\end{equation*}
the ratio $G_{eq}/G_N$ can then be expressed in terms of $\Gamma$ and $\alpha$ as
\begin{equation*}
\frac{G_{eq}}{G_N}=\frac{e^{\Gamma_{eq}}}{e^{\Gamma_0}\sqrt{1+\alpha_0^2}}.
\end{equation*}
Assuming that $\Gamma_0$ and $\ln (1+\alpha_0)$ are negligible
compared to $\Gamma_{eq}$ (i.e. the Universe is close to GR
today, \cite{Bert}) we can write
\begin{align*}
\ln \left( \frac{G_{eq}}{G_N} \right) \simeq \Gamma_{eq} &=
\frac{1}{2}\beta (\psi_{eq}-\psi_{\infty})^2 \\ &=\frac{2 \pi}{(2
  \omega_{eq}+3)} \frac{1}{\beta}.
\end{align*}
This allows us to constrain $\omega_{eq}$ in terms of $\beta$ and
$a_2$, as shown in figure \ref{STresult1}.  Constraints imposed upon
$\omega$ and $a_2$ have the same consequences for the evolution of $G$
  as previously discussed.  The parameter $\beta$ controls the
  evolution of $\omega$, as can be seen from (\ref{alpha}).  In the
  limit $\beta \rightarrow 0$ it can be seen that $\omega$ becomes constant
  and this class of scalar-tensor theories becomes indistinguishable from the
  Brans-Dicke theory.  Constraints upon $\beta$ therefore correspond to
  constraints on the allowed variation of $\omega$ and hence $G$.

The parameter $\beta$ is
taken to be small here so that the ``kick'' on the scalar field during
electron-positron annihilation can be neglected.  These effects
have been explored by Damour and Pichon in \cite{Dam}, for the case
$a_2=0$, and are expected to have the same result in this more general
scenario; we will not repeat their analysis of this effect here.
\begin{figure}
\begin{center}
\epsfig{file=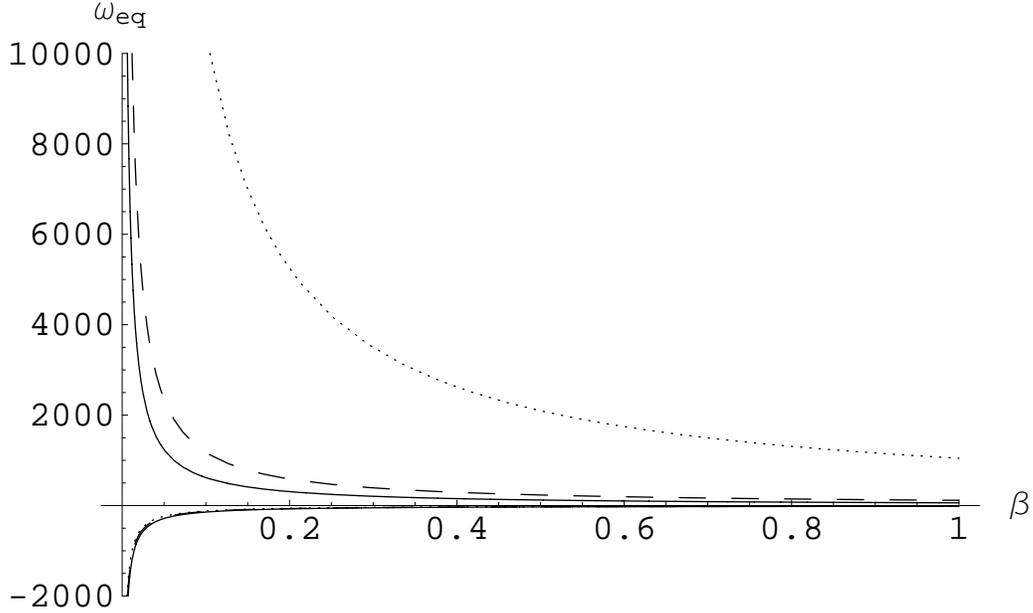,height=8.5cm}
\end{center}
\caption{\textit{The allowed parameter space in this graph is the region above the
    line in the region $\omega>0$ and below the line in the region
    $\omega<0$, to 2$\sigma$.  The solid lines correspond to
    $a_2=10^{-11}$ the dashed lines to $a_2=10^{-13}$ and the dotted
    lines to $a_2=-10^{-11}$.  $\omega_{eq}$ is the value of
    $\omega$ at matter-radiation equality and $\beta$ is a parameter
    of the theory, defined in (\ref{alpha}).}}
\label{STresult1}
\end{figure}

\subsubsection{Constraints on $\omega_0$ for a universe containing
    matter and radiation}

The scalar field can be evolved forward in time from the time of
matter-radiation equality to the present, using the solution
(\ref{solution}).  The two arbitrary constants in this expression can be
fixed using our limiting values of $G_{eq}$ derived above and by assuming that
the evolution of the scalar field has effectively ceased by this time,
i.e. $\psi'_{eq}=0$, as is the case for the models considered here.

In order to gain a quantitative limit on $\psi_1-\psi_{\infty}$, and
hence on $\omega_0$, it is necessary to calculate $N_{eq}$.  From the
definition of $N$, we can write
\begin{align*}
N_{eq}&=-\ln (1+z_{eq})-\Gamma_{eq} + \Gamma_0\\
&\simeq -\ln (1+z_{eq})-\frac{1}{2}\beta(\psi_{eq}-\psi_{\infty})^2
\end{align*}
where $z_{eq}$ is the redshift at $t_{eq}$ and we have
assumed, as before, the term $\ln A_0$ to be negligible.

It now remains to determine 
$1+z_{eq}=\rho_{m0}/\rho_{r0}$ for the case $\omega=\omega(\phi)$.  If
we now assume $\omega_0$ to be moderately large, and recall
our assumption of spatial flatness, we can write
\begin{equation*}
\rho_{tot0}=\rho_{m0}+\rho_{r0} = \frac{3H_0^2}{8\pi G_N}.
\end{equation*}
For $\rho_{r0}=7.84 \times 10^{-34}g \; cm^{-3}$ this gives
$\rho_{m0} \simeq 1.87 \times 10^{-29} \; h^2 g \; cm^{-3}$.  We
have $1+z_{eq} \simeq 2.4 \times 10^4 \; h^2$ and so finally obtain, for $h=0.7$,
\begin{equation*}
N_{eq} \simeq -9.37-\frac{1}{2}\beta(\psi_{eq}-\psi_{\infty})^2.
\end{equation*}

Evolving $\psi-\psi_{\infty}$ from $N_{eq}$ to $N_0=0$ and using
\begin{equation}
\label{w(t)}
\omega_0=\frac{2\pi}{\beta^2(\psi_0-\psi_{\infty})^2}-\frac{3}{2}
\end{equation}
we obtain the results shown in figure \ref{STresult2}.
\begin{figure}[ht]
\begin{center}
\epsfig{file=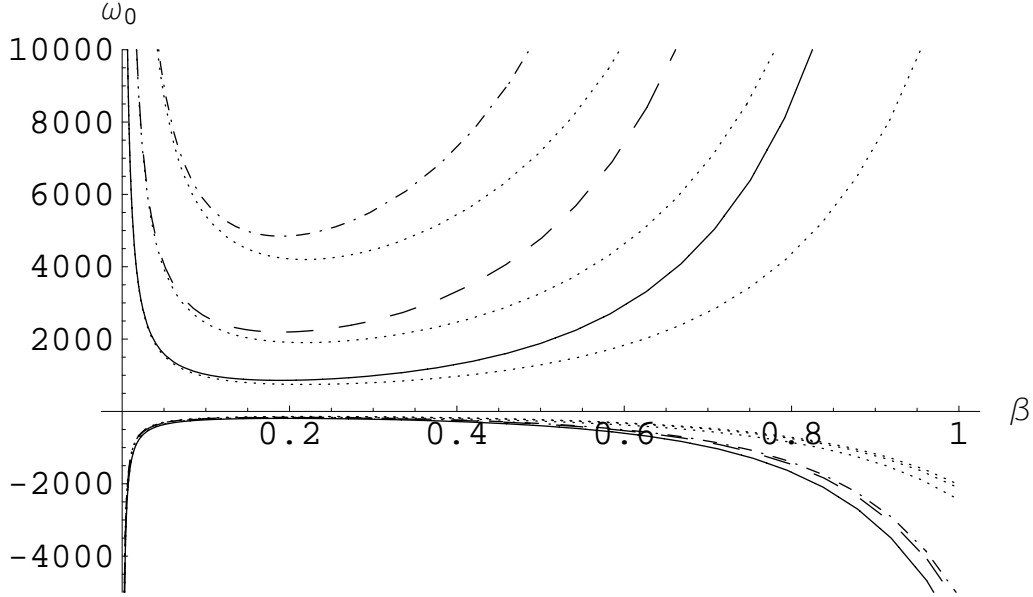,height=8.5cm}
\end{center}
\caption{\textit{The allowed parameter space in this graph is the region above the
    line in the region $\omega>0$ and below the line in the region
    $\omega<0$, to 2$\sigma$ for a universe containing matter and radiation only.  The solid lines correspond to
    $a_2=10^{-11}$, the dashed lines to $a_2=-10^{-11.523}$, and the dot-dashed
    lines to $a_2=-10^{-11.155}$.  The dotted
    lines show the corresponding results for a universe with a
    non-zero vacuum energy.}}
\label{STresult2}
\end{figure}

\subsubsection{Constraints on $\omega_0$ for a universe containing
    matter, radiation and nonzero vacuum energy}

We can repeat the previous analysis for the more realistic case of a
universe with a period of late-time vacuum domination.  We now need
the time of matter-radiation equality, $N_{eq1}$, and
the time of matter-vacuum equality, $N_{eq2}$.

Assuming spatial flatness, a large $\omega_0$, $h=0.7$,
$\rho_{\Lambda0}/\rho_{m0}=2.7$, and $\rho_{r0}$ as above gives
\begin{equation*}
N_{eq1} \simeq -8.06-\frac{1}{2}\beta(\psi_{eq1}-\psi_{\infty})^2,
\end{equation*}
and
\begin{equation*}
N_{eq2} \simeq -0.33-\frac{1}{2}\beta(\psi_{eq2}-\psi_{\infty})^2.
\end{equation*}

We can now evolve $\psi(N)-\psi_{\infty}$ from $N_{eq1}$ to $N_{eq2}$
using (\ref{solution}) and the same boundary conditions as before.
(In finding $N_{eq2}$ we used an iterative method to evaluate
$\psi_{eq2}-\psi_{\infty}$).  We then use the solution (\ref{solution2})
to evolve the field from $N_{eq2}$ to $N_0$; the constants $C$ and
$D$ in (\ref{solution2}) are set by matching
$\psi(N)-\psi_{\infty}$ and its first derivative with the solution (\ref{solution})
at $N_{eq2}$.  Finally, we can calculate $\omega_0$ using
(\ref{w(t)}).  The results of this procedure are shown in figure \ref{STresult2}.

\subsection{Discussion}

Using the framework provided by scalar-tensor theories of gravity we have investigated
the effect of a time-varying $G$ during primordial nucleosynthesis.
We determined the effect on primordial nucleosynthesis numerically, using a
modified version of the Kawano code \cite{Kaw92}, and constrained the
parameters of the underlying theory using these results.  Our results
are consistent with the interpretation that the abundance
of $^4$He is primarily only sensitive to the value of $G$ at the time
when weak interactions freeze out, for the class of scalar-tensor models studied.

Using our numerically determined constraints on the evolution of $G$, we
imposed the 2$\sigma$ upper and lower bounds shown in figure
\ref{BDresult} on the Brans-Dicke parameter $\omega$.  For
a constant $G$ (i.e. $a_2=0$) we get the bounds $\omega \gtrsim 332$ or
$\omega \lesssim -37$, for a universe containing matter and radiation
only, and the bounds  $\omega \gtrsim 277$ or $\omega \lesssim -31$,
for a universe containing matter, radiation and a nonzero vacuum
energy.  As the parameter $a_2$ is increased, the strength of the upper bound decreases
whilst that of the lower bound increases; the opposite behaviour occurs
as $a_2$ is decreased.  This is because the strength of the
bound on $\omega$ is essentially due to the allowed value of $G$ at
matter-radiation equality.  If primordial nucleosynthesis allows this value to be vastly
different from $G_N$ then a significant evolution of $G$ during matter
domination, and hence a low $\vert \omega \vert$, is permitted.  A value of
$G_{eq}$ close to $G_N$ means that only a very slow variation of $G$,
and hence large $\vert \omega \vert$ is permitted.  As $a_2>0$ corresponds to an
increasing $G$ during primordial nucleosynthesis this corresponds to a higher value
of $G_{eq}$ and hence a looser upper bound on $\omega$ and a tighter
lower bound (as $G$ decreases during matter domination for $\omega >-3/2$ and
increases for $\omega <-3/2$); $a_2<0$ corresponds to $G$ decreasing
during radiation domination, and so has the opposite effect.  

The more stringent effect of a nonzero value of
$a_2$ on the upper bounds is due to the current observational determinations of the
$^4$He abundance \cite{Bar03} disfavouring $G>G_N$ during primordial
nucleosynthesis (see e.g. \cite{Bar04}).

The interpretation of the constraints on the more general ST theories
is a little more complicated, due to the increased complexity of the
theories.  The bounds on $\omega(\phi)$ at matter-radiation equality,
shown in figure \ref{STresult1}, are seen to be stronger for smaller
$\beta$ and weaker for larger for $\beta$ (assuming $\beta$ to be
small enough to safely ignore the effect of the $e^-e^+$ kick
analysed by \cite{Dam}).  This should be expected as $\omega \sim \beta^{-2}$ in the models
we are studying.  We see from the constraints on $\omega$
at the present day, shown in figure \ref{STresult2},
that the bounds on $\omega$ become tighter as $\beta$ gets
very small and large, with an apparent minimum in the
bound at $\beta \sim 0.2$.  The tight bounds for very small $\beta$
are due to the tight bounds on $\omega_{eq}$ for small $\beta$. The
tight bounds at large $\beta$ are due to the
attraction towards GR at late times that occurs for this class of scalar-tensor
theories (see e.g. \cite{Dam93}).  This attractor mechanism is
more efficient for larger values of $\beta$, as can be seen from the
solution (\ref{master}), and at late times so $\omega$ is drawn to a larger
value for a larger $\beta$.

The effect of including a late-time vacuum-dominated stage of the Universe's
evolution is to weaken slightly the bounds that can be placed on
$\omega$ at the present day, as can be seen from figures
\ref{BDresult} and \ref{STresult2}.  This weakening of the bounds is due to
a shortening of the matter-dominated period of the Universe's history
which is essentially the only period, after the effects of the free
scalar-dominated phase become negligible, during which $G$ evolves.

We find that the constraints that can be imposed upon the present day
value of $\omega$ from primordial nucleosynthesis are, for most of the allowed parameter space, considerably
weaker than those obtained from observations within the solar system.
To date, the tightest constraint upon $\omega_0$ are imposed by
Bertotti, Iess and Tortora \cite{Bert} who find $\vert \omega_0 \vert
\gtrsim 40000$, to $2\sigma$.  This constraint is obtained from observations of the
Shapiro delay of radio signals from the Cassini spacecraft as it
passes behind the Sun.  We consider
the constraints imposed upon $\omega$ here to be complementary to
these results as they probe different length and time scales, as well
as different epochs of the Universe's history.

\section{Fourth-Order Theories}

The modified cosmological dynamics for fourth-order gravity lead to
different predictions for the outcomes of primordial nucleosynthesis
and microwave background formation, compared to the standard
general relativistic model. The relevant modifications to these physical
processes, and the bounds that they can impose upon the theory, will be
discussed in this section. We will use the solutions (\ref{power}) as they
have been shown to be the generic attractors as $t\rightarrow \infty $
(except for the case $-\frac{1}{4}<\delta <0$ when $\gamma =1$, which has
been excluded as physically unrealistic on the grounds of structure
formation).

\subsection{Primordial Nucleosythesis}

We find that the temperature-time adiabat during radiation domination for
the solution (\ref{power}) is given by the exact relation 
\begin{equation}
t^{2(1+\delta)}=\frac{A}{T^{4}}  \label{T(t)}
\end{equation}%
where, as usual (with units $\hbar =c=1=k_{B}$), 
\begin{equation*}
\rho =\frac{g\pi ^{2}}{15}T^{4}
\end{equation*}%
where $g$ is the total number of relativistic spin states at temperature $T$%
. The constant $A$ can be determined from the generalised Friedmann equation
(\ref{Friedman2}) and is dependent on the present day value of the Ricci
scalar, through equation (\ref{chi}). (This dependence is analogous to the
dependence of scalar-tensor theories on the evolution of the non-minimally
coupled scalar, as may be expected from the relationship between these
theories \cite{Mag94}). As a first approximation, we assume the universe to
have been matter dominated throughout its later history; this allows us to
write 
\begin{equation}
A=\left( \frac{45(1-2\delta )(1-2\delta -5\delta ^{2})}{32(1-\delta )g\pi
^{3}G}\right) \left( \frac{2(1+\delta )}{3H_{0}}\right) ^{2\delta }
\label{A}
\end{equation}%
where $H_{0}$ is the value of Hubble's constant today and we have used the
solution (\ref{power}) to model the evolution of $a(t)$. Adding a recent
period of accelerated expansion will refine the constant $A$, but in the
interests of brevity we exclude this from the current analysis.  As
usual, the weak-interaction time is given by 
\begin{equation*}
t_{wk}\propto \frac{1}{T^{5}}.
\end{equation*}%
The freeze-out temperature, $T_{f}$, for neutron-proton kinetic
equilibrium is then defined by 
\begin{equation*}
t(T_{f})=t_{wk}(T_{f}).
\end{equation*}%
Hence the freeze-out temperature in this theory, with $\delta \neq 0$, is
related to that in the general relativistic case with $\delta =0$, $%
T_{f}^{GR}$:
\begin{equation}
T_{f}=C(T_{f}^{GR})^{\frac{3(1+\delta )}{(3+5\delta )}}  \label{freeze}
\end{equation}%
where 
\begin{equation}
C=\left( \frac{(1-\delta )}{(1-2\delta )(1-2\delta -5\delta ^{2})}\right) ^{%
\frac{1}{2(2+5\delta )}}\left( \frac{45}{32g\pi ^{3}G}\right) ^{\frac{\delta
(1+\delta )}{2(3+5\delta )}}\left( \frac{3H_{0}}{2(1+\delta )}\right) ^{%
\frac{\delta }{(3+5\delta) }}.  \label{freeze2}
\end{equation}%
The neutron-proton ratio, $n/p$, is now determined at temperature $T$ when
the equilibrium holds by 
\begin{equation*}
\frac{n}{p}=\exp \left( -\frac{\Delta m}{T}\right) .
\end{equation*}%
where $\Delta m$ is the neutron-proton mass difference. Hence the
neutron-proton ratio at freeze-out in the $R^{1+\delta }$ early universe is
given by 
\begin{equation*}
\frac{n}{p}=\exp \left( -\frac{\Delta m}{C(T_{f}^{GR})^{1-\varepsilon }}%
\right) ,
\end{equation*}%
where 
\begin{equation*}
\varepsilon \equiv \frac{2\delta }{3+5\delta }.
\end{equation*}

The frozen-out $n/p$ ratio in the $R^{1+\delta }$ theory is given by a
power of its value in the general relativistic case, $\left( \frac{n}{p}%
\right) _{GR}\approx 1/7$, by 
\begin{equation*}
\frac{n}{p}=\left( \frac{n}{p}\right) _{GR}^{C(T_{f}^{GR})^{\varepsilon }}.
\end{equation*}%
It is seen that when $C(T_{f}^{GR})^{\varepsilon }>1$ ($\delta <0$) there is
a smaller frozen-out neutron-proton ratio that in the general relativistic
case and consequently a lower final helium-4 abundance than in the standard
general-relativistic early universe containing the same number of
relativistic spin states. This happens because the freeze-out temperature is
lower than in GR. The neutrons remain in equilibrium to a
lower temperature and their slightly higher mass shifts the number balance
more towards the protons the longer they are in equilibrium. Note that a
reduction in the helium-4 abundance compared to the standard model of
GR is both astrophysically interesting and difficult to
achieve (all other variants like extra particle species \cite{shv, nus},
anisotropies \cite{ht, jb1, jb2, jb32}, magnetic fields \cite{mag, YM},
gravitational waves \cite{jb1, jb2, skew}, or varying $G$ \cite{bmaeda,
newt, clift}, lead to an \textit{increase} in the expansion rate and in the
final helium-4 abundance). Conversely, when $C(T_{f}^{GR})^{\varepsilon }<1$
($\delta >0$) freeze-out occurs at a higher temperature than in general
relativity and a higher final helium-4 abundance fraction results. The final
helium-4 mass fraction $Y$ is well approximated by 
\begin{equation}
Y=\frac{2n/p}{(1+n/p)}.
\end{equation}

It is now possible to constrain the value of $\delta $ using observational
abundances of the light elements. In doing this we will use the results of
Carroll and Kaplinghat \cite{Car02} who consider nucleosynthesis with a
Hubble constant parametrised by 
\begin{equation*}
H(T)=\left( \frac{T}{1MeV}\right) ^{\alpha }H_{1}.
\end{equation*}%
Our theory can be cast into this form by substituting 
\begin{equation*}
\alpha =\frac{2}{(1+\delta )}
\end{equation*}%
and 
\begin{equation*}
H_{1}=\frac{(1+\delta )}{2}A^{-\frac{1}{2(1+\delta )}}(1MeV)^{\frac{2}{%
(1+\delta )}},
\end{equation*}%
so, taking $g=43/8$, $G=6.72\times 10^{-45}MeV^{-2}$ and $H_{0}=1.51\times
10^{-39}MeV$ \cite{Ben03}, this can be rewritten as 
\begin{equation*}
H_{1}=\frac{(1+\delta )}{2}\left( \frac{7.96\times 10^{-43}(1-\delta )}{%
(1-2\delta )(1-2\delta -5\delta ^{2})}\right) ^{\frac{1}{2(1+\delta )}%
}\left( \frac{2.23\times 10^{-39}}{(1+\delta )}\right) ^{\frac{\delta }{%
(1+\delta )}}MeV.
\end{equation*}%
Carroll and Kaplinghat use the observational abundances inferred by Olive
et. al. \cite{Oli00}, 
\begin{align*}
0.228\leqslant Y_{P}& \leqslant 0.248 \\
2\leqslant 10^{5}\times & \frac{D}{H}\leqslant 5 \\
1\leqslant 10^{10}\times & \frac{^{7}Li}{H}\leqslant 3,
\end{align*}%
to impose the constraint 
\begin{equation*}
H_{1}=H_{c}\left( \frac{T_{c}}{MeV}\right) ^{-\alpha }
\end{equation*}%
where $H_{c}=2.6\pm 0.9\times 10^{-23}MeV$ at $T_{c}=0.2MeV$ for $%
0.5\leqslant \eta _{10}\leqslant 50$, or $H_{c}=2.0\pm 0.3\times 10^{-23}$
for $1\leqslant \eta _{10}\leqslant 10$ and $\eta _{10}$ is $10^{10}$ times
the baryon to photon ratio.

These results can now be used to impose upon $\delta$ the constraints 
\begin{equation*}
-0.017 \leqslant \delta \leqslant 0.0012,
\end{equation*}
for $0.5 \leqslant \eta_{10} \leqslant 50$, or 
\begin{equation}
-0.0064 \leqslant \delta \leqslant 0.0012,
\end{equation}
for $1 \leqslant \eta_{10} \leqslant 10$.

\subsection{Microwave Background Formation}

The horizon size at the epoch of matter-radiation equality is of great
observational significance. During radiation domination cosmological
perturbations on sub-horizon scales are effectively frozen. Once matter
domination commences, however, perturbations on all scales are allowed to
grow and structure formation begins. The horizon size at matter-radiation
equality is therefore frozen into the power spectrum of perturbations and is
observable. Calculation of the horizon sizes in this theory proceeds in a
similar way to that in Brans-Dicke theory \cite{cmb1}.

In making an estimate of the horizon size in $R^{1+\delta }$ theory we will
use the generalised Friedmann equation, (\ref{Friedman2}), in the form 
\begin{equation}
H^{2}+\delta H\frac{\dot{R}}{R}-\frac{\delta R}{6(1+\delta )}=\frac{8\pi
G(1-\delta )}{3(1-2\delta )}\frac{R_{0}^{\delta }}{R^{\delta }}\rho .
\label{Fred}
\end{equation}
Again, we assume the form (\ref{power}) to model the evolution of the scale
factor during the epoch of matter domination. This gives 
\begin{align*}
a(t)& =a_{0}\left( \frac{t}{t_{0}}\right) ^{\frac{2(1+\delta )}{3}} \\
H_{0}& =\frac{2(1+\delta )}{3t_{0}} \\
\rho _{m}& =\frac{3H_{0}^{2}}{16\pi G}\frac{(1-2\delta )(2-3\delta -8\delta
^{2})}{(1-\delta )(1+\delta )^{2}}\frac{a_{0}^{3}}{a^{3}} \\
R(t)& =\frac{4(1+5\delta +4\delta ^{2})}{3t^{2}}
\end{align*}%
during the matter-dominated era. In order to simplify matters, we assume the
above solutions to hold exactly from the time of matter-radiation equality
up until the present (neglecting the small residual radiation effects and
the late time acceleration). Substituting them into (\ref{Fred}), along
with $\rho _{eq}=2\rho _{meq}$ at equality, we can then solve for $H_{eq}$ to
first order in $\delta $ to find 
\begin{equation}
\frac{a_{eq}H_{eq}}{a_{0}H_{0}}\simeq \sqrt{2}\sqrt{1+z_{eq}}^{\frac{%
1-2\delta }{1+\delta }}(1-2.686\delta )+O(\delta ^{2})  \label{Heq}
\end{equation}%
where $z_{eq}$ is the redshift at matter radiation equality and $H$ has been
treated as an independent parameter. The value of $1+z_{eq}$ can now be
calculated in this theory as 
\begin{equation}
1+z_{eq}=\frac{\rho _{r0}}{\rho _{m0}}.  \label{zeq}
\end{equation}%
Taking the present day temperature of the microwave background as $%
T=2.728\pm 0.004K$ \cite{Fix96} gives 
\begin{equation}
\rho _{r0}=3.37\times 10^{-39}MeV^{4}  \label{rhor0}
\end{equation}%
where three families of light neutrinos have been assumed at a temperature
lower than that of the microwave background by a factor $(4/11)^{\frac{1}{3}%
} $. Using the same values for $G$ and $H_{0}$ as above we than find from
the above expression for $\rho _{m}$ that 
\begin{equation}
\rho _{m0}=2.03\times 10^{-35}\frac{(1-2\delta )(2-3\delta -8\delta ^{2})}{%
(1-\delta )(1+\delta )^{2}}MeV^{4}.  \label{rhom0}
\end{equation}%
Substituting (\ref{rhor0}), (\ref{rhom0}) and (\ref{zeq}) into (\ref{Heq})
then gives the expression for the horizon size at equality, to first order
in $\delta $, as 
\begin{equation}
\frac{a_{eq}H_{eq}}{a_{0}H_{0}}\simeq 155(1-19\delta )+O(\delta ^{2}).
\end{equation}%
This expression shows that the horizon size at matter-radiation equality
will be shifted by $\sim 1\%$ for a value of $\delta \sim 0.0005$. This
shift in horizon size should be observable in a shift of the peak of the
power spectrum of perturbations, compared to its position in the standard
general relativistic cosmology. Microwave background observations,
therefore, allow a potentially tight bound to be derived on the value of $%
\delta $. This effect is analogous to the shift of power-spectrum peaks in
Brans-Dicke theory (see e.g. \cite{cmb1}, \cite{cmb2}).

A full analysis of the spectrum of perturbations in this theory requires a
knowledge of the evolution of linearised perturbations as well as a
marginalisation over other parameters which can mimic this effect (e.g.
baryon density). Such a study is beyond the scope of the present work.
\clearemptydoublepage
\chapter{Weak-Field Constraints}
\label{Weak-Field Constraints}

\bigskip

It is the weak-field limit of gravity that is most readily accessible
to us for experiment and observation.  By performing gravitational
experiments in the solar system, and comparing the results of these
experiments to the predictions from different gravitational theories,
it is possible to place constraints on the theory under
consideration.  To date, it is weak field gravitational experiments
that have so far offered the best constraints on the theory (though
with astrophysical and cosmological observations becoming ever more
accurate there is a possibility of the situation changing in the not
too distant future).

The consideration of gravitational experiments in the weak-field, low
velocity limit is usually performed within the frame-work of the
parameterised post-Newtonian (PPN) formalism.  In this chapter we will
give a brief explanation of the PPN formalism, and explain how it can
be used to constrain scalar-tensor and fourth-order theories of
gravity.  We will also point out the areas in which this formalism
cannot be applied so easily, and provide alternative ways to proceed.

This chapter is based on the work of Clifton and Barrow \cite{Cli}
and Clifton \cite{Clifton}.

\section{Parameterised Post-Newtonian Approach}

We will provide in this section a brief explanation of the PPN
approach (for a more detailed explanation of this formalism the
reader is referred to Will, \cite{will}).

The PPN formalism does not deal with exact solutions, but instead uses
perturbative expansions about a Minkowski background.  In the
weak-field, low velocity limit we need only deal with the lowest orders of
the perturbative expansion.  Solving the field equations order by
order, we can then find approximate solutions which can be compared
with observation.

Following Will we define an ``order of smallness'' such that
\begin{equation*}
U \sim v^2 \sim \frac{p}{\rho} \sim \Pi \sim O(2)
\end{equation*}
where $U$ is the Newtonian potential, $v$ is the velocity of a body in
the space-time, $p$ is pressure, $\rho$ is energy density and $\Pi$ is
specific energy density (ratio of energy density to rest mass
density) and $O(2)$ means second order in ``smallness''.  In the solar
system we have that $U \leqslant 10^{-5}$ is a small perturbation to flatness, and
all of the other above quantities are less that $U$ and so are also
small (see Will \cite{will} for more details).  It is also
assumed that the time evolution of the solar system is slow
\begin{equation*}
\frac{\vert \partial / \partial t \vert}{\vert \partial / \partial x
  \vert} \sim O(1).
\end{equation*}
A perturbative expansion about the Minkowski background can now be
performed in the order of smallness that has just been defined.

To obtain the Newtonian limit for particles following
time-like geodesics we now need to know only the $O(2)$ term
in the $g_{0 0}$ component of the metric.  The Newtonian limit of light rays are straight
lines, so for null geodesics the Newtonian limit is given simply by
the Minkowski background.  The post-Newtonian limit is then given by
the next non-zero order in the perturbative expansion of the
metric components.  For time-like particles this is given by $g_{0 0}$
to $O(4)$, $g_{0 \mu}$ to $O(3)$ and $g_{\mu \nu}$ to $O(2)$.  For
null particles it is given by $g_{0 0}$ to $O(2)$ and $g_{\mu \nu}$ to $O(2)$.

The usual approach at this point is to construct a metric containing a
number of the possible functionals of the matter variables, to the
required post-Newtonian order.  Giving the different functionals
(often referred to as potentials) arbitrary coefficients it is then
possible to construct a generalised metric.  The values of the
coefficients are then given by solving the field equations of the
theory being considered, order by order in perturbation.  To restrict
the number of potentials in the metric a number of conditions are
usually imposed such as asymptotic flatness and simplicity of the
functional form.  After specifying a particular gauge the generalised metric can
then be written down explicitly.

The appeal of this approach is that the generalised metric can be used
by experimenters to place constraints on the coefficients of the
metric potentials, independent of any further considerations of the
specific gravitational theory being tested.  For any particular
gravitational theory (that fits into this framework) the values of the
coefficients can then be expressed in terms of the parameters of the
theory.  In this way the results of gravitational experiments can be
applied to any gravitational theory that fits into the PPN
frame-work.

A simplified version of the post-Newtonian metric is given by
\begin{align}
\label{PPN}
g_{00} &=-1+2 U-2 \beta U^2+(2+2 \gamma+\zeta_1) \Phi_1 \\ & \qquad +2 (1-2 \beta +3
\gamma+\zeta_2) \Phi_2 +2 (1+\zeta_3) \Phi_3 +2 (3 \gamma +3 \zeta_4) \nonumber
\Phi_4\\ \nonumber
g_{\mu \nu} &= (1+2 \gamma U) \delta_{\mu \nu}\\ \nonumber
g_{0 \mu} &= -\frac{1}{2} (3+4 \gamma+\zeta_1 +\alpha_1-\alpha_2)
V_{\mu} - \frac{1}{2} (1- \zeta_1+\alpha_2) W_{\mu})
\end{align}
where $\beta$, $\gamma$, $\zeta_1$, $\zeta_2$, $\zeta_3$, $\zeta_4$,
$\alpha_1$ and $\alpha_2$ are some of the possible post-Newtonian
coefficients and $\Phi_1$, $\Phi_2$, $\Phi_3$, $\Phi_4$, $V_{\mu}$
and $W_{\mu}$ are post-Newtonian potentials.  (See \cite{will} for a more comprehensive version of
the post-Newtonian metric, and definitions of the above post-Newtonian
potentials).  For an idealised spherical and non-rotating massive object at the origin, in an
otherwise empty space-time, this metric reduces to
\begin{equation*}
ds^2=-\left( 1-\frac{2 G m}{r}+ \beta \frac{2 G^2 m^2}{r^2} \right)
dt^2+\left( 1+\gamma \frac{2 G m}{r} \right) \delta_{\mu \nu}
dx^{\mu} dx^{\nu}
\end{equation*}
which is exactly the metric considered by Eddington \cite{edd},
Robertson \cite{Rob} and Schiff \cite{Sch}.

The PPN approach has been shown to work very well for a number of
gravitational theories, but it does have some drawbacks.  From the
point of view of the present study the principle drawback is the
assumption of the Minkowski background.  Whilst the Minkowski
background is very well justified in GR, it is less well justified in
some alternative theories.  This is principally due to the lack of any
equivalent to Birkhoff's theorem.  In fact, in a previous chapter we
have shown explicitly the existence of solutions which are spherically
symmetric, non-static vacuum solutions.  This breakdown of Birkhoff's
theorem can be understood quite simply in terms of the conformal
equivalence of these theories in vacuum to GR in the presence of a
scalar field.  Just as it conceivable that a scalar field in GR could
cause a cosmological evolution (even in the absence of any other
matter sources), so it is conceivable here that the vacuum solutions
of these modified theories could evolve cosmologically.

These problems are especially evident in the $R^n$ theories.  In these
theories, even in the static case, it has already been shown that at asymptotically large
spatial distances the general solution is not attracted towards
flatness.  This strange asymptotic behaviour means that the theories
do not fit straightforwardly into the PPN frame-work.  We develop new
ways of dealing with this strange behaviour below, and apply the
results of gravitational experiments to these theories.

\section{Scalar-Tensor Theories}

The post-Newtonian limit of the scalar-tensor theories of gravity has
been well studied numerous times in the past.  We present here a brief
illustration of how the post-Newtonian coefficients are obtained, and
how they can be used to constrain the theory.

In order to fit into the PPN formalism the metric and scalar field
must be subjected to a perturbative expansion of the form
\begin{align*}
g_{\mu \nu} &= \eta_{\mu \nu} + h_{\mu \nu}\\
\phi &= \phi_0 + \varphi
\end{align*}
where
\begin{align*}
h_{0 0} &\sim O(2) + O(4)\\
h_{\mu \nu} &\sim O(3)\\
h_{0 \mu} &\sim O(2)\\
\varphi &\sim O(2) +O(4).
\end{align*}
These expansions can then be substituted into the scalar-tensor field equations
(\ref{STfields}) and (\ref{STfields2}), which can be solved order by
order in perturbations to obtain (after an appropriate gauge choice) \cite{will}
\begin{align}
\label{STPPN}
g_{00} &=-1+2 U-2 (1+\Lambda) U^2+4 \left( \frac{3+2 \omega}{4+2
  \omega} \right) \Phi_1 \\ 
& \qquad +4 \left( \frac{1+2 \omega}{4+2 \omega} -\Lambda \right)
  \Phi_2 +2 \Phi_3 +6 \left( \frac{1+\omega}{2+\omega} \right) \Phi_4 \nonumber \\ \nonumber
g_{\mu \nu} &= \left(1+2 \left( \frac{1+\omega}{2+\omega} \right) U \right) \delta_{\mu \nu}\\ \nonumber
g_{0 \mu} &= -\frac{1}{2} \left( \frac{10+7 \omega}{2+\omega} \right)
V_{\mu} - \frac{1}{2} W_{\mu}
\end{align}
where
\begin{equation*}
\Lambda \equiv \frac{d\omega/d\phi}{(4+2 \omega) (3+2 \omega)^2}
\end{equation*}
and Newton's constant has been set to
\begin{equation*}
G \equiv \left( \frac{4+2 \omega}{3+2 \omega} \right)
\frac{1}{\phi_0}.
\end{equation*}
A direct comparison of (\ref{PPN}) and (\ref{STPPN}) allows the PPN
coefficients to be read off as
\begin{align*}
\gamma &= \frac{1+\omega}{2+\omega}\\
\beta &= 1+\Lambda\\
\zeta_1 &= \zeta_2 = \zeta_3 = \zeta_4 = \alpha_1 = \alpha_2 = 0.
\end{align*}
The Brans-Dicke theory can be particularly well constrained by
observations constraining the PPN coefficients.  For the Brans-Dicke
theory $\omega =$constant, so $\Lambda=0$, and the constraints on the
coefficient $\gamma$,
\begin{equation*}
\gamma = 1+(2.1 \pm 2.3) \times 10^{-5},
\end{equation*}
from observations of the Shapiro time delay of radio signals from the
Cassini space-probe \cite{Bert}, gives the constraint
\begin{equation*}
\omega > 40 000.
\end{equation*}
This constraint is very limiting indeed for the Brans-Dicke theory,
but less so for theories in which $\omega$ is allowed to be
a function of $\phi$.  For these theories, whilst the present day
value of $\omega$ in the solar system is tightly constrained, the
value of $\omega$ at other points in the Universe's history
(particularly in the early Universe) is less restricted.  To
better constrain these theories we need to obtain limits on the
parameter $\beta$ or, even better, use cosmological constraints from
the early Universe, such as those which were discussed in the last chapter.

\section{Fourth-Order Theories}

In order to calculate the classical tests of metric theories of gravity
(i.e. bending and time-delay of light rays and the perihelion precession of
Mercury) we require a spherically symmetric solution to the $\mathcal{L}=R^{1+\delta}$
field equations (\ref{field}).  Due to the complicated form of these
equations we are unable to find the general solution; instead we propose to
use first-order solutions around attractors as $r\rightarrow
\infty $ or $t \rightarrow \infty$.  This method should be applicable to gravitational experiments
performed in the solar system as the gravitational field in this region can
be considered weak and we will be considering experiments performed at large 
$r$ (in terms of the Schwarzschild radius of the massive objects in the
system) and at late times.

\subsection{Static Space-Time}

In this section we will use the static and spherically symmetric
solution found previously (\ref{Chan}).  We choose to arbitrarily set the
oscillatory parts of the solution to zero, and
hence ensure that the gravitational force is always attractive. This
considerable simplification of the solution also allows a straightforward
calculation of both null and time-like geodesics which can be used to compute
the outcomes of the classical tests in this space-time.

\subsubsection{Solution in isotropic coordinates}

Having removed the oscillatory parts of the solution we are left with the
part corresponding to the exact solution (\ref{Chan}). Making the coordinate
transformation 
\begin{equation*}
r^{(1-2 \delta+4 \delta^2)/(1-d)}= \left(1-\frac{C}{4 \hat{r}^{\sqrt{\frac{%
(1-2 \delta+4 \delta^2)}{(1-2 \delta-2 \delta^2)}}}} \right)^2 \hat{r}^{%
\sqrt{\frac{(1-2 \delta+4 \delta^2)}{(1-2 \delta-2 \delta^2)}}}
\end{equation*}
the solution (\ref{Chan}) can be transformed into the isotropic coordinate
system 
\begin{equation}  \label{iso}
ds^2=-A(\hat{r}) d t^2+B(\hat{r}) ( d\hat{r}^2+ \hat{r}^2 (d
\theta^2+\sin^2\theta d \phi^2))
\end{equation}
where 
\begin{equation*}
A(\hat{r})= \hat{r}^{\frac{2 \delta (1+2 \delta)}{\sqrt{(1-2
\delta-2\delta^2)(1-2\delta+4 \delta^2)}}} \left(1+\frac{C}{4 \hat{r}^{\sqrt{%
\frac{(1-2 \delta+4 \delta^2)}{(1-2 \delta-2 \delta^2)}}}} \right)^2 \left(1-%
\frac{C}{4 \hat{r}^{\sqrt{\frac{(1-2 \delta+4 \delta^2)}{(1-2 \delta-2
\delta^2)}}}} \right)^{-\frac{2 (1+4 \delta)}{(1-2 \delta+4 \delta^2)}}
\end{equation*}
and 
\begin{equation*}
B(\hat{r})= \hat{r}^{-2+2\frac{(1-\delta)}{\sqrt{(1-2
\delta-2\delta^2)(1-2\delta+4 \delta^2)}}} \left(1-\frac{C}{4 \hat{r}^{\sqrt{%
\frac{(1-2 \delta+4 \delta^2)}{(1-2 \delta-2 \delta^2)}}}} \right)^{\frac{4
(1-\delta)}{(1-2 \delta+4 \delta^2)}}.
\end{equation*}
This is, to linear order in $C$, 
\begin{equation*}
A(\hat{r})= \hat{r}^{\frac{2 \delta (1+2 \delta)}{\sqrt{(1-2
\delta-2\delta^2)(1-2\delta+4 \delta^2)}}} \left(1+ \frac{(1-\delta) (1-2
\delta)}{(1-2 \delta+4 \delta^2)}\frac{C}{\hat{r}^{\sqrt{\frac{(1-2 \delta+4
\delta^2)}{(1-2 \delta-2 \delta^2)}}}} \right)
\end{equation*}
and 
\begin{equation*}
B(\hat{r})= \hat{r}^{-2+2\frac{(1-\delta)}{\sqrt{(1-2
\delta-2\delta^2)(1-2\delta+4 \delta^2)}}} \left(1- \frac{(1-\delta)}{(1-2
\delta+4 \delta^2)}\frac{C}{\hat{r}^{\sqrt{\frac{(1-2 \delta+4 \delta^2)}{%
(1-2 \delta-2 \delta^2)}}}} \right).
\end{equation*}

\subsubsection{Newtonian limit}

We first investigate the Newtonian limit of the geodesic equation in order
to set the constant $C$ in the solution (\ref{iso}) above. As usual, we have 
\begin{equation*}
\Phi _{,\mu }=\Gamma _{\;00}^{\mu }
\end{equation*}%
where $\Phi $ is the Newtonian gravitational potential. Substituting in the
isotropic metric (\ref{iso}) this gives 
\begin{align}
\nabla \Phi & =\frac{\nabla A(\hat{r})}{2B(\hat{r})}  \label{motion} \\
& =\frac{\delta (1+2\delta )\hat{r}^{1-2\sqrt{\frac{1-2\delta -2\delta ^{2}}{%
1-2\delta +4\delta ^{2}}}}}{\sqrt{(1-2\delta -2\delta ^{2})(1-2\delta
+4\delta ^{2})}} \nonumber \\ & \qquad -\frac{(1-\delta )(1-8\delta +4\delta ^{2})C\hat{r}^{1-\frac{%
3(1-2\delta )}{\sqrt{(1-2\delta -2\delta ^{2})(1-2\delta +4\delta ^{2})}}}}{2%
\sqrt{(1-2\delta -2\delta ^{2})(1-2\delta +4\delta
  ^{2})^{3}}}+O(C^{2}). \nonumber
\end{align}

The second term in the expression goes as $\sim \hat{r}^{-2+O(\delta ^{2})}$
and so corresponds to the Newtonian part of the gravitational force. The
first term, however, goes as $\sim \hat{r}^{-1+O(\delta ^{2})}$ and has no
Newtonian counterpart. In order for the Newtonian part to dominate over the
non-Newtonian part we must impose upon $\delta $ the requirement that it is
at most 
\begin{equation*}
\delta \sim O\left( \frac{C}{r}\right) .
\end{equation*}%
If $\delta $ were larger than this then the non-Newtonian part of the
potential would dominate over the Newtonian part, which is clearly
unacceptable at scales over which the Newtonian potential has been measured
and shown to be accurate.

This requirement upon the order of magnitude of $\delta$ allows (\ref{motion}%
) to be written 
\begin{equation}  \label{force}
\nabla \Phi = \frac{\delta}{\hat{r}^{1+O(C^2)}}-\frac{C}{2 \hat{r}^{2+O(C^2)}%
}+O(C^2)
\end{equation}
where expansions in $C$ have been carried out separately in the coefficients
and the powers of $\hat{r}$ of the two terms.

Comparison of (\ref{force}) with the Newtonian force law 
\begin{equation*}
\nabla \Phi_N=\frac{G m}{r^2}
\end{equation*}
allows the value of $C$ to be read off as 
\begin{equation*}
C=-2 G m +O(\delta).
\end{equation*}

\subsubsection{Post-Newtonian limit}

We now wish to calculate, to post-Newtonian order, the equations of motion
for test particles in the metric (\ref{iso}). The geodesic equation can be
written in its usual form 
\begin{equation*}
\frac{d^{2}x^{\mu }}{d\lambda ^{2}}+\Gamma _{\;ij}^{\mu }\frac{dx^{i}}{%
d\lambda }\frac{dx^{j}}{d\lambda }=0,
\end{equation*}%
where $\lambda $ can be taken as proper time for a time-like geodesic or as
an affine parameter for a null geodesic. In terms of coordinate time this
can be written 
\begin{equation}
\frac{d^{2}x^{\mu }}{dt^{2}}+\left( \Gamma _{\;ij}^{\mu }-\Gamma _{\;ij}^{0}%
\frac{dx^{\mu }}{dt}\right) \frac{dx^{i}}{dt}\frac{dx^{j}}{dt}=0.
\label{motion2}
\end{equation}%
We also have the integral 
\begin{equation}
g_{ij}\frac{dx^{i}}{dt}\frac{dx^{j}}{dt}=S  \label{motion3}
\end{equation}%
where $S=-1$ for particles and $0$ for photons.

Substituting (\ref{iso}) into (\ref{motion2}) and (\ref{motion3}) gives, to
the relevant order, the equations of motion 
\begin{multline}
\frac{d^{2}\mathbf{x}}{dt^{2}}=-\frac{Gm}{r^{2}}\left( 1+\left\vert \frac{d%
\mathbf{x}}{dt}\right\vert ^{2}\right) \mathbf{e_{r}}+4\frac{G^{2}m^{2}}{%
r^{3}}\mathbf{e_{r}}+4\frac{Gm}{r^{2}}\mathbf{e_{r}}\cdot \frac{d\mathbf{x}}{%
dt}\frac{d\mathbf{x}}{dt}  \label{geo1} \\
-\frac{\delta }{r}\left( 1-\left\vert \frac{d\mathbf{x}}{dt}\right\vert
^{2}\right) \mathbf{e_{r}}-4\frac{\delta ^{2}}{r}\mathbf{e_{r}}+\delta \frac{%
Gm}{r^{2}}\mathbf{e_{r}}+O(G^{3}m^{3})
\end{multline}%
and 
\begin{equation}
\left\vert \frac{d\mathbf{x}}{dt}\right\vert ^{2}=1-4\frac{Gm}{r}+\frac{S}{%
r^{2\delta }}-2S\frac{Gm}{r^{1+2\delta }}+O(G^{2}m^{2}).  \label{geo2}
\end{equation}%
(In the interests of concision we have excluded the $O(\delta ^{2})$ terms
from the powers of $r$; the reader should regard them as being there
implicitly). The first three terms in equation (\ref{geo1}) are identical to
their general relativistic counterparts. The next two terms are completely
new and have no counterparts in general relativity. The last term in
equation (\ref{geo1}) can be removed by rescaling the mass term by $%
m\rightarrow m(1+\delta )$; this has no effect on the Newtonian limit of the
geodesic equation as any term $Gm\delta $ is of post-Newtonian order.

\subsubsection{The bending of light and time delay of radio signals}

From equation (\ref{geo2}) it can be seen that the solution for null
geodesics, to zeroth order, is a straight line that can be parametrised by 
\begin{equation*}
\mathbf{x}=\mathbf{n}(t-t_{0})
\end{equation*}%
where $\mathbf{n}\cdot \mathbf{n}=1$. Considering a small departure from the
zeroth order solution we can write 
\begin{equation*}
\mathbf{x}=\mathbf{n}(t-t_{0})+\mathbf{x}_{1}
\end{equation*}%
where $\mathbf{x}_{1}$ is small. To first order, the equations of motion (%
\ref{geo1}) and (\ref{geo2}) then become 
\begin{equation}
\frac{d^{2}\mathbf{x}}{dt^{2}}=-2\frac{Gm}{r^{2}}\mathbf{e_{r}}+4\frac{Gm}{%
r^{2}}(\mathbf{n}\cdot \mathbf{e_{r}})\mathbf{n}  \label{geo3}
\end{equation}%
and 
\begin{equation}
\mathbf{n}\cdot \frac{d\mathbf{x}}{dt}=-2\frac{Gm}{r}.  \label{geo4}
\end{equation}

Equations (\ref{geo3}) and (\ref{geo4}) can be seen to be identical to the
first-order equations of motions for photons in GR. We
therefore conclude that any observations involving the motion of photons in
a stationary and spherically symmetric weak field situation cannot tell any
difference between GR and this $R^{1+\delta }$ theory, to
first post-Newtonian order. This includes the classical light bending and
time delay tests, which should measure the post-Newtonian parameter $\gamma $ to be one
in this theory, as in GR.

\subsubsection{Perihelion precession}

In calculating the perihelion precession of a test particle in the geometry (%
\ref{iso}) it is convenient to use the standard procedures for computing the
perturbations of orbital elements (see \cite{Sma53} and \cite{Rob68}). In
the notation of Robertson and Noonan \cite{Rob68} the measured rate of
change of the perihelion in geocentric coordinates is given by 
\begin{equation}
\frac{d\tilde{\omega}}{dt}=-\frac{p\mathcal{R}}{he}\cos \phi +\frac{\mathcal{%
J}(p+r)}{he}\sin \phi  \label{per}
\end{equation}%
where $p$ is the semi-latus rectum of the orbit, $h$ is the
angular-momentum per unit mass, $e$ is the eccentricity and $\mathcal{R}$
and $\mathcal{J}$ are the components of the acceleration in radial and
normal to radial directions in the orbital plane, respectively. The radial
coordinate, $r$, is defined by 
\begin{equation}
r\equiv \frac{p}{(1+e\cos \phi )}  \label{r}
\end{equation}%
and $\phi $ is the angle measured from the perihelion. We have, as usual,
the additional relations 
\begin{equation*}
p=a(1-e^{2})
\end{equation*}%
and 
\begin{equation}
h\equiv \sqrt{Gmp}\equiv r^{2}\frac{d\phi }{dt}.  \label{dr}
\end{equation}

From (\ref{geo1}), the components of the acceleration can be read off as 
\begin{equation}
\mathcal{R}=-\frac{Gm}{r^{2}} -\frac{Gm}{r^2} v^2 +4\frac{Gm}{r^{2}}v_{\mathcal{R}}^{2}+4\frac{%
G^{2}m^{2}}{r^{3}}-\frac{\delta }{r}+\frac{\delta }{r}v^{2}-4\frac{\delta
^{2}}{r}  \label{Raccel}
\end{equation}%
and 
\begin{equation}
\mathcal{J}=4\frac{Gm}{r^{2}}v_{\mathcal{R}}v_{\mathcal{J}}  \label{Jaccel}
\end{equation}%
where we now have the radial and normal-to-radial components of the velocity
as 
\begin{align*}
v_{\mathcal{R}}& =\frac{eh}{p}\sin \phi \\
v_{\mathcal{J}}& =\frac{h}{p}(1+e\cos \phi )
\end{align*}%
and $v^{2}=v_{\mathcal{R}}^{2}+v_{\mathcal{J}}^{2}$. In writing (\ref{Raccel}%
), the last term of (\ref{geo1}) has been absorbed by a rescaling of $m$, as
mentioned above.

The expressions (\ref{Raccel}) and (\ref{Jaccel}) can now be substituted
into (\ref{per}) and integrated from $\phi=0$ to $2 \pi$, using (\ref{r})
and (\ref{dr}) to write $r$ and $dr$ in terms of $\phi$ and $d\phi$. The
perihelion precession per orbit is then given, to post-Newtonian accuracy,
by the expression 
\begin{equation}  \label{precession}
\Delta \tilde{\omega} = \frac{6 \pi G m}{a(1-e^2)}-\frac{2\pi\delta}{e^2}
\left( e^2-1-\frac{(1+4 \delta) a (1-e^2)}{Gm} \right).
\end{equation}
The first term in (\ref{precession}) is clearly the standard general
relativistic expression. The second term is new and contributes to leading
order the term 
\begin{equation*}
\frac{2 \pi a}{Gm} \left( \frac{1-e^2}{e^2} \right) \delta.
\end{equation*}

Comparing the prediction (\ref{precession}) with observation is a
non-trivial matter. The above prediction is the highly idealised precession
expected for a time-like geodesic in the geometry described by (\ref{iso}).
If we assume that the geometry (\ref{iso}) is a good approximation to the
weak field for a static Schwarzschild-like mass then it is not trivial to
assume that the time-like geodesics used to calculate the rate of perihelion
precession (\ref{precession}) are the paths that material objects will
follow. Whilst we are assured from the generalised Bianchi identities
\cite{Mag94} of the covariant conservation of energy-momentum,
${T^{ab}}_{; b}=0 $, and hence of the geodesic motion of an ideal fluid of
pressureless dust, $U^{i}U_{\;;i}^{j}=0$, this does not ensure the geodesic
motion of extended bodies. This deviation from geodesic motion is known as
the Nordvedt effect \cite{Nor68} and, whilst being zero for GR, is
generally non-zero for extended theories of gravity. From
the analysis so far it is also not clear how orbiting matter and other
nearby sources (other than the central mass) will contribute to the geometry
(\ref{iso}).

In order to make a prediction for a physical system such as the solar
system, and in the interests of brevity, some assumptions must be made. It
is firstly assumed that the geometry of space-time in the solar system can
be considered, to first approximation, as static and spherically symmetric.
It is then assumed that this geometry is determined by the Sun, which can be
treated as a point-like Schwarzschild mass at the origin, and is isolated
from the effects of matter outside the solar system and from the background
cosmology. It is also assumed that the Nordvedt effect is negligible and
that extended massive bodies, such as planets, follow the same time-like
geodesics of the background geometry as neutral test particles.

In comparing with observation it is useful to recast (\ref{precession}) in
the form 
\begin{equation*}
\Delta \tilde{\omega}=\frac{6\pi Gm}{a(1-e^{2})}\lambda
\end{equation*}%
where 
\begin{equation*}
\lambda =1+\frac{a^{2}(1-e^{2})^{2}}{3G^{2}m^{2}e^{2}}\delta .
\end{equation*}%
This allows for easy comparison with results which have been used to
constrain the standard post-Newtonian parameters, for which 
\begin{equation*}
\lambda =\frac{1}{3}(2+2\gamma -\beta ).
\end{equation*}%
The observational determination of the perihelion precession of Mercury is
not clear cut and is subject to a number of uncertainties; most notably the
quadrupole moment of the Sun (see e.g. \cite{Pir03}). We choose to use the
result of Shapiro et. al. \cite{Sha76},
\begin{equation}
\lambda =1.003\pm 0.005,  \label{Shapiro}
\end{equation}%
which for standard values of $a$, $e$ and $m$ \cite{All63} gives us the
constraint 
\begin{equation}
\delta =2.7\pm 4.5\times 10^{-19}.
\end{equation}%
In deriving (\ref{Shapiro}) the quadrupole moment of the Sun was assumed to
correspond to uniform rotation.  For more modern estimates of the
anomalous perihelion advance of Mercury see \cite{Pir03}. 

\subsection{Non-Static Space-Time}

We will now proceed to calculate the equations of motion of particles
following geodesics of the space-time (\ref{Fon}), to post-Newtonian order.  It will be shown that not
only are the terms due to the linear perturbations different to the
static case (as should be expected as the perturbations themselves have been shown to
be  background dependant) but that the background itself contributes
an extra term to the post-Newtonian equation of motion.

\subsubsection{Equation of motion}

We will now proceed to calculate the geodesics of the non-static
solution (\ref{lin5}).  In doing this we will neglect the
contribution of the $r^2$ mode to (\ref{lin5}) so that we only take
into account the modes which go as $r^{-1}$, in the limit $\delta
\rightarrow 0$.  This is the mode corresponding to the
linearisation of the exact solution (\ref{Fon}).

The geodesic equation can be written, as usual, in the form
\begin{equation*}
\frac{d^{2}x^{\mu }}{d\lambda ^{2}}+\Gamma _{\;ij}^{\mu }\frac{dx^{i}}{%
d\lambda }\frac{dx^{j}}{d\lambda }=0,
\end{equation*}
where $\lambda $ can be taken as proper time for a time-like geodesic, or as
an affine parameter for a null geodesic. In terms of coordinate time this
can be re-written as
\begin{equation}
\label{geo9}
\frac{d^{2}x^{\mu }}{dt^{2}}+\left( \Gamma _{\;ij}^{\mu }-\Gamma _{\;ij}^{0}%
\frac{dx^{\mu }}{dt}\right) \frac{dx^{i}}{dt}\frac{dx^{j}}{dt}=0.
\end{equation}
Substituting the linearised solution into this equation will then give the
equations of motion for test particles in this space-time.  

Substituting (\ref{Fon}) into (\ref{geo9}) gives, to
post-Newtonian order, the equation of motion
\begin{multline}
\label{eqnn}
\frac{d^2\textbf{x}}{d\tau^2} = -\frac{Gm}{r^2} \textbf{e}_r+4
(1-\delta) \frac{G^2m^2}{r^3} \textbf{e}_r -(1-2 \delta) \frac{G
  m}{r^2} \left( \frac{d\textbf{x}}{d \tau} \right)^2 \\+4 (1-\delta)
\frac{G m}{r^2} \textbf{e}_r \cdot \frac{d\textbf{x}}{d \tau}
\frac{d\textbf{x}}{d \tau} -\mathcal{H} \frac{d\textbf{x}}{d \tau}
\left(1- \left(\frac{d\textbf{x}}{d \tau} \right)^2 \right)
\end{multline}
where
\begin{equation*}
\mathcal{H} \equiv \frac{1}{a} \frac{da}{d \tau}.
\end{equation*}
The conformal time coordinate $\tau$ is defined by $dt \equiv a d\tau$
and $c_4$ has been set by the appropriate Newtonian limit.  The equations
of motion for this solution are considerably simpler than those of the
static solution (\ref{Chan}), but still differ from those of GR in significant ways.
All of the terms except the last in equation (\ref{eqnn}) have GR
counterparts and the powers of $r$ in these terms are all the same as
in the general relativistic case.  The premultiplicative factors of these
terms are, however, modified and can be described adequately within
the frame-work of the PPN approach by assigning
\begin{equation*}
\beta = 1 \qquad \text{and} \qquad \gamma = 1-2 \delta.
\end{equation*}
By making this identification, the constraints on $\gamma$ from
observations of the Shapiro time delay of radio signals from the
Cassini space probe \cite{Bert} can be used to
impose upon $\delta$ the constraint
\begin{equation}
\delta = -1.1 \pm 1.2 \times 10^{-5}.
\end{equation}
As well as the usual effects associated with $\gamma-1$ being non-zero
there are extra effects in this space-time due to the last term in
(\ref{eqnn}).  This term is proportional to the velocity of the
test-particle (when $v << c$) and is zero for photons.  For this reason we
identify it as a friction term, with the friction coefficient being
given by $\mathcal{H}$.  This `friction' is a purely gravitational
effect and is not due to any non-gravitational interaction of test particles with
any other matter.

It is clear that the constraints which can be imposed on this theory
depend crucially on whether or not the space-time of the solar system can be assumed to be
static.  This is a non-trivial matter, as Birkhoff's theorem is
no longer valid.  Taking the usual assumption of the geometry of the
solar system being static leads us to a solution with a non-trivial
asymptotic dependence on $r$, as $r \rightarrow \infty$.  These
unusual asymptotics can be removed, but this appears to require that the
usual assumption of the solar system having a static geometry be
abandoned.  Further study is required on this subject to determine
which of these solutions is the most appropriate for application to
the solar system.
\clearemptydoublepage
\chapter{Conclusions}
\label{Conclusions}

\bigskip

The objective of this work has been to develop the framework of
alternative theories of gravity, to investigate the solutions to their
field equations and to identify the principle ways in which they deviate from
GR.  This framework has then been used to study the ways in which GR can be
considered to be special or unique, to identify new
behaviour that was not previously present and to place observational
constraints on deviations from the standard theory.  In
achieving these goals a number of approaches have been used in both
cosmological and weak-field environments.

The focus of the analysis of cosmological solutions has been on,
but not limited to, studies of FRW universes.  These universes
are appealing for a number of reasons.  As well as having the same symmetries as our own observable
Universe (on the largest scales) their high degree of symmetry is
particularly useful for finding simple exact solutions to the
complicated field equations of the theory.
These solutions have been used to model a variety of physical processes
such as structure formation, primordial nucleosynthesis and microwave
background formation.  All of these processes have been seen to behave
differently from the corresponding cases in the standard model.  In
this way we have been able to see explicitly what
effect such modifications to GR would have on the evolution of the
Universe, as well as on the physical processes occurring within it.

Further to studying FRW universes we have also considered a
variety of other cosmological solutions.  In particular we have found
the conditions for the existence of G\"{o}del, Einstein static and de
Sitter universes in a wide variety of theories.  For the case of
the G\"{o}del universe the existence of closed time-like curves, and
hence time travel into the past, has been analysed in detail.  Furthermore, anisotropic solutions of
Bianchi type I have been found and used to study the approach
to anisotropic singularities in this extended framework.  Using these
solutions it has been found that an infinite sequence of chaotic mixmaster oscillations
does not occur in all theories, as it does in GR.  As well as
studying homogeneous cosmologies, investigation has also been made of the effects of
inhomogeneity on cosmological models within this framework.  We have
used FRW universes to model spherical collapse and have found
exact inhomogeneous, spherically symmetric and time-dependent vacuum
solutions.  The very existence of these solutions is forbidden in GR
by Birkhoff's theorem, and so they are of particular interest in
looking for new behaviour which is not available in the standard
model.

The tightest constraints currently available on gravitational theory
are from experiments in the solar system and astrophysical tests in
the proximity of (almost) spherical massive bodies.  With this in mind
the spherically symmetric vacuum solutions to the field equations are of
particular interest.  We have found exact solutions for this
situation, and analysed their behaviour.  These solutions have been
shown to exhibit entirely new behaviour not previously obtainable in
GR, including time dependence and non-asymptotic flatness.  The weak
field limit of these solutions has been found and the
geodesics calculated to Post-Newtonian order.  Comparison of these
solutions with observational results has allowed tight constraints to
be placed on the theory.

In summary, the principle new results in this work are that it has:
\begin{itemize}
\item{Investigated the spatially flat FRW solutions
  of theories with Lagrangian density $\mathcal{L}=R^n$.}
\item{Found the conditions for the existence of G\"{o}del, Einstein
  static and de Sitter universes in a wide class of fourth-order
  theories.}
\item{Investigated the phenomenon of closed time-like curves
  in these theories.}
\item{Found exact Bianchi type I anisotropic solutions for a wide
  class of scale invariant fourth-order gravity theories.}
\item{Investigated the approach to anisotropic singularities in these
  theories.}
\item{Found exact static and spherically symmetric solutions in vacuum
  for $R^n$ theories and investigated their stability.}
\item{Found exact inhomogeneous cosmological solutions for
  scalar-tensor and $R^n$ theories.}
\item{Explicitly shown the lack of validity of Birkhoff's theorem in
  scalar-tensor and $R^n$ theories.}
\item{Investigated the spherical collapse model in scalar-tensor
  theories, and the spatial variations in $G$ that can result from it.}
\item{Investigated the possibility of generalising the usual
  scalar-tensor theories by allowing a transfer of energy between
  scalar and matter fields.}
\item{Used primordial nucleosynthesis to impose constraints upon
  scalar-tensor and $R^n$ theories, and their resulting cosmologies.}
\item{Considered microwave background formation in $R^n$ theories.}
\item{Imposed constraints on $R^n$ theories by modelling the solar
  system as either a static or non-static spherically symmetric vacuum
  space-time.}
\end{itemize}

This work has confirmed the excellent agreement between GR and
gravitational experiment and observation.  It has uncovered new
effects of modifying the standard theory and it has shown that such
modifications are very strongly constrained, particularly in the low
curvature regime of the solar system.

There are a variety of directions in which future research on this
subject can proceed.  Firstly, new effects and behaviours that have
been found in this work can be used to model other physical processes.
Particular processes of interest may be particle production
near anisotropic singularities, modelling of
anomalous observations (such as the Pioneer anomaly) or effects on
structure formation and the dynamics of astrophysical bodies.  A
second avenue of investigation would be to further investigate the
effects that can be found to arise in theories of gravity which do not
allow Birkhoff's theorem to be formulated.  A start has been made on
this by identifying exact vacuum solutions in Brans-Dicke and $R^n$ theories
which are spherically symmetric and either static or non-static.  More
work is required to fully understand how these solutions should be
interpreted and what their physical effects would be.  Other future
research may focus on the effects of
self-interacting scalar fields in some of the contexts that have been
explored here.  The gravitational scalars that have been
considered in this work have all been massless.  Self-interactions are of
direct interest for problems such as the late-time acceleration of the
Universe as well as early-time behaviour, such as that which occurs
during inflation.  Understand these theories, and their
physical effects, would be of direct interest for considerations of
these scenarios.
\clearemptydoublepage

% -- APPENDICES --

\appendix
\clearemptydoublepage

% -- BIBLIOGRAPHY --

{\small
\addcontentsline{toc}{chapter}{Bibliography}
\bibliographystyle{unsrt}
\bibliography{references}
}

% -- END --

\end{document}